\documentclass[a4paper,12pt]{report}
\usepackage{listings}
\usepackage{amsmath,amssymb,epsfig,bm}
\usepackage[greek,english]{babel}
\usepackage{graphicx}
\usepackage[utf8x]{inputenc}
\usepackage{color}
\usepackage{float}
\usepackage{kerkis}
\usepackage{subcaption}
\usepackage{lineno}
\usepackage[normalem]{ulem}
\usepackage{enumitem}
\usepackage[hyphens]{url}
\usepackage{hyperref}
\usepackage{pdflscape}
\usepackage{array}

\newlist{stages}{enumerate}{1}
\setlist[stages,1]{font={\bfseries},label={Stage \arabic*:},leftmargin=*,labelindent=1em}

\graphicspath{ {images/} }

\newcommand{\sen}[1]{\foreignlanguage{english}{#1}}

\begin{document}

\begin{titlepage}
    \begin{center}
        \vspace*{1cm}

        \Large
        \textbf{Study of the PICOSEC-Micromegas Detector with Test Beam Data and Phenomenological Modelling of its Response}

        \vspace{0.5cm}
        \normalsize
        \vspace{1.5cm}
        \textbf{Konstantinos Paraschou}

        \vfill

        A diploma thesis submitted in partial fulfillment of 

        \vspace{0.2cm}

        the requirements for the degree of

        \vspace{1.0cm}

        Master of Science in Computational Physics,

        \vspace{0.5cm}

        under the supervision of \\
        Professor Spyridon Eust. Tzamarias

        \vspace{0.5cm}

        \includegraphics[width=0.4\textwidth]{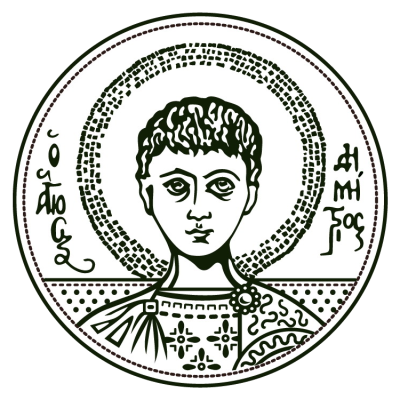}
        
        School of Physics\\
        Aristotle University of Thessaloniki\\
        Greece\\
        March 2018

    \end{center}
\end{titlepage}

\newpage
\thispagestyle{empty}
\mbox{}
\newpage

\begin{abstract}
In this work, a part of the Research and Development effort of the PICOSEC detector is presented.
The PICOSEC detector is a novel gas-filled detector, based on the Micromegas detector,
which has been developed by the RD51-PICOSEC collaboration\footnote{
    \tiny
    \textit{
        $^a$ IRFU, CEA, Universit\'e Paris-Saclay, F-91191 Gif-sur-Yvette, France 
    }
 
    \quad\quad \textit{ $^b$ European Organization for Nuclear Research (CERN), CH-1211 Geneve 23, Switzerland }

 \quad\quad\textit{$^c$ State Key Laboratory of Particle Detection and Electronics, University of Science and
 Technology of China, Hefei 230026, China }

 \quad\quad\textit{$^d$ Department of Physics, Aristotle University of Thessaloniki, Thessaloniki, Greece }

 \quad\quad    \textit{$^e$ Institute of Nuclear Physics, NCRS Demokritos, 15310 Aghia Paraskevi, Athens, Greece }

 \quad\quad    \textit{$^f$  National Technical University of Athens, Athens, Greece }

  \quad\quad   \textit{$^g$ Laborat\'orio de Instrumenta\c c \~ao e F\'isica Experimental de Part\'iculas, Lisbon, Portugal }

 \quad\quad\textit{$^h$ RD51 collaboration, European Organization for Nuclear Research (CERN), CH-1211
 Geneve 23, Switzerland }

 \quad\quad\textit{$^i$  Instituto Galego de F\'isica de Altas Enerx\'ias (IGFAE), Universidade de Santiago de
 Compostela, Spain
}
}.
Instead of relying on traditional direct ionization, the PICOSEC detector takes advantage of the prompt timing characteristics of Cherenkov radiation
by converting the Cherenkov photons into electrons through the use of a photocathode.
The detector has been put into two type of tests where experimental data are collected.
One involves a laser beam and a single photoelectron response at CEA-SACLAY, while the other involves a test beam of $150\,GeV$ muons at the CERN SPS H4 secondary beamline, with multiple photoelectrons.
The methods employed in the statistical analysis of the detector's timing properties are summarized and in their application an optimal time resolution of $76$ ps is achieved for single photoelectrons.
A strange dependence of the mean timing on the size of the electron peak is observed,
which mimics the behaviour of the ``time walk'' effect.
A simulation of the detector is developed to study the timing properties, 
which is also the main scope of the thesis.
After a detailed investigation,
it is found that the electrons in the gaseous mixture move with a different drift velocity before and after the first multiplication,
introducing this ``time walk'' effect,
and a phenomenological explanation of this effect is provided.
Finally, a maximum likelihood statistical method for the estimation of the mean number of photoelectron extracted per particle is developed and an optimal time resolution of $24$ ps is achieved with a mean number of photoelectrons per particle equal to $10.4$.
In this work, a deep understanding is acquired on a new detector who has brought upon unprecedented results in the field of gas-filled detectors. 
\end{abstract}

\newpage
\thispagestyle{empty}
\mbox{}
\newpage

\selectlanguage{greek}
\begin{abstract}
    Στην παρούσα εργασία παρουσιάζεται ένα μέρος της συλλογικής προσπάθειας Έρευνας και Ανάπτυξης του ανιχνευτή \sen{PICOSEC}.
    Ο ανιχνευτής \sen{PICOSEC} είναι ένας καινοτόμος ανιχνευτής αερίου γεμίσματος βασιζόμενος \break στον ανιχνευτή \sen{Micromegas},
    ο οποίος έχει αναπτυχθεί από την σύμπραξη \sen{RD51-PICOSEC}.
    Aντί να στηρίζεται στον παραδοσιακό τρόπο ιονισμού, ο ανιχνευτής \sen{PICOSEC} εκμεταλλεύεται τα γρήγορα χαρακτηριστικά χρονισμού της ακτινοβολίας 
    \sen{Cherenkov} μετατρέποντας τα \sen{Cherenkov} φωτόνια σε ηλεκτρόνια μέσω της χρήσης μιας φωτοκαθόδου.
    Ο ανιχνευτής έχει υποβληθεί σε δύο είδη ελέγχων όπου πειραματικά δεδομένα έχουν συλλεχθεί.
    Ο ένας περι-\break λαμβάνει μια δέσμη λέιζερ με μια απόκριση ενός φωτοηλεκτρονίου στο \sen{CEA-SACLAY}, 
    ενώ ο άλλος περιλαμβάνει μια δέσμη ελέγχου από μιόνια των $150\,GeV$ στην \sen{H4} δευτερεύουσα γραμμή δέσμης
    στον \sen{SPS} του \sen{CERN} και πολλαπλά φωτοηλεκτρόνια.
    Παρουσιάζονται συνοπτικά οι μέθοδοι που χρησιμοποιήθη- καν στην στατιστική ανάλυση των ιδιοτήτων χρονισμού του ανιχνευτή, και με την εφαρμογή τους 
    τα αποτελέσματα δείχνουν ότι η βέλτιστη χρονική διακριτική ικανότητα που επιτυγχάνεται στην ανίχνευση μονήρων φωτοηλεκτρονίων φτάνει στα $76$ \sen{ps}. 
    Παρατηρείται μια παράξενη εξάρτηση του μέσου χρονισμού στο μέγεθος του σήματος, η οποία μιμείται την συμπεριφορά του φαινομένου \sen{``time walk''}.
    Μια προσομοίωση του ανιχνευτή αναπτύσσεται για την μελέτη των ιδιοτήτων του χρονισμού, η οποία είναι και το κύριο θέμα της εργασίας.
    Μετά από μια λεπτομερή έρευνα, προκύπτει πως τα ηλεκτρόνια στο αέριο μίγμα κινούνται με διαφορετική ταχύτητα ολίσθησης πριν και μετά τον πρώτο τους πολλαπλασιασμό,
    εισάγοντας έτσι το φαινόμενο \sen{``time walk''}, και προτείνεται μια φαινομενολογική εξήγηση.
    Τέλος, αναπτύσσεται μια στατιστική μέθοδος μέγιστης πιθανοφάνειας για τον προσδιορισμό του μέσου αριθμού φωτοηλεκτρονίων που εξάγονται ανά σωματίδιο,
    και η βέλτιστη χρονική διακριτική ικανότητα που επιτυγχάνεται είναι στά $24$ \sen{ps} με κατά μέσο όρο $10.4$ φωτοηλεκτρόνια ανά σωματίδιο.
    Στην παρούσα εργασία αποκτήθηκε μια βαθιά κατανόηση ενός νέου ανιχνευτή ο οποίος έχει επιφέρει αποτελέσματα άνευ προηγουμένων στο πεδίο των ανιχνευτών αερίου γεμίσματος.
\end{abstract}
\selectlanguage{english}

\newpage
\thispagestyle{empty}
\mbox{}
\newpage

\thispagestyle{empty}
\chapter*{Acknowledgements} 

I would first like to thank my advisor Prof. Spyros Tzamarias.
In his guidance, I acquired a great deal of knowledge, but more importantly, he taught me how to research and that the scientific method requires attention to detail.
I am indebted to the RD51-PICOSEC collaboration which accepted me as a researcher and offered me the opportunity and the means to contribute in the development of a very exciting novel detector.
I can't not thank the rest of my supervisors who guided and helped me in my work, Prof. Chara Petridou and Prof. Kostas Kordas.
Of equal importance were the fruitful discussions with Prof. Christos Eleftheriadis.

In the context of my thesis work I got to collaborate with colleagues who are exceptional people and have become my friends, Vasilis and Ioannis, as well as the new members of our team, Orestis, Anastasia and Vaggelis.

In the simulation side of the thesis, Rob Veenhof's help and comments were invaluable and I cannot thank him enough.
I also want to express my gratitude to Ioannis Katsioulas for his help and friendship.

Furthermore, I had the pleasure to come in direct and frequent contact with colleagues of the RD51 collaboration, who offered me encouragement and their friendship, including Paco, Thomas, Eraldo, Lukas, Yi, Jona, Michael and others.

Last but not least, I want to thank my parents and the rest of my friends who are not mentioned here.
The support they provided me with and the fact that they believed in me were catalytic in my development as a person.

\newpage
\thispagestyle{empty}
\mbox{}
\newpage

\cleardoublepage
\phantomsection
\addcontentsline{toc}{chapter}{\listtablename}
\listoftables

\cleardoublepage
\phantomsection
\addcontentsline{toc}{chapter}{\listfigurename}
\listoffigures

\tableofcontents

\chapter{Introduction}

Both nuclear and particle physics rely on particle detectors to measure the properties of individual particles as well as collections of them, such as the cascades that are endlessly being formed in the Earth's atmosphere.
These particle detectors are built taking into account the way particles interact with matter.
One of the first particle detectors, the cloud chamber, was a vessel filled with supersaturated vapor of water or alcohol.
A charged particle passing through it would ionize the vapor and ultimately, cloud tracks would form in the trail of the particle's passage.
These cloud tracks would be visible to the naked eye and the storage of these measurements was by the means of photographs.

This process of using photographs to archive measurements is understandably very tedious and vulnerable to human error.
Thankfully, technology has advanced up to a point that electronics are used not only to store measurements from detectors,
but are also a crucial part of the detector that enables it to ``see'' signals which would otherwise be unmeasurable.
In fact, this advancement of technology allows us to simulate detectors using computers.
The simulation of detectors is important because it confirms (or rejects) theoretical models, enhances the validity of experimental measurements and, finally, assists in the process of developing a detector in that the effect of a variation in the detector's design can be examined in the simulation without building a prototype.

The aim of most modern detectors is to measure the position and time at which a particle passes through the detector, as well as the energy it has.
Usually, energy is measured with the technique of calorimetry, the particle interacts with the material of the ``calorimeter'', deposits all of its energy and gets absorbed. 
Time is measured simply by measuring the time at which the detector sends its signal.
Position can be measured indirectly by placing several small detectors and by knowing which detector got hit.
As precision becomes better and better, it is difficult and impractical to construct detectors with a size in the order of micrometers.
To reach such a precision, micropattern detectors employ very finely segmented electronic circuits in their internal volume.
Two prominent examples of these micropattern technologies are the Micromegas\cite{Giomataris1996} and the GEM\cite{Sauli1997} detectors.

This thesis is part of the Research and Development effort on the PICOSEC Micromegas detector.
The PICOSEC detector is a variation of the Micromegas detector for the purpose of reaching time resolution in the order of picoseconds.
The main motivation to reach such time resolution was to distinguish between the many interaction vertices that will be present in the High Luminosity Large Hadron Collider.
Other applications to take advantage of this achievement have been proposed and are being evaluated in terms of their feasibility.
Both analysis of experimental data and their simulation is presented in this work, in addition to the phenomenology of the observed effects.

In the second chapter a general and very brief introduction into the workings of gas-filled detectors is presented.
More details are given for the special case of the Micromegas detector.
In the third chapter, the PICOSEC Micromegas detector and the experimental tests that were carried out to characterize its performance are explained.
In the fourth chapter, the signal processing and statistical techniques which are used to analyze the experimental data collected in the Laser Test are shown, in addition to their application.
The fifth chapter consists of the simulation, which is the primary objective of this work and includes its description, its verification and its microscopic investigation. 
The sixth chapter concerns the analysis of experimental data collected with relativistic muons, and the best achieved results are shown.
The conclusions of this study consist the seventh, and final, chapter.


\chapter{Gas-Filled Detectors}

Particle detectors of the gaseous kind are closed vessels filled with a gaseous mixture, which is biased with an electric field through an applied voltage difference in the boundaries of the vessel.
The process of detection is a strictly electromagnetic interaction.
A charged particle that will pass through the gaseous mixture of the detector may ionize it through the electromagnetic interaction and a pair of a positively charged ion and an electron will be formed.
The ion and the electron will move under the influence of the electric field and the electron will cause an ``electron avalanche'' similarly to how a relatively small disturbance in the snowy layer of a mountain accelerates rapidly, drags more and more snow in its wake and ultimately creates an avalanche of immense power.
The movement of electrons will cause an electric signal which can be recorded by the electronics.
The theoretical basis of gas-filled detectors is briefly summarized in this chapter and adapted from the much more analytical references \cite{sauli}, \cite{riegler-drift}.

\section{Kinetic Processes in the Gas}

\subsection{Diffusion of Charges}
Consider a charged particle, ion or electron, freely moving in the gas.
In the absence of electric fields and inelastic interactions, the charged particle behaves like a neutral molecule \cite{sauli} and the classic kinetic theory of gases describes the phenomenon. 
The probability of an atom/molecule to have an energy $E$ at an absolute temperature $T$, follows the Maxwell-Boltzmann distribution:

\begin{equation}
    F(E) = 2 \sqrt{\frac{E}{\pi (kT)^3}} e^{-\frac{E}{kT}}
\end{equation}
    
in which $k = 1.38 \cdot 10^{-16} erg/ ^o K = 8.617 \cdot 10^{-5} eV/ ^o K$ is Boltzmann's constant.
Notice that this distribution is independent of the mass of the particle.
In fact, the only variable that affects the probability distribution of the particle's energy is the absolute temperature and the average energy $\bar{E}$ is equal to:

\begin{equation}
    \bar{E} = k T
\end{equation}

Through a change in variables, one can easily arrive to the Maxwell-Boltzmann distribution of the velocity:

\begin{equation}
    f(v) = 4\pi \sqrt{\left(\frac{m}{2\pi kT}\right)^3} v^2 e^{-\frac{mv^2}{2kT}}
\end{equation}

where $m$ is the mass of the particle and the average velocity $\bar{v}$ is equal to:

\begin{equation}
    \bar{v} = \sqrt{\frac{8kT}{\pi m}}
\end{equation}

With respect to its position, the particle can be considered to perform a Brownian motion.
Solving the Boltzmann equation, the probability to find the particle at a distance $x$ from its initial position at a time $t$ follows the Gaussian distribution:

\begin{equation}
    G(x;t) = \frac{1}{\sqrt{4\pi Dt}}e ^{-\frac{x^2}{4Dt}}
\end{equation}

where $D$ is the diffusion coefficient.
The mean value is equal to zero and the root mean square of the distribution, or standard deviation is equal to:
\begin{equation}
    \sigma_x = \sqrt{2Dt}
\end{equation}
for linear diffusion and
\begin{equation}
     \sigma_V = \sqrt{6Dt}
\end{equation}
for 3D-volume diffusion.
In other words, the variance of the displacement, which is the square of the standard deviation, grows linearly with time.

\subsection{Drift of Charges}

When an electric field is applied, the movement of charges is no longer isotropically random.
The electric field will drive the electrons/ions along its electric field lines but colliding with the gas medium, they will instantaneously change direction (elastic collisions) or even lose energy as well (inelastic collisions).
As a result, they will move ``back and forth'' but on average they will follow the direction of the electric field lines with an average speed that is much smaller than the instantaneous microscopic velocity.
This rate of this movement per unit time is called the drift velocity $v_d$.
An interesting phenomenon is the fact that the faster an electron is microscopically moving, the slower it will drift.
Suppose that in each collision, all of its energy is lost and begins again from rest.
Any direction of motion after the collision is irrelevant since it is at rest.
Therefore, it can only follow the drift lines of the electric field directly.
In this case, the drift velocity is maximal.
If all of the collisions are elastic, the particle will always bounce around and the electric field will spend time to recover a forward direction of motion.

The drift velocity appears in the above Gaussian distribution through a mean value that is linearly changing with time.
For the 1D case:
\begin{equation}
    G(x;t) = \frac{1}{\sqrt{4\pi Dt}}e ^{-\frac{(x-v_d t)^2}{4Dt}}
\end{equation}

A quantity which relates the drift velocity and the electric field $\mathcal{E}$ is the mobility $\mu$:

\begin{equation}
    \mu = \frac{v_d}{\mathcal{E}}
\end{equation}

For the case of ions, this mobility is found to be constant even for very high electric fields.
This reflects the fact that the average energy of the ions barely changes with the electric field.
However, this is not the case for electrons and the relation between their drift velocity and the electric field is far from linear.

In the presence of an electric field, it has been observed that diffusion is not isotropic,
it separates into a longitudinal diffusion $\sigma_L = \sqrt{2D_Lt}$ along the direction of the electric field lines,
where $D_L$ is the corresponding longitudinal diffusion coefficient,
and a transverse diffusion $\sigma_T = \sqrt{4D_T t}$ perpendicular to the electric drift lines,
where $D_T$ is the transverse diffusion coefficient.
Both diffusions are still proportional to the square root of the time $t$.

\subsection{Side Note on Brownian Motion with Drift}
To avoid confusion in the following, the inverse distribution of another shall be defined.
If the probability distribution of a positive random variable $X$ is $f(x)$ and is continuous,
then the inverse distribution is the distribution of the reciprocal of the random variable, i.e. of $Y = \frac{1}{X}$.
The inverse distribution $g(y)$ is then given by:

\begin{equation}
    g(y) = \frac{1}{y^2} f\left(\frac{1}{y}\right)
\end{equation}

When a particles performs a Brownian Motion with positive drift,
then the first time at which the particle will reach a fixed level $a$ in the direction of the drift movement,
follows the inverse Gaussian (or Wald) distribution:

\begin{equation}\label{inverse-gaus}
    IG(t;\mu,\lambda) = \sqrt{ \frac{\lambda}{2\pi x^3} } \exp \left( \frac{-\lambda(x-\mu)^2}{2\mu^2x}\right)
\end{equation}
where $\mu$ is the mean, or expectation value of the distribution, and $\lambda$ is the shape parameter of the distribution.
The variance of this distribution is given by
\begin{equation}
    V[X] = \frac{\mu ^3}{\lambda}
\end{equation}

This has been proven by Schr\"odinger\cite{schro} but an additional proof can be found in Appendix \ref{app-igaus}.
The movement of an electron or an ion in the gas and in a constant electric field can be approximated by a Brownian Motion and a Drift component and as such, the inverse-Gaussian distribution will describe the time it takes for the electron or ion to arrive at fixed level with respect to its initial position.
For a drift velocity $v_d$ and a longitudinal diffusion coefficient $D_L$, the time of arrival $t$ on a fixed level $a$ follows the inverse-Gaussian $T \sim IG\left(\frac{a}{v_d},\frac{a^2}{2D_L}\right)$.

\subsection{Electron Avalanche}

Microscopically, a free electron in the gas consistently gains energy from the electric field and consistently loses some of it in collisions with the gas molecules.
If the electric field is high enough the energy gain will be larger than the collisional losses and if the electron's energy becomes high enough, it will ionize the gas molecule and a new electron-ion pair will be formed.
Both of the electrons now will again begin to acquire energy in the electric field and each of them will again ionize the gas.
This process will keep repeating itself forming an avalanche of electrons that is ionizing more and more atoms/molecules in its wake until all of them finally reach the anode.
The initial electron that caused the avalanche is called the primary electron,
and the electrons produced in the avalanche are called secondary electrons.
The mean number of electrons $\bar{n}$ in the avalanche is governed by the familiar differential equation: 
\begin{equation}
    d\bar{n} = \alpha \bar{n} dx - \eta \bar{n} dx
\end{equation}
where $x$ is the distance from the initial position of the electron that began the avalanche, $\alpha$ is the first Townsend coefficient and $\eta$ is the attachment coefficient.
The first Townsend coefficient expresses the number of electron-ion pairs formed per unit of distance by one electron and naturally depends on the electric field.
The attachment coefficient on the other hand, expresses the number of electrons that re-attach with the gas mixture.
The solution of this differential equation for a constant electric field is:
\begin{equation}
    \bar{n} = e^{(\alpha - \eta)x}
\end{equation}

where the difference $\alpha-\eta$ is what can be practically measured in an experiment and is called the effective Townsend coefficient.
This avalanche is the main process of amplifying the response of an ionization in the gas for most detectors.
Some gas-filled detectors, called ionization chambers, operate in such conditions that this electron multiplication is not happening.
These chambers, however, require much more sophisticated electronics as the low signals are susceptible to noise.

The formation of the avalanche is a statistical phenomenon and naturally the number of electrons in the avalanche fluctuate.
This fact is actually the limiting factor to the energy resolution of gas-filled detectors.
In the case of small rate of multiplications the distribution of the number of electrons follows an exponential distribution.
At higher rates, a ``hump'' begins to appear at very low multiplicities and its shape resembles that of a ``P\'{o}lya'' distribution.
Whether this ``P\'{o}lya'' distribution is the actual underlying distribution is an ongoing debate 
and more details in this subject can be found in \cite{heinrich-thesis},\cite{alkhazov},\cite{legler}.
Nevertheless, the ``P\'{o}lya'' distribution has proven to be a very good approximation to the observed spectra and its properties makes it a very attractive fit function for the purposes of describing these fluctuations.

\begin{figure}[t]
    \centering
    \includegraphics[width = 0.49\textwidth]{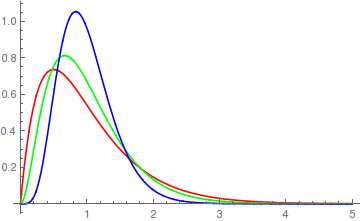}
    \caption[P\'olya distributions.]{P\'olya distributions with $\bar{n}=1$ and (red) $\theta = 1$, (green) $\theta =2$, and (blue) $\theta = 3$.}
    \label{polya-gas}
\end{figure}

Although the distribution is called ``P\'{o}lya'', what is meant in the context of gas-filled detectors is the Gamma distribution:
\begin{equation}\label{eq:polya}
    P(n; \bar{n}, \theta) = \frac{ (\theta+1)^{(\theta+1)}}{\bar{n}\Gamma(\theta+1)} \left( \frac{n}{\bar{n}}\right)^\theta e^{-(\theta+1)n/\bar{n}}
\end{equation}
where $\theta$ is the shape parameter of the distribution. The expectation value is equal to:
\begin{equation}
    E[n] = \bar{n}
\end{equation}
and the variance is:
\begin{equation}
    V[n] = \frac{\bar{n}^2}{\theta+1}
\end{equation}

Examples of the P\'olya distribution are shown in Figure \ref{polya-gas}.
This distribution describes the multiplicity of electrons in an avalanche which was caused by a single primary electron.
If instead $N$ primary electrons, all beginning from the same position, go on to form the avalanche, the number of electrons in the avalanche will be given by what shall be called the N-P\'{o}lya distribution:
\begin{equation}\label{npolya}
    P_N(n; \bar{n}, \theta, N) =  \frac{ (\theta+1)^{N(\theta+1)}}{\bar{n}\Gamma(N(\theta+1))} \left( \frac{n}{\bar{n}}\right)^{N(\theta+1)-1} e^{-(\theta+1)n/\bar{n}}
\end{equation}
where $\bar{n}$ is the mean number of electrons in an avalanche caused by a single primary electron,
$N$ is the number of primary electrons which caused the avalanche 
and $\theta$ is the shape parameter which is same for any number of primary electrons.
The expectation value of the N-P\'{o}lya is equal to: 
\begin{equation}
    E[n] = N\bar{n}
\end{equation}
and its variance is:
\begin{equation}
    V[n] = \frac{N\bar{n}^2}{\theta+1}
\end{equation}

The usefulness of this generalization from the P\'{o}lya distribution to the N-P\'{o}lya one is the fact that, given the calibration to a single primary electron avalanche (i.e. estimation of $\bar{n}$ and $\theta$), one can predict the spectrum of any number of primary electrons.
Additionally, these N-P\'{o}lya spectrums can be superimposed for different $N$ to take into account the fluctuations in the number of primary electrons.

\section{Penning Transfer}

Collisions with a gas molecule have many types.
The most dominant one in low energies is the elastic collision.
If a gas is molecular, it can absorb energy and store it in the form of a vibration mode.
Also, most gases have many excitation levels, in which energy is can be again absorbed from collisions.
Then above all excitation levels lies the ionization level in which an electron is ejected from the atom/molecule.

When there is a mixture of different types of gases, where the excitation level of a gas component is higher than the ionization level of another, then the Penning transfer effect takes place.
In some cases it can even account for a big part of the gain (number of secondary electrons produced per primary electron).
The Penning transfer effect is a collection of different reactions.
The most dominant reactions in this collection are the ``collision with ionizing energy transfer''
$$A^* + B \rightarrow A + B^+ + e^-$$ 
in which an excited state $A^*$ of an atom/molecule of type $A$ collides with an atom/molecule of type $B$ and ionizes it by transfering its excitation energy,
and the ``radiative photo-ionization''
$$A^* + B \rightarrow A + \gamma + B \rightarrow A + B^+ + e^-$$ 
in which a similar reaction happens in an indirect way, i.e. the excited molecule $A^*$ de-excites, radiating a photon $\gamma$, which photon excites atom/molecule $B$.
Usually this reaction is parameterized in terms of a transfer rate $r$ which expresses the probability that an excitation in the above form ionizes another atom/molecule through the above reactions, provided that it has sufficient energy. 

\section{Signal Induction}

A signal is induced on an electrode when a charge is moving inside the electric field.
There is very often the misconception that the signal is created when a charge is collected by the electrode.
Altough this (incorrect) picture can provide equivalent results in some cases under some assumptions, it generally fails when there are more than two electrodes in the equivalent circuit of the detector.
A useful theorem to calculate the induced signal on any of the electrodes in the detector is Ramo's theorem\cite{ram39} which states:
\begin{quote}
    The current induced on a grounded electrode by a point charge $q$ moving along a trajectory $x(t)$ is 
    \begin{equation}\label{eq:ramo}
        I^{ind}_n(t) = \frac{-q \mathbf{E_n}\left(x\left(t\right)\right)\cdot \mathbf{v(t)}}{V_W}
    \end{equation}
    where $E_n(x)$ is the electric field in the case where the charge $q$ is removed, electrode $n$ is set to voltage $V_W$, and all other electrodes are grounded.
\end{quote}

In cases where the electric field cannot be analytically derived but demands a numerical computation, usually realized with Finite-Element-Method commercial software, the weighting field should also be calculated.
Practically, this is done by following the guidelines of Ramo's theorem, i.e. repeating the calculation for each electrode of interest, each time setting it to a voltage equal to $1$ and grounding the rest.
Detailed discussion on the induction of signals can be found in \cite{riegler-drift} and \cite{alexopoulos-signal-formation}, where rigorous derivations of Ramo's theorem are also shown.

\section{Micromegas Detector}

In most gas-filled detectors, the charged particle to be detected passes through the active volume of the detector and ionizes the gas atoms/molecules and primary electrons are created.
The primary electrons are then drifting and diffusing in the gas and if the electric field allows it,
they will create electron avalanches for the purpose of signal amplification.
Ultimately, the motion of the ions and electrons will induce an electrical signal on the electrodes which will be recorded by the electronics. 

%


\begin{figure}[t]
    \centering
    \includegraphics[width = 0.49\textwidth]{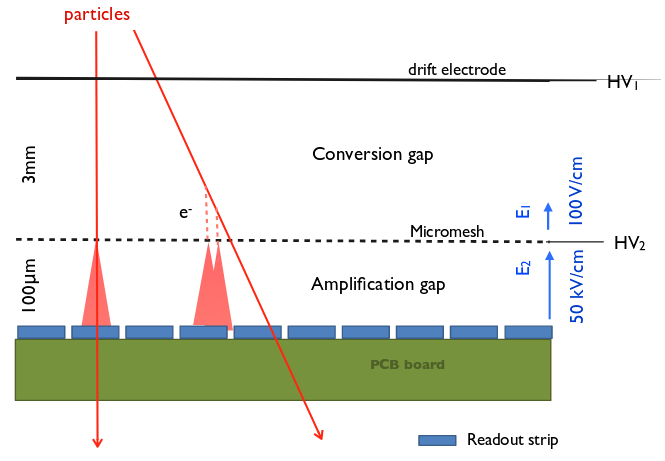}
    \caption[Micromegas detector concept.]{Micromegas detector concept. Image adapted from \cite{micropic}.}
    \label{mm-gas}
\end{figure}

The Micromegas (MICRO-MEsh GASeous) detector (as illustrated in Figure \ref{mm-gas}) is one very similar to a chamber with two parallel plates.
It is separated in two stages and it consists of a cathode, a micro-mesh with holes in the order of tens of micrometers and
an anode with parallel micro-strips whose size is in the order of hundreds of micrometers.
The separation between the cathode and the micro-mesh is several millimeters while the separation between the micro-mesh and the anode strips is around one hundred millimeters.

The first stage, called conversion or drift gap, consists of the active volume between the cathode and the micro-mesh.
The primary ionization happens in this gap, where because of its ``large'' volume, plenty of primary electrons are produced randomly along the incident particle's track.
A voltage difference is applied between the cathode and the micro-mesh such that the electric field is large but not enough so that the electron avalanche does not develop, or if it does, it is small.
The electrons drift until the mesh and according to their position and the electric field, they will either get absorbed by the mesh or continue and enter the second stage, the amplification gap.
The fraction of electrons that manage to pass through the mesh is called the electron transparency of the mesh.
Another voltage difference is applied between the micro-mesh and the anode which is very high such that a considerable electron avalanche is formed.
In the amplification gap, the electrons that manage to pass through the mesh begin an electron avalanche until they reach the anode.
The ions will then drift until the micro-mesh where the majority of them will be absorbed by it.
The fraction of the ions that pass through the mesh, called ion backflow, will induce a negligible signal and will produce negligible space charge (if the ion backflow is small).
This makes the Micromegas suitable for high rate applications and its scalability is limited only by the uniformity of the electric field in its drift gap.

The time resolution of the Micromegas detector is limited because of
\begin{enumerate}
    \item the longitudinal diffusion of charges in the drift gap and,
    \item the randomness in which primary electrons are produced in the charged particle's track.
\end{enumerate}

\subsection{The PICOSEC variant}

\begin{figure}[t]
    \centering
    \includegraphics[width = 0.79\textwidth]{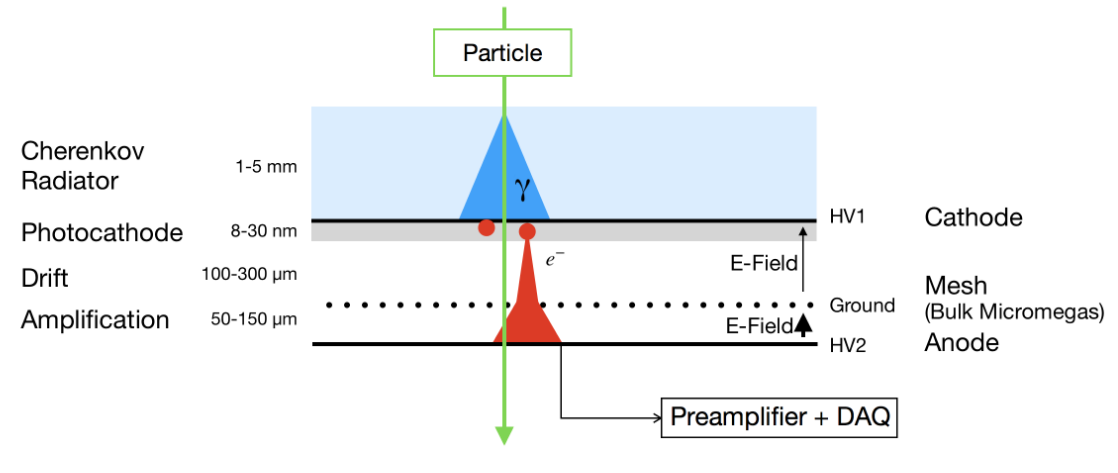}
    \caption[PICOSEC detector concept.]{PICOSEC detector concept. Image adapted from \cite{picosec}.}
    \label{pico-gas}
\end{figure}

The PICOSEC detector (concept illustrated in Figure \ref{pico-gas}) is a Micromegas-based detector invented for addressing the challenge of reaching a time resolution in the order of tens of picoseconds.
It borrows the structure of a Micromegas with a very small conversion gap $(200\mu m)$,
which in fact is so small that it is no longer a conversion gap;
a muon has very slim chances of producing a primary electron that will cause an avalanche big enough to be detected.
Therefore the gap will be referred to as drift gap, and
the voltage difference applied between the cathode and the (micro-)mesh will be referred to as drift voltage.
The voltage difference is high enough that electron avalanches form in this gap;
these will be called pre-amplification avalanches.
    
So far in this discussion, the detector is almost blind to charged particles.
However, by equipping it with a photocathode instead of a normal cathode, it can act as a photodetector with very good time resolution.
Relativistic charged particles can become visible again to the detector by adding a Cherenkov radiator on top of the photocathode.
This has been realized with a $3\, mm$ thick $MgF_2$ crystal.
This way, when a relativistic charge particle passes through the radiator and has a speed greater than the speed of light in the medium, 
the radiator itself will emit Cherenkov photons.
These Cherenkov photons will interact with the photocathode, become photoelectrons and enter the gaseous volume.

There the field in the drift gap (drift field) is large enough that it allows multiplication.
This avalanche multiplication will continue until the micro-mesh where a fraction of the electron will be absorbed.
Those that transmit through will continue the avalanche multiplication in the much higher (amplification) field, inducing a current on the anode.
This current can be readout with appropriate electronics.

\chapter{Experimental Tests}
In this chapter the Picosec detector is described in more detail with focus on its geometrical characteristics and operating conditions.
Most of the details are also summarized in \cite{picosec}.
The first section describes the structural details of the detector while
in the last two sections the two types of experimental campaigns are described, the ``Laser Test'' setup and the ``Testbeam'' setup. 
Although both experimental tests are considered ``test beams'', when referring to the ``Testbeam'', the campaing described in Section \ref{sec:testbeam} is assumed.
A summary of the analysis on the Laser Test is mentioned in Chapter \ref{chap:laser} while a summary of the additional methods used for data collected in the Testbeam is summarized in Chapter \ref{chap:muons}.

\section{Details on the Picosec Prototype}

The prototype's readout is a bulk Micromegas detector \cite{bulk} consisting of a single circular anode 
(without micro-strips) which has a diameter of $1\,cm$.
As part of the bulk Micromegas detector, 
the center of the (micro-)mesh sits above the anode at a distance of $128\,\mu m$
\footnote{It is believed that during the fabrication process, 
the photoresist material laminated between the anode and the mesh is slightly compressed 
and the real distance between the center of the mesh and the anode plane is $\sim 120\,\mu m$.}.
The wires of the mesh are $18\,\mu m$ thick and the openings between the wires form squares with sides equal to $45\, \mu m$,
 acounting for a total of $63\,\mu m$ pitch (distance between the centers of each hole).
These particular characteristics correspond to an optical transparency of the mesh equal to $51 \%$, 
which should not be confused with the electron transparency. 
Optical transparency refers to the fraction of visible photons (for infinite plane waves parallel to the plane of the mesh) which are transmitting through the mesh.
The electron transparency refers to the fraction of electrons transmitting through the mesh and is heavily dependent on the electric field configuration.
Furthermore, the mesh is plainly woven and as such its maximum thickness would have been $36\,\mu m$ if it wasn't calendered, 
i.e. compressed, and therefore its maximum thickness is approximately equal to $30\,\mu m$.
The mesh is held into place by 6 cylindrical pillars with diameters of $200\,\mu m$, arranged in a regular hexagonal 
around the center of the anode.
The height of these pillars is no longer than $128\,\mu m$ and thus they do not protrude above the mesh.

\begin{figure}[t]
    \centering
    \includegraphics[width=0.8\textwidth]{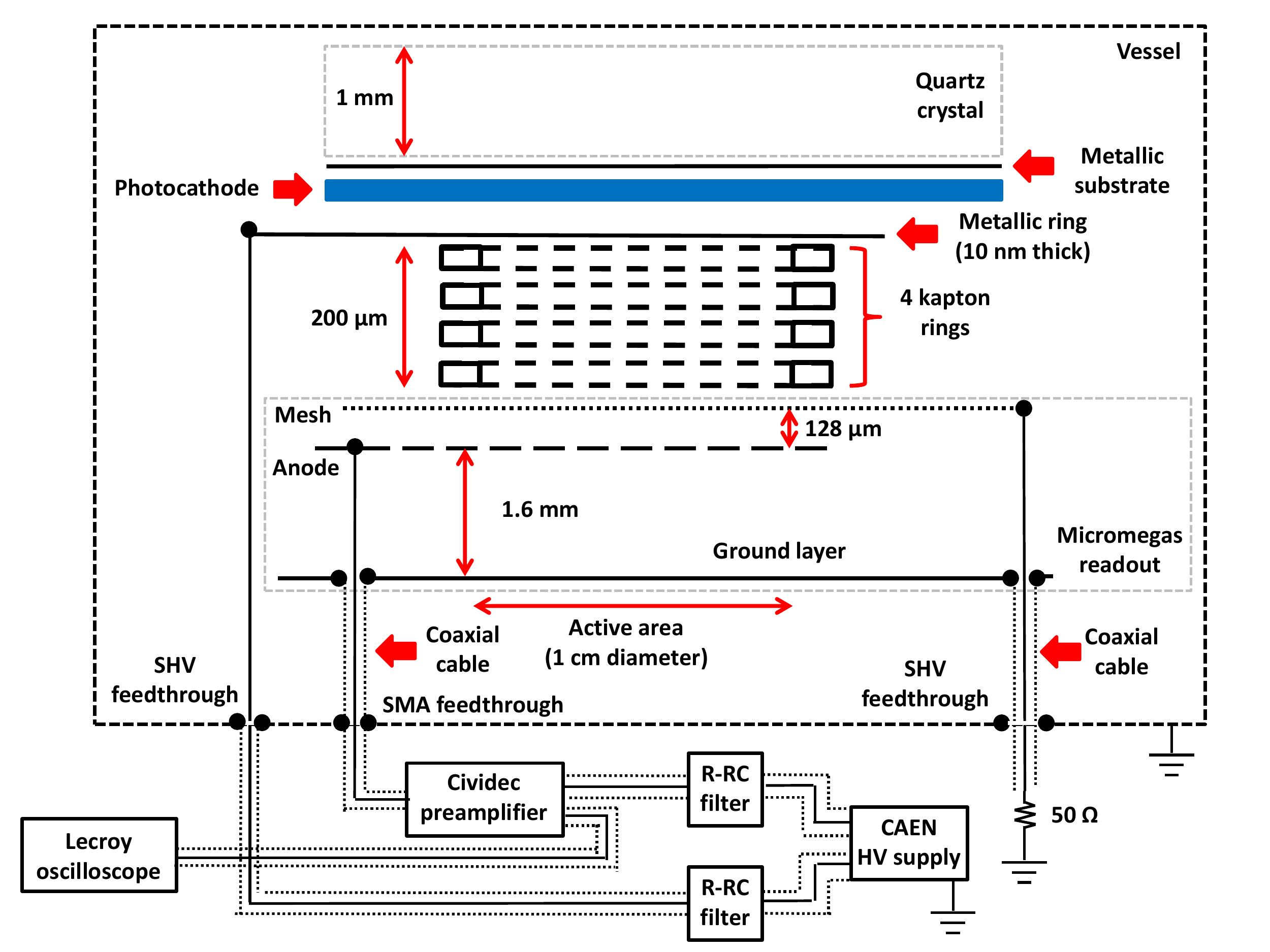}
    \caption[Sketch of PICOSEC prototype.]{Sketch of the PICOSEC prototype. The sketch is not to scale for visual clarity. Image is adapted from \cite{picosec}}
    \label{sketch}
\end{figure}

To define the drift gap, four kapton rings, each of $50\, \mu m$ thickness are placed on top of each other 
which are then placed on top of the insulating layers (during the lamination process) that define the amplification gap.
In other words, the drift gap is extending at $200\, \mu m$ from the center of the mesh.
On top of the kapton rings sits the arrangement of the photocathode and the Cherenkov radiator.
The Cherenkov radiator is a $3\,mm$-thick $MgF_2$ crystal, on whose whole surface is deposited a $5.5\,nm$ chromium metallic substrate.
In the center of the radiator, an $18\,nm$-thick layer of CsI with a diameter of $1\,cm$ is deposited to act as a photocathode\cite{csi}.
In contact with the metallic substrate layer comes a metallic ring of $10\, nm$ thickness, on which the ``drift'' high voltage is applied.

This setup is enclosed in a steel vessel which is completely filled with a gas mixture.
There are two gas mixtures which are used.
The first, called COMPASS gas, is composed of $80\%\,Ne$, $10\%\,CF_4$ and $10\%\,C_2H_6$.
The second one is composed of $80\%\,CF_4$ and $20\%\,C_2H_6$.
This function will hereafter be referred to as the CF$_4$ gas.
Both mixtures were used in the Laser Tests while only the COMPASS mixture was used in the Beam Tests and only the COMPASS mixture was simulated.

To prevent unwanted reflections, the mesh connection, exposed with a SHV feedthrough, 
is grounded through a long coaxial cable with a BNC connector and termination resistance of $50$ Ohm.
Another SHV feedthrough connects the cathode with a CAEN High Voltage Supply and an indermediate lowpass filter to reduce noise fluctuations.
Finally, an SMA feedthrough connects the anode to a CIVIDEC C2-HV Broadband Diamond Amplifier\cite{cividec}
with an analog bandwidth of $40\,GHz$,a $43\,dB$, 
an equivalent input noise of $0.4\,\mu A$, and a guaranteed linearity in the output range of $\pm 1\,V$.
The sketch of the prototype is illustrated in Figure \ref{sketch}.

\section{Laser Test}\label{laser-test-section}

\begin{figure}[t]
    \centering
    \includegraphics[width=0.8\textwidth]{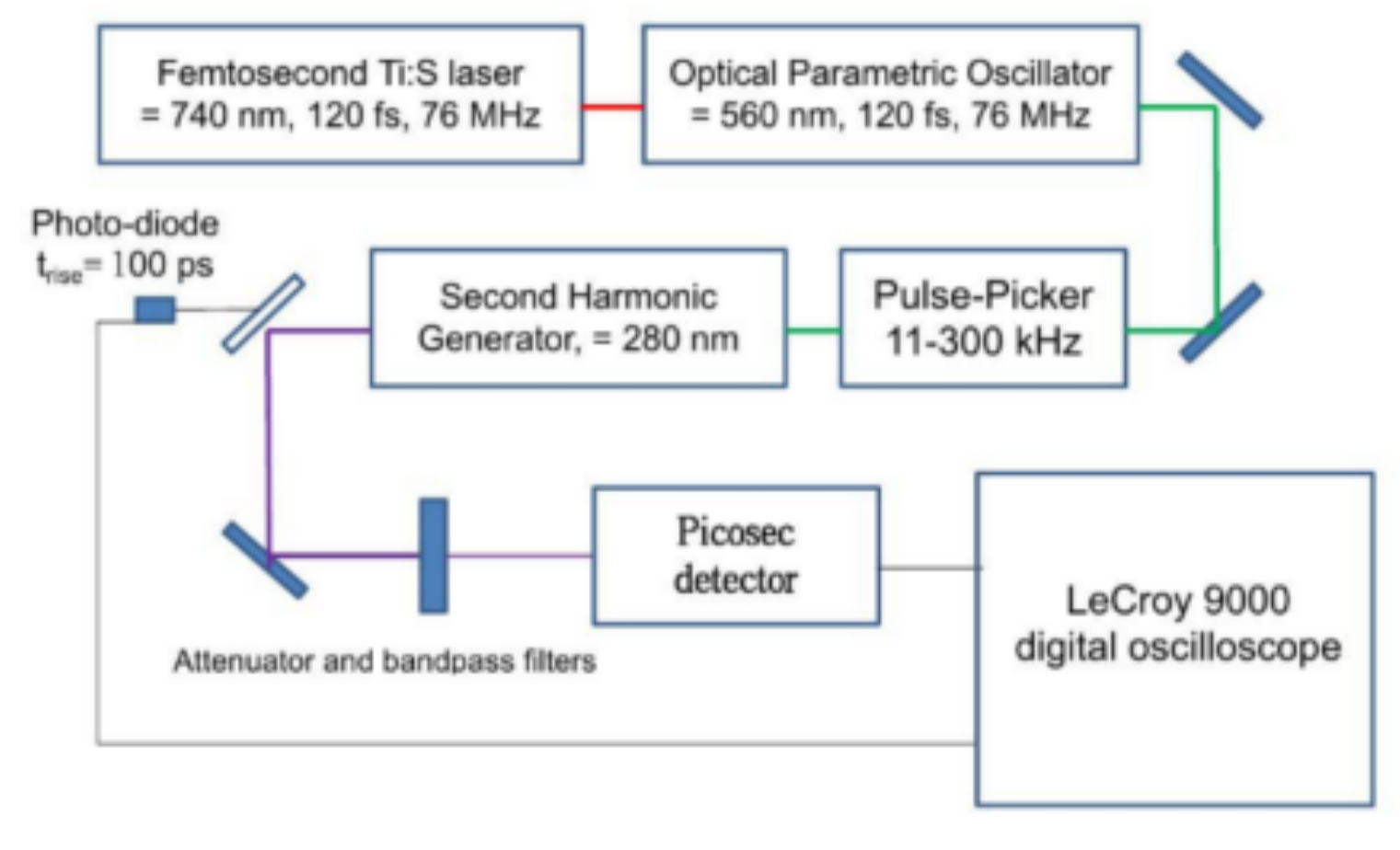}
    \caption[Laser Test setup sketch.]{Experimental setup employed to study the PICOSEC's response to single photoelectrons. Details for the setup is included in the text. Image is adapted from \cite{picosec}.}
    \label{sketch-laser}
\end{figure}

To investigate the detector's response to single photoelectrons, it was put to test at the Saclay Laser-matter Interaction Center (IRAMIS/SLIC,CEA).
The goal of the PICOSEC detector is to reach a time resolution in the order of picoseconds.
For this reason, a femtosecond laser was used with a pulse rate in the range of $[9\,kHz,\  4.7\,MHz]$ with wavelengths at $267-278\, nm$ and a focal length in the order of $\sim 1\,mm$.
A fast photodiode was used to find a time reference.
The laser beam from the femtosecond laser is split into two beams where one the beams is sent directly to the photodiode while the other beam is sent to the PICOSEC detector.
Before reaching the PICOSEC, the laser beam transmits through a number of electroformed nickel micro-meshes with $100-2000\,LPI$ (Lines Per Inch) which have optical transmissions between $10\%$ and $25\%$, 
totalling to attuenation factors of $4,5,10$,
and any other combination of these.
The attenuation of the beam is adjusted such that the number of photoelectrons that are finally extracted from the photocathode is at most one. 
A pictorial representation of the setup is shown in Figure \ref{sketch-laser}.
The signals from both the PICOSEC detector (equipped with the CIVIDEC amplifier mentioned before) 
and the photodiode are recorded with an oscilloscope with a bandwidth of $2.5\,GHz$ 
and a rate of digitizations of $20\,GSamples/s$ which translates to a time step of $50\,ps$ between two adjacent samples. 

This campaign of measurements, hereafter reffered to as the Laser Test, consisted of voltage scans with both the CF$_4$ and the COMPASS gas mixtures.
The anode voltage was scanned between $600\,V$ and $650\,V$ for the CF$_4$ mixture while it was scanned between $450\,V$ and $525\,V$ withe voltage steps of $25\,V$.
They drift voltages were scanned between $350\,V$ and $500\,V$ for the CF$_4$ gas and between $200\,V$ and $425\,V$ for the COMPASS gas, both in steps of $25\,V$.
The full range of the drift voltage scan was not made for every anode voltage. 
Each combination of anode and drift voltages had to first 
output at measureable signal (distinguishable from the noise)
and exhibit a stable operation of the detector, i.e. that it was not sparking.

In order to achieve a single photoelectron response there had to be many events that would not produce any photoelectrons.
Because of this, the PICOSEC had to be included in the trigger chain
(through a constant threshold discrimination) to make the data collection efficient.

\section{Testbeam Campaign}\label{sec:testbeam}

\begin{figure}[t]
    \centering
    \includegraphics[width=0.8\textwidth]{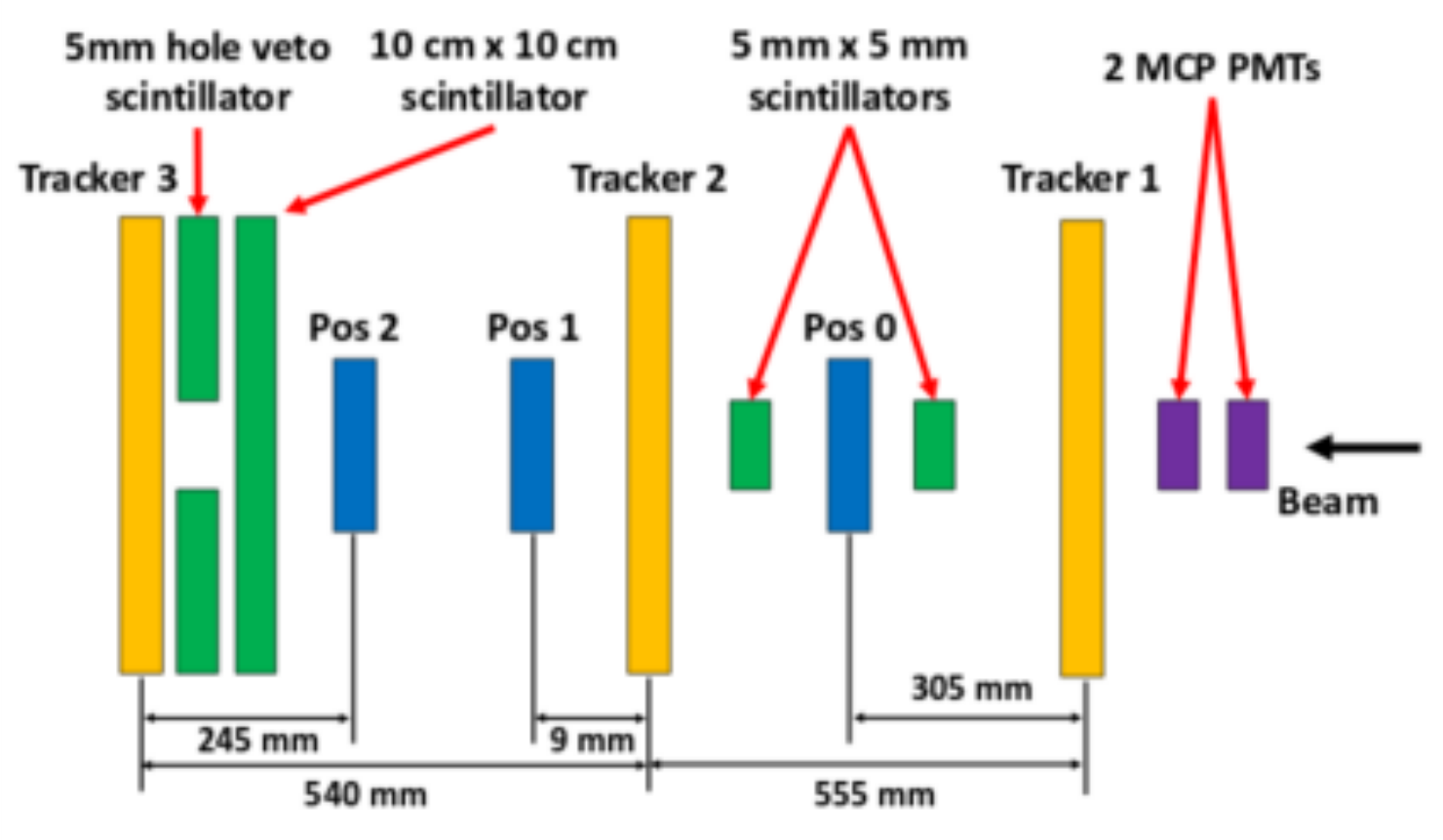}
    \caption[Testbeam setup sketch.]{
    Experimental setup employed to study the PICOSEC's response to relativistic charge particles. Details for the setup is included in the text. Image is adapted from \cite{picosec}.}
    \label{beam}
\end{figure}

The more important test is the measurement of (relativistic) charged particle to take advantage of the Cherenkov radiation.
This test is realized at the CERN SPS H4 secondary beamline using muons with a momentum of $150\,GeV$.
A series of measurements in the same beamline included the use of pions.
Results of measurements with pions however are not presented in this work.
The results with muons that will be presented were collected in the Testbeam campaign of July 2017.
Figure \ref{beam} shows the scheme of the experimental setup.
A series of scintillators was used to trigger the charged particles.
Either a ``large'' $(30\times 20 mm^2)$ or two ``small'' scintillators $(5\times 5\, mm^2)$ are used.
The large scintillator allows the study of the PICOSEC across its whole active area while the small scintillator allows all tracks to be depositing all of the photons on the photocathode and the active area of the PICOSEC. 
More details on this subject in Chapter \ref{chap:muons}.
To supress muons that undergo scattering and particle showers coming from hadronization interactions, a large scintillator with a hole $(5\times 5\,mm^2)$ in its center is used.
Obviously, this is only possible when using the small scintillator.

To measure the time resolution, a time reference is need.
For this purpose, two Hamamatsu MicroChannel Plate PhotoMultiplier Tubes (MCP-PMTs)\cite{mcp} model R3809U-50, are used.
The MCP-PMTs are photodetectors, however a quartz entrance window of $3\,mm$ that is attached on its photocathode is used to provide a detection mechanism for relativistic charged particles
similar to the detection mechanism of PICOSEC (through Cherenkov radiation).
The signal of the MCP-PMTs is measurement with the signal from the PICOSEC at the same time from the same oscilloscope with a bandwidth of $2.5\,GHz$ and a rate of $20\,GSamples/s$.
The time resolution of the MCP-PMTs was measured be under $5\,ps$.

To provide spatial information, a tracker consisting of three gem detectors is used which are triggered from the scintillators mentioned previously and are readout with an APV25 and SRS Data AcQuisiton (DAQ) system\cite{srs}.  
Because the data acquisition system of the gem is different than the data acquisition system of the PICOSEC (i.e. the oscilloscope), the event number of the SRS is broadcast to a channel of the oscilloscope in the form of a digital bitstream.
This allows the alignment and correspondence of events recorded by the SRS DAQ and the oscilloscope, providing this way the track impact point of the charge particle.

In between the Testbeam measurements, a UV LED was used to take single photoelectron calibration data.
In these measurements, no timing information could be provided and the detector's waveform was triggered directly with a constant threshold leading edge discrimination.

\section{Cross Sections of COMPASS and CF$_4$ Gases}

\begin{figure}[H]
    \centering
    \begin{subfigure}[h]{0.75\textwidth}
        \includegraphics[width=0.95\textwidth]{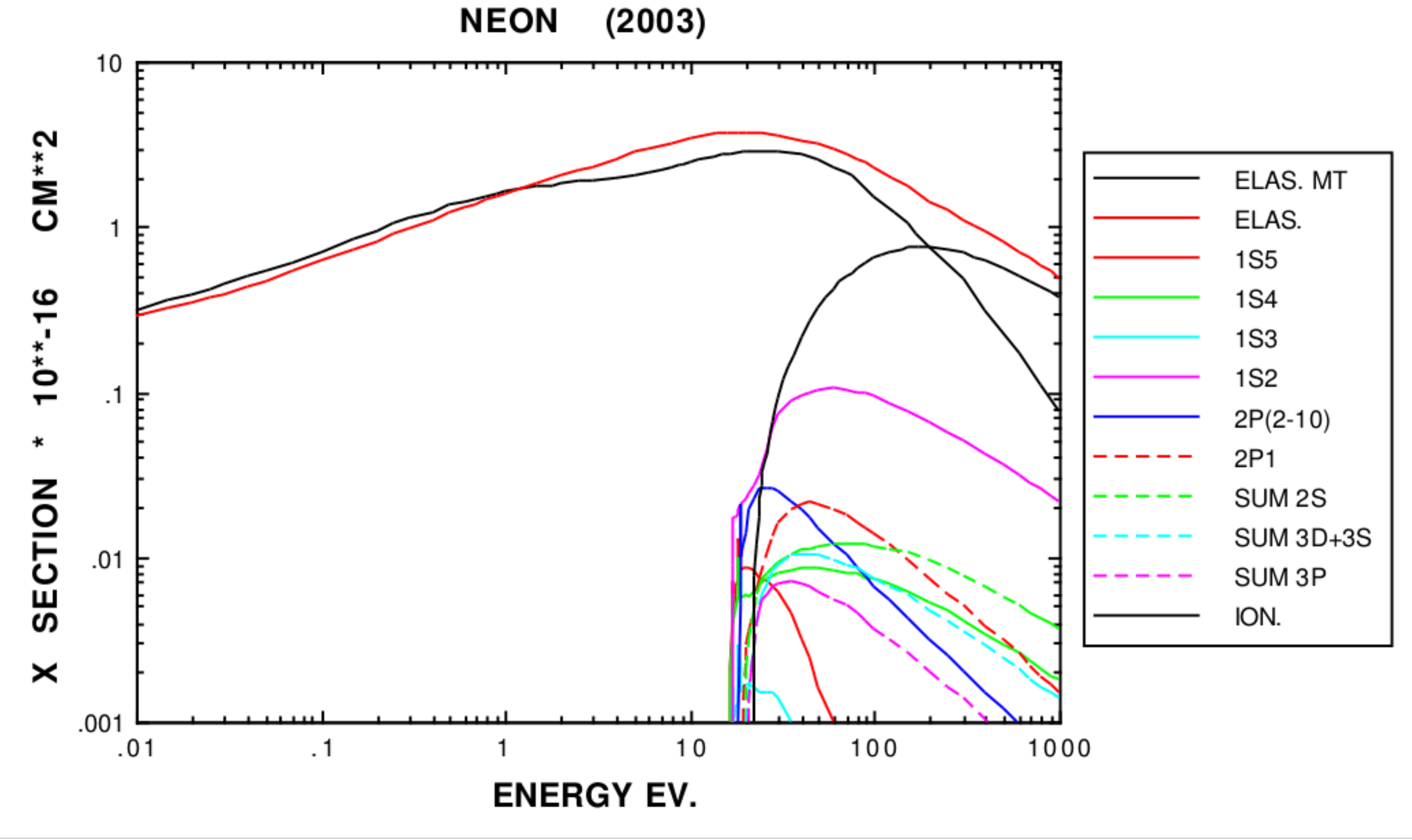}
        \caption{}
    \end{subfigure}\\
    \begin{subfigure}[h]{0.75\textwidth}
        \includegraphics[width=0.95\textwidth]{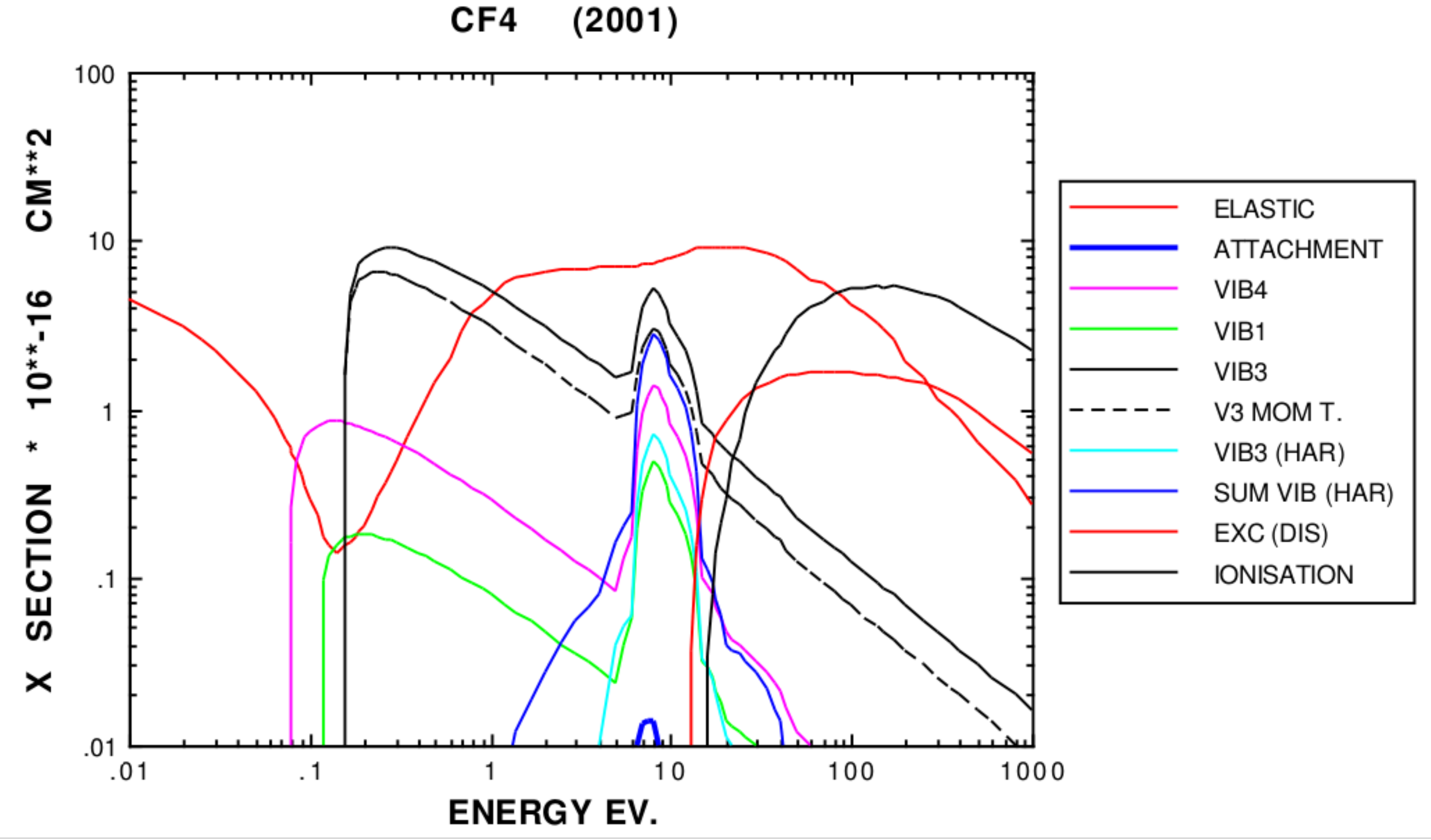}
        \caption{}
    \end{subfigure}\\
    \begin{subfigure}[h]{0.75\textwidth}
        \includegraphics[width=0.95\textwidth]{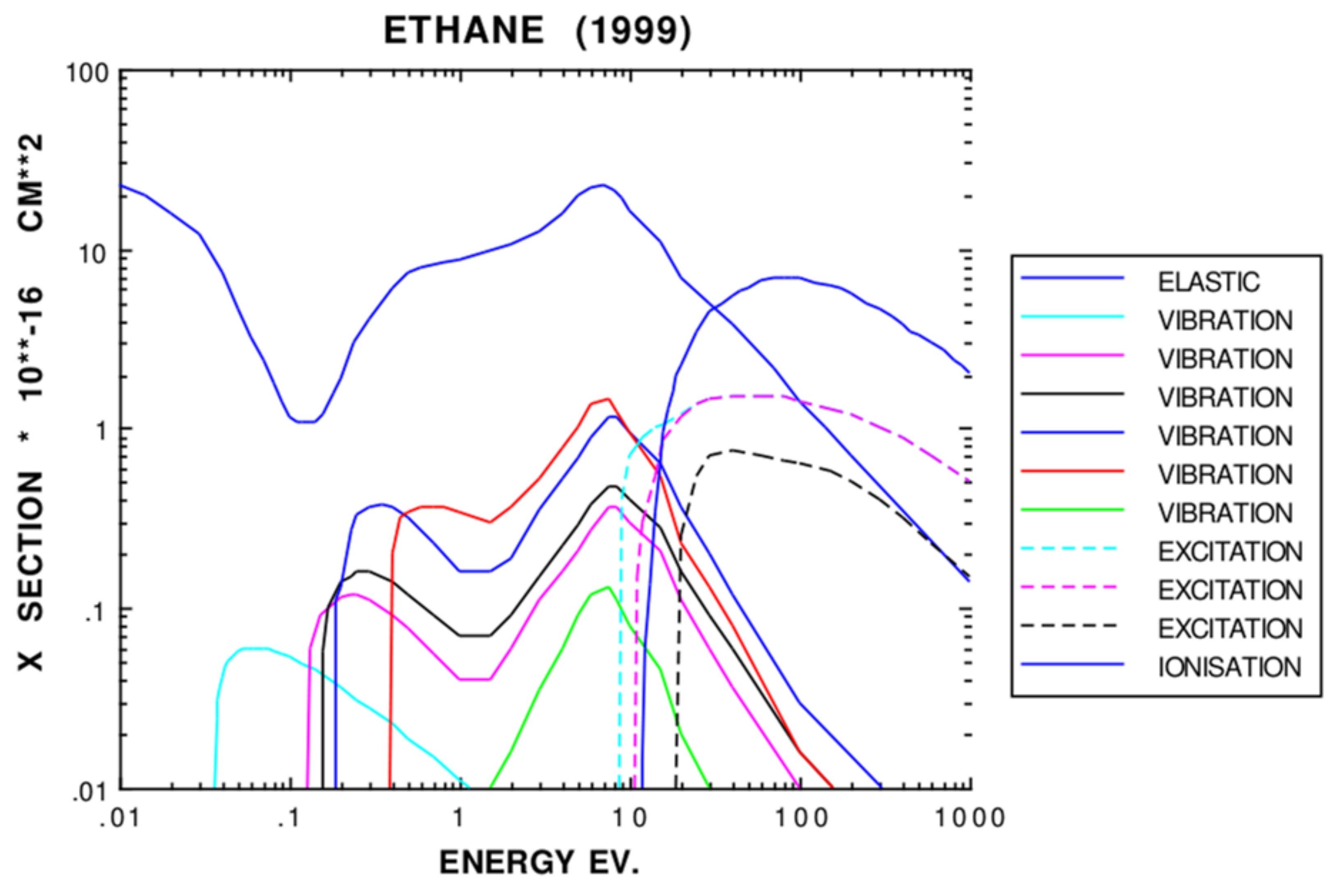}
        \caption{}
    \end{subfigure}
\caption[Cross sections of interactions in the gas mixtures.]{
       Cross sections versus the energy of the electron for the (a) $Ne$, (b) CF$_4$ and (c) $C_2H_6$ gases. Adapted from Magboltz\cite{magboltz}.
    }
    \label{xs}
\end{figure}

The cross sections of the individual components of the COMPASS and the CF$_4$ gases are included and presented in Figure \ref{xs}.
The ionisation thresholds of the three gas components are $I = 21.56,\ 15.9,\ 11.52$ for $Ne$, CF$_4$ and $C_2H_6$, respectively.
Because $Ne$ has many excitations that are above the ionisation thresholds of the two other gas components, it is expected that there will be a considerable Penning effect in the COMPASS mixture.

\chapter{Analysis of the Laser Test}\label{chap:laser}
\section{Waveforms and Signal Processing} \label{sec:sp}

\begin{figure}[t]
    \centering
    \begin{subfigure}[h]{0.49\textwidth}
        \includegraphics[width=0.95\textwidth]{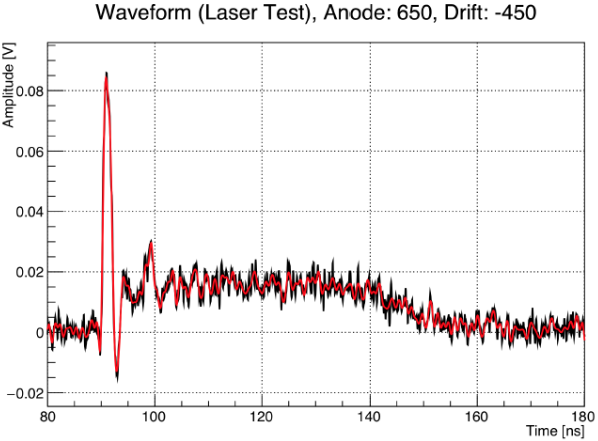}
        \caption{}
    \end{subfigure}
    \begin{subfigure}[h]{0.49\textwidth}
        \includegraphics[width=0.95\textwidth]{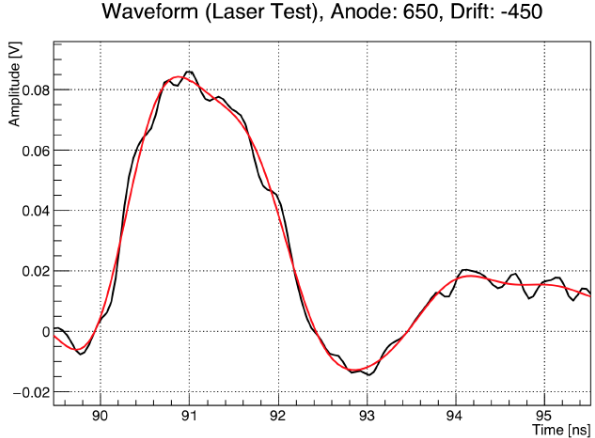}
        \caption{}
    \end{subfigure}
    \caption[Typical waveform of the detector.]{A typical waveform of the detector which is amplitude as a function of time. (a) shows the entirety of the waveform while (b) is focused on the electron peak. The black line corresponds to the measured, digitized signal whilst the red line corresponds to the same signal after it has been processed through Fourier-based filters.
}
    \label{typ-mm-wave}
\end{figure}

Every time the experimental setup is triggered a single object is recorded for each detector, its electrical signal (sometimes referred to as waveform).
The signal is essentially a time series of voltage amplitudes in regularly spaced time intervals. 
This time interval is equal to $\Delta t = 50\,ps$ for Laser and Muon data,
or equal to $\Delta t = 25\,ps$ in the case of Muon-Calibration data.
A typical waveform\footnote{It should be noted that the real waveform is negative but all negative waveforms are inverted in the software for clarity and to avoid confusion.} 
is shown in Figure \ref{typ-mm-wave} with black, where a clear separation between the electron and the ion components can be made.
\footnote{The analysis is based on the ROOT\cite{ROOT} Data Analysis Framework.}
The electron component (or electron peak) is the fast signal (width at baseline $\sim 2\,ns$) which is induced by the movement of the electrons.
The ion component (or ion tail) is the slow tail (width at baseline $\sim 60\,ns$) and is induced by the movement of the positive ions that drift from the anode to the micro-mesh and then from the micro-mesh to the cathode (those that manage to transmit through the micro-mesh).
The waveform is digitized by an 8-bit oscilloscope which means that each point in the signal takes a discrete value from $2^8 = 256$ regularly spaced amplitudes. 
The red line in Figure \ref{typ-mm-wave} shows an example of waveform processed with a Fourier based filter.
Although this filter successfully damps a significant part of the noise, it has been studied that it also reduces the timing information of the signal in small pulses (with small amplitudes), i.e. the time resolution of small pulses worsens.
More details in the Fourier based filters used can be found in \cite{vasilis}.

The waveforms do not necessarily have their baseline equal to zero amplitude.
For this reason, a baseline correction is applied on event-by-event basis.
In each waveform, the baseline characteristics (mean and RMS) are calculated from the first few points lying in the window of the first $~75\,ns$.
Then, the baseline mean is subtracted on a point-by-point basis and is restored to zero amplitude.
The RMS of the baseline is an estimation of the noise RMS in the waveform.

\subsection{Electron Peak Size}
All of the triggered events are assumed to be caused by a single photoelectron, i.e. of all the incident photons in the cathode only one is converted to an electron, as explained in Section \ref{laser-test-section}.
One of the most important characteristics of the waveforms is the electron peak amplitude, which is the amplitude of the point with the highest amplitude in the waveform.
It would be unnecessary and misleading to interpolate between the tallest points as it is clear in Figure \ref{typ-mm-wave} that noise fluctuations are much faster than the electron peak.
Interpolating between the tallest points would be equivalent to interpolating the noise.
The electron peak amplitude is one of the two variables that expresses the size of the electron peak 
and is related to the number of secondary electrons produced in the detector, or the gain of the detector.
The other variable expressing the size of the electron peak is the charge of it.
The charge of the electron peak is equal to the integral of the electron peak, from its start until its end, 
transformed to (pico)Coulombs assuming a termination resistance of $50\,\Omega$.
This charge is a much more reliable expression of the electron peak's size because the noise contribution is averaged out in the integral.

\begin{figure}[H]
    \centering
        \includegraphics[width=0.49\textwidth]{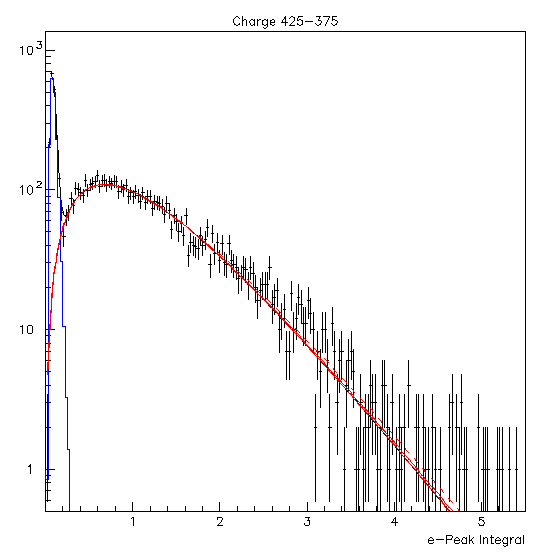}
        \caption[Electron peak charge distribution with noise.]{
        Distribution of the electron peak charge in waveforms accumulated in June laser tests with voltage settings of $425\,V$ on the anode, and $375\,V$ on the cathode. Black points correspond to the experimental distribution, while the red line corresponds to a Polya fit and the blue line corresponds to the noise contribution.
    }
    \label{mm-anal-polya}
\end{figure}

\begin{figure}[H]
    \centering
        \includegraphics[width=0.49\textwidth]{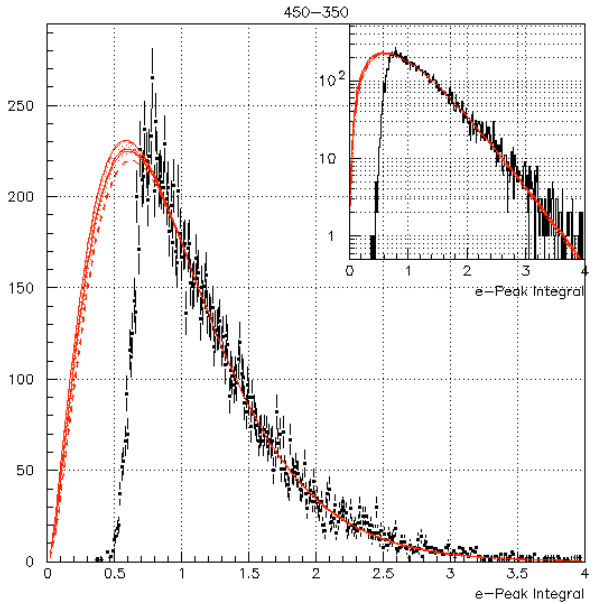}
        \caption[Truncated distribution of electron peak charge.]{
        Distribution of the electron peak charge with black points in waveforms accumulated with voltage settings of $450\,V$ on the anode, and $350\,V$ on the cathode. P\'olya fits with varying boundaries or bin sizes are shown with red.}
    \label{polya-cut}
\end{figure}

Both the electron peak charges and amplitudes are parameterized using the P\'olya distribution in Equation \ref{eq:polya}.
The distribution of the electron peak charges is shown in Figure \ref{mm-anal-polya}.
The black points correspond to the experimental distribution collection in a June laser test with an anode voltage of $425\,V$ and a cathode voltage of $375\,V$.
The red line corresponds to the P\'olya parameterizations while the contribution of the noise are shown with blue.
The usefulness of the P\'olya parameterization lies in the fact that it provides a simple and reliable formula to express the distribution of the electron peak size and also to extrapolate the distribution in unobserved regions.
The need for this parameterization to extrapolate the distribution is clear in Figure \ref{polya-cut} where the charge distribution is shown with black points for an anode voltage of $450\,V$ and a drift voltage of $350\,V$ with the chamber filled with COMPASS gas.
Red lines correspond to different P\'olya fits with varying parameters (fit boundaries, bin size, etc.).
It is clear that the distribution is heavily truncated because of the trigger setup and as a result a lot of the photoelectrons go unobserved.
Calculating the mean value of ``amplification'' would be impossible (or biased) without the use of such a parameterization.

For the charge estimation, a start point and an endpoint is needed for the integration.
The start point of the electron peak is defined as the first point of the leading edge with an amplitude higher than the RMS of the noise, as calculated from the baseline.
The end point of the electron peak is defined as the local minimum lying between the electron peak and the ion tail.
In some cases, the waveform components are not clearly separated such as in Figure \ref{typ-mm-wave} and this local minimum does not exist causing an artificial end point to be selected.
For this reason, it is much more reliable to use the electron peak amplitude in place of the electron peak charge in these cases for fast analyses.
This issue can be remedied by fitting the whole electron peak, as will be described later in Section \ref{section:ept}.

\subsection{Electron Peak Timing}\label{section:ept}

One of the most important characteristic (especially in this study) of the waveform is its timing.
There exist two dominant factors that limit the resolution of timing a signal.

\begin{figure}[H]
    \centering
    \begin{subfigure}[h]{0.49\textwidth}
        \includegraphics[width=0.95\textwidth]{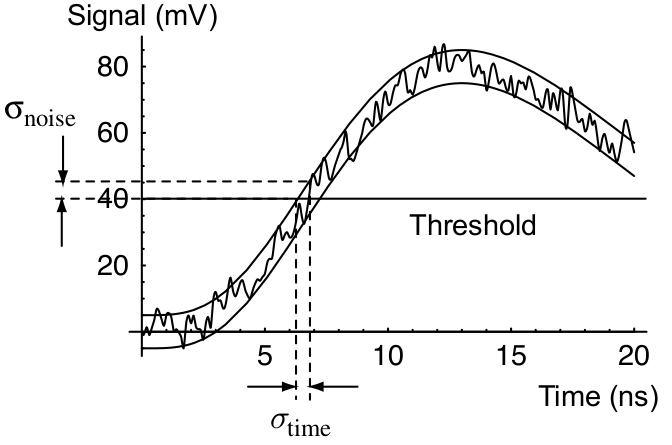}
        \caption{}
    \end{subfigure}
    \begin{subfigure}[h]{0.49\textwidth}
        \includegraphics[width=0.95\textwidth]{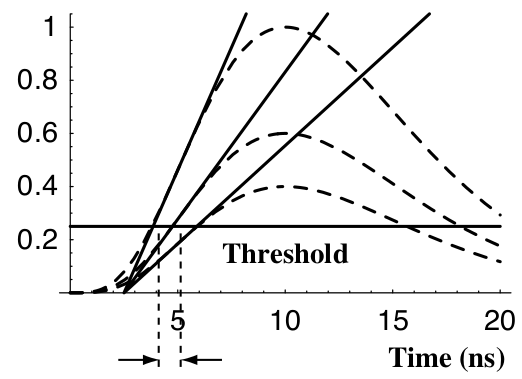}
        \caption{}
    \end{subfigure}
    \caption[Time jitter and time walk.]{ (a) The effect of ``time jitter''; a superimposed noise of RMS $\sigma_{\text{noise}}$ in the waveform will lead to a crossing time of a certain threshold with RMS $\sigma_{\text{time}}$. (b) The effect of ``time walk''; signals with different pulse heights but same intrinsic timing cross a fixed threshold at different times. Both images are adapted from \cite{riegler-drift}.
            }
    \label{anal-time-noise}
\end{figure}

The first one is what is known as ``time jitter''. 
When measuring a waveform, it is unavoidable that electronic noise will be superimposed on the signal.
Through a simple geometrical projection, electronic noise in amplitude $\sigma_{\text{noise}}$ will translate in a noise in time $\sigma_{\text{time}}$, or time jitter, with $\sigma_{\text{time}} = \left|\frac{d f(t)}{d t}\right|^{-1}\sigma_{\text{noise}}$ where $f(t)$ is the signal.
This effect is illustrated in Figure \ref{anal-time-noise}\,a.
Assuming that all signals have a fixed amplitude and shape, and that discrimination is happening with the crossing of a fixed threshold, the time resolution of this timing technique coincides with the time jitter $\sigma_{\text{time}}$.

If the signal is now not of a fixed amplitude, then a taller signal (with a higher amplitude) will cross a fixed threshold in an earlier time than a smaller signal would.
This dependence of the timing with the amplitude of the signal is called ``time walk'' or ``slewing''.
Ignoring this correlation would cause a timing resolution worse than the time jitter.
A time walk effect is not catastrophic and can easily be corrected in an off-line analysis.
Actually this correction can be avoided by using a timing technique that is more sophisticated than the constant threshold discrimination, for example constant fraction discrimination.
In constant fraction discrimination, the timing of the waveform is defined as the time corresponding to a specific fraction of the signals amplitude, e.g. the time corresponding to the $20\%$ of the peak amplitude.
By using this technique, as long as the shape of the signal is not changing, there should be no time walk effect.

Most of the timing techniques which are being employed are based on constant fraction discrimination to avoid this time walk effect
\footnote{As will be shown later, there still exists a time walk/slewing effect in constant fraction timing that is not caused by the change of the signal's shape.
In fact, it is caused by the physical processes in the detector and the primary purpose of the simulation was to study this effect and the mechanism that produces it.}.
A number of techniques are used to time the signal to make sure that all results are consistent with each other.
The techniques are as follows:
\begin{enumerate}
    \item The most naive application of constant fraction discrimination is to find the two successive points whose first point's amplitude is less than the $a\%$ of the electron peak amplitude while the second point's amplitude is larger.
        Then a linear interpolation is applied to find the timing that corresponds to the $a\%$ discrimination.
    \item A more sophisticated approach is to include more points in this interpolation, i.e. two from the left of the $a\%$ and two from the right.
        Then a cubic interpolation is applied to find the timing that corresponds to the $a\%$ discrimination.
    \item Another technique uses the fact the derivative of the pulse is maximum at the point of inflection. 
        A cubic interpolation is performed similarly to above technique where the points are on the left and the right of the time with the maximum derivative.
        The tangent line to the cubic interpolating polynomial is found and the point where it crosses the zero amplitude is defined as the timing of the waveform.
    \item Another family of techniques is to use parametric functions to fit and model the whole leading edge and then employ the constant fraction discrimination on the parametric function.
        The logistic function is a convenient option to use as the parametric function.
        Its functional form is:
        \begin{equation}
        f_4(x;p_0,p_1,p_2,p_3) = \frac{p_0}{1 + e^{-(x-p_1)p_2}} + p_3
        \end{equation}
        Its convenience comes from the fact that the constant fraction discrimination can be analytically employed.
        Assuming a $a\%$ constant fraction and an amplitude of $V_{\text{max}}$ the timing is equal to:
        \begin{equation}\label{t4eq}
            t_4= p_1 - \frac{1}{p_2}\ln \left(\frac{p_0}{a\cdot V_{\text{max}}-p_3}-1\right)
        \end{equation}

    \item Similarly to the previous one, the generalized logistic function is used to model the leading edge.
        The functional form of the generalized logistic function that is used is:
        \begin{equation}
            f_5(x;p_0,p_1,p_2,p_3) = \frac{p_0}{\left(1 + e^{-(x-p_1)p_2}\right)^{p_3}} 
        \end{equation}
        Solving for the time of constant fraction in the same spirit as before gives:
        \begin{equation}\label{t5eq}
            t_5= p1 - \frac{1}{p_2}\ln \left(\left(\frac{p_0}{a\cdot V_{\text{max}}}\right)^{1/p_3}-1\right)
        \end{equation}

    \item In the two previous techniques it is clear that the electron peak amplitude is used in Equations \ref{t4eq} and \ref{t5eq}.
        Therefore, no matter how good these techniques are, they are limited from the estimation of the electron peak's amplitude.
        Simply picking the tallest point to estimate the amplitude is slightly biased when noise is superimposed to the real signal.
        An improvement to these techniques is to find a parameteric function to model the whole electron peak, so as to estimate its maximum amplitude.
        The obvious choice, allowing for plenty of degrees of freedom in the fit, is to use the difference of two generalized logistic functions.
        This function would take the form of
        \begin{equation}\label{dif-logi}
            f_6(x;p_0,p_1,\dots,p_6) = \frac{p_0}{\left(1 + e^{-(x-p_1)p_2}\right)^{p_3}} - \frac{p_0}{\left(1 + e^{-(x-p_4)p_5}\right)^{p_6}}
        \end{equation}
        Unfortunately, this function cannot be analytically solved for the constant fraction timing or for the electron peak amplitude (maximum).
        It is trivial though to employ numerical techniques to solve it, such as the method of Newton-Raphson, and acquire estimates for the timing of the waveform, 
        the electron peak maximum, as well as the electron peak charge through the integration of the function.
\end{enumerate}

All techniques give similar results and unless otherwise noted, the technique using the simple logistic function is used (Technique 4.).

\section{Timing Properties} \label{sec:tim}

The purpose of the PICOSEC detector is to provide precise timing information in the passage of a particle.
The way to demonstrate, as well as quantify, the precision of this timing information is by repeating the same measurement over and over again, having a reference time with a much better precision than the precision of the PICOSEC.
Then, the temporal distance of the detector's signal timing to the reference timing will follow a distribution.
This distribution can be approximated through the histogram of the many measurements and its spread is equal to the time resolution of the detector.

\begin{figure}[H]
    \centering
        \includegraphics[width=0.49\textwidth]{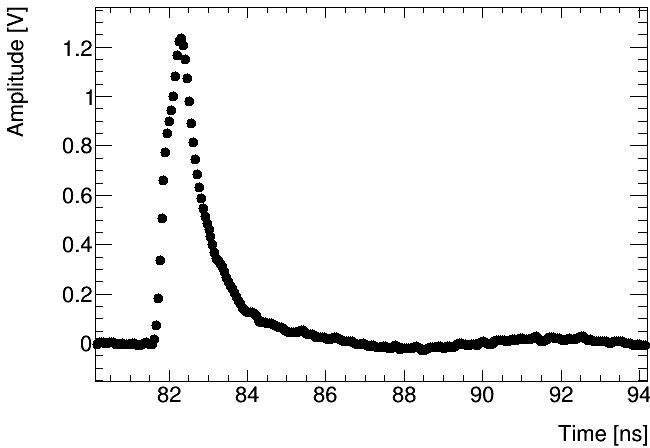}
        \caption[Typical photocathode waveform.]{
        A typical waveform of an event recorded by the photodiode whose timing (estimated with $20\%$ CFD) is recorded and used as the reference time on an event-by-event basis.}
    \label{mm-anal-pd}
\end{figure}

For the reference time, the signal of a photodiode is recorded and timed.
This signal has a time resolution that is less than $4\,ps$ and is suitable to study the time resolution of the PICOSEC detector.
An example of such a signal can be seen in Figure \ref{mm-anal-pd}.
The technique that is used to time the photodiode waveform is the one based on cubic interpolation and $20\%$ Constant Fraction Discrimination.

\begin{figure}[H]
    \centering
        \includegraphics[width=0.49\textwidth]{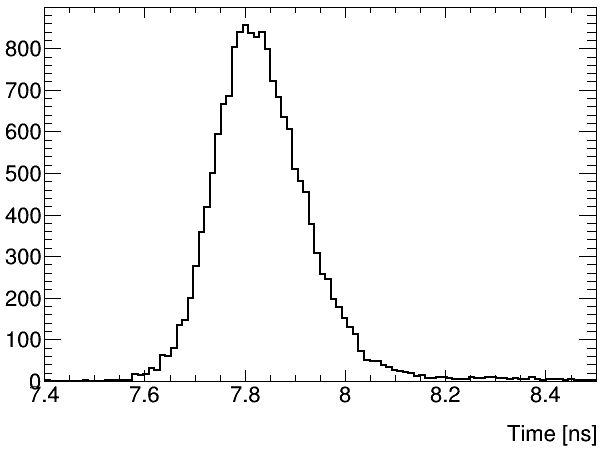}
        \caption[Temporal distance distribution.]{
    Distribution of the temporal distance between the timing of the PICOSEC and the reference time. The data were collected with the $\text{CF}_4$ gas mixture and anode and drift voltage settings of $650\,V$ and $450\,V$, respectively.}
    \label{mm-anal-resex}
\end{figure}

By timing the reference signal of the photodiode and subtracting it from the timing of the PICOSEC waveform, hereafter called Signal Arrival Time $SAT$,
it is shown in Figure \ref{mm-anal-resex} that the distribution of $\Delta T/SAT$ is highly asymmetric, 
for data collected with the $\text{CF}_4$ gas mixture and anode and drift voltages set at $650\,V$ and $450\,V$, respectively.
An asymmetric distribution is a very alarming indication that there is probably a systematic effect that is at play.
In the case of timing, this systematic effect is usually the slewing effect, which is the correlation of the mean signal arrival time with the size of the electron peak.

\begin{figure}[H]
    \centering
    \begin{subfigure}[h]{0.49\textwidth}
        \includegraphics[width=0.95\textwidth]{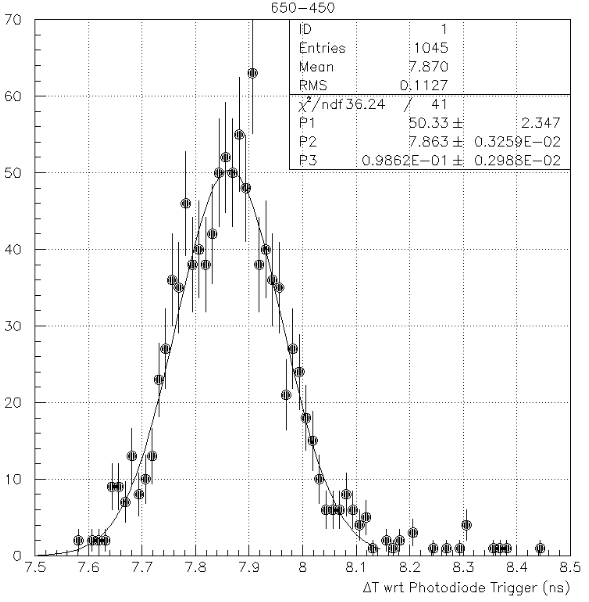}
        \caption{}
    \end{subfigure}
    \begin{subfigure}[h]{0.49\textwidth}
        \includegraphics[width=0.95\textwidth]{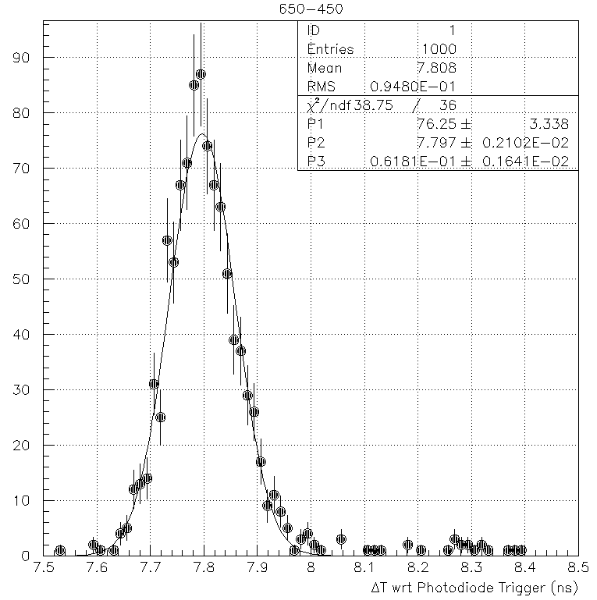}
        \caption{}
    \end{subfigure}
    \caption[Distribution of SAT for different electron peak charges.]{ 
    Distribution of SAT for events with electron peak charge in the range of (a) $[1.14\,pC,\,\,1.69\,pC]$ and (b) $[7.96\,pC,\,\,10.40\,pC]$. The data were collected with the $\text{CF}_4$ gas mixture and anode and drift voltage settings of $650\,V$ and $450\,V$, respectively.
    }
    \label{mm-anal-res12}
\end{figure}

To quantify and correct the mean SAT dependence on the size of the electron peak, the collection of events must be separated in narrow bins of the electron peak charge.
In the two distributions of Figures \ref{mm-anal-res12}\,a and b, the SAT from events in different bins of charges are selected. 
Figure \ref{mm-anal-res12}\,a corresponds to an electron peak charge in the range $[1.14\,pC,\,\,1.69\,pC]$, while b corresponds to a range of $[7.96\,pC,\,\,10.40\,pC]$.
The solid lines represent Gaussian fits.
The mean value of the Gaussian is the mean SAT while the standard deviation of the Gaussian represents the time resolution for the specific electron peak charge.

\begin{figure}[H]
    \centering
    \begin{subfigure}[h]{0.4\textwidth}
        \includegraphics[width=0.95\textwidth]{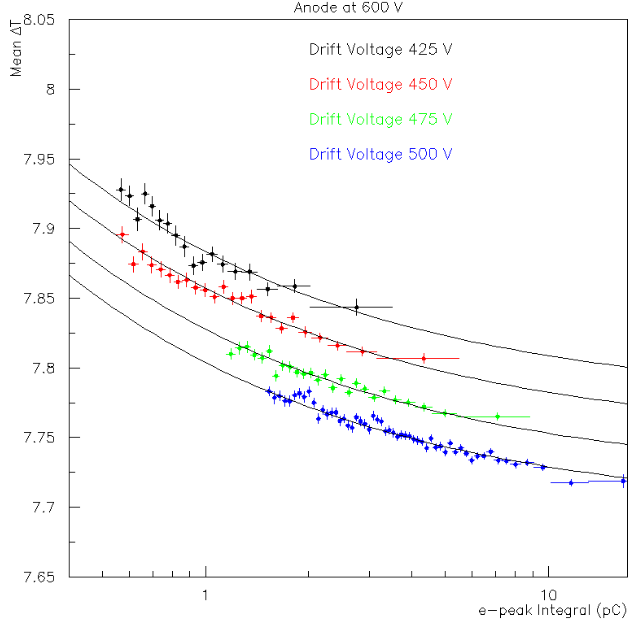}
        \caption{}
    \end{subfigure}
    \begin{subfigure}[h]{0.4\textwidth}
        \includegraphics[width=0.95\textwidth]{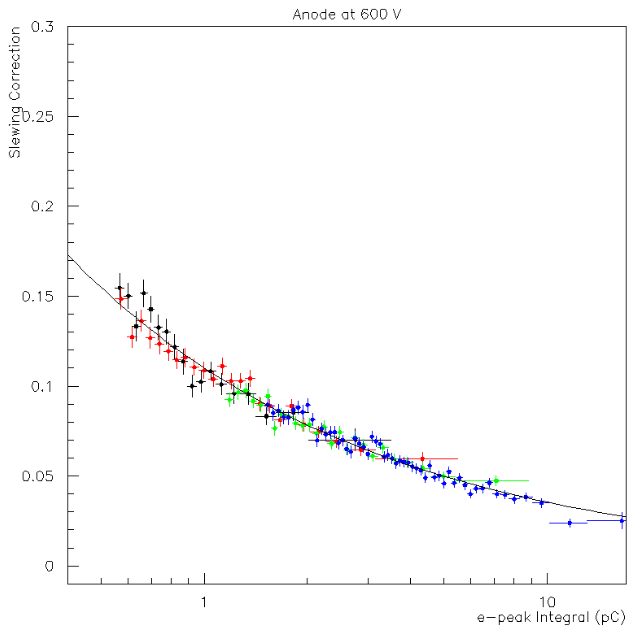}
        \caption{}
    \end{subfigure}
    \caption[Mean SAT as a function of the electron peak charge in CF$_4$ gas.]{ 
        Mean SAT as a function of the electron peak charge. The data were collected with the $\text{CF}_4$ gas mixture and anode and drift voltage settings of $650\,V$ and $\text{black: }425\,V,\text{ red: }450\,V,\text{ green: }475\,V,\text{ and blue: }500\,V$, respectively.
        (a) corresponds to the mean SAT versus the electron peak charge, while the same data are presented in (b) but with the related constant term subtracted for each data set. 
        Dashed lines represent fits with the function of Equation \ref{eq-power}.
    }
    \label{anal-slew600}
\end{figure}

After dividing the spectrum in many different narrow bins and repeating this procedure of fitting Gaussian distributions, the mean SAT and time resolution dependence on the electron peak size is found. 
For all voltage settings or choice of gaseous mixture, both the mean SAT and time resolution was found to be decreasing with increasing electron peak charge (and or voltage).
This procedure is again repeated for the different collected set of events with different anode and drift voltages.
The mean SAT as a function of the electron peak charge, with a anode voltage of $600\,V$ 
and drift voltages of $\left\{425\,V,\ 450\,V,\ 475\,V,\ 500\,V\right\}$, using the $\text{CF}_4$ gas, is presented in Figure \ref{anal-slew600}\,a.
It was found that the mean SAT versus the electron peak size (voltage or charge) are following a curve that can be approximated with a power law and a constant term (Equation \ref{eq-power}).
\begin{equation}\label{eq-power}
    g(x;a,b,w) = a + \frac{b}{x^w}
\end{equation}
where $x$ is either the electron peak charge or the electron peak amplitude, $a$ is the constant term and $b,w$ are the variables of the power law term. The $a,b,w$ free parameters are estimated through fits.
It was also found that for a fixed anode voltage, all correlations between mean SAT and electron peak size are described by the same parameters for the power law, while a different constant term was required for each drift voltage.
For this reason, a global fit was performed for all drift voltages with fixed anode voltages.
In this global fit, the parameters $b,w$ are common across all drift voltages, while the constant term $a$ is unique to each drift voltage $(a_1,a_2,a_3,...)$.
The fit is shown with dashed lines.
In Figure \ref{anal-slew600}\,b, the related constant term is subtracted from each mean SAT according to the drift voltage.
The fact that the difference between the drift voltages is only between the constant terms is clear.

\begin{figure}[H]
    \centering
    \begin{subfigure}[h]{0.49\textwidth}
        \includegraphics[width=0.95\textwidth]{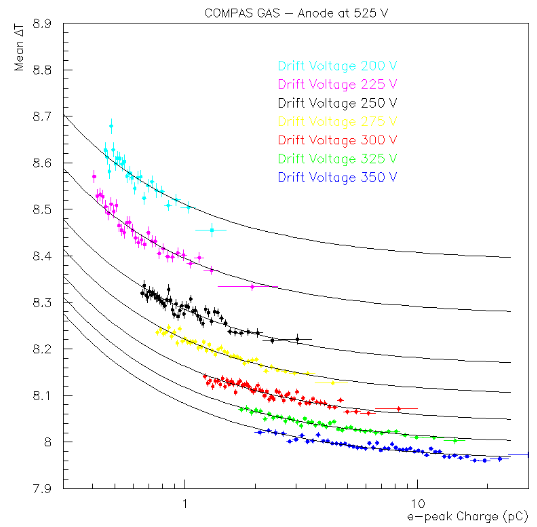}
        \caption{}
    \end{subfigure}
    \begin{subfigure}[h]{0.49\textwidth}
        \includegraphics[width=0.95\textwidth]{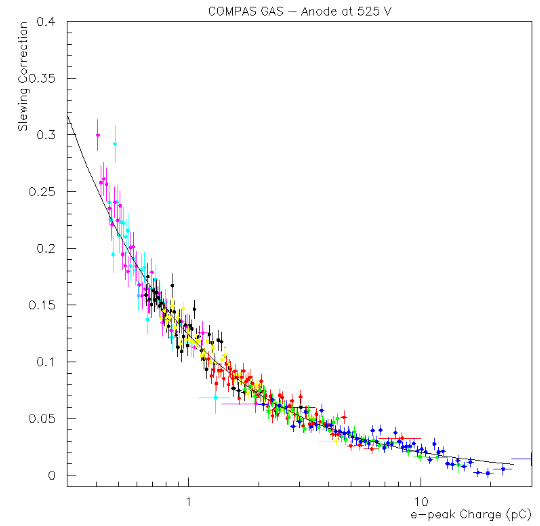}
        \caption{}
    \end{subfigure}
    \caption[Mean SAT as a function of the electron peak charge in COMPASS gas.]{ 
        Mean SAT as a function of the electron peak charge. The data were collected with the $\text{COMPASS}$ gas mixture and anode and drift voltage settings of $525\,V$ and $\text{cyan: }200\,V,$ $\text{magenta: }225\,V,$ $\text{black: }250\,V,$ $\text{yellow: }275\,V,$ $\text{red: }300\,V,$ $\text{green: }325\,V,\text{ and blue: }350\,V$, respectively.
        (a) corresponds to the mean SAT versus the electron peak charge, while the same data are presented in (b) but with the related constant term subtracted for each data set. 
        Dashed lines represent fits with the function of Equation \ref{eq-power}.
    }
    \label{anal-slew525compass}
\end{figure}

Similarly, the same effect is present in the data collected using the COMPASS gas mixture.
In Figure \ref{anal-slew525compass}\,a the mean SAT versus the electron peak charge is shown for an anode voltage of $525\,V$ and drift voltages of $200\,V,$ $225\,V,$ $250\,V,$ $275\,V,$ $300\,V,$ $325\,V,$ $350\,V$, using the COMPASS gas mixture.
In Figure \ref{anal-slew525compass}\,b, the related constant term is subtracted from each mean SAT according to the drift voltage.
The fact that the difference between the drift voltages lies only in the constant terms is clear.
The curves shown in Figures \ref{anal-slew600}\,b and\ref{anal-slew525compass}\,b constitute the time which has to be subtracted from each individual event's SAT, in order to correct for this ``slewing''.

\begin{figure}[H]
    \centering
    \begin{subfigure}[h]{0.49\textwidth}
        \includegraphics[width=0.95\textwidth]{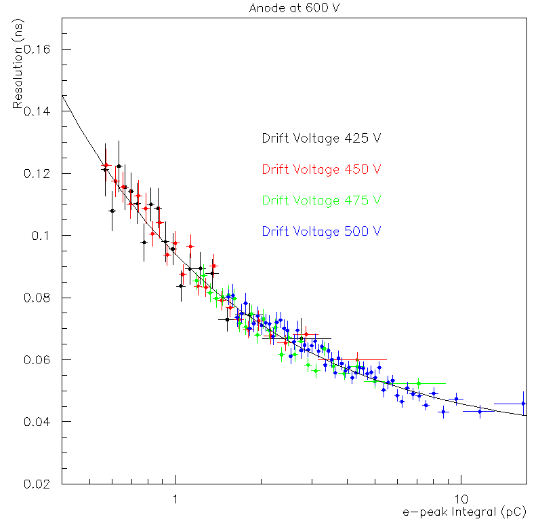}
        \caption{}
    \end{subfigure}
    \begin{subfigure}[h]{0.49\textwidth}
        \includegraphics[width=0.95\textwidth]{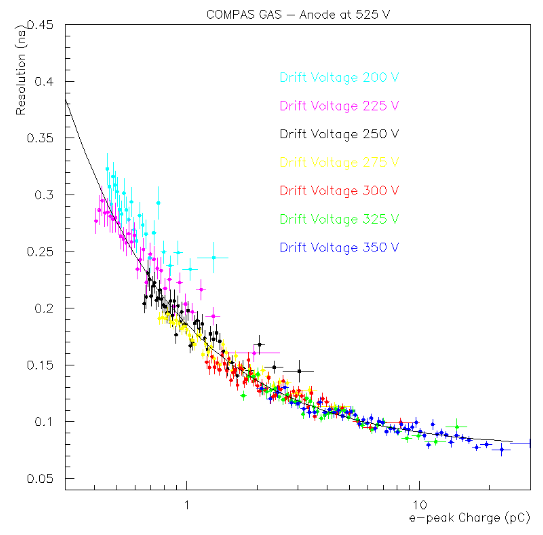}
        \caption{}
    \end{subfigure}
    \caption[Time resolution as a function of the electron peak charge.]{ 
    Time resolution as a function of the electron peak charge. 
The data were collected with the (a) $\text{CF}_4$ and (b) $\text{COMPASS}$ gas mixtures. 
The color codes and voltage settings are identical to ones presented in Figures \ref{anal-slew600} and \ref{anal-slew525compass}, respectively. 
        Solid lines represent fits with the function of Equation \ref{eq-power}.
    }
    \label{anal-rescf4comp}
\end{figure}

The time resolution, not suffering from ``slewing'', is illustrated in Figure \ref{anal-rescf4comp}, 
with (a) corresponding to the data collected with the $\text{CF}_4$ mixture which are shown in Figure \ref{anal-slew600} and (b) corresponding to the data collected with the COMPASS mixture as shown in Figure \ref{anal-slew525compass}. 
The time resolution versus the electron peak charge was found to follow a single curve for a fixed anode voltage, independently of the drift voltage.
Even though the functional form of the time resolution with respect to the electron peak charge is not changing with different drift voltages, this does not mean that the total time resolution is independent of the drift voltage.
Different drift voltages with the same anode voltage obviously gives a different amplification and larger drift voltages correspond to larger electron peaks.
In Figures \ref{anal-rescf4comp} it is clear how measurements at different drift voltages occupy a different region in the curve.
Equation \ref{eq-power} was chosen to parameterize the dependence of the time resolution versus the size of the electron peak.
This time all parameters are common to all drift voltage settings in the global fit, as shown with solid lines in Figures \ref{anal-rescf4comp}.

\begin{figure}[H]
    \centering
    \begin{subfigure}[h]{0.49\textwidth}
        \includegraphics[width=0.95\textwidth]{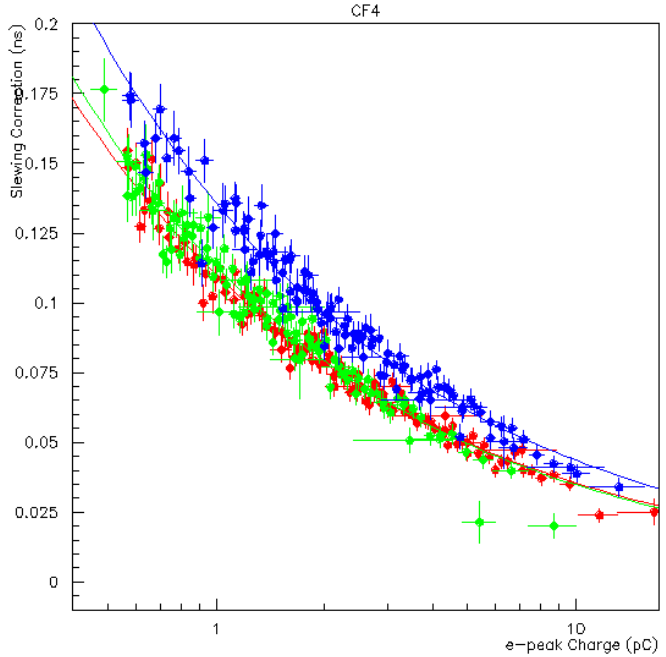}
        \caption{}
    \end{subfigure}
    \begin{subfigure}[h]{0.49\textwidth}
        \includegraphics[width=0.95\textwidth]{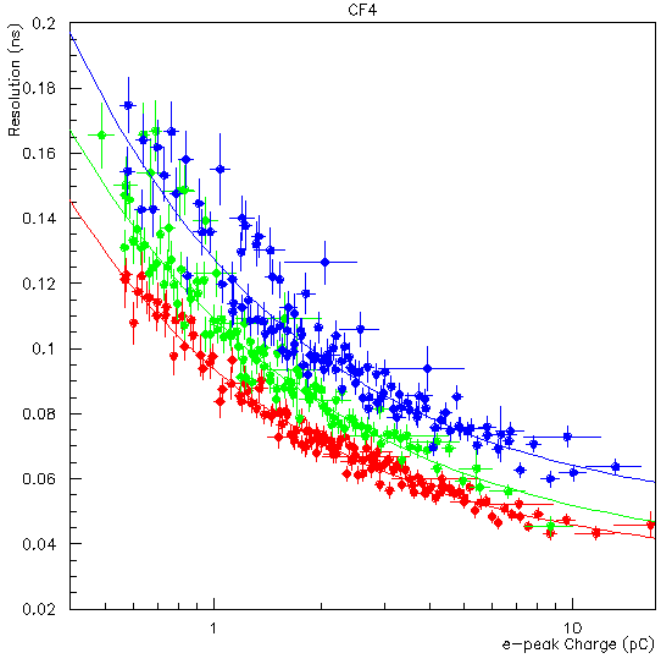}
        \caption{}
    \end{subfigure}
    \begin{subfigure}[h]{0.49\textwidth}
        \includegraphics[width=0.95\textwidth]{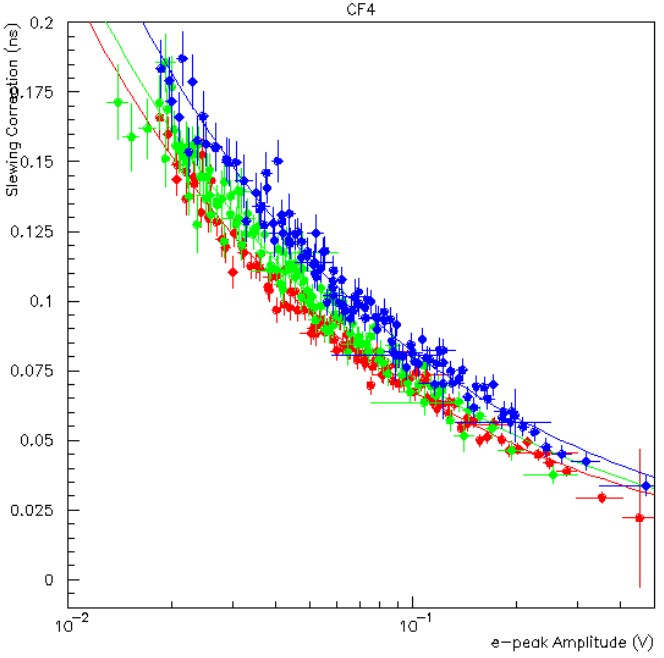}
        \caption{}
    \end{subfigure}
    \begin{subfigure}[h]{0.49\textwidth}
        \includegraphics[width=0.95\textwidth]{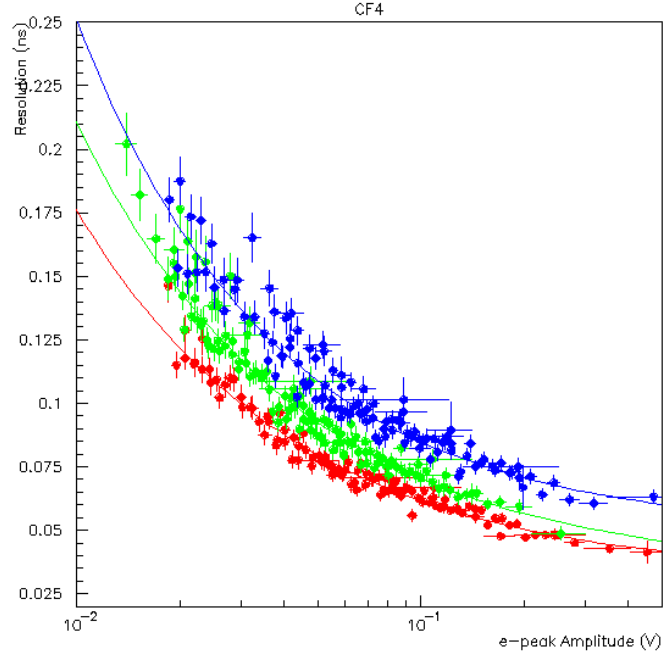}
        \caption{}
    \end{subfigure}
    \caption[Mean SAT correction and time resolution versus electron peak charge and amplitude for CF$_4$ gas.]{ 
        (a) Mean SAT correction versus electron peak charge. (b) Time resolution versus electron peak charge. 
        (c) Mean SAT correction versus electron peak amplitude. (d) Time resolution versus electron peak amplitude. 
        All results correspond to the $\text{CF}_4$ gas mixture with red corresponding to an anode voltage of $600\,V$,
        green to $625\,V$, and blue to $650\,V$.
    }
    \label{totcor-cf4-ch}
\end{figure}

\begin{figure}[H]
    \centering
    \begin{subfigure}[h]{0.49\textwidth}
        \includegraphics[width=0.95\textwidth]{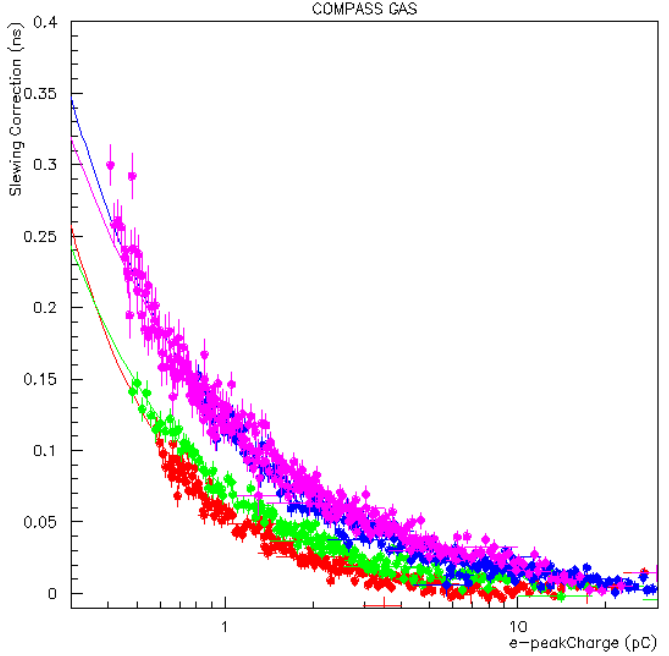}
        \caption{}
    \end{subfigure}
    \begin{subfigure}[h]{0.49\textwidth}
        \includegraphics[width=0.95\textwidth]{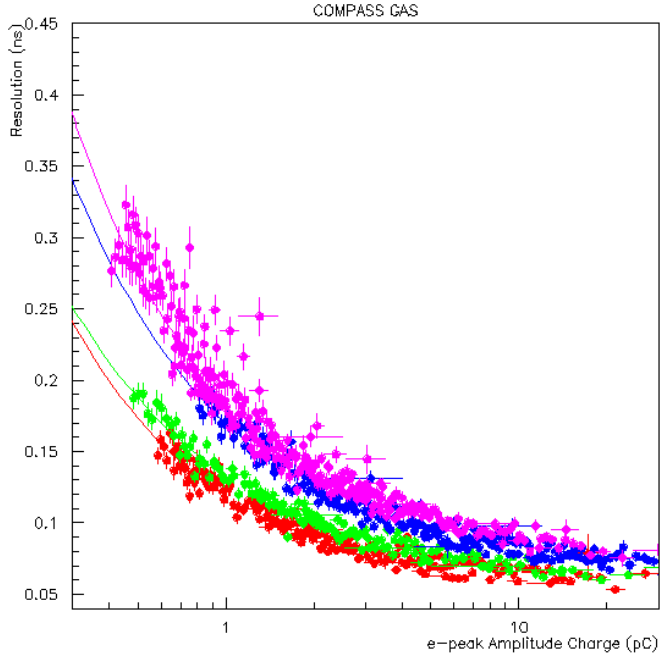}
        \caption{}
    \end{subfigure}
    \begin{subfigure}[h]{0.49\textwidth}
        \includegraphics[width=0.95\textwidth]{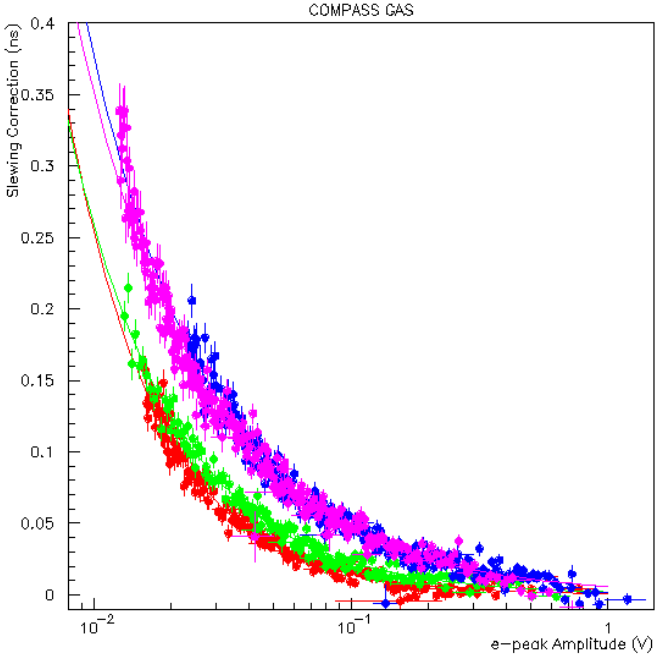}
        \caption{}
    \end{subfigure}
    \begin{subfigure}[h]{0.49\textwidth}
        \includegraphics[width=0.95\textwidth]{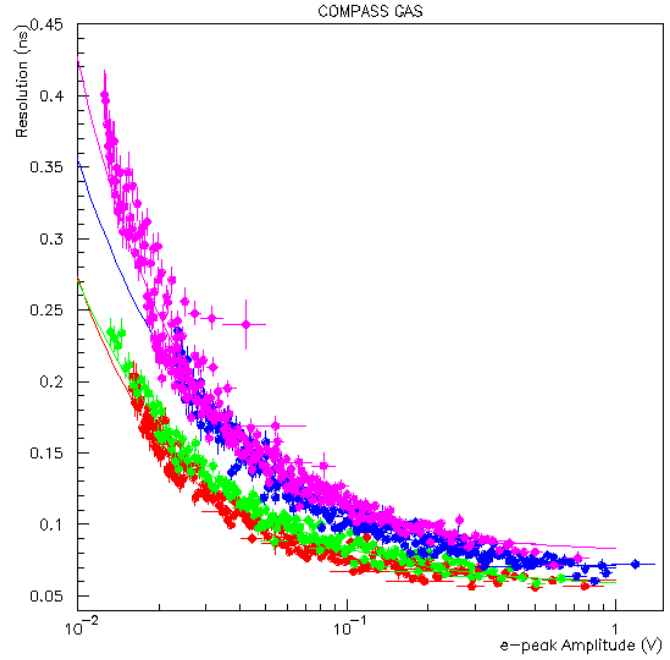}
        \caption{}
    \end{subfigure}
    \caption[Mean SAT correction and time resolution versus electron peak charge and amplitude for COMPASS gas.]{ 
        (a) Mean SAT correction versus electron peak charge. (b) Time resolution versus electron peak charge. 
        (c) Mean SAT correction versus electron peak amplitude. (d) Time resolution versus electron peak amplitude. 
        All results correspond to the COMPASS gas mixture with red corresponding to an anode voltage of $450\,V$,
        green to $475\,V$, blue to $650\,V$ and magenta to $525\,V$.
    }
    \label{totcor-compass-ch}
\end{figure}

It is now the turn for the anode voltage to be investigated.
It has been shown that the mean SAT correction (Slewing Correction) follows a single curve for a fixed anode voltage.
The same has been shown for the time resolution.
For the $\text{CF}_4$ mixture, the mean SAT correction versus the electron peak charge is shown in Figure \ref{totcor-cf4-ch}\,a while the time resoluiton versus the electron peak charge is shown in Figure \ref{totcor-cf4-ch}\,b.
The same is shown versus electron peak amplitude in Figures \ref{totcor-cf4-ch}\,c and d, respectively.
For the COMPASS mixture, the mean SAT correction versus the electron peak charge is shown in Figure \ref{totcor-compass-ch}\,a while the time resoluiton versus the electron peak charge is shown in Figure \ref{totcor-compass-ch}\,b.
The same is shown versus electron peak amplitude in Figures \ref{totcor-compass-ch}\,c and d, respectively.
Different colors correspond to different anode voltages, whilst all drift voltages associated to a specific anode voltage have the same color.

On a first look, it seems like that by reducing the anode voltage, the time resolution gets better.
Altough this sounds very strange, the reason that this happens is that by lowering the anode voltage, the drift voltage can be furtherly increased making the pre-amplification region faster before the detector becomes unstable and begins forming sparks.
Because of this, the pre-amplification region is the dominant source that contributes in the time resolution.

All of the parameterizations are shown in Tables \ref{tab:1}-\ref{tab:6}, where both parameterizations in terms of electron peak charge and amplitude are presented.
Table \ref{tab:1} corresponds to the power law component of the parameterization of the mean SAT for data collected with the CF$_4$ gas.
Table \ref{tab:2} corresponds to the constant component of the parameterization of the mean SAT for data collected with the CF$_4$ gas.
Table \ref{tab:3} corresponds to the parameterization of the time resolution for data collected with the CF$_4$ gas.
Table \ref{tab:4} corresponds to the power law component of the parameterization of the mean SAT for data collected with the COMPASS gas.
Table \ref{tab:5} corresponds to the constant component of the parameterization of the mean SAT for data collected with the COMPASS gas.
Table \ref{tab:6} corresponds to the parameterization of the time resolution for data collected with the CF$_4$ gas.
\newpage
\begin{table}[H]
  \centering
  \footnotesize
  \caption{CF$_4$ Slewing Correction}
  \begin{tabular}{p{1.5cm}|cccc}\hline
      Anode Voltage & \multicolumn{2}{p{3.5cm}}{Electron peak charge parameterization.} & \multicolumn{2}{p{3.5cm}}{Electron peak amplitude parameterization.} \\     \cline{2-5}  
      $[V]$ & b $[ns/pC^w]$  & w & b $[ns/pC^w]$  & w \\ \hline
    600 &0.11012 &0.49206 &0.021513     &     0.49925\\
    625 &0.11308 &0.51168 &0.023235     &     0.49479\\
    650 &0.13580 &0.49375 &0.026131     &     0.49536\\ \hline

  \end{tabular}
  \label{tab:1}
\end{table}

\vspace{-0.2cm}
\begin{table}[H]
    \centering
    \footnotesize
    \caption{CF$_4$ Slewing Constant Term.}
    \begin{tabular}{p{1.5cm}p{1.5cm}|p{3.0cm}p{3.0cm}}\hline
        Anode Voltage $[V]$ & Drift Voltage $[V]$ & a $[ns]$ (in electron peak charge parameterization) & a $[ns]$ (in electron peak amplitude parameterization) \\     \hline  
        600  & 425 & $7.7733 \pm 0.0014 $ & $7.7686 \pm 0.0044 $ \\
             & 450 & $7.7471 \pm 0.0019 $ & $7.7414 \pm 0.0032 $ \\
             & 475 & $7.7176 \pm 0.0006 $ & $7.7112 \pm 0.0023 $ \\
             & 500 & $7.6936 \pm 0.0004 $ & $7.6892 \pm 0.0012 $ \\ \hline
        625  & 350 & $7.9007 \pm 0.0030 $ & $7.8984 \pm 0.0030 $ \\
             & 375 & $7.8677 \pm 0.0034 $ & $7.8615 \pm 0.0044 $ \\
             & 400 & $7.8185 \pm 0.0011 $ & $7.8099 \pm 0.0011 $ \\ 
             & 425 & $7.7869 \pm 0.0008 $ & $7.7759 \pm 0.0010 $ \\
             & 450 & $7.7532 \pm 0.0007 $ & $7.7427 \pm 0.0010 $ \\ \hline
        650  & 350 & $7.9092 \pm 0.0026 $ & $7.9076 \pm 0.0024 $ \\
             & 375 & $7.8677 \pm 0.0017 $ & $7.8621 \pm 0.0017 $ \\
             & 400 & $7.8185 \pm 0.0012 $ & $7.8144 \pm 0.0012 $ \\
             & 425 & $7.7846 \pm 0.0008 $ & $7.7797 \pm 0.0008 $ \\
             & 450 & $7.7536 \pm 0.0007 $ & $7.7490 \pm 0.0007 $ \\ \hline
    \end{tabular}
    \label{tab:2}
\end{table}

\vspace{-0.2cm}
\begin{table}[H]
  \centering
  \footnotesize
  \caption{CF$_4$ Resolution}
  \begin{tabular}{p{1.5cm}|cccccc}\hline
      Anode Voltage & \multicolumn{3}{p{4.5cm}}{Electron peak charge parameterization.} & \multicolumn{3}{p{4.5cm}}{Electron peak amplitude parameterization.} \\     \cline{2-7}  
       $[V]$ & b $[ns/pC^w]$  & w & a $[ns]$ & b $[ns/pC^w]$  & w & a $[ns]$ \\ \hline
      600  & 0.061460  &  0.66176 &0.032447    &    0.0062345  & 0.68148 &0.031709 \\
      625  & 0.075729  &  0.61689 &0.033636    &    0.0076818  & 0.68053 &0.033543 \\
      650  & 0.080632  &  0.67428 &0.047075    &    0.0059885  & 0.76212 &0.05013  \\ \hline
  \end{tabular}
  \label{tab:3}
\end{table}

\vspace{-0.2cm}
\begin{table}[H]
  \centering
  \footnotesize
  \caption{COMPASS Slewing Correction}
  \begin{tabular}{p{1.5cm}|cccc}\hline
      Anode Voltage & \multicolumn{2}{p{3.5cm}}{Electron peak charge parameterization.} & \multicolumn{2}{p{3.5cm}}{Electron peak amplitude parameterization.} \\     \cline{2-5}  
      $[V]$ & b $[ns/pC^w]$  & w & b $[ns/pC^w]$  & w \\ \hline
      450& 0.055703  &  1.2617  &0.00083297    &      1.2422   \\ 
      475& 0.074144  &  0.98339 &0.0020207     &      1.0538   \\
      500& 0.11789   &  0.89270 &0.0056577     &      0.91195  \\
      525& 0.12367   &  0.78459 &0.005936      &      0.88623  \\ \hline
  \end{tabular}
  \label{tab:4}
\end{table}

\begin{table}[H]
    \centering
    \footnotesize
    \caption{COMPASS Slewing Constant Term.}
    \begin{tabular}{p{1.5cm}p{1.5cm}|p{3.0cm}p{3.0cm}}\hline
        Anode Voltage $[V]$ & Drift Voltage $[V]$ & a $[ns]$ (in electron peak charge parameterization) & a $[ns]$ (in electron peak amplitude parameterization) \\     \hline
        450   &  300 &$8.0174 \pm 0.0019 $& $8.0396   \pm   0.0153$ \\
              &  325 &$7.9871 \pm 0.0013 $& $7.9891   \pm   0.0032$ \\
              &  350 &$7.9567 \pm 0.0011 $& $7.9534   \pm   0.0019$ \\
              &  375 &$7.9174 \pm 0.0009 $& $7.9135   \pm   0.0015$ \\
              &  400 &$7.8878 \pm 0.0007 $& $7.8844   \pm   0.0015$ \\
              &  425 &$7.8551 \pm 0.0006 $& $7.8539   \pm   0.0008$ \\ \hline
        475   &  300 &$8.0482 \pm 0.0017 $& $8.0521   \pm   0.0015$ \\
              &  325 &$7.9927 \pm 0.0014 $& $7.9905   \pm   0.0014$ \\
              &  350 &$7.9499 \pm 0.0010 $& $7.9481   \pm   0.0012$ \\
              &  375 &$7.9175 \pm 0.0008 $& $7.9151   \pm   0.0009$ \\
              &  400 &$7.8894 \pm 0.0006 $& $7.8891   \pm   0.0008$ \\ \hline
        500   &  275 &$8.0998 \pm 0.0030 $& $8.0966   \pm   0.0054$ \\
              &  300 &$8.0529 \pm 0.0030 $& $8.0472   \pm   0.0033$ \\
              &  325 &$7.9956 \pm 0.0025 $& $7.9874   \pm   0.0045$ \\
              &  350 &$7.9611 \pm 0.0022 $& $7.9559   \pm   0.0040$ \\
              &  375 &$7.9212 \pm 0.0015 $& $7.9182   \pm   0.0024$ \\
              &  400 &$7.8866 \pm 0.0014 $& $7.8840   \pm   0.0015$ \\ \hline
        525   &  200 &$8.3870 \pm 0.0043 $& $8.4182   \pm   0.0068$ \\
              &  225 &$8.2708 \pm 0.0025 $& $8.2845   \pm   0.0043$ \\
              &  250 &$8.1609 \pm 0.0030 $& $8.1640   \pm   0.0031$ \\
              &  275 &$8.0969 \pm 0.0025 $& $8.0934   \pm   0.0020$ \\
              &  300 &$8.0389 \pm 0.0023 $& $8.0390   \pm   0.0018$ \\
              &  325 &$7.9938 \pm 0.0020 $& $7.9945   \pm   0.0015$ \\
              &  350 &$7.9583 \pm 0.0017 $& $7.9593   \pm   0.0010$ \\ \hline
    \end{tabular}
    \label{tab:5}
\end{table}

\begin{table}[H]
  \centering
  \footnotesize
  \caption{COMPASS Resolution}
  \begin{tabular}{p{1.5cm}|cccccc}\hline
      Anode Voltage & \multicolumn{3}{p{4.5cm}}{Electron peak charge parameterization.} & \multicolumn{3}{p{4.5cm}}{Electron peak amplitude parameterization.} \\     \cline{2-7}  
       $[V]$ & b $[ns/pC^w]$  & w & a $[ns]$ & b $[ns/pC^w]$  & w & a $[ns]$ \\ \hline
      450 &  0.063099  &  0.89335 &0.056088   & 0.0013546  &     1.0977  &0.05942  \\
      475 &  0.076703  &  0.77578 &0.056345   & 0.0037606  &     0.87790 &0.0555   \\
      500 &  0.10352   &  0.81002 &0.06586    & 0.0056683  &     0.85500 &0.063736  \\
      525 &  0.11017   &  0.85942 &0.075582   & 0.0027259  &     1.0500  &0.080534  \\ \hline
  \end{tabular}
  \label{tab:6}
\end{table}

\section{Total Time Resolution} \label{sec:tot}

The total resolution is not trivial in its estimation.
The standard procedure would be to simply correct all events for their ``time walk'', and compute the standard deviation of the distribution of SAT.
However, there exists a bias that prevents this from being possible.
This bias is because the PICOSEC participates in the trigger setup and there is a threshold on the amplitude of the signal.
Because of this threshold, the spectrum of the electron peak amplitudes is truncated and, indirectly, the spectrum of the electron peak charge looks as if truncated as well.
For this reason, the time resolution has to be extrapolated to an unobserved region by using the parameterization of the electron peak charge size.

Suppose that the SAT $\Delta T$ follows a Gaussian distribution $G(\Delta T)$ for a fixed electron peak size $Q$, with a mean value given by the mean SAT correction $W(Q)$ and the time resolution $R(Q)$ which are both functions of $Q$, and the spectrum of the electron peak size is given by the distribution $A(Q)$.

\begin{equation}
    G\left(\Delta T; W(Q),R(Q) \right)
    =
    \frac{1}{\sqrt{2\pi} R(Q)}
    \exp \left(
        -\frac{(\Delta T - W(Q)^2}{2R^2(Q)}
    \right)
\end{equation}

In a region of electron peak size $[Q_1,Q_N]$, the distribution of SAT is then $f(\Delta T)$:

\begin{equation}
    f(\Delta T) = 
    \int_{Q_1}^{Q_N}
    A(Q)\cdot G\left(\Delta T; W(Q), R(Q)\right)
    dQ
\end{equation}

This integral can be approximated by:

\begin{equation}
    f(\Delta T) \approx 
    \sum_{i=1}^{N}
    A(Q_i)\cdot G\left(\Delta T; W(Q_i), R(Q_i)\right)
    \Delta Q
\end{equation}

or

\begin{equation}
    f(\Delta T) \approx 
    \sum_{i=1}^{N}
    a_i\cdot g_i\left(\Delta T\right)
\end{equation}

where $N$ is a large number, $g_i(\Delta T) = G\left(\Delta T; W(Q_i), R(Q_i)\right)$, $V_i = W^2(Q_i)$, and $a_i = A(Q_i)\cdot \Delta Q$ such that $\sum_{i=1}^N a_i = 1$. 

The mean $E[\Delta T]$ and the variance $V[\Delta T]$ values of this distribution are equal to:

\begin{equation}
    E[\Delta T] = \sum_{i=1}^N a_i \cdot W(Q_i) 
\end{equation}
\begin{equation}
    V[\Delta T] = \sum_{i=1}^N a_i^2 \cdot V_i + \sum_{i=1}^N a_i a_j \left(V_i+V_j + (\mu_i-\mu_j)^2\right) 
\end{equation}

\begin{figure}[H]
    \centering
    \begin{subfigure}[h]{0.49\textwidth}
        \includegraphics[width=0.95\textwidth]{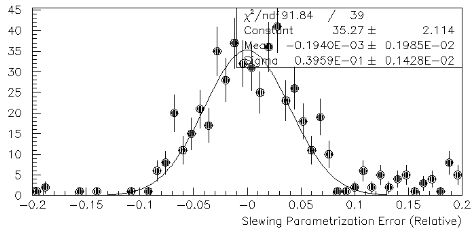}
        \caption{}
    \end{subfigure}
    \begin{subfigure}[h]{0.49\textwidth}
        \includegraphics[width=0.95\textwidth]{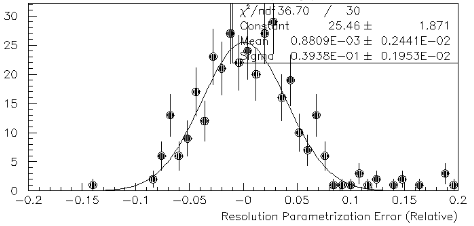}
        \caption{}
    \end{subfigure}
    \caption[Distributions of relative errors for the CF$_4$ gas.]{ 
        Distribution of relative errors for all data points of the CF$_4$ gas mixture of the (a) mean SAT correction and (b) time resolution.
        Solid lines represent Gaussian fits.
    }
    \label{errors-cf4}
\end{figure}

\begin{figure}[H]
    \centering
    \begin{subfigure}[h]{0.49\textwidth}
        \includegraphics[width=0.95\textwidth]{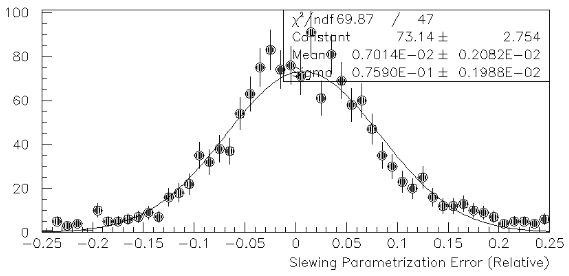}
        \caption{}
    \end{subfigure}
    \begin{subfigure}[h]{0.49\textwidth}
        \includegraphics[width=0.95\textwidth]{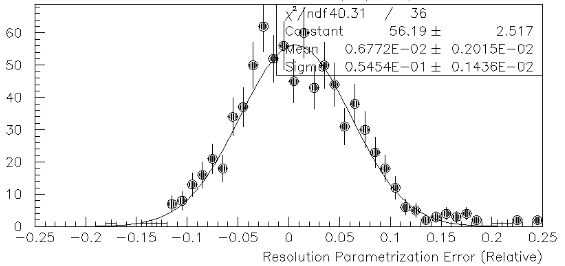}
        \caption{}
    \end{subfigure}
    \caption[Distributions of relative errors for the COMPASS gas.]{ 
        Distribution of relative errors for all data points of the COMPASS gas mixture of the (a) mean SAT correction and (b) time resolution.
        Solid lines represent Gaussian fits.
    }
    \label{errors-comp}
\end{figure}

By applying error propagation on this formula, one gets:

\begin{equation}
    \delta_{V[\Delta T]} = 
    \sqrt{
        \sum_{i=1}^N a_i\cdot \delta_i^2 + 4 \sum_{i=1}^N a_i^2 \cdot \left( \mu_i - \sum_{j=1}^N a_j \cdot\mu_j  \right)^2 w_i^2
    }
\end{equation}

where $\delta_i$ is the error with which the function $W(Q)$ is known and $w_i$ is the error with which the function $W(Q)$ is known.
Because all the estimated variables are highly correlated with each other, unless the full covariance matrix was taken into account, the errors would have been misleading.
To avoid using the full covariance matrix, the relative error between the data points and the parameterizations is found to express $\delta_i$ and $w_i$.
These relative errors, i.e. the difference between parameterization and data points, divided by the parameterization, are histogrammed in Figures \ref{errors-cf4} for the $\text{CF}_4$ data, where (a) corresponds to the mean SAT correction error and (b) corresponds to the resolution error.
The same applies for Figures \ref{errors-comp} which are the corresponding distributions for the COMPASS data.
Solid lines represent Gaussian Fits.
The mean of all the Gaussians are consistent with zero.
This is logical as all parameterizations are performed by minimizing the difference between the fit function and the data points.
The standard deviations of the Gaussian fits are used to express the $\delta_i$ and $w_i$ errors.
If $\sigma$ is the standard deviation of the Gaussian fit in Figure \ref{errors-cf4}\,a, then $d_i$ equals to:

\begin{equation}
    \delta_i = \sigma\cdot W(Q_i)
\end{equation}

In summary, for the CF$_4$ data:

\begin{equation}
    \delta_i = 0.03959\cdot W(Q_i) 
\end{equation}
\begin{equation}
    w_i = 0.03938\cdot R(Q_i) 
\end{equation}

While for the COMPASS data:

\begin{equation}
    \delta_i = 0.07590\cdot W(Q_i) 
\end{equation}
\begin{equation}
    w_i = 0.05454\cdot R(Q_i) 
\end{equation}

When the electron peak size $Q_i$ is less than the threshold of the corresponding dataset and thus belongs in the unobserved region, the above errors are doubled.
Because there is no way to observe the spectrum in that region, this conservative choice is made for the uncertainties.

In order to take into account the uncertainty of the electron peak size spectrum, many P\'olya fits are performed on each voltage settings, each time with different fit range or bin sizes.
Thus, several values are obtained for the parameters of the electron peak size spectrum.
Then for each P\'olya parameterization, the total time resolution and its error is computed.
Of the total time resolutions with their errors, their weighted average and standard deviation is evaluated to express finally the total time resolution with its uncertainty.

The final results are compiled in Tables \ref{tab:7}-\ref{tab:10}, where the total time resolution, both with and without the slewing correction, for a) the observed region, b) the prediction of the statistical technique in the observed region and c) the prediction of the statistical technique in the unobserved region.
Table \ref{tab:7} corresponds to the total time resolution achieved using the CF$_4$ gas and a parameterization in terms of the electron peak amplitude. 
Table \ref{tab:8} corresponds to the total time resolution achieved using the CF$_4$ gas and a parameterization in terms of the electron peak charge. 
Table \ref{tab:9} corresponds to the total time resolution achieved using the COMPASS gas and a parameterization in terms of the electron peak amplitude. 
Table \ref{tab:10} corresponds to the total time resolution achieved using the COMPASS gas and a parameterization in terms of the electron peak charge. 

\begin{figure}[H]
    \centering
    \begin{subfigure}[h]{0.49\textwidth}
        \includegraphics[width=0.95\textwidth]{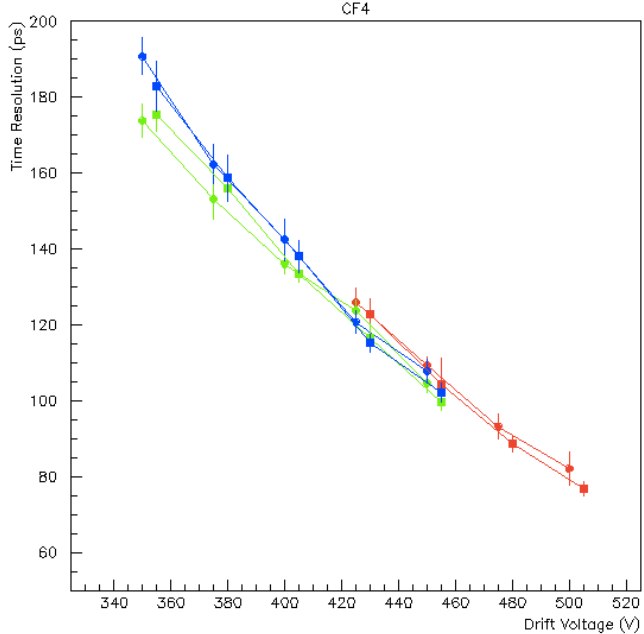}
        \caption{}
    \end{subfigure}
    \begin{subfigure}[h]{0.49\textwidth}
        \includegraphics[width=0.95\textwidth]{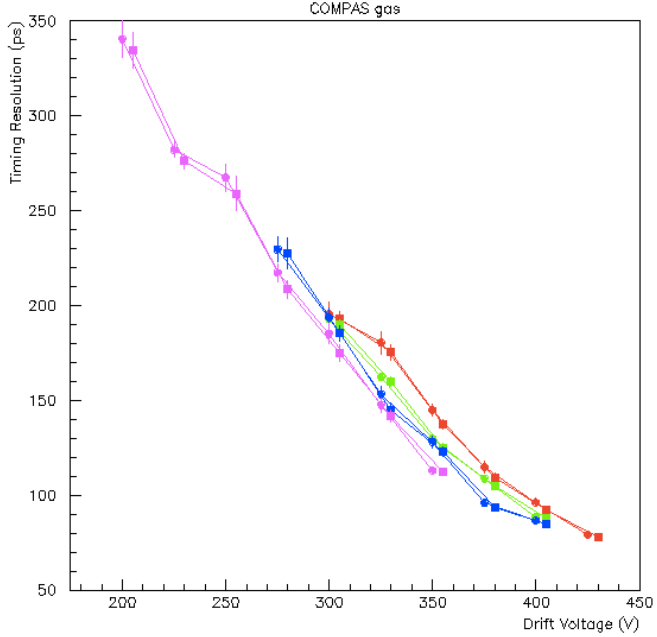}
        \caption{}
    \end{subfigure}
    \caption[Total time resolution as a function of the drift voltage.]{ 
        Total time resolution as a function of the applied drift voltage.
        (a) concerns data collected with the CF$_4$ gas mixture and anode voltages of (red) $600\,V$, (green) $625\,V$ and (blue) $650\,V$.
        (b) concerns data collected with the COMPASS gas and anode voltages of (red) $450\,V$, (green) $475\,V$, (blue) $500\,V$ and (magenta) $525\,V$.
        Circles correspond to the total time resolution without including the mean SAT correction while the correction is included for points denoted with squares.
        The square points are artificially shifted $10\,V$ horizontally to the right for better visual clarity.
    }
    \label{tot_res}
\end{figure}

Figure \ref{tot_res} presents the final prediction of the total time resolution, extrapolated to the unobserved region of electron peaks.
Figure \ref{tot_res}\,a corresponds to data collected with the CF$_4$ gas filling while b corresponds to data collected with the COMPASS gas.
Interestingly, the total time resolution seems almost independent of the anode voltage.
The best time resolution is reached with the CF$_4$ gas mixture, anode voltage of $600\,V$ and drift voltage of $500\,V$.
It is clear that by increasing the drift voltage the time resolution gets better.
Based on this, it is quite likely that the drift voltage should be furtherly increased even at the cost of a reduced anode voltage\footnote{This is in fact what happened in a later Testbeam campaign. Results of this are shown in Chapter \ref{chap:muons}}.

\begin{landscape}
\newcolumntype{L}{>{\centering\arraybackslash}m{2cm}}
\newcolumntype{R}{>{\centering\arraybackslash}m{1.5cm}}
\begin{table}[H]
    \centering
    \footnotesize
    \caption{CF$_4$ Total time resolution based on electron peak amplitude.}
    \begin{tabular}{RR|LLLLLLL}\hline
        & & \multicolumn{3}{c}{Experimental measurements} & \multicolumn{2}{c}{Convolution up to threshold} & \multicolumn{2}{p{5.0cm}}{Predicted resolution (electron peak amplitude $>0.01\,V$) }\\
\hline
Anode Voltage $[V]$ & Drift Voltage $[V]$ & Analysis Threshold $[V]$ & Raw resolution $[ns]$ & Slew. Corr. Resolution $[ns]$ & 
 Raw resolution $[ns]$ & Slew. Corr. Resolution $[ns]$ & Raw resolution $[ns]$ & Slew. Corr. Resolution $[ns]$ \\ \hline 
        600 &425 &0.018 &  111.0 $\pm$ 3.2   & 107.3  $\pm$ 1.7   & 110.2   $\pm$  1.2  &  108.3   $\pm$ 0.5  &  134.7  $\pm$ 2.3 &   130.4   $\pm$ 1.0 \\ 
            &450 &0.018 &  \ 96.8  $\pm$  4.9  &\  94.8  $\pm$  1.2  &\  97.9   $\pm$  1.2  & \ 94.6    $\pm$ 0.5  &  115.5  $\pm$ 1.5 &   109.3   $\pm$ 0.7 \\ 
            &475 &0.034 &  \ 74.8  $\pm$  1.0  &\  73.2  $\pm$  2.4  &\  75.2   $\pm$  0.7  &\  73.1    $\pm$ 0.3  &\  98.0   $\pm$ 1.5 & \  91.0    $\pm$ 0.9 \\ 
            &500 &0.045 &  \ 67.8  $\pm$  0.6  &\  65.5  $\pm$  1.2  &\  67.7   $\pm$  0.7  &\  65.7    $\pm$ 0.3  &\  86.9   $\pm$ 1.6 & \  80.0    $\pm$ 1.0 \\ \hline 
        625 &350 &0.017 &  153.1 $\pm$  4.3  &  153.4 $\pm$  4.0  &  153.3  $\pm$  1.0  &  152.3   $\pm$ 0.7  &  192.5  $\pm$ 2.2 &   190.0   $\pm$ 1.5 \\ 
            &375 &0.014 &  150.8 $\pm$  6.3  &  150.0 $\pm$  6.0  &  153.6  $\pm$  0.7  &  151.5   $\pm$ 0.7  &  174.3  $\pm$ 1.4 &   171.1   $\pm$ 1.2 \\ 
            &400 &0.017 &  121.8 $\pm$  1.8  &  121.8 $\pm$  1.8  &  123.6  $\pm$  1.2  &  121.1   $\pm$ 0.7  &  148.1  $\pm$ 2.4 &   143.2   $\pm$ 1.8 \\ 
            &425 &0.018 &  108.2 $\pm$  1.4  &  106.6 $\pm$  1.1  &  111.8  $\pm$  1.2  &  108.5   $\pm$ 0.7  &  131.7  $\pm$ 2.4 &   125.5   $\pm$ 1.0 \\ 
            &450 &0.035 & \ 82.4  $\pm$  1.9  & \ 81.6  $\pm$  1.4  & \ 84.9   $\pm$  1.2  & \ 82.8    $\pm$ 0.5  &  112.8  $\pm$ 2.2 &   105.8   $\pm$ 1.0 \\ \hline
        650 &350 &0.017 & 158.7  $\pm$  5.0  &  158.4 $\pm$  5.0  &  164.0  $\pm$  2.0  &  161.9   $\pm$ 1.0  &  202.3  $\pm$ 2.2 &   198.1   $\pm$ 2.4 \\ 
            &375 &0.020  &  135.5 $\pm$  3.0  &  135.3 $\pm$  5.0  &  139.0  $\pm$  2.0  &  138.6   $\pm$ 1.0  &  177.9  $\pm$ 2.3 &   172.4   $\pm$ 2.3 \\ 
            &400 &0.035 & 105.4  $\pm$  2.5  &  105.4 $\pm$  1.0  &  108.6  $\pm$  0.6  &  107.0   $\pm$ 0.5  &  150.3  $\pm$ 2.7 &   143.5   $\pm$ 2.3 \\ 
            &425 &0.035 &\ 95.0   $\pm$  2.0  & \ 94.2  $\pm$  1.8  &\  98.9   $\pm$  0.7  &\  96.5    $\pm$ 0.5  &  130.1  $\pm$ 2.0 &   122.9   $\pm$ 1.5 \\ 
            &450 &0.050  & \ 83.7  $\pm$  1.5  & \ 82.7  $\pm$  1.2  &\  85.6   $\pm$  1.0  &\  83.4    $\pm$ 0.5  &  113.8  $\pm$ 3.0 &   105.4   $\pm$ 1.4 \\ \hline
    \end{tabular}
    \label{tab:7}
\end{table}

\newcolumntype{L}{>{\centering\arraybackslash}m{2cm}}
\newcolumntype{R}{>{\centering\arraybackslash}m{1.5cm}}
\begin{table}[H]
    \centering
    \footnotesize
    \caption{CF$_4$ Total time resolution based on electron peak charge.}
    \begin{tabular}{RR|LLLLLLL}\hline
        & & \multicolumn{3}{c}{Experimental measurements} & \multicolumn{2}{c}{Convolution up to threshold} & \multicolumn{2}{p{5.0cm}}{Predicted resolution (electron peak charge $>0.333\,pC$) }\\
        \hline
        Anode Voltage $[V]$ & Drift Voltage $[V]$ & Analysis Threshold $[pC]$& Raw resolution $[ns]$ & Slew. Corr. Resolution $[ns]$ & 
        Raw resolution $[ns]$ & Slew. Corr. Resolution $[ns]$ & Raw resolution $[ns]$ & Slew. Corr. Resolution $[ns]$ \\ \hline 
        600 & 425 &0.55  &  107.5   $\pm$2.5 &   106.0     $\pm$3.0 & 107.3   $\pm$0.7 &   105.5 $\pm$0.5  &  125.7  $\pm$ 1.5 &   122.2   $\pm$1.3 \\ 
            & 450 &0.55  &  95.9    $\pm$1.3 &   93.0     $\pm$6.0 & 97.1    $\pm$0.7 &   94.2   $\pm$0.4  &  110.7  $\pm$ 1.0 &   105.7   $\pm$0.5 \\ 
            & 475 &1.15  &  72.0      $\pm$1.8 &   71.5    $\pm$1.2 & 71.9    $\pm$0.5 &   70.3  $\pm$0.3  &  93.1   $\pm$ 1.1 &   87.2    $\pm$0.5 \\ 
            & 500 &1.50   &  64.5    $\pm$3.0 &   63.5    $\pm$1.2 & 64.1    $\pm$0.5 &   62.4    $\pm$0.2  &  81.6   $\pm$ 1.1 &   75.5    $\pm$0.5 \\ \hline 
        625 & 350 &0.55  &  144.8   $\pm$2.5 &   146.8   $\pm$2.5 & 148.7   $\pm$0.6 &   147.8   $\pm$0.5  &  178.5  $\pm$ 1.1 &   176.5   $\pm$1.0 \\ 
            & 375 &0.34  &  152.0     $\pm$5.0 &   154.9   $\pm$1.0 & 157.9   $\pm$1.0 &   155.0 $\pm$0.8  &  159.0  $\pm$ 1.0 &   156.0   $\pm$0.9 \\ 
            & 400 &0.55  &  117.7   $\pm$1.0 &   116.9   $\pm$1.0 & 119.8   $\pm$0.8 &   117.7   $\pm$0.5  &  138.5  $\pm$ 1.2 &   134.4   $\pm$0.8 \\ 
            & 425 &0.55  &  109.5   $\pm$2.0 &   104.8   $\pm$3.0 & 108.8   $\pm$0.8 &   105.7   $\pm$0.5  &  123.0  $\pm$ 1.0 &   117.7   $\pm$0.6 \\ 
            & 450 &1.10   &  82.4    $\pm$0.5 &   81.3    $\pm$0.5 & 83.1    $\pm$0.6 &   81.3    $\pm$0.3  &  105.5  $\pm$ 1.2 &   99.5    $\pm$0.5 \\ \hline
        650 & 350 &0.55  &  161.7   $\pm$3.0 &   156.4   $\pm$5.0 & 157.3   $\pm$0.8 &   155.3   $\pm$0.7  &  185.6  $\pm$ 1.7 &   181.7   $\pm$1.6 \\ 
            & 375 &0.55  &  140.7   $\pm$4.0 &   139.3   $\pm$5.0 & 143.3   $\pm$0.7 &   140.5   $\pm$0.5  &  165.3  $\pm$ 1.0 &   160.1   $\pm$0.8 \\ 
            & 400 &1.10   &  107.0     $\pm$3.0 &   107.0     $\pm$2.0 & 107.1   $\pm$0.5&105.6   $\pm$0.4  &  142.6  $\pm$ 1.4 &   136.3   $\pm$1.0 \\ 
            & 425 &1.10   &  94.7    $\pm$1.1 &   94.0      $\pm$1.1 & 96.3    $\pm$0.6 &   94.0  $\pm$0.3  &  122.7  $\pm$ 1.2 &   115.2   $\pm$0.7 \\ 
            & 450 &1.50   &  84.2    $\pm$0.7 &   83.2    $\pm$1.2 & 84.9    $\pm$0.7 &   82.3    $\pm$0.3  &  108.7  $\pm$ 1.6 &   101.0   $\pm$1.0 \\ \hline
    \end{tabular}
    \label{tab:8}
\end{table}

\newcolumntype{L}{>{\centering\arraybackslash}m{2cm}}
\newcolumntype{K}{>{\centering\arraybackslash}m{2.5cm}}
\newcolumntype{R}{>{\centering\arraybackslash}m{1.5cm}}
\begin{table}[H]
    \centering
    \footnotesize
    \caption{COMPASS Total time resolution based on electron peak amplitude.}
    \begin{tabular}{RR|LLLLLKK}\hline
        & & \multicolumn{3}{c}{Experimental measurements} & \multicolumn{2}{c}{Convolution up to threshold} & \multicolumn{2}{p{5.0cm}}{Predicted resolution (electron peak amplitude $>0.01\,V$) }\\
\hline
Anode Voltage $[V]$ & Drift Voltage $[V]$ & Analysis Threshold $[V]$ & Raw resolution $[ns]$ & Slew. Corr. Resolution $[ns]$ & 
 Raw resolution $[ns]$ & Slew. Corr. Resolution $[ns]$ & Raw resolution $[ns]$ & Slew. Corr. Resolution $[ns]$ \\ \hline 
        450& 300 &0.016 &  166.2  $\pm$ 6.0  &  165.2  $\pm$ 6.0  &  159.3  $\pm$ 1.5  &  158.3  $\pm$ 1.1  &  216.0  $\pm$ 5.0   & 211.5  $\pm$ 3.0    \\ 
           & 325 &0.018 &  144.2  $\pm$ 3.1  &  143.6  $\pm$ 2.5  &  144.2  $\pm$ 0.7  &  142.1  $\pm$ 0.5  &  190.8  $\pm$ 2.2   & 183.0  $\pm$ 1.6  \\ 
           & 350 &0.020 &  120.7  $\pm$ 1.7  &  119.5  $\pm$ 1.4  &  123.2  $\pm$ 0.7  &  121.0  $\pm$ 0.5  &  160.2  $\pm$ 3.0   & 151.0  $\pm$ 1.5  \\ 
           & 375 &0.035 &   88.6  $\pm$ 0.9  &  88.8   $\pm$ 0.7  &  89.5   $\pm$ 0.5  &  88.7   $\pm$ 0.5  &  122.4  $\pm$ 12.5  & 114.3  $\pm$ 9.5  \\ 
           & 400 &0.050 &   77.1  $\pm$ 0.4  &  78.0   $\pm$ 0.7  &  77.6   $\pm$ 0.4  &  77.2   $\pm$ 0.3  &  97.4   $\pm$ 7.7   & 92.3   $\pm$ 5.6  \\ 
           & 425 &0.100 &   70.8  $\pm$ 0.9  &  70.8   $\pm$ 0.6  &  69.3   $\pm$ 0.3  &  69.2   $\pm$ 0.3  &  78.7   $\pm$ 3.7   & 76.7   $\pm$ 2.7  \\ \hline 
        475& 300 &0.0125&  184.8  $\pm$ 2.4  &  184.3  $\pm$ 2.3  &  194.1  $\pm$ 1.0  &  190.0  $\pm$ 0.5  &  211.8  $\pm$ 7.0   & 205.1  $\pm$ 6.0    \\ 
           & 325 &0.017 &  139.0  $\pm$ 3.3  &  139    $\pm$ 1.4  &  145.0  $\pm$ 0.6  &  141.6  $\pm$ 0.5  &  171.6  $\pm$ 10.0  & 163.0  $\pm$ 8.0    \\ 
           & 350 &0.034 &  100.6  $\pm$ 0.6  &  100.1  $\pm$ 0.6  &  101.1  $\pm$ 0.5  &  100.0  $\pm$ 0.4  &  136.4  $\pm$ 13.3  & 127.8  $\pm$ 10.2 \\ 
           & 375 &0.035 &   90.0  $\pm$ 0.7  &  91.0   $\pm$ 0.7  &  92.0   $\pm$ 0.5  &  90.7   $\pm$ 0.3  &  112.5  $\pm$ 7.9   & 105.9  $\pm$ 0.6  \\ 
           & 400 &0.050 &   79.1  $\pm$ 0.5  &  80.5   $\pm$ 0.5  &  79.4   $\pm$ 0.4  &  78.7   $\pm$ 0.2  &  91.1   $\pm$ 4.7   & 87.4   $\pm$ 3.3  \\ \hline
        500& 275 &0.022 &  179.3  $\pm$ 3.1  &  180.7  $\pm$ 1.2  &  183.5  $\pm$ 1.0  &  180.8  $\pm$ 0.8  &  242.7  $\pm$ 22.2  & 231.4  $\pm$ 18.8 \\ 
           & 300 &0.025 &  151.9  $\pm$ 4.1  &  153.2  $\pm$ 1.3  &  158.6  $\pm$ 0.7  &  155.3  $\pm$ 0.8  &  207.9  $\pm$ 18.4  & 194.7  $\pm$ 14.6 \\ 
           & 325 &0.034 &  117.6  $\pm$ 3.5  &  118.8  $\pm$ 0.8  &  121.8  $\pm$ 0.7  &  119    $\pm$ 0.5  &  164.8  $\pm$ 16.3  & 150.6  $\pm$ 11.7 \\ 
           & 350 &0.050 &   99.5  $\pm$ 1.5  &  100.6  $\pm$ 0.5  &  102.0  $\pm$ 0.5  &  100.3  $\pm$ 0.3  &  132.3  $\pm$ 11.6  & 121.6  $\pm$ 8.0    \\ 
           & 375 &0.100 &   87.5  $\pm$ 0.5  &  88.2   $\pm$ 0.8  &  86.7   $\pm$ 0.4  &  86.2   $\pm$ 0.3  &  101.0  $\pm$ 5.6   & 96.6   $\pm$ 4.0    \\ 
           & 400 &0.120 &   80.1  $\pm$ 0.5  &  80.8   $\pm$ 1.2  &  79.6   $\pm$ 0.3  &  79.1   $\pm$ 0.3  &  87.4   $\pm$ 3.2   & 84.7   $\pm$ 2.2  \\ \hline
        525& 200 &0.012 &  330.3  $\pm$ 3.5  &  326.1  $\pm$ 3.5  &  325.2  $\pm$ 2.1  &  321.9  $\pm$ 2.0  &  358.8  $\pm$ 13.8  & 354.1  $\pm$ 13.3 \\ 
           & 225 &0.013 &  279.8  $\pm$ 3.7  &  281.8  $\pm$ 8.8  &  282.1  $\pm$ 1.1  &  278.0  $\pm$ 1.1  &  321.3  $\pm$ 15.2  & 314.7  $\pm$ 14.2 \\ 
           & 250 &0.018 &  218.1  $\pm$ 3.6  &  202.8  $\pm$ 4.5  &  211.6  $\pm$ 1.0  &  208.8  $\pm$ 1.0  &  278.1  $\pm$ 24.7  & 269.0  $\pm$ 22.6 \\ 
           & 275 &0.018 &  175.2  $\pm$ 3.5  &  176.3  $\pm$ 3.3  &  185.5  $\pm$ 1.0  &  180.2  $\pm$ 0.7  &  234.4  $\pm$ 18.4  & 222.4  $\pm$ 15.7 \\ 
           & 300 &0.035 &  133.3  $\pm$ 1.5  &  133.7  $\pm$ 1.0  &  135.8  $\pm$ 0.8  &  133.8  $\pm$ 0.8  &  194.5  $\pm$ 21.7  & 182.8  $\pm$ 18.0   \\ 
           & 325 &0.048 &  111.7  $\pm$ 1.1  &  112.5  $\pm$ 1.0  &  113.8  $\pm$ 0.6  &  112.2  $\pm$ 0.4  &  157.0  $\pm$ 16.0  & 146.3  $\pm$ 12.7 \\ 
           & 350 &0.050 &  100.9  $\pm$ 0.8  &  102.3  $\pm$ 0.7  &  101.9  $\pm$ 0.5  &  100.5  $\pm$ 0.5  &  118.2  $\pm$ 6.4   & 112.2  $\pm$ 4.5  \\ \hline
    \end{tabular}
    \label{tab:9}
\end{table}

\newcolumntype{L}{>{\centering\arraybackslash}m{2cm}}
\newcolumntype{R}{>{\centering\arraybackslash}m{1.5cm}}
\begin{table}[H]
    \centering
    \footnotesize
    \caption{COMPASS Total time resolution based on electron peak charge.}
    \begin{tabular}{RR|LLLLLLL}\hline
        & & \multicolumn{3}{c}{Experimental measurements} & \multicolumn{2}{c}{Convolution up to threshold} & \multicolumn{2}{p{5.0cm}}{Predicted resolution (electron peak charge $>0.333\,pC$) }\\
        \hline
        Anode Voltage $[V]$ & Drift Voltage $[V]$ & Analysis Threshold $[pC]$& Raw resolution $[ns]$ & Slew. Corr. Resolution $[ns]$ & 
        Raw resolution $[ns]$ & Slew. Corr. Resolution $[ns]$ & Raw resolution $[ns]$ & Slew. Corr. Resolution $[ns]$ \\ \hline 
        450 &   300 &   0.50 &   164.0  $\pm$ 4.2    &   165.0  $\pm$ 4.5    &   158.9  $\pm$ 0.7    &   157.5  $\pm$ 0.6  &   189.6  $\pm$ 1.6    &   184.6   $\pm$1.4    \\
            &   325 &   0.55 &   147.0  $\pm$ 4.0    &   147.0  $\pm$ 4.0    &   144.0  $\pm$ 0.6    &   142    $\pm$ 0.5  &   176.6  $\pm$ 1.5    &   169.5   $\pm$1.1    \\
            &   350 &   0.60 &   121.6  $\pm$ 1.8    &   119.6  $\pm$ 1.7    &   124.2  $\pm$ 0.6    &   121.9  $\pm$ 0.4  &   148    $\pm$ 1.2    &   140.1   $\pm$1.0  \\
            &   375 &   1.20 &   88.4   $\pm$ 0.5    &   88.8   $\pm$ 0.5    &   88.8   $\pm$ 0.4    &   88.3   $\pm$ 0.3  &   115.5  $\pm$ 1.2    &   108.7   $\pm$0.6    \\
            &   400 &   1.70 &   77.0   $\pm$ 0.5    &   77.8   $\pm$ 0.5    &   76.3   $\pm$ 0.3    &   75.9   $\pm$ 0.3  &   95.2   $\pm$ 1.0    &   90.3    $\pm$0.5    \\ \hline
        475 &   425 &   3.80 &   69.5   $\pm$ 0.6    &   69.9   $\pm$ 0.4    &   68.1   $\pm$ 0.2    &   68.1   $\pm$ 0.2  &   77.7   $\pm$ 0.7    &   76.0    $\pm$0.4    \\
            &   300 &   0.40 &   180.0  $\pm$ 6.0    &   179.0  $\pm$ 2.0    &   180.5  $\pm$ 1.0  &   176.7    $\pm$ 0.6  &   193.3  $\pm$ 1.0    &    187.8  $\pm$0.8    \\
            &   325 &   0.60 &   140.0  $\pm$ 1.0    &   139.0  $\pm$ 1.0    &   140.5  $\pm$ 0.6    &   139.2  $\pm$ 0.5  &   162.8  $\pm$ 1.1    &   160.3   $\pm$0.7    \\
            &   350 &   1.18 &   99.7   $\pm$ 0.6    &   100.3  $\pm$ 0.6    &   100.3  $\pm$ 0.5    &   99.5   $\pm$ 0.5  &   130.7  $\pm$ 1.5    &   123.8   $\pm$1.0  \\
            &   375 &   1.25 &   89.0   $\pm$ 0.6    &   89.8   $\pm$ 0.5    &   91.0   $\pm$ 0.5    &   90.0   $\pm$ 0.3  &   111.2  $\pm$ 1.0    &   105.3   $\pm$0.5    \\
            &   400 &   1.60 &   79.1   $\pm$ 0.5    &   80.6   $\pm$ 0.5    &   79.4   $\pm$ 0.4    &   78.8   $\pm$ 0.5  &   89.0   $\pm$ 0.7    &   86.0    $\pm$0.3    \\\hline

        500 &   275 &   0.75 &   175.0  $\pm$ 3.1    &   175.0  $\pm$ 4.0    &   179.0  $\pm$ 1.2    &   177.0  $\pm$ 1.0  &   235.0  $\pm$ 3.0    &   230.0   $\pm$3.0  \\
            &   300 &   0.80 &   150.8  $\pm$ 1.8    &   150.0  $\pm$ 1.2    &   153.3  $\pm$ 1.0  &   150.6    $\pm$ 0.6  &   197.0  $\pm$ 2.3    &   186.0   $\pm$2.0  \\
            &   325 &   1.28 &   115.8  $\pm$ 1.2    &   115.8  $\pm$ 0.8    &   117.8  $\pm$ 0.5    &   116.1  $\pm$ 0.4  &   155.9  $\pm$ 2.0    &   145.5   $\pm$1.0  \\
            &   350 &   1.80 &   98.3   $\pm$ 0.9    &   100.3  $\pm$ 1.2    &   99.9   $\pm$ 0.5    &   98.7   $\pm$ 0.4  &   130.4  $\pm$ 1.5    &   121.25  $\pm$1.0  \\
            &   375 &   4.30 &   85.3   $\pm$ 0.5    &   85.8   $\pm$ 0.5    &   85.2   $\pm$ 0.3    &   84.9   $\pm$ 0.3  &   95.8   $\pm$ 1.5    &   92.6    $\pm$0.6    \\
            &   400 &   5.00 &   78.8   $\pm$ 0.5    &   79.6   $\pm$ 0.5    &   78.7   $\pm$ 0.3    &   78.5   $\pm$ 0.2  &   86.5   $\pm$ 0.5    &   83.8    $\pm$0.3    \\\hline
        525 &   200 &   0.45 &   290.0  $\pm$ 7.0    &   286.0  $\pm$ 7.0    &   290.0  $\pm$ 2.0  &   288.5    $\pm$ 1.0  &   340.5  $\pm$ 2.0    &   337.5   $\pm$2.0  \\

           &   225 &   0.40 &   261.8  $\pm$ 3.0    &   257.3  $\pm$ 3.0    &   262.7  $\pm$ 1.1    &   259.0  $\pm$ 1.0  &   283.1  $\pm$ 1.2    &   278.0   $\pm$1.2    \\
            &   250 &   0.60 &   210.3  $\pm$ 3.0    &   208.7  $\pm$ 6.0    &   205.5  $\pm$ 1.2    &   204.8  $\pm$ 1.0  &   261.3  $\pm$ 2.5    &   254.2   $\pm$2.0  \\
            &   275 &   0.60 &   180.2  $\pm$ 2.0    &   177.2  $\pm$ 3.0    &   180.0  $\pm$ 0.6    &   177.1  $\pm$ 0.5  &   217.2  $\pm$ 1.3    &   208.4   $\pm$1.0  \\
            &   300 &   1.20 &   133.0  $\pm$ 1.4    &   133.5  $\pm$ 0.8    &   135.0  $\pm$ 0.7    &   133.4  $\pm$ 0.5  &   188.0  $\pm$ 1.5    &   174.8   $\pm$1.0  \\
            &   325 &   1.70 &   111.9  $\pm$ 0.8    &   112.5  $\pm$ 0.6    &   113.6  $\pm$ 0.7    &   112.3  $\pm$ 0.5  &   149.8  $\pm$ 1.0    &   141.6   $\pm$0.7    \\
            &   350 &   2.00 &   100.0  $\pm$ 0.9    &   100.7  $\pm$ 0.9    &   102.0  $\pm$ 1.0  &   99.0     $\pm$ 1.0  &   115.3  $\pm$ 1.0    &   110.5   $\pm$0.5    \\\hline
    \end{tabular}
    \label{tab:10}
\end{table}

\end{landscape}

\section{Average Waveforms}

It has been claimed that the cause of the correlation between the mean SAT and the electron peak's size is not because of a weakness in the timing technique.
Instead, it is believed that it comes from the physical processes taking place inside the detector.
To exclude the possibility that the shape of the waveform is changing with different electron peak sizes, causing this ``time walk'' effect, the average waveform is found.
The average waveform is found by taking all waveforms of a dataset, belonging in a region of electron peak size $[Q_i,\,Q_f]$, synchronizing all of the selected waveforms and normalizing them to unity electron peak size, and then averaging them point-by-point.

\begin{figure}[H]
    \centering
    \begin{subfigure}[h]{0.49\textwidth}
        \includegraphics[width=0.95\textwidth]{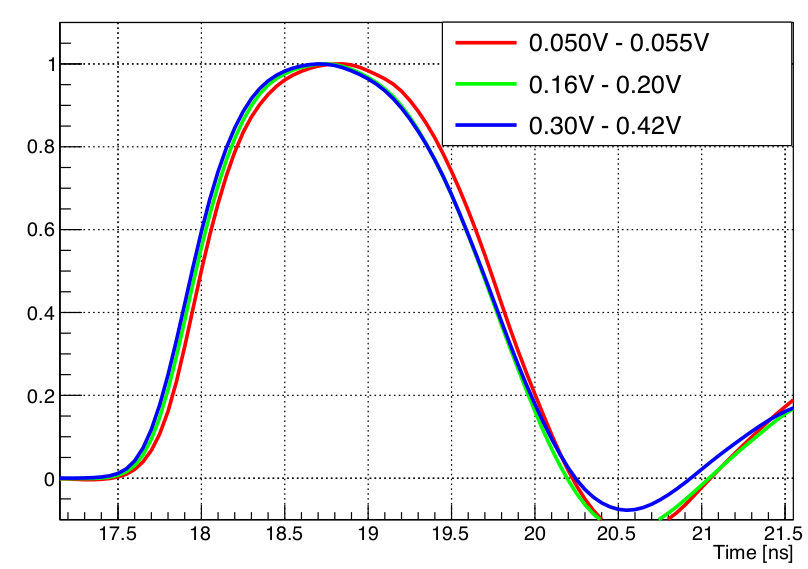}
        \caption{}
    \end{subfigure}
    \begin{subfigure}[h]{0.49\textwidth}
        \includegraphics[width=0.95\textwidth]{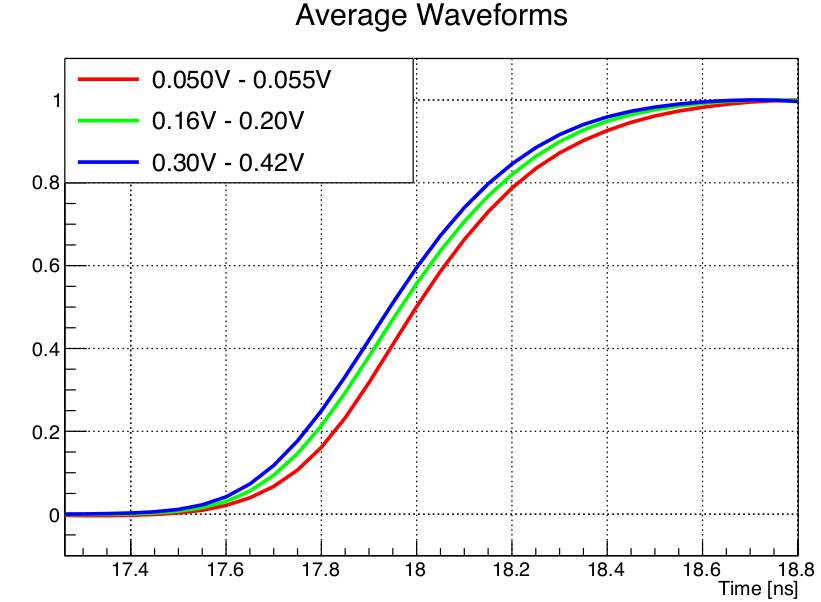}
        \caption{}
    \end{subfigure}
    \caption[Average waveforms for CF$_4$ gas.]{ 
        Average waveform for data collected with anode voltage of $650\,V$, drift voltage of $450\,V$ and the CF$_4$ gas filling. (a) corresponds to the entirety of the electron peak while (b) is focused on the leading edge. With red is presented the average waveform of pulses with an electron peak amplitude in the range $[0.05\,V,\ 0.055\,V]$, green in $[0.16\,V,\ 0.20\,V]$ and blue in $[0.30\,V,\ 0.42\,V]$.
    }
    \label{aver-cf4}
\end{figure}

\begin{figure}[H]
    \centering
    \begin{subfigure}[h]{0.49\textwidth}
        \includegraphics[width=0.95\textwidth]{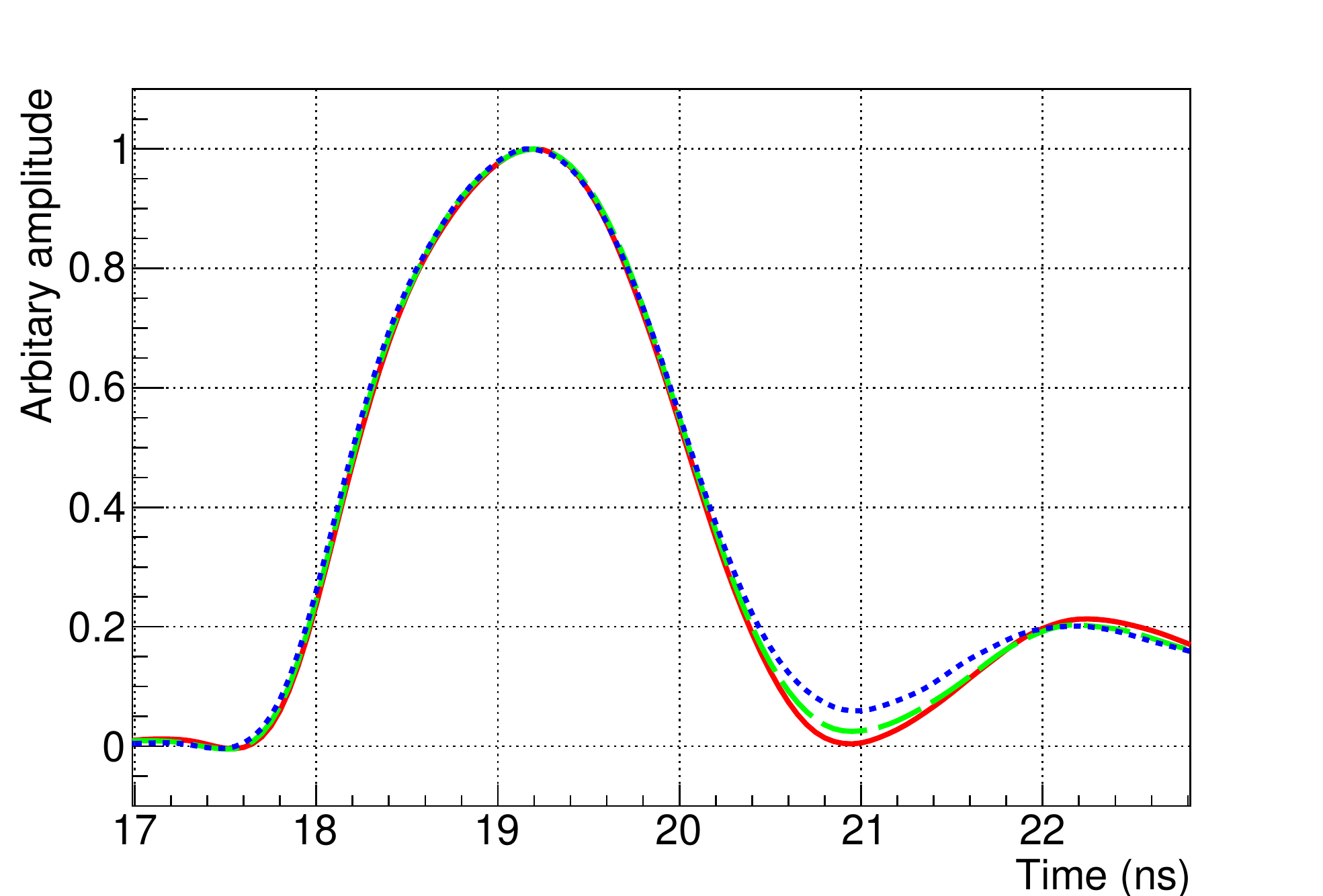}
        \caption{}
    \end{subfigure}
    \begin{subfigure}[h]{0.49\textwidth}
        \includegraphics[width=0.95\textwidth]{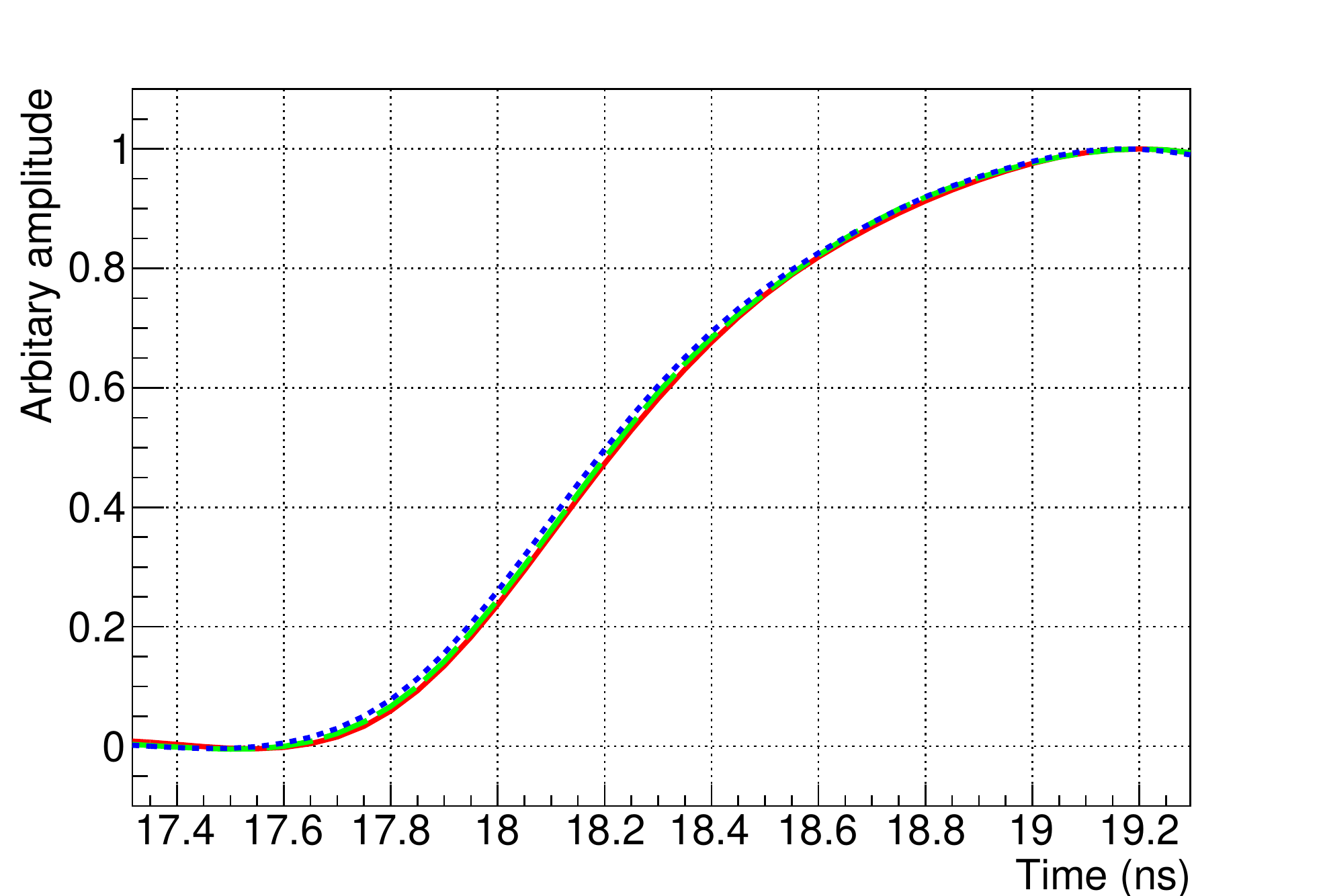}
        \caption{}
    \end{subfigure}
    \caption[Average waveforms for COMPASS gas.]{ 
        Average waveform for data collected with anode voltage of $450\,V$, drift voltage of $350\,V$ and the COMPASS gas filling. (a) corresponds to the entirety of the electron peak while (b) is focused on the leading edge. With red is presented the average waveform of pulses with an electron peak charge in the range $[1.0\,pC,\ 1.1\,pC]$, green in $[2.0\,pC,\ 2.5\,pC]$ and blue in $[3\,pC,\ 4\,pC]$.
    }
    \label{aver-comp}
\end{figure}

The average waveform for data collected with CF$_4$ gas, anode voltage of $650\,V$ and drift voltage of $450\,V$ is shown in Figure \ref{aver-cf4}, with (a) corresponding to the whole electron peak while (b) is focused in the leading edge of the electron peak.
Similarly, the average waveform for data collected with COMPASS gas, anode voltage of $450\,V$ and drift voltage of $350\,V$ is shown in Figure \ref{aver-comp}, with (a) corresponding to the whole electron peak while (b) is focused in the leading edge of the electron peak.

It is obvious that the waveforms seems to be shifting in time with a change in electron peak size.
The shape of the leading edge is barely changing neither for the CF$_4$ nor for the COMPASS data.
This is a very strong indication that there exists a physical process that causes this particular correlation.
This indication was the primary reason for making the simulation, in order to understand why and how this effect emerges.

\section{Summary}

A number of effects have been discovered in the analysis of the Laser Calibration data.
The most important conclusion is the fact that by increasing the drift voltage, the total time resolution improves.
A decrease in the anode voltage barely changes the time resolution but
to preserve the stability of the detector, the anode voltage should be decreased.

Two more interesting effects were discovered that have no simple explanation.
These are:
\begin{enumerate}
    \item For a fixed anode voltage, the functional form of the mean SAT with respect to the size of the electron peak changes only by a constant term with different drift voltages.
    \item For a fixed anode voltage, the functional form of the time resolution with respect to the size of the electron peak does not change with different drift voltages.    
\end{enumerate}
These phenomena were the primary reason for which the simulation study was initiated.
It should be emphasized that more details concerning these results can be found in references \cite{noteA} and \cite{noteB}.

\chapter{Simulation}

\section{Description of the Simulation}

\begin{figure}[t]
    \centering
    \includegraphics[width=0.95\textwidth]{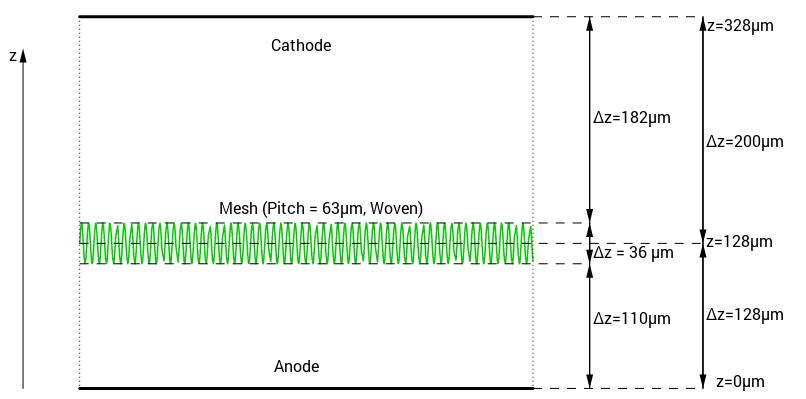}
    \caption[Diagram of the simulated detector model.]{Diagram of the simulated detector model. The geometry is periodic in the plane that is perpendicular to the z-axis (electric field). The reference $z=0$ is the anode.} 
    \label{simu:fig:dia}
\end{figure}

The simulation is based on the Garfield++\cite{garfieldpp}
software, which in turn is based on Garfield\cite{garfield} using Magboltz's technique\cite{magboltz}. 
The ANSYS\cite{ansys} software is used to model the electric field. 
Although the real micro-mesh is woven and calendered, its calendering is not included in the electric field modelling.
The diagram of Figure \ref{simu:fig:dia} shows the structure of the chamber in the simulation.
Its specifications are:

\begin{enumerate}
\item The reference $z=0$ begins at the anode.
\item The woven micro-mesh is placed such that its center lies at $z = 128\mu m$.
\item The wires of the micro-mesh are of $18\mu m$ diameter, providing a total thickness of $36\,\mu m$.
\item The distance between each hole, or the ``pitch'', is $63\,\mu m$.
\item The cathode lies at $ z = 328\,\mu m$.
\item An ``anode voltage'' is applied on the anode.
\item The micro-mesh is set to ground.
\item A ``drift voltage'' is applied on the cathode.
\end{enumerate}

It is impractically time-consuming to make a full simulation.
Moreover, the electronic chain's response is not known and must be measured or estimated separately.
It was for these reasons that the simulation is separated in three stages.

\begin{stages}
\item Simulation of the pre-amplification region\footnote{The pre-amplification region is the region defined between the cathode and the micro-mesh.}.
\item Simulation of the amplification region\footnote{The amplification region is the region defined between the micro-mesh and the anode}.
\item Combination of Stage 1 and Stage 2, and convolution with the electronic chain's response for the generation of signal waveforms.
\end{stages}

The motivation to separate the simulation in these stages was to study the mechanism that produces the effects observed in the analysis. 
Furthermore, the required CPU time is greatly reduced which makes it easier to change input parameters (like the penning transfer rate
\footnote{The penning transfer rate refers to the net probability for an excited molecule to ionize another molecule in the mixture.}
or the photoelectron's initial energy and direction).

It has to be mentioned that in the simulation

\begin{itemize}
    \item Only electrons were simulated while the ions were ignored.
    \item Initially, the penning transfer was but was taken into account in later simulations.
        It is known that the penning transfer accounts for a considerable amount of the total gain for the COMPASS gas mixture
    \item The impulse response of the electronics chain' is estimated a posteriori in a data-driven way.
        More details in Section \ref{sec:response}.
    \item Real data are digitized with 8-bit precision.
        This fact is ignored and amplitudes take continuous values in the simulation.
\end{itemize}

These approximations are completely justified and do not affect the results of the simulation.

\subsection{Stage 1 - Pre-Amplification}\label{sec:preamp}

The simulation of the effects in the pre-amplification region, whose input parameters are:

\begin{itemize}
    \item At $z = 328\,\mu m$ and $t = 0$, a photoelectron is created with uniformly random coordinates in the transverse xy-plane.
    \item In most cases, the photoelectron begins with a kinetic energy of $0.1\, eV$ with an isotropically random direction.
        In special cases, these values are varied to check the influence of these variables.
    \item Electrons are tracked microscopically.
    \item The gaseous mixture used is the COMPASS one.
    \item Environment conditions are $293.15\,K^o$ temperature at $1010\,mbar$ pressure. 
\end{itemize}

To avoid redundant computing time, the region of the simulation is reduced such that all electrons are stopped right after they pass through the micro-mesh region at the plane with $z = 110\,\mu m$.
Meanwhile, their whole drift lines are stored, i.e. their position in space and time in each of their step.
This makes for an enormous amount of information which is filtered through by a processing script extracting the following variables.

\begin{enumerate}
    \item The photoelectron's initial coordinates.
    \item The coordinates in space and time of the first ionization.
    \item The number of scatterings of the photoelectron until the first ionization.
    \item The times at which each electron of the avalanche pass through a fixed level of $z$.
        \footnote{This level is usually set at $z=146\,\mu m$ to ``ignore'' the micro-mesh's influence, or at $z = 128\,\mu m$ to include it.
        In other cases it is a collection of levels to ``monitor'' the evolution of the avalanche.}.
    \item The number of electrons passing through the corresponding fixed level.
\end{enumerate}

\begin{figure}[H]
    \centering
    \begin{subfigure}[h]{0.49\textwidth}
        \includegraphics[width=0.95\textwidth]{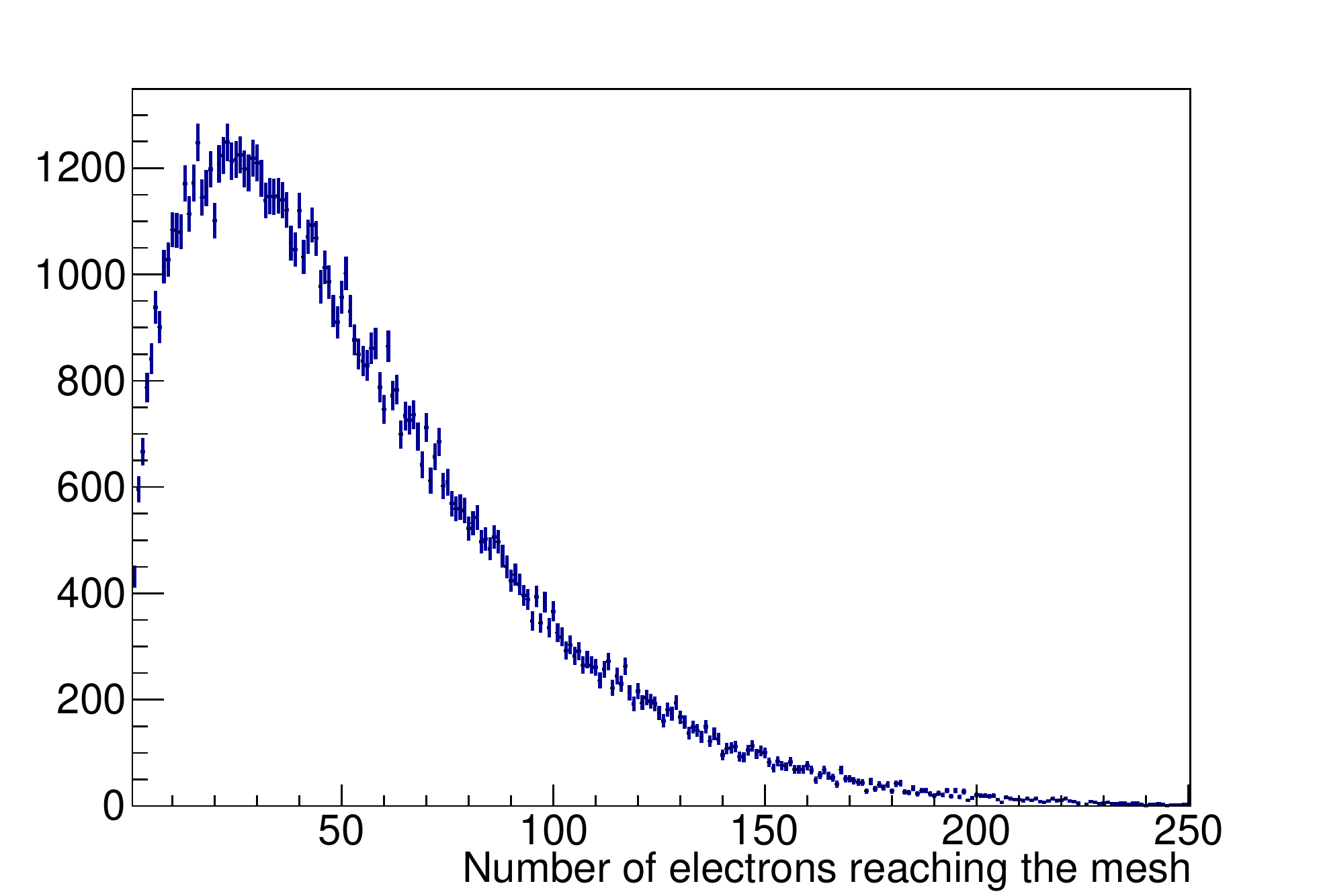}
        \caption{}
    \end{subfigure}
    \begin{subfigure}[h]{0.49\textwidth}
        \includegraphics[width=0.95\textwidth]{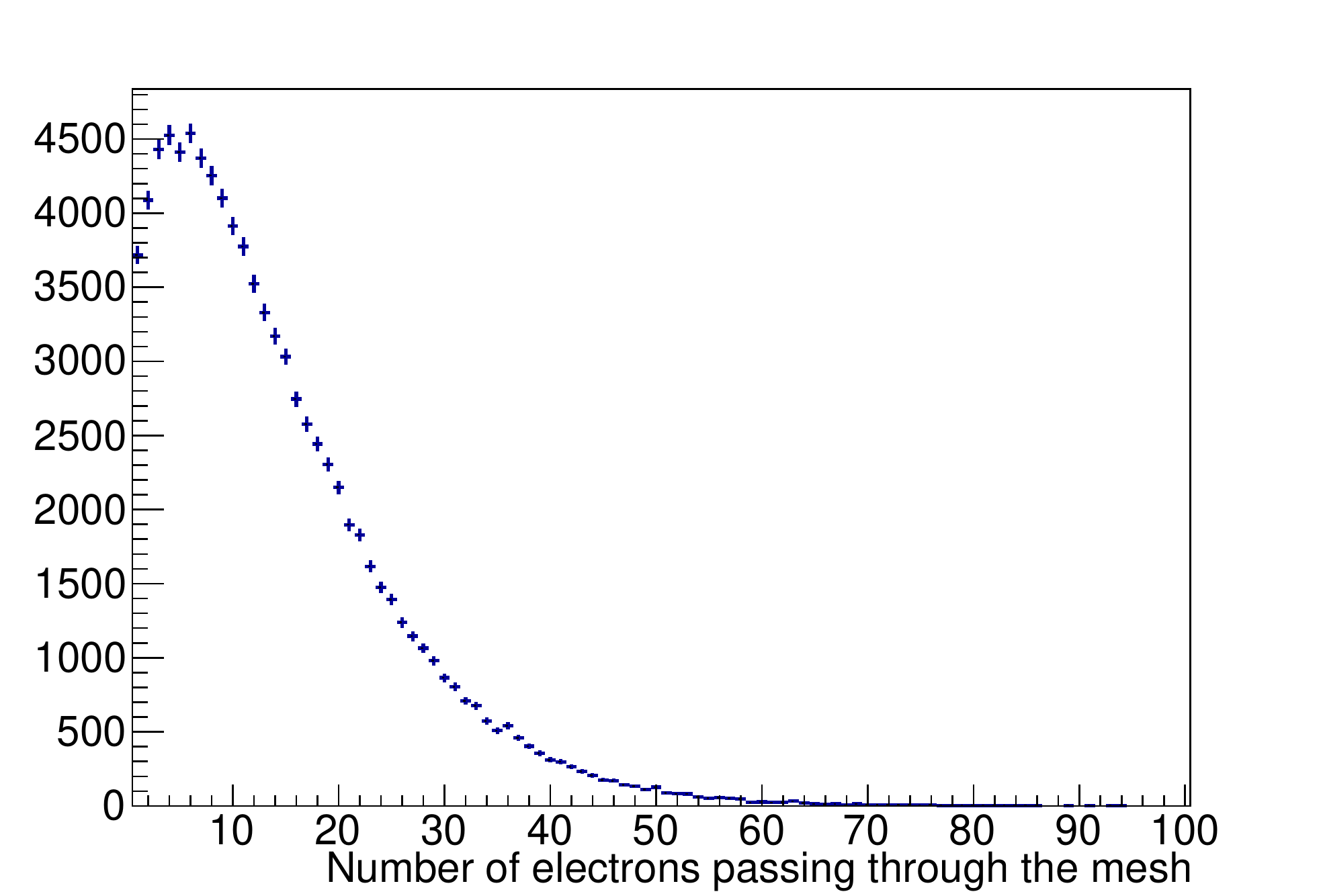}
        \caption{}
    \end{subfigure}
    \begin{subfigure}[h]{0.49\textwidth}
        \includegraphics[width=0.95\textwidth]{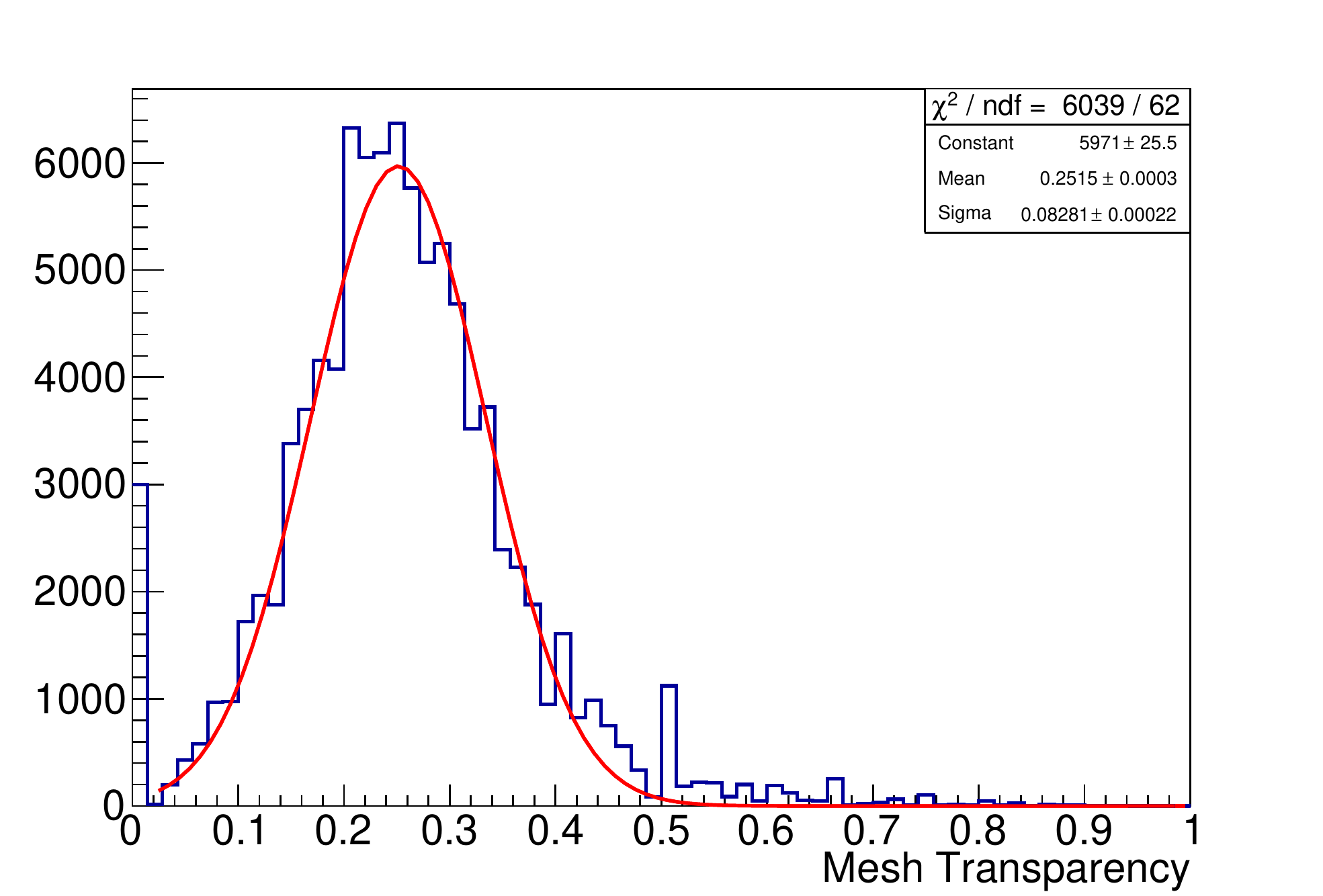}
        \caption{}
    \end{subfigure}
    \caption[Number of electrons and electron mesh transparency.]{(a) The distribution of the number of electrons produced in the pre-amplification region reaching the mesh. (b) The distribution of the number of electrons produced in the pre-amplification region passing through the mesh. (c) The distribution of the fraction of electrons produced in the pre-amplification region which pass through the mesh (transparency).
    The structure shown in (c) is an artifact of the division between small integer numbers.
    } 
    \label{simu:fig:example}
\end{figure}

\begin{figure}[H]
    \centering
    \begin{subfigure}[h]{0.49\textwidth}
        \includegraphics[width=0.95\textwidth]{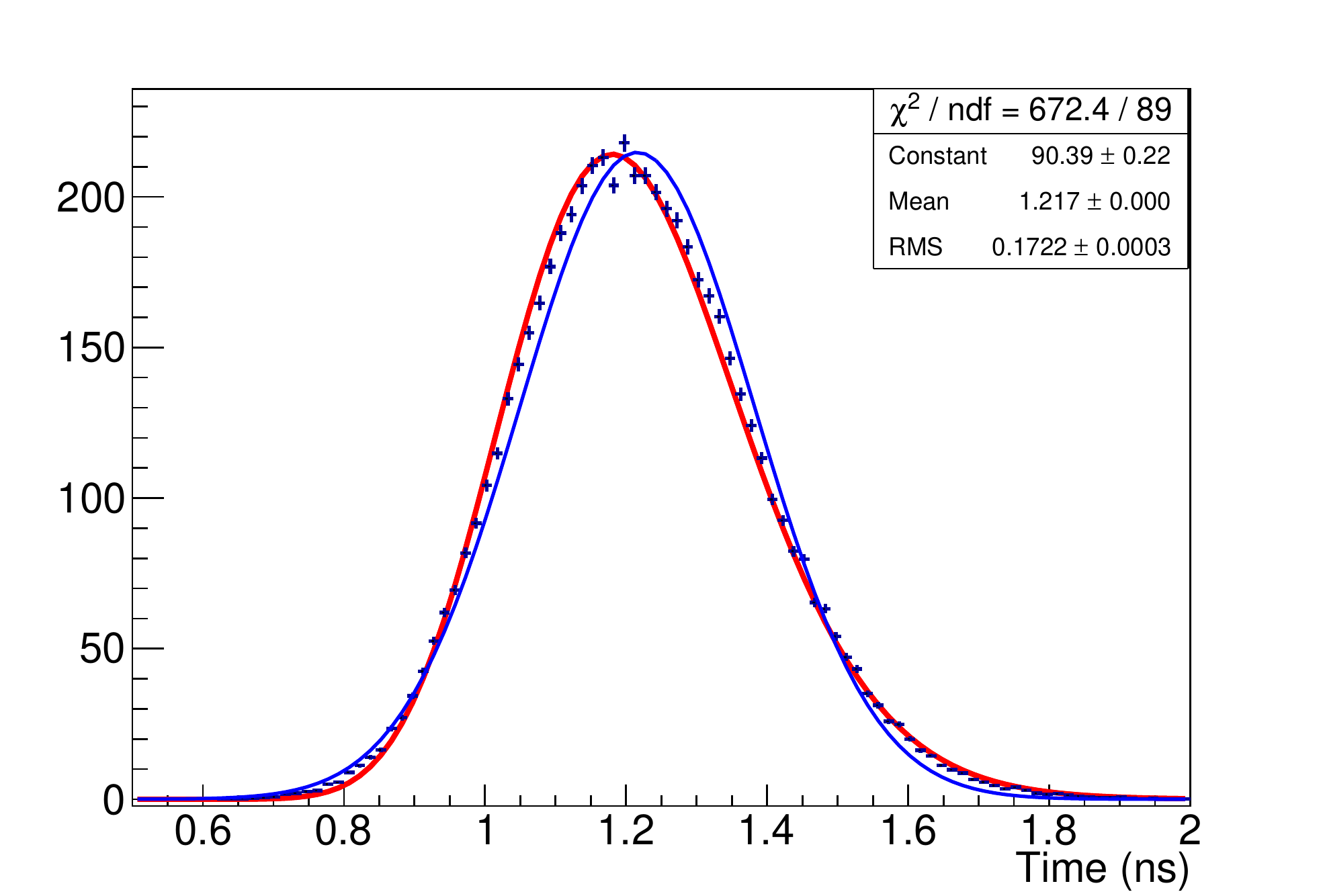}
        \caption{}
    \end{subfigure}
    \begin{subfigure}[h]{0.49\textwidth}
        \includegraphics[width=0.95\textwidth]{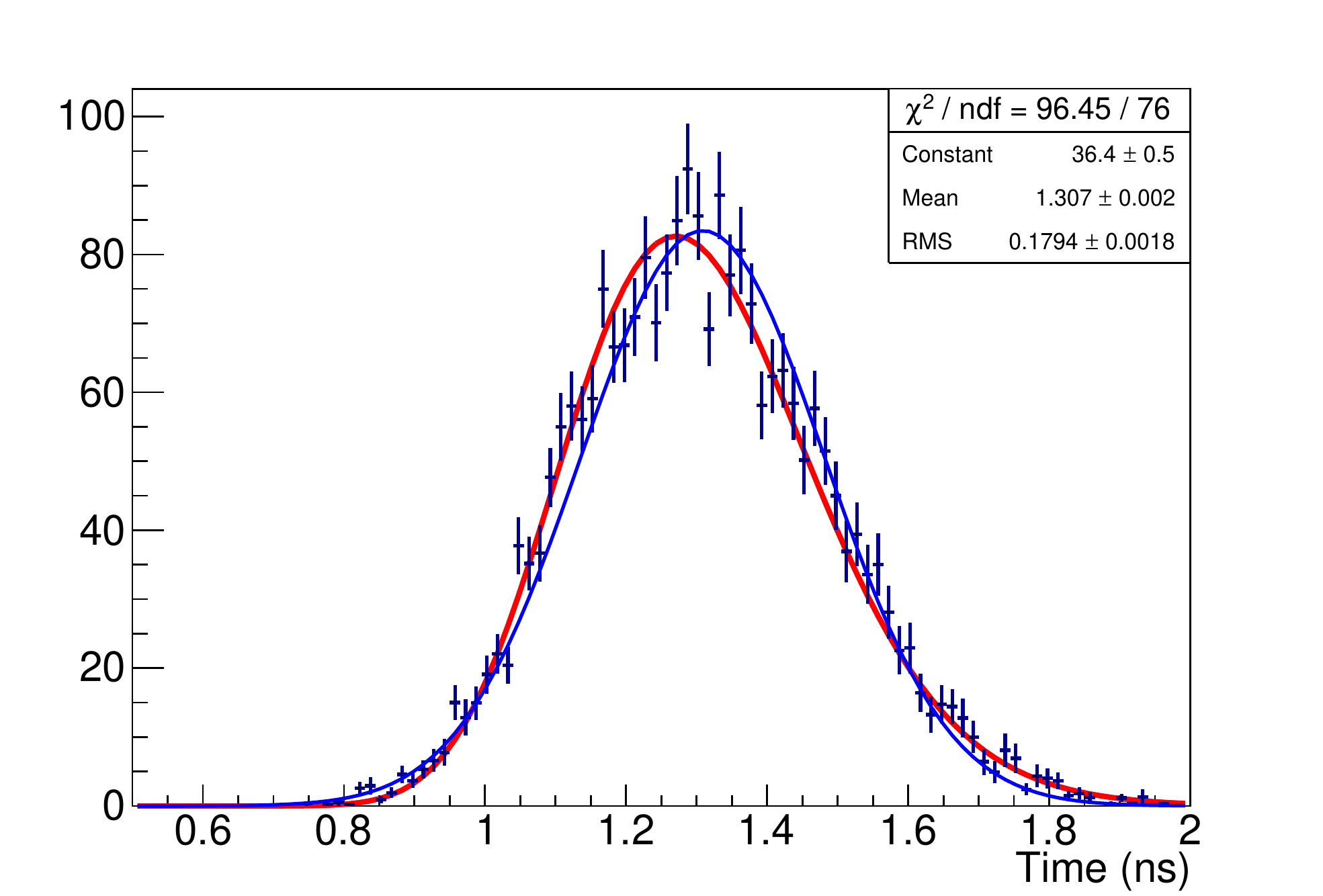}
        \caption{}
    \end{subfigure}
    \caption[Time of arrival distributions.]{Points represent the distribution of the time an electron, produced in the pre-amplification region, takes to reach the middle of the mesh.
    The shown plots represent distributions of cases where the number of electrons produced in the pre-amplification are between (a) $[25,30]$ and (b) $[1,5]$.
    The blue lines represent (unsuccesful) fits using  Gaussian functions, whilst the red lines represent fits using the inverse Gaussian distribution of Equation \ref{inverse-gaus}.  
    } 
    \label{simu:fig:example2}
\end{figure}

For example, Figure \ref{simu:fig:example} shows the number of electrons produced in the pre-amplification region by a single photoelectron (a) before and (b) after the micro-mesh, and (c) the fraction of the electrons passing through the micro-mesh (electron mesh transparency) for an anode and drift voltage equal to $450\,V$ and $350\,V$, respectively.
Figure \ref{simu:fig:example2} shows the electrons' time of arrival on the micro-mesh when their multiplicity (of the electrons in the pre-amplification avalanche) is between (a) $[25,\,30]$ and (b) $[1,\,5]$.

\subsection{Stage 2 - Amplification} \label{sec:ampli}

\begin{figure}[H]
    \centering
    \begin{subfigure}[h]{0.49\textwidth}
        \includegraphics[width=0.95\textwidth]{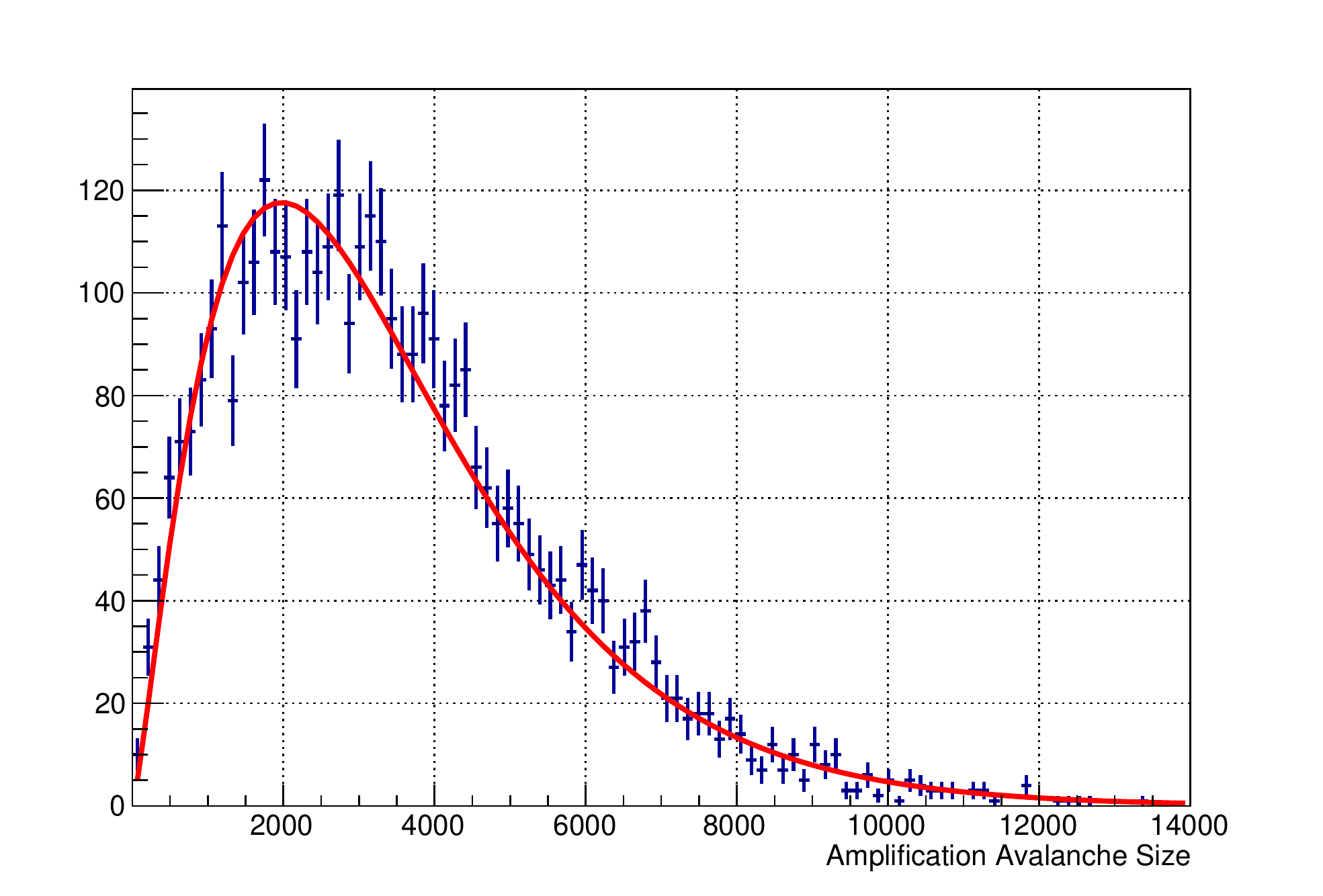}
        \caption{}
    \end{subfigure}
    \begin{subfigure}[h]{0.49\textwidth}
        \includegraphics[width=0.95\textwidth]{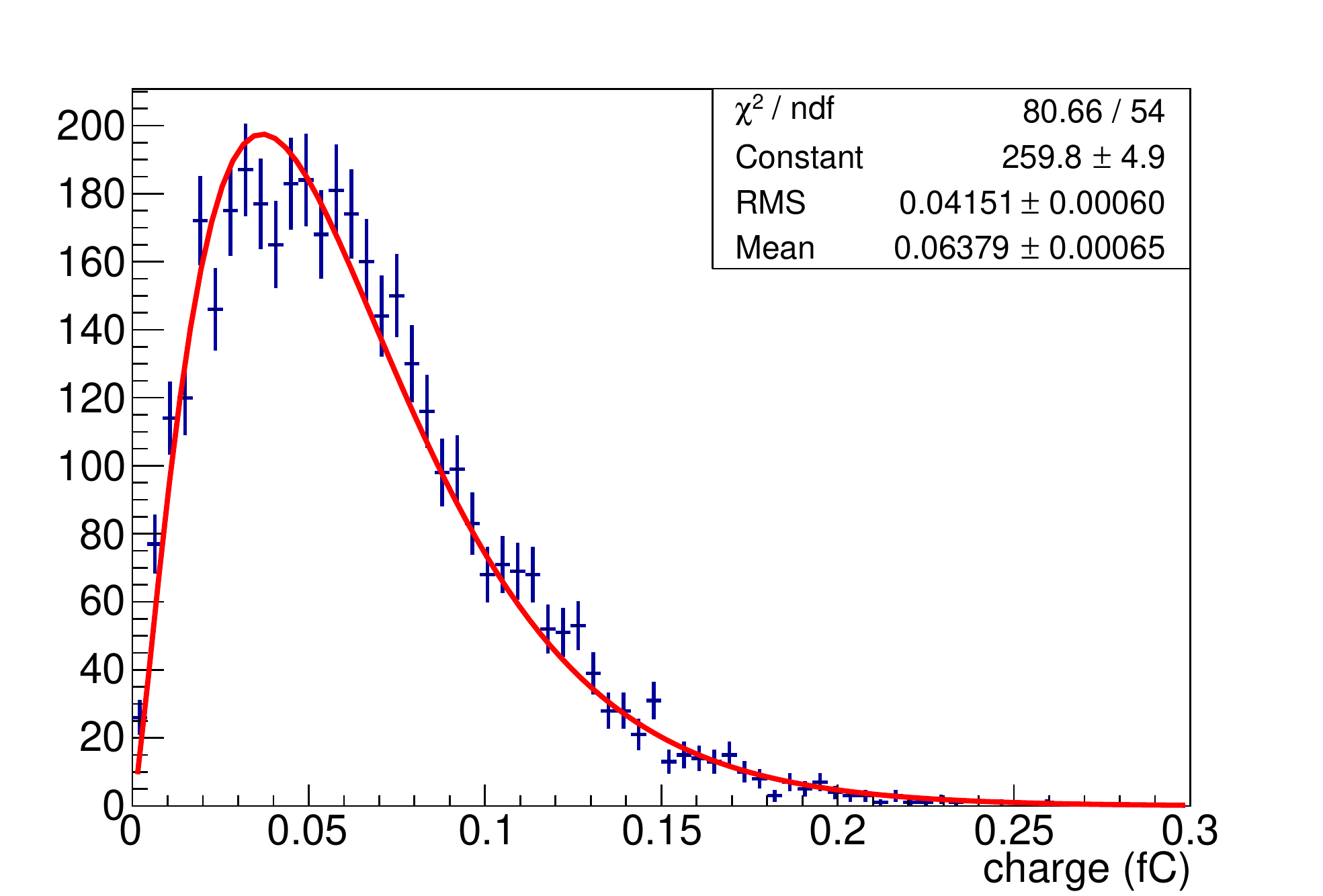}
        \caption{}
    \end{subfigure}
    \begin{subfigure}[h]{0.49\textwidth}
        \includegraphics[width=0.95\textwidth]{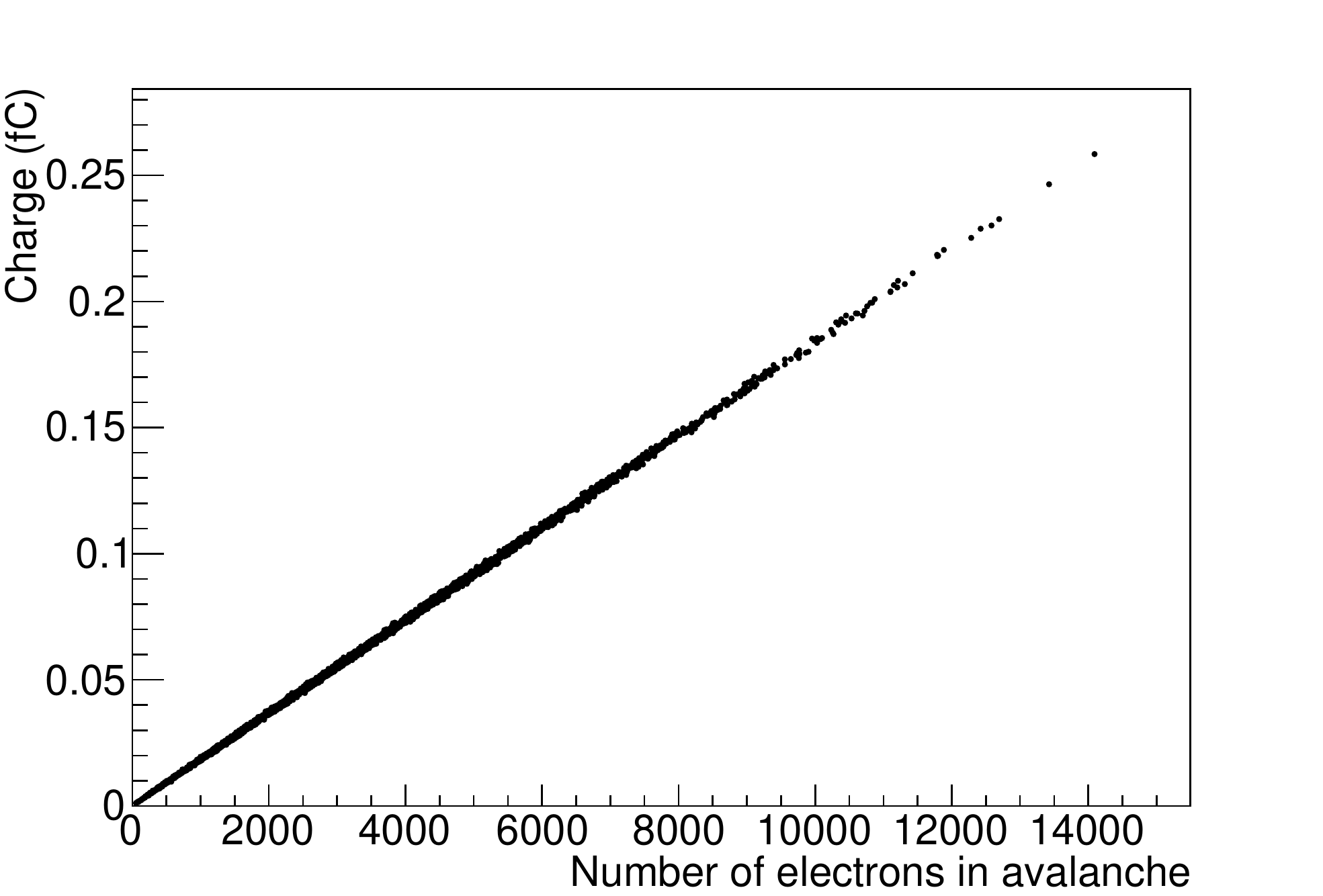}
        \caption{}
    \end{subfigure}
    \caption[Number of electrons and electron peak charge distribution.]{(a) Distribution of the amplification avalanche size (blue). The red line corresponds to the realization of a single Polya fit. The estimated parameters for the fit are $RMS = 2279 \pm 35$ and $Mean = 3473 \pm 36$.
    (b)Distribution of the amplification induced total charge (blue). The red line corresponds to the realization of a single Polya fit. The estimated parameters for the fit are $RMS = 0.042 \pm 0.0006$ and $Mean = 0.064 \pm 0.0007$.
    (c) Scatter plot of induced total charge vs the number of electrons in the amplification avalanche.
    In all figures, anode voltage is $450\,V$.
    } 
    \label{simu:fig:ampli}
\end{figure}

Simulating each and every one of the electrons passing through the micro-mesh and into the amplification region would be impractically CPU time-consuming.
To reduce computational time, the microscopic modelling of Garfield++ was used to simulate $10000$ single electron avalanches in the amplification region, beginning on the mesh.
For each of the $10000$ electron avalanches, the detector's electrical response was calculated, as defined by the Ramo's theorem (Equation \ref{eq:ramo}), i.e. the current as a function of time which is induced on the anode, as well as the multiplicity of the secondary electrons in the amplification avalanche.
The total charge induced on the anode (integral of the induced current) from a single electron beginning on the micro-mesh is presented on the distribution of Figure \ref{simu:fig:ampli}\,b.
Electrons that passed through the micro-mesh but re-attached with the gaseous mixture before initiating an avalanche produce a negligible amount of current and charge.
These events are not shown in Figure \ref{simu:fig:ampli}.
and the probability for such an event to occur is approximately $1.5\%$ for this case ($450\,V$ anode voltage and COMPASS gas).
The red line of the same Figure represents a fit with a P\'{o}lya distribution function.
In the stages to follow, this P\'{o}lya parameterization is used to simulate the induced charge caused by a single electron passing through the micro-mesh.

If the electronics chain' response function was known, the procedure of the signal generation would have gone as follows:
In an event where $k$ electrons are transmitting through the mesh, $k$ single electron avalanches are randomly sampled.
Because the electrical response is governed by linear differential equations, and assuming the $k$ single electron amplification avalanches are independent of each other in their formation, the $k$ corresponding induced current signals are superimposed, taking into account the time of each electron's transmission through the micro-mesh.
Finally, the total induced current as a function of time is transformed to the observed voltage signal using the characteristic transfer function of the electronics chain.

However, this characteristic transfer function is not known and another strategy was followed, taking advantage of the linearity of the response of the electronics.
As explained in detail in the next section, real PICOSEC waveforms were used to estimate the shape of the signal (voltage versus time) corresponding to an amplification avalanche starting from a single electron transmitting through the micro-mesh.
So, if $k$ electrons started an amplification avalanche, each would contribute with the standard signal described previously.
The size of each of these $k$ signals should follow the P\'{o}lya distribution shown in Figure \ref{simu:fig:ampli}\,b.
As such, for the $i-$th electron $(i=1,2,...,k)$ passing through the micro-mesh, a charge $(Q_i)$ is sampled from the P\'{o}lya parameterization and is assigned to the corresponding signal, i.e. the standard signal is normalized such that its integral corresponds to the sampled charge $Q_i$.
The probability that an electron does not produce an avalanche is also taken into account.
Each $i-$th electron's corresponding pulse is also shifted in time according to its time of passing through the mesh.
The superposition of all these pulses constitutes the simulation of the detector and the electronics chain response to a single photoelectron beginning from the cathode.

An additional jitter in time that is not included in this simulation is the time jitter of the single electron amplification avalanche starting from the mesh.
This should only barely affect the time resolution of events with low secondary electron multiplicities in the pre-amplification region.
It has been studied, though, that this time (the temporal distance between the passing through the mesh and the arrival on the anode) jitters with an RMS of around $20\,ps$.
This is a much smaller effect than the RMS of pre-amplification electrons' arrival times on the micro-mesh, especially in events with a low secondary electron multiplicity in the pre-amplification region.
It has to be emphasized that the time jitter caused by the micro-mesh transmission is not ignored but is included.

\subsection{Stage 3 - Electronics Chain Response}\label{sec:response}

\begin{figure}[H]
    \centering
    \begin{subfigure}[h]{0.49\textwidth}
        \includegraphics[width=0.95\textwidth]{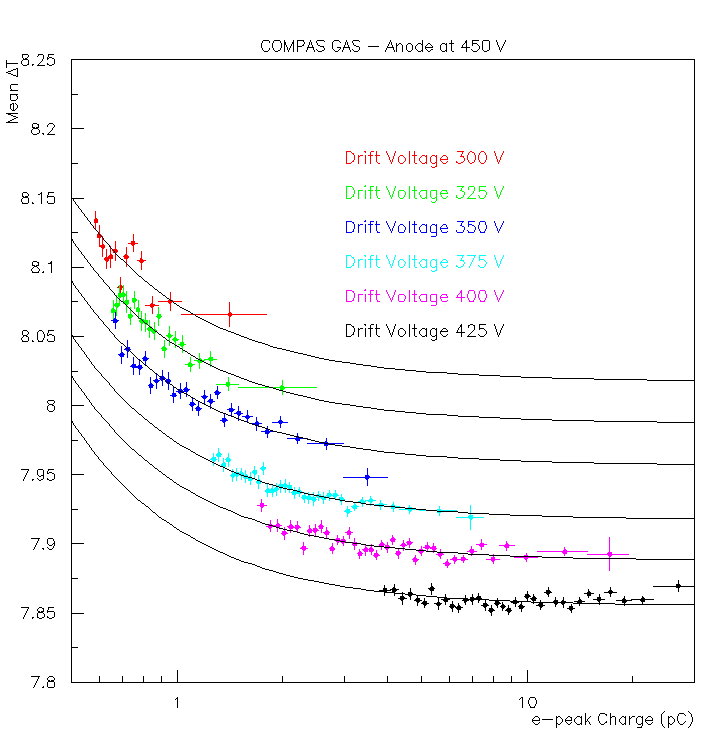}
        \caption{}
    \end{subfigure}
    \begin{subfigure}[h]{0.49\textwidth}
        \includegraphics[width=1.05\textwidth]{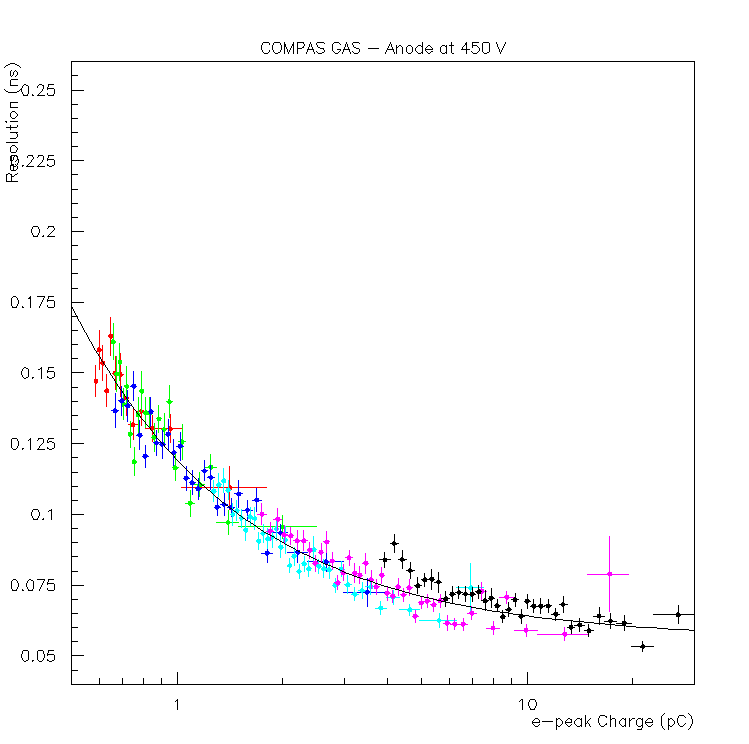}
        \caption{}
    \end{subfigure}
    \caption[Signal arrival time and time resolution with respect to electron peak charge.]{In the left plot, the SAT dependence is presented with respect to electron peak charge for experimental runs with anode voltage of $450\,V$, while in the right plot, the time resolution dependence is presented. }  
    \label{simu:fig:postconv}
\end{figure}

To determine the electronics chain impulse response, the experimental data collected with anode voltage of $450\,V$ and drift voltage of $425\,V$ is used in combination with the simulation.
The choice of voltages is such that signal arrival time dependencies are minimized.
Indeed as shown in Figure \ref{simu:fig:postconv}, 
the mean and the spread of the signal arrival times in the experimental signals depend weakly on the size of the signal (amplitude or charge).
The procedure described in the previous Section \ref{sec:ampli} 
requires the electrical signal response of a single electron amplification avalanche in order to simulate the experimental readout.
Hereafter, the function $S(t)$ will symbolize this which according to the procedure can be expressed as:

\begin{equation}
    S(t) = \sum_{i=1}^k f(t-\tau_i)\cdot q_i
\end{equation}

where $f(x)$ is the function of time describing the shape of an electronic pulse that is produced by a single electron avalanche, such as $\int_0^\infty f(x) dx = 1$.
Therefore, $f(t-\tau_i)\cdot q_i$ is the contribution of the $i$-th avalanching electron to the electronic signal and $\tau_i$ is the time at which the single electron avalanche started.
The variable $k$ is the number of pre-amplification electrons transmitting through the micro-mesh and initiate avalanches in the amplification region.
This follows a probability distribution function (PDF) $R(k)$, which is evaluated in the simulation as described in Section \ref{sec:preamp} (e.g. see Figure \ref{simu:fig:example}).
The variables $\tau_i$, $i=1,2,..,k$, are treated as random variables following the PDF $\Phi(\tau;k)$ which depends on the parameter $k$ and can be evaluated from the simulation.
As demonstrated in Fig. \ref{simu:fig:example2}, $\Phi(\tau;k)$ is well described by an Inverse Gaussian with mean and RMS values depending on the number of the avalanching electrons, $k$. 
Accordingly, $q_i$ is the charge induced in the anode by the $i$-th single electron avalanche and follows the PDF $G(q)$, which is well described by a P\'{o}lya distribution, as shown in Figure \ref{simu:fig:ampli}. 

The likelihood to get a pulse $S(t)$ produced by $k$ single electron avalanches, each contributing with pulse of charge $q_i$ $(i=1,...,k)$ and starting at time $\tau_i$ is 

\begin{equation}
    P(\tau_1,...,\tau_k,q_1,...,q_k;k) = R(k)\cdot \prod_{i=1}^k \Phi(\tau_i;k) \cdot \prod_{i=1}^k G(q_i)
\end{equation}

Obviously, the total charge of the pulse $S(t)$ equals to $\sum_{i=1}^k q_i$.
The average pulse which corresponds to the whole region of observed charges is expressed as:

\begin{multline}
    \bar{S}(t) = \\
    = \sum_{k=1}^\infty R(k) \int_0^\infty\cdots \int_0^\infty \sum_{i=1}^k q_i f(t-\tau_i) \prod_{j=1}^k \Phi(\tau_j;k) \cdot \prod_{j=1}^k G(q_j) dq_1...dq_kd\tau_1...d\tau_k
\end{multline}

In the case where we are interested in the average pulse with total charge equal to $Q$, we express the likelihood of observing such a pulse by the function

\begin{multline}\label{four}
    P_{_{Q_{tot}= Q}} (\tau_1,...,\tau_k,q_1,...,q_k;k) = \\
    = \frac{\delta\left(Q-\sum_{i=1}^k q_i\right) \cdot R(k)\cdot \prod_{i=1}^k \Phi(\tau_i;k) \cdot \prod_{i=1}^k G(q_i)}{N(k)}
\end{multline} 

where

\begin{equation}\label{eqNk}
    N(k)
    =\int_0^\infty\cdots \int_0^\infty \delta\left(Q-\sum_{i=1}^k q_i\right) \cdot R(k)\cdot \prod_{i=1}^k G(q_i)dq_1...dq_k
\end{equation}

The variables $q_i$ $(i=1,2,...,k)$ and $k$ are not independent anymore.
The average pulse with the requirement $Q_{tot} = Q$ is now expressed as 

\begin{multline}\label{barS}
    \left<S(t)\right>_{_{Q_{tot}=Q}} = \\
    = \sum_{k=1}^\infty R(k) \int_0^\infty\cdots \int_0^\infty \sum_{i=1}^k q_i f(t-\tau_i) \cdot\delta\left(Q-\sum_{i=1}^k q_i\right)\cdot \\
    \cdot \prod_{j=1}^k \Phi(\tau_j;k) \cdot \prod_{j=1}^k G(q_j) dq_1...dq_kd\tau_1...d\tau_k
\end{multline}

After some simple manipulations Equation \ref{barS} becomes

\begin{multline}\label{barS2}
    \left<S(t)\right>_{_{Q_{tot}=Q}} 
    = \sum_{k=1}^\infty R(k) \sum_{i=1}^k \int_0^\infty f(t-\tau_i) \Phi(\tau_i;k) d\tau_i\cdot \\ 
    \cdot \int_0^\infty\cdots \int_0^\infty q_i  \cdot\delta\left(Q-\sum_{i=1}^k q_i\right)\cdot \prod_{j=1}^k G(q_i) dq_1...dq_k
\end{multline}

Taking into account Equation \ref{four}, the expectation value of the charge induced to the anode by the $i$-th (out of $k$) single electron avalanche, when the total induced charge equals $Q$, is expressed as

\begin{equation}\label{meanq}
    \left<q_i\right>_{_{Q=\sum_{j=1}^k q_j}} = \frac{\int_0^\infty\cdots \int_0^\infty  q_i \cdot\delta\left(Q-\sum_{i=1}^k q_i\right)\cdot \prod_{j=1}^k G(q_i) dq_1...dq_k}{N(k)}
\end{equation}

By multiplying and dividing Equation \ref{barS2} with $N(k)$, the average pulse is expressed as

\begin{multline}\label{barS3}
    \left<S(t)\right>_{_{Q_{tot}=Q}} 
    = \sum_{k=1}^\infty R(k) N(k) \sum_{i=1}^k \int_0^\infty f(t-\tau_i) \Phi(\tau_i;k) d\tau_i \cdot \\
\cdot \int_0^\infty\cdots \int_0^\infty \frac{q_i \cdot\delta\left(Q-\sum_{i=1}^k q_i\right)\cdot \prod_{j=1}^k G(q_i)}{N(k)} dq_1...dq_k
\end{multline}

Substituting Equation \ref{meanq} into Equation \ref{barS3}, the above expression is simplified as

\begin{equation}\label{barS4}
    \left<S(t)\right>_{_{Q_{tot}=Q}} = \sum_{k=1}^\infty R(k) N(k) \cdot
     \sum_{i=1}^k     
          \left<q_i\right>_{_{Q=\sum_{j=1}^k q_j}}   
\int_0^\infty f(t-\tau_i) \Phi(\tau_i;k) d\tau_i 
\end{equation}

The integral in the above equation constitutes the average pulse shape (i.e. the electrical pulse normalized to unity charge) when the pulse is produced by $k$ single electron avalanches.
Notice that the integral $\int_0^\infty f(t-\tau_i) \Phi(\tau_i;k) d\tau_i$ is actually independent of $i$, i.e. 
\begin{equation}
\int_0^\infty f(t-\tau_i) \Phi(\tau_i;k) d\tau_i
=\int_0^\infty f(t-\tau) \Phi(\tau;k) d\tau
\end{equation}
Therefore, Equation \ref{barS4} does not depend explicitly on any of the arrival times $\tau_1,\tau_2,...,\tau_k$ and Equation \ref{barS4} becomes

$$
    \left<S(t)\right>_{_{Q_{tot}=Q}} = \sum_{k=1}^\infty R(k) N(k)\int_0^\infty f(t-\tau) \Phi(\tau;k) d\tau
\sum_{i=1}^k     
    \left<q_i\right>_{_{Q=\sum_{j=1}^k q_j}}   
$$

\begin{equation}
    \hspace{-0.3cm}=Q\sum_{k=1}^\infty R(k) N(k) \int_0^\infty f(t-\tau) \Phi(\tau;k) d\tau
\end{equation}

\begin{equation}\label{barS5}
    \hspace{0.4cm}=Q\int_0^\infty f(t-\tau) \left\{\sum_{k=1}^\infty R(k) N(k) \Phi(\tau;k)\right\} d\tau
\end{equation}

where we used 
$\sum_{i=1}^k q_i = Q$, which means that $\sum_{i=1}^k \left< q_i\right> = \left< Q \right> = Q$.

As described by Equation \ref{eqNk}, $N(k)$ expresses the probability that $k$ single electron amplification avalanches are producing a pulse with total charge $Q$.
Thus, $R(k)\cdot N(k)$ expresses the probability that in a sample of pulses with total charge $Q$, a pulse has been produced by $k$ single electron avalanches.
Consequently, the term inside the brackets of Equation \ref{barS5} expresses the average time distribution of an electron passing through the micro-mesh and initiates a single electron avalanche where the total induced charge of all the avalanches in the event is equal to $Q$.

\begin{equation} \label{averF}
    \left<\Phi(\tau)\right>_{_{Q_{tot} = Q}} =
    \sum_{k=1}^\infty R(k) N(k) \Phi(\tau;k)
\end{equation}
With this definition, Equation \ref{barS5} is written

\begin{equation}\label{barS6}
    \frac{\left<S(t)\right>_{_{Q_{tot}=Q}}}{Q}
    = \int_0^\infty f(t-\tau) 
    \left<\Phi(\tau)\right>_{_{Q_{tot} = Q}} d\tau
\end{equation}

    Equation \ref{barS6} relates the detector's electrical response to a single electron avalanche, $f(t)$, which we want to estimate, to the observed average shape of experimentally observed pulses.  
The left hand side, $\frac{\left<S(t)\right>_{_{Q_{tot}=Q}}}{Q}$, is evaluated by averaging experimental signals with electron peak charge equal to $Q$.
    The function $\left<\Phi(\tau)\right>_{_{Q_{tot} = Q}}$ describes the average electrons' time of transmission through the mesh and is taken from the simulation.
The single electron avalanche response, $f(t)$, is expressed as a flexible parametric function, and the parameters are estimated by fitting convolution of the right hand side to the normalized average shape of experimental pulses of the left hand side.

    However, it is impractical to find enough pulses with electron peak charge exactly equal to $Q$ to calculate the left hand side of Eq. \ref{barS6}.
As mentioned before, we concentrate our analysis in the high values tail of the experimental charge spectrum where the experimental pulse shape remains constant and the mean SAT and resolution effects are negligible, i.e. in this analysis
    $\frac{\left<S(t)\right>_{_{Q_{tot}=Q_\lambda}}}{Q_\lambda} = const.$ for $Q_\lambda > 15pC$.
The left hand side of Equation \ref{barS6} is then estimated as

\begin{equation}\label{barS7}
    \frac{\left<S(t)\right>_{_{Q_{tot}=Q}}}{Q}
    \rightarrow \frac{1}{N} \sum_{\lambda=1}^N 
    \frac{\left<S(t)\right>_{_{Q_{tot}=Q_\lambda}}}{Q_\lambda}
    , \hspace{2cm}Q_\lambda > 15pC
\end{equation}

\begin{figure}[H]
    \centering
    \begin{subfigure}[h]{0.49\textwidth}
        \includegraphics[width=0.95\textwidth]{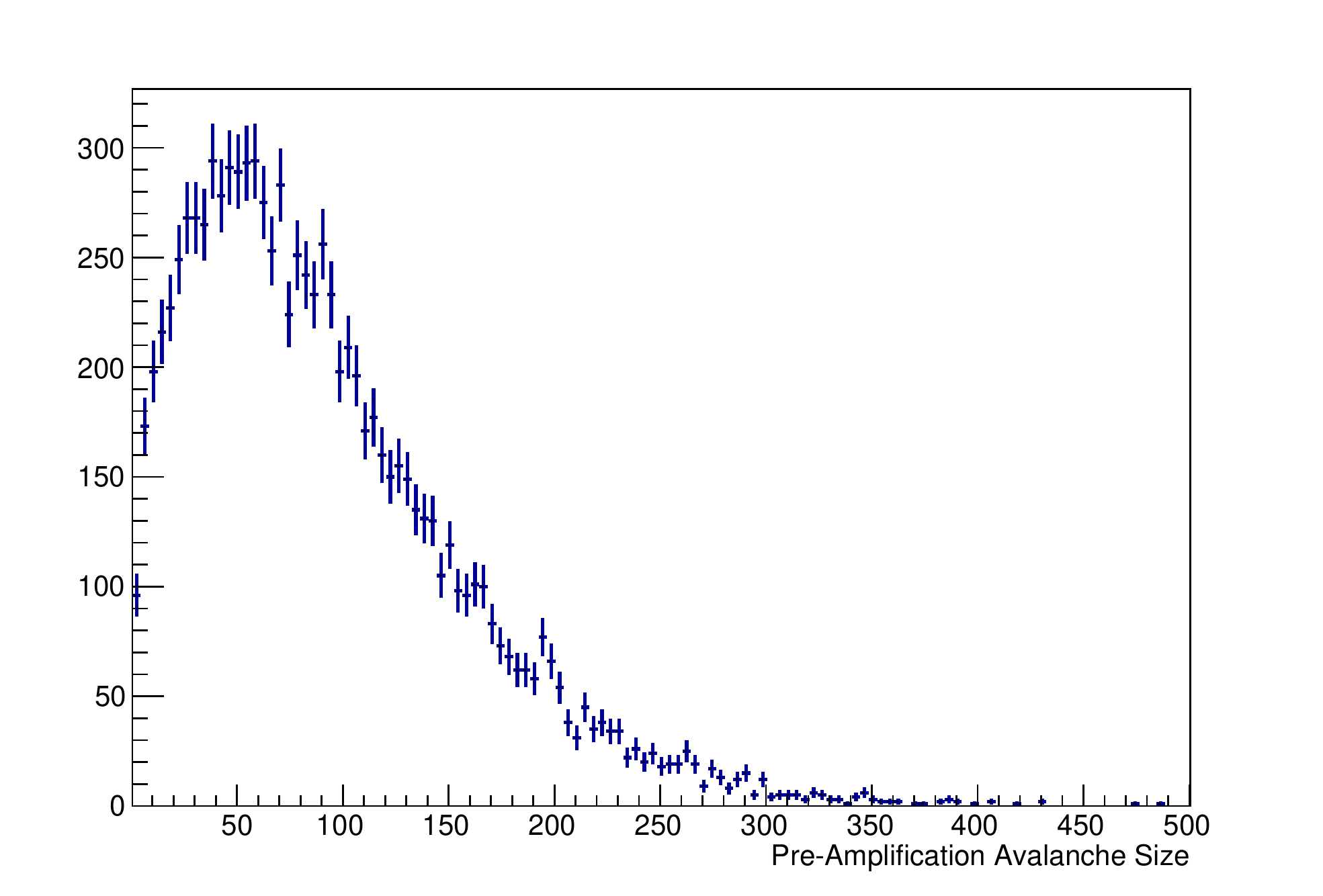}
        \caption{}
    \end{subfigure}
    \begin{subfigure}[h]{0.49\textwidth}
        \includegraphics[width=0.95\textwidth]{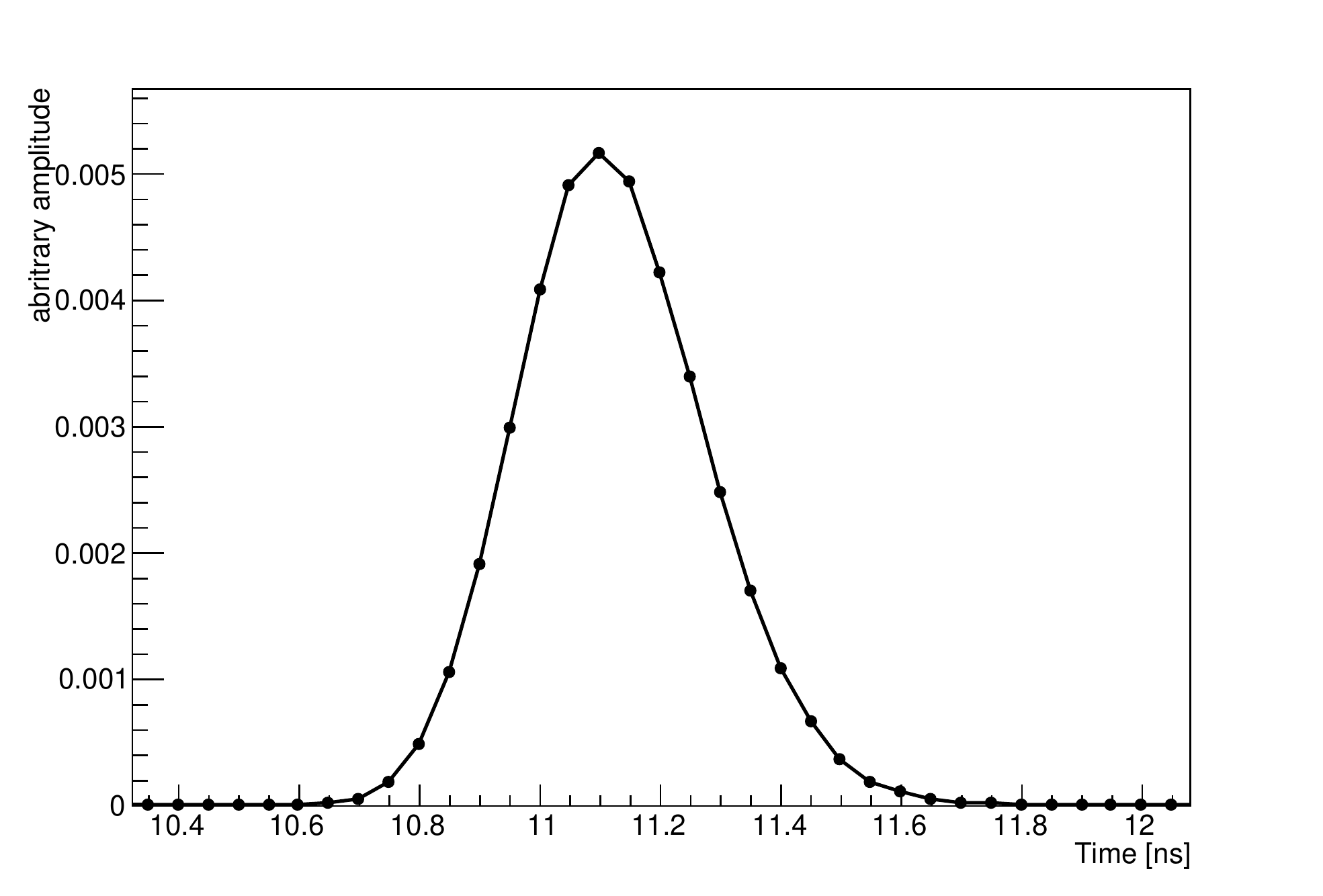}
        \caption{}
    \end{subfigure}
    \caption[Multiplicity of electrons and mean arrival time distributions.]{The distribution of the multiplicity of electrons passing through the mesh is shown on the left plot. To ensure that no SAT is affecting the result, only the events that belong in the tail of the distribution are considered $(>200)$. The right plot corresponds to the arrival time distribution of these selected events.}  
    \label{simu:fig:preconv}
\end{figure}

\begin{figure}[H]
    \centering
    \begin{subfigure}[h]{0.49\textwidth}
        \includegraphics[width=1.05\textwidth]{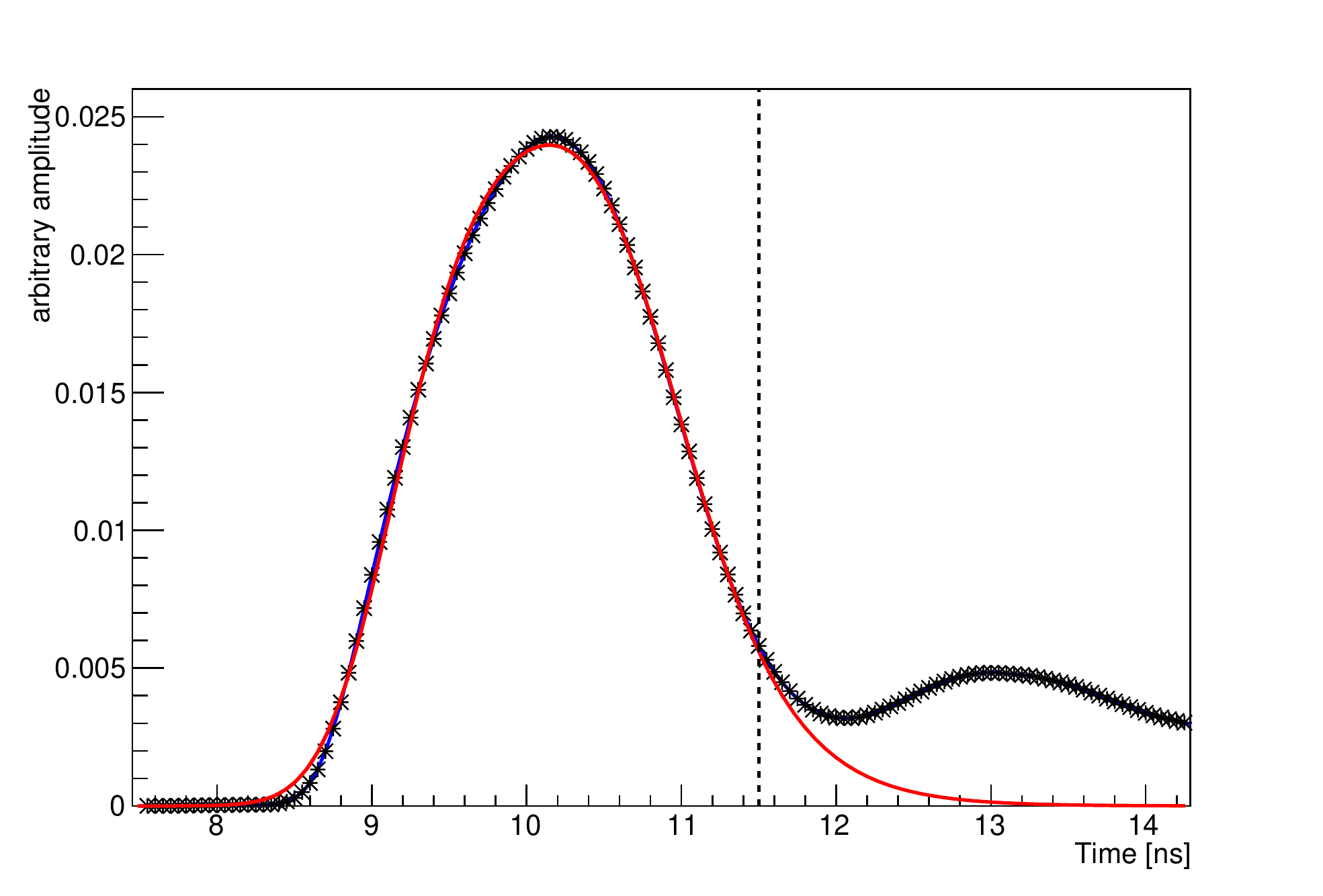}
        \caption{}
    \end{subfigure}
    \begin{subfigure}[h]{0.49\textwidth}
        \includegraphics[width=1.05\textwidth]{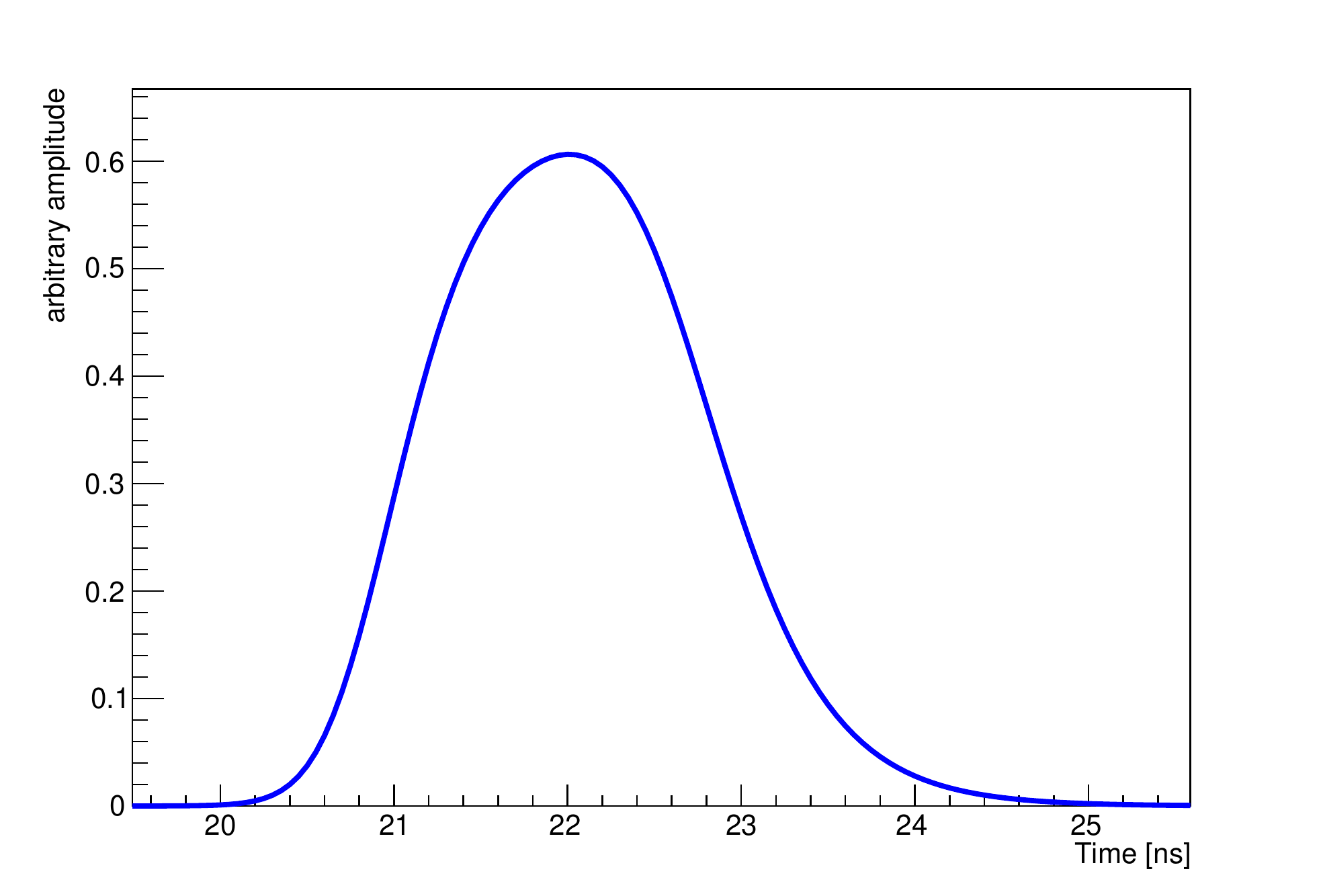}
        \caption{}
    \end{subfigure}
    \caption[Average waveforms of simulation and experiment and the final impulse response.]{(a) The black points in the right plot correspond to the experimental average waveform for events with an electron peak charge above $15.\,pC$. This region was chosen to prevent the SAT dependence from affecting the result. The fit result is shown in the red line of the plot. The dashed line corresponds to the right limit of the fit region. (b) The final impulse response as a function of time is shown.
    }  
    \label{simu:fig:resp}
\end{figure}

Specifically:
\begin{enumerate}[label=\alph*)]
    \item Waveforms at the high tail of the charge spectrum of the experimental pulses are selected. As an example, in the case of anode voltage of $450\,V$ and drift voltage of $425\,V$, waveforms with electron peak charge greater than $15\,pC$ were selected, which as shown in Figure \ref{simu:fig:postconv} have an almost stable SAT and timing resolution.
    \item  Equation \ref{barS7} is used to define the average experimental shape. 
        Before averaging, the normalized pulses have to be synchronized, using the reference time as found from the reference timing detector. 
        As an example, the black points of Figure \ref{simu:fig:resp} represent the mean shape electron peak shape for electron peak charge $> 15\,pC$, when the chamber was operated with $450\,V$ and $425\,V$ anode and drift voltages, respectively. 
    \item The detector response to a single electron avalanche, $f(t)$, is parameterized as
\begin{equation}\label{simu:eq:glogi}
    f(t;p_0,p_1,p_2,p_3,p_4,p_5,p_6) = \frac{p_0}{\left( 1+ e^{-\left(t-p_1\right)p_2}\right)^{p_3}}
    - \frac{p_0}{\left( 1+ e^{-\left(t-p_4\right)p_5}\right)^{p_6}}
\end{equation}
        where $p_0,p_1,...,p_6$ are parameters estimated by fitting the mean experimental electron peak pulse shape, using the convolution defined in Equation \ref{barS6}.
     
     \item The time distribution $\left<\Phi(\tau)\right>_{_{Q_{tot} = Q}}$ is evaluated using the simulation of the electrons produced in the pre-amplification region passing through the micro-mesh. 
         As shown in Figure \ref{simu:fig:example2}, according to the simulation, the time of an electron passing through the micro-mesh follows an Inverse Gaussian distribution with mean and sigma values which are in general dependent on the number of electrons ($k$) produced in the pre-amplification region.
         However, for high values of $k$ this time distribution remains practically constant and high values of $k$ naturally correspond to high total charge $Q$.
         Therefore, following the same arguments as in b), $\left<\Phi(\tau)\right>_{_{Q_{tot} = Q}}$ is evaluated using simulated events at which the number of electrons that have passed through the mesh are on the tail of the distribution $( \left<\Phi(\tau)\right>_{_{Q_{tot} = Q}} \rightarrow \Phi(\tau)_{_{Q_{tot} > Q_c}} )$, i.e. $Q_c>200$ electrons, as shown in Figure \ref{simu:fig:preconv}\,a.
    \item To fit the mean experimental electron peak pulse shape, the convolution in Equation \ref{barS6} is approximated as 
        \begin{equation}
            \int_0^\infty f(t-\tau) \Phi(\tau)_{_{Q_{tot} > Q_c}} d\tau \approx \sum_{i=1}^N f(t-\tau_i) \Phi(\tau_i)_{_{Q_{tot} > Q_c}} \Delta\tau.
        \end{equation}
        for (large) $N$ steps, of a (small) size $\Delta\tau$. 
        The red solid curve shown in Figure \ref{simu:fig:resp}\,a represents the result of such a fit where the mean experimental pulse shape has been fitted up to a point shown by the black dashed line in the Figure. 
        By restricting the limits of the fit the contribution of the ion component of the waveform is avoided, as well as the electronic reflections shown at the right of the dashed line in Figure \ref{simu:fig:resp}\,a.
\end{enumerate}

The resulting fit parameters are $p_0 = 2.973,\,p_1 = 0.0358,\,p_2= 18.676,\,p_3=3.376,\,p_4=2.727,\,p_5=20.297,\,p_6=3.683$. 
The fit is shown in Figure \ref{simu:fig:resp}\,a
in red.
The graphical representation of the detector's response to single electron amplification avalanches $f(t)$, is shown in Figure \ref{simu:fig:resp}\,b.

\subsection{Generation of Waveforms}
After the estimation of the parameters in Equation \ref{simu:eq:glogi}, the response of a single electron amplification avalanche is described and the procedure in Section \ref{sec:ampli}
can be followed to generate simulated waveforms with the same format as the observed one.
For example, assuming that a single photoelectron beginning from the cathode produced secondary electrons in the pre-amplification region and $N$ of them transmitted through the micro-mesh and entered the amplification region.
The $i$-th of $N$ electrons passed through at time $\tau_i$. For each of these, a random uniform number is sampled to decide whether it attached. If the electron is decided to survive and initiate an avalanche, a random amplification value $q_i$ is randomly sampled from the P\'{o}lya parameterization $G(q)$ described in Section \ref{sec:ampli}.
If the electron re-attached in the gaseous mixture, the ``amplification'' is set to $q_i= 0$.
The analog signal $S(t)$ that would be readout from the detector is the superposition of the $N$ electron responses:

\begin{equation}\label{simu:analog}
    S(t) = \sum_{i=1}^N q_i\cdot f(t-\tau_i)
\end{equation}
where $f(t)$ is the shape of the single electron avalanche response.
Lastly, this analog expression in Equation \ref{simu:analog} takes a digital form by sampling the expression every $50\,ps$.
The digital form is then stored in the same format and analyzed in the same way as the experimental data.

Because of the several normalization in the averaging of the waveforms, the absolute amplification of the electronics is lost.
There must a exist, though, an absolute scaling factor which includes this absolute amplification and can transform the arbitrary amplitude of the simulated waveforms to the real voltage that would be measured.
This factor can be recovered by scaling the distribution of the simulated electron peak charge to the distribution of the experimental electron peak charge.
To find it, a P\'{o}lya distribution $P(x;a_0,a_1,a_2)$ is fitted to the charge spectrum of the simulated electron peak. 
The P\'{o}lya $P(x;\hat{a_0},\hat{a_1},\hat{a_2})$, where parameters $\hat{a_0},\hat{a_1},\hat{a_2}$ are the estimations of parameters $a_0,a_1,a_2$, must describe the experimental charge spectrum when it is transformed as

\begin{equation}
P_{exp}(x;b_0;b_1) = b_0\cdot P(b_1 \cdot x |  \hat{a_0}, \hat{a_1},\hat{a_2})
\end{equation}
where $b_0$ is the normalizing factor between the two distributions and $b_1$ is the scaling factor between the values of the waveforms' charge.
By fitting the function $P_{exp}(x;b_0,b_1)$ on the experimental charge distribution, the scaling factor $b_1$ is found.
Obviously, if $I$ is the integral of a simulated electron peak waveform, its corresponding charge is $Q = \frac{I}{b_1}$.

\begin{figure}[H]
    \centering
        \includegraphics[width=1.05\textwidth]{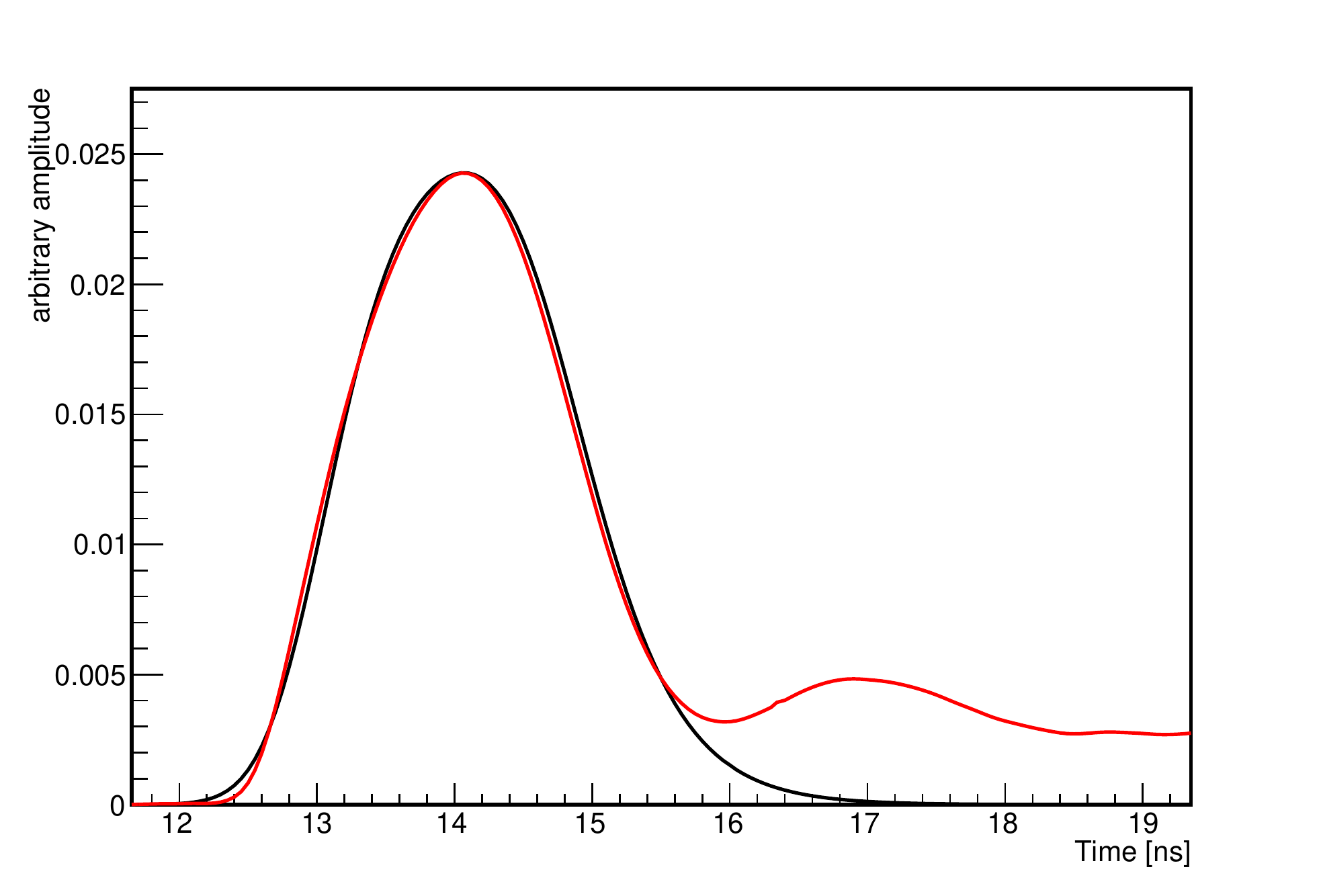}
        \caption[Average waveforms of experimental and simulated signals.]{The average waveforms of the experimental and simulated waveforms are illustrated with red and black, respectively, for pulses with charge in the region $[5\,pC,7\,pC]$}  
    \label{simu:fig:resp2}
\end{figure}

Even though the detector response to a single electron amplification avalanche is estimated, only events from the high tail of the charge spectrum were used.
It has to be verified that the shape of the electron peak is similar in all regions of charge.
As an example, a comparison between simulated (black) and experimental signal (red) is shown in Figure \ref{simu:fig:resp2},
for pulses with a charge in the range $[5\,pc,7\,pc]$.
The similarity of the two shapes demonstrates the reliability of this technique.

\begin{figure}[H]
    \centering
    \begin{subfigure}[h]{0.49\textwidth}
        \includegraphics[width=0.95\textwidth]{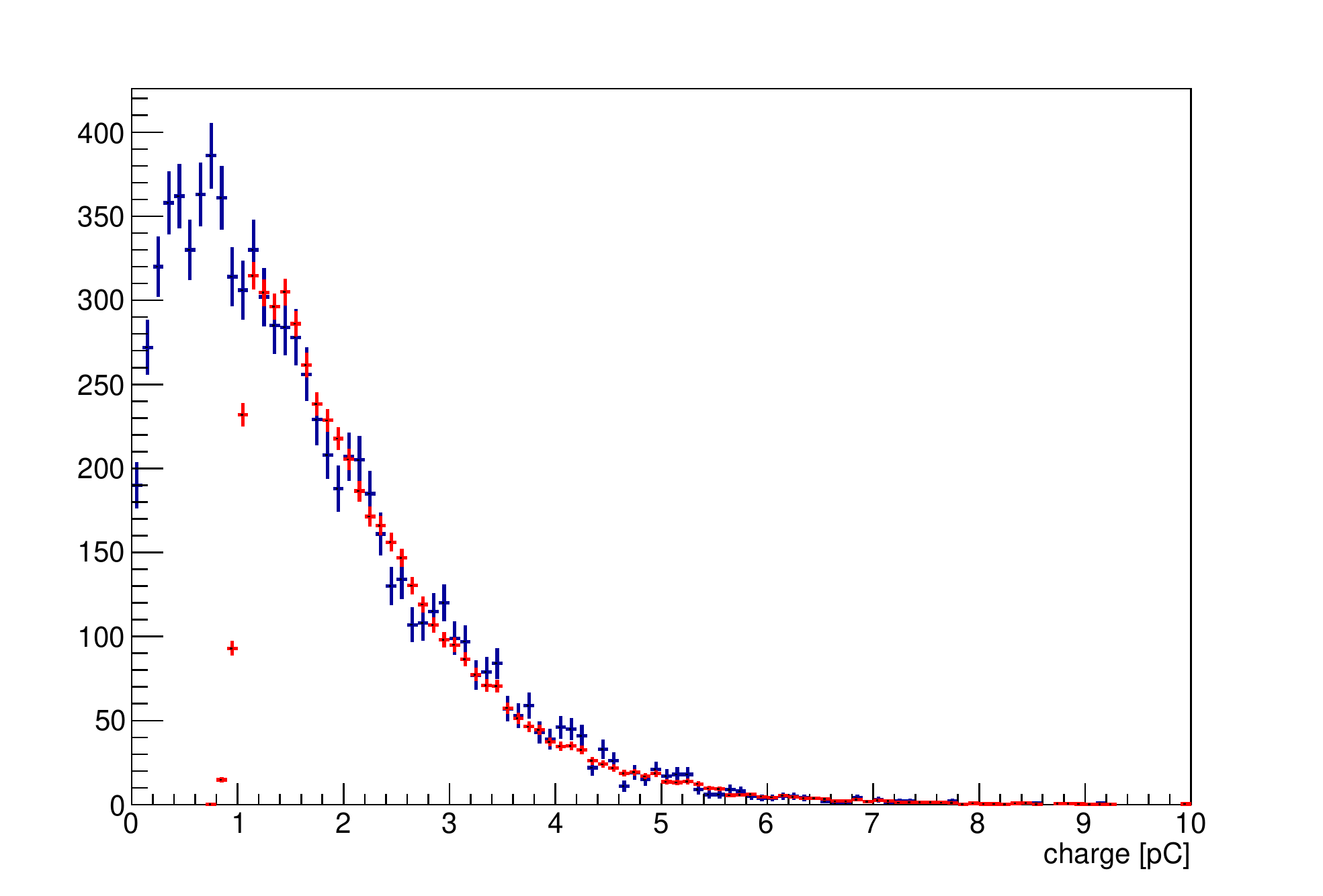}
        \caption{}
    \end{subfigure}
    \begin{subfigure}[h]{0.49\textwidth}
        \includegraphics[width=0.95\textwidth]{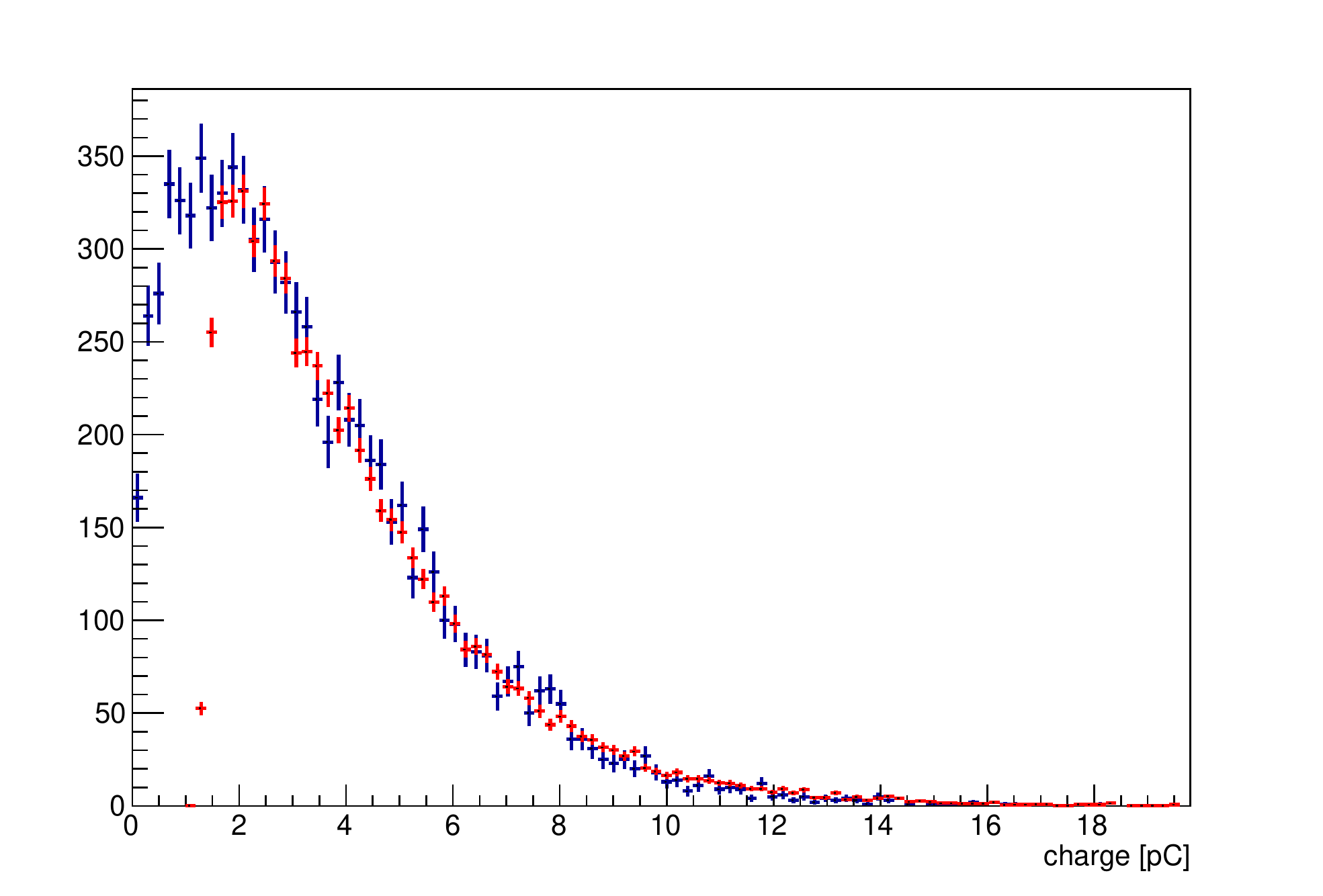}
        \caption{}
    \end{subfigure}
    \begin{subfigure}[h]{0.49\textwidth}
        \includegraphics[width=0.95\textwidth]{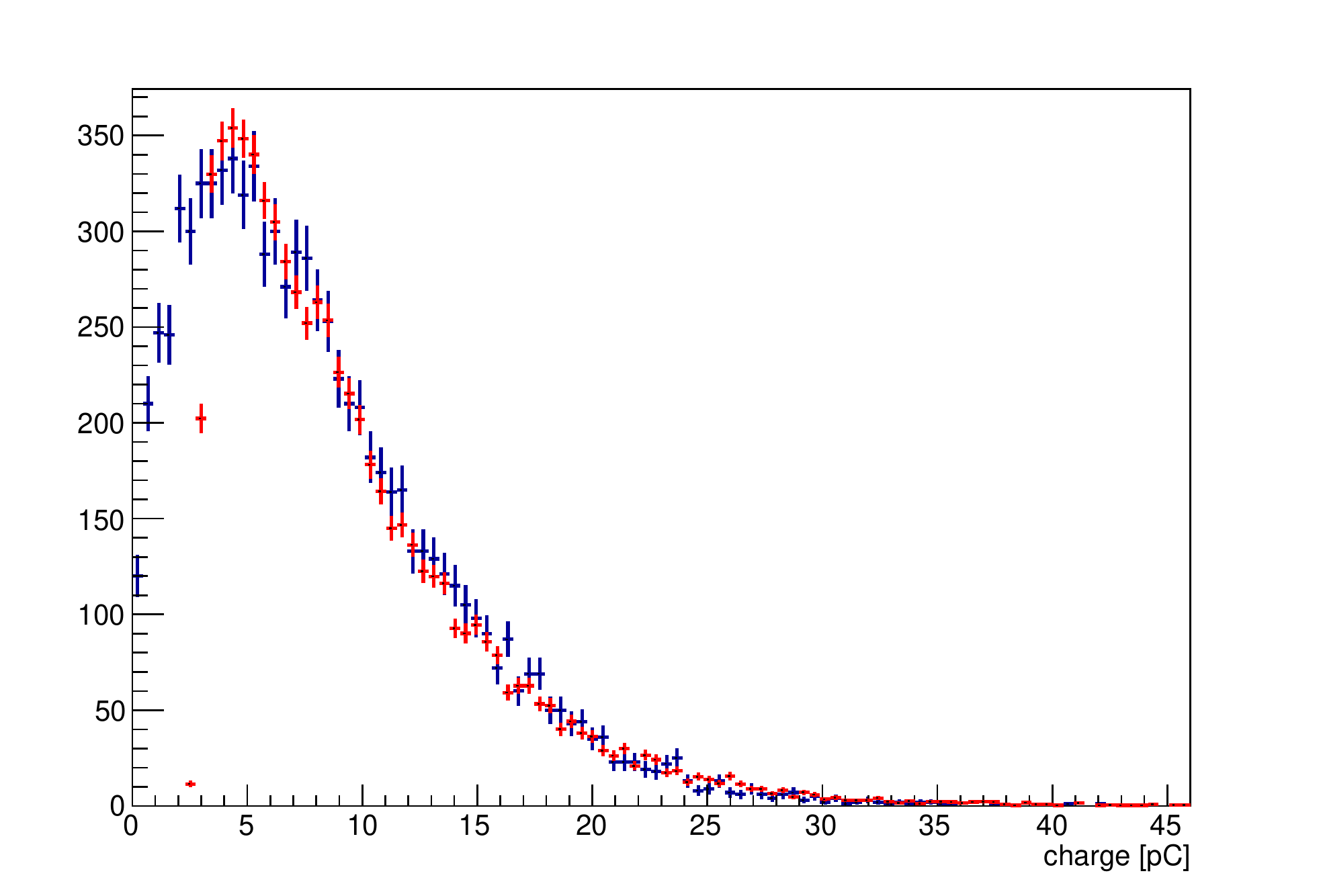}
        \caption{}
    \end{subfigure}
    \caption[Simulated electron peak charge distribution.]{Distribution of the simulated electron peak charge (blue), scaled with a factor $G$, such that the distribution matches the experimental one in shape (red). Anode voltage is $450\,V$ and drift voltage is (a) $375\,V$, (b) $400\,V$, and (c) $425\,V$. The scaling factors for each case are (a) $G = 30.2$, (b) $G = 27.8$, (c) $G = 21.9$.}  
    \label{simu:fig:scales}
\end{figure}

Furthermore, the scaling factor $b_1$ should be independent of the operating voltages, i.e. it should be expected that the value of the scale factor $b_1$ is the same across different drift voltage settings without any extra fine-tuning.
The detector response to single photoelectrons was simulated with the anode voltage set to $450\,V$ and the drift voltages at $325\,V,350\,V$, $375\,V,4    00\,V$ and $425\,V$. 
Fits to estimate the scaling factor are shown in Figure \ref{simu:fig:scales} for cases where the drift voltage was set to $375\,V, 400\,V$ and $425\,V$. 
In spite of the expectation of it to be the same, the values found for the scaling factors in the three cases were significantly different, namely $b_1 = 30.2,\,27.8\,21.9$, respectively for the three different drift voltages. 

\begin{figure}[H]
    \centering
    \includegraphics[width=0.95\textwidth]{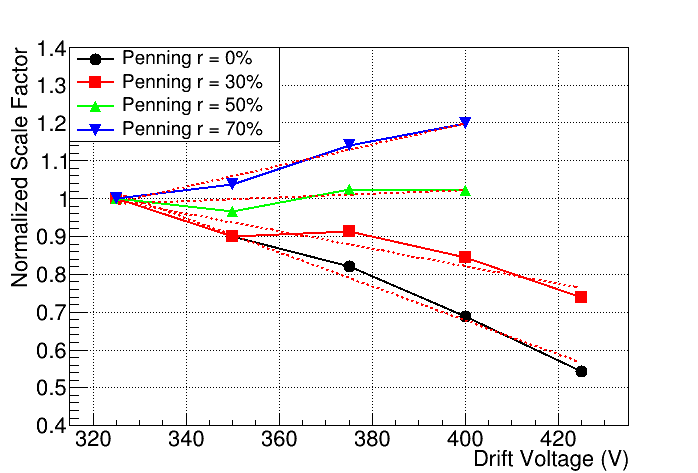}
    \caption[Scaling factor with respect to the drift voltage for different Penning transfer rates.]{
        Normalized scaling factor with respect to the scaling factor at drift voltage of $325\,V$, versus the drift voltage.
        Different colors correspond to different transfer rates with (black) $r=0$, (red) $r=0.3$, (green) $r=0.5$ and (blue) $r=0.7$, normalized to the scaling factor at the drift voltage of $325\,V$. 
    }  
    \label{simu:fig:penning}
\end{figure}

The simplest explanation for this dependence on the drift voltage is because of the ignorance of the Penning transfer effect.
Indeed, by repeating the simulation with different values for the Penning transfer rate $r$, specifically for $r = 0.3,\,0.5$ and $0.7$, the dependence appears to vanish for approximately $r=0.5$. 
Figure \ref{simu:fig:penning} shows how the scaling factor changes with respect to the drift voltage for the different Penning transfer rates (black) $r=0$, (red) $r=0.3$, (green) $r=0.5$ and (blue) $r=70$, normalized to the scaling factor at the drift voltage of $325\,V$.
Because the experimental charge distributions are truncated, there is a large uncertainty on the scaling factors and a better estimation of the transfer rate is very difficult be made.

\section{Macroscopic Analysis of the Simulation}
\subsection{SAT and Time Resolution Dependence on the Electron Peak Size}

A variety of anode and drift voltage settings were used to simulate the single photoelectron response of the PICOSEC.
The digitized signals were analyzed by the same analysis code that the real experimental data are analyzed with and the electron peak arrival time has been estimated by the same fitting procedures.
In this section, the SAT and time resolution dependence on the size of the electron peak is compared between in real and simulated data.
After calculating and adjusting for the gain factors, the SAT and resolution dependencies on the electron peak charge can be compared between simulation and experiment. 

\begin{figure}[H]
    \centering
    \begin{subfigure}[h]{0.49\textwidth}
        \includegraphics[width=0.95\textwidth]{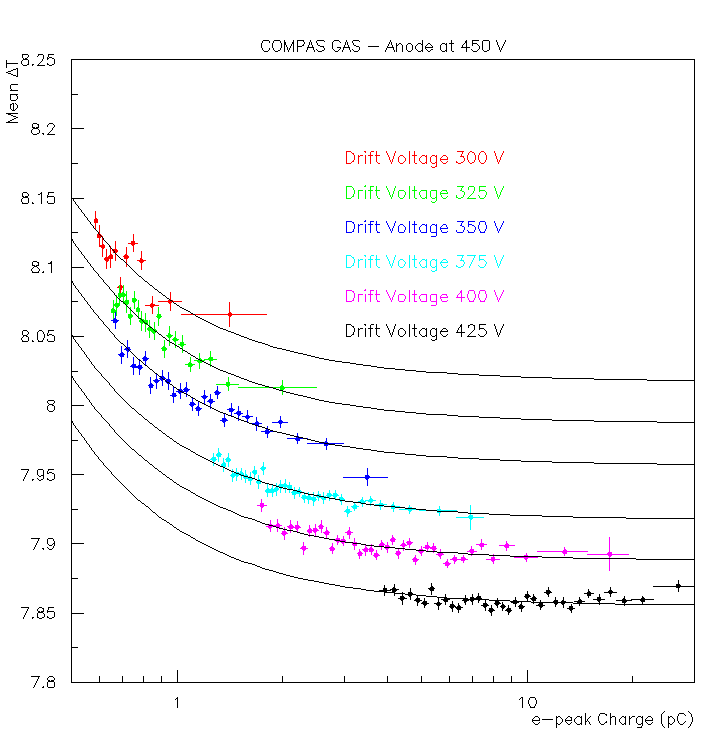}
        \caption{}
    \end{subfigure}
    \begin{subfigure}[h]{0.49\textwidth}
        \includegraphics[width=0.95\textwidth]{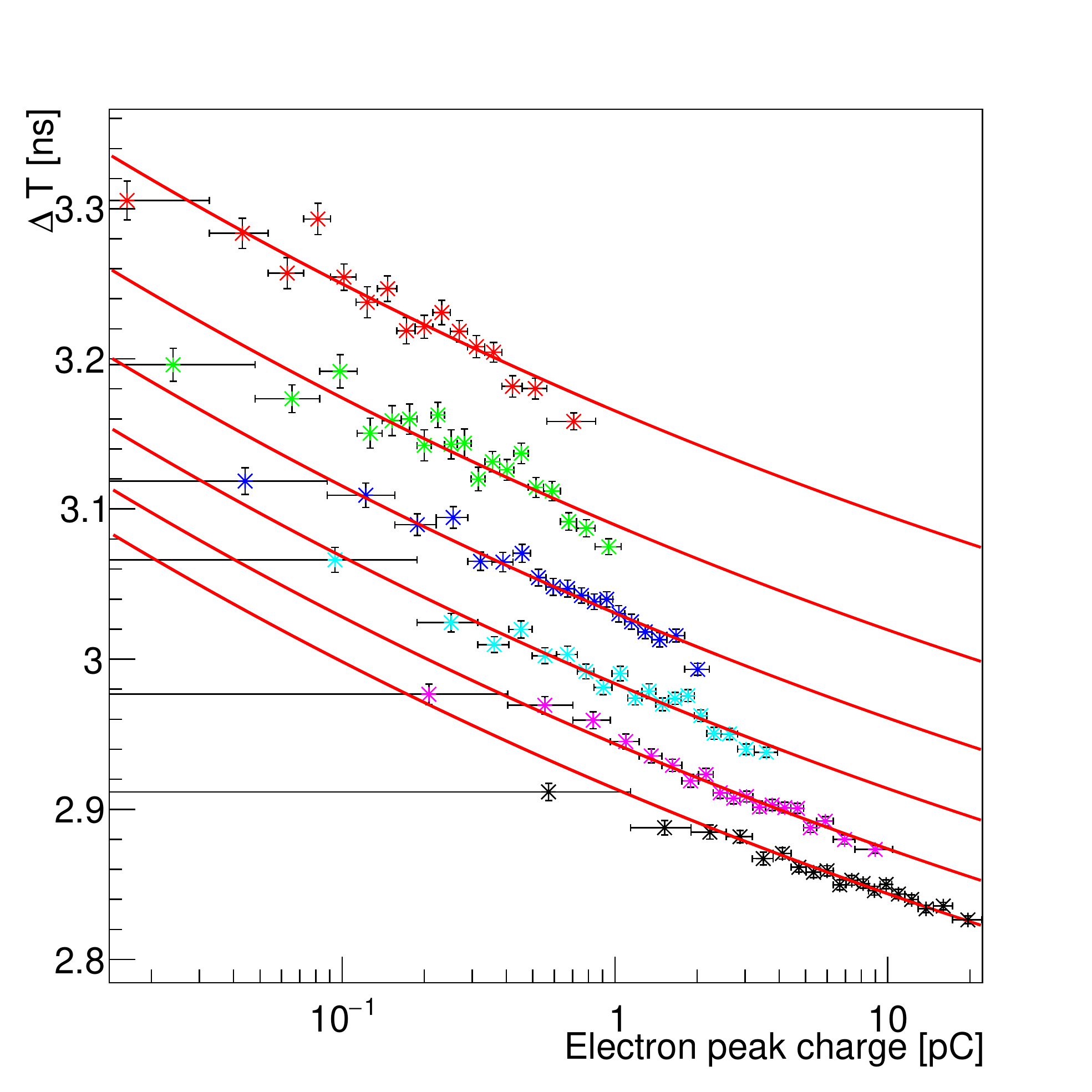}
        \caption{}
    \end{subfigure}
    \caption[Mean signal arrival time for experiment and simulation without noise.]{Both figures show the mean temporal distance with respect to the electron peak charge for different drift voltages and an anode voltage of $450\,V$, with COMPASS gas filling.
    The left figure corresponds to the experimental data, whilst the right figure corresponds to the simulated data.
    Solid lines indicate the corresponding parameterizations through a constant term for each drift voltage and a global power law, found to be common to all drift voltages (Equation \ref{eq-power}).}  
    \label{simu:fig:slewing}
\end{figure}

The dependence of SAT on the electron peak charge is shown in Figure \ref{simu:fig:slewing} for the different drift voltages in both experimental and simulated data.
Experimental data are presented in Figure \ref{simu:fig:slewing}\,a, while simulated data are presented in Fig. \ref{simu:fig:slewing}\,b.
As explained before, the experimental data points have been fitted with a power law term plus a constant where the parameters of the power law are common to all the different drift voltage settings, while a different constant term is assigned to each drift voltage setting.
The same fit constraints are imposed successfully to the simulated data.
This qualitative similarity demonstrates the fact that the simulation supports the claim that besides the different drift velocities at different drift fields (which is expressed by the different values of the constant term at different drift voltages), the SAT dependence on the electron peak size is almost independent of the drift voltage.
The resulting fits are illustrated as solid lines in these figures.
However, comparing the SAT dependence on the electron peak size between experiment and simulation, it is clear that the experimental SAT dependence is much steeper than the one predicted by the simulation.
It has to be emphasized that the Penning effect barely affects these dependencies and mostly increases only the number of secondary electrons and as such, the total amplification.

\begin{figure}[H]
    \centering
    \begin{subfigure}[h]{0.49\textwidth}
        \includegraphics[width=0.95\textwidth]{simunote/simu-exp-res.png}
        \caption{}
    \end{subfigure}
    \begin{subfigure}[h]{0.49\textwidth}
        \includegraphics[width=0.95\textwidth]{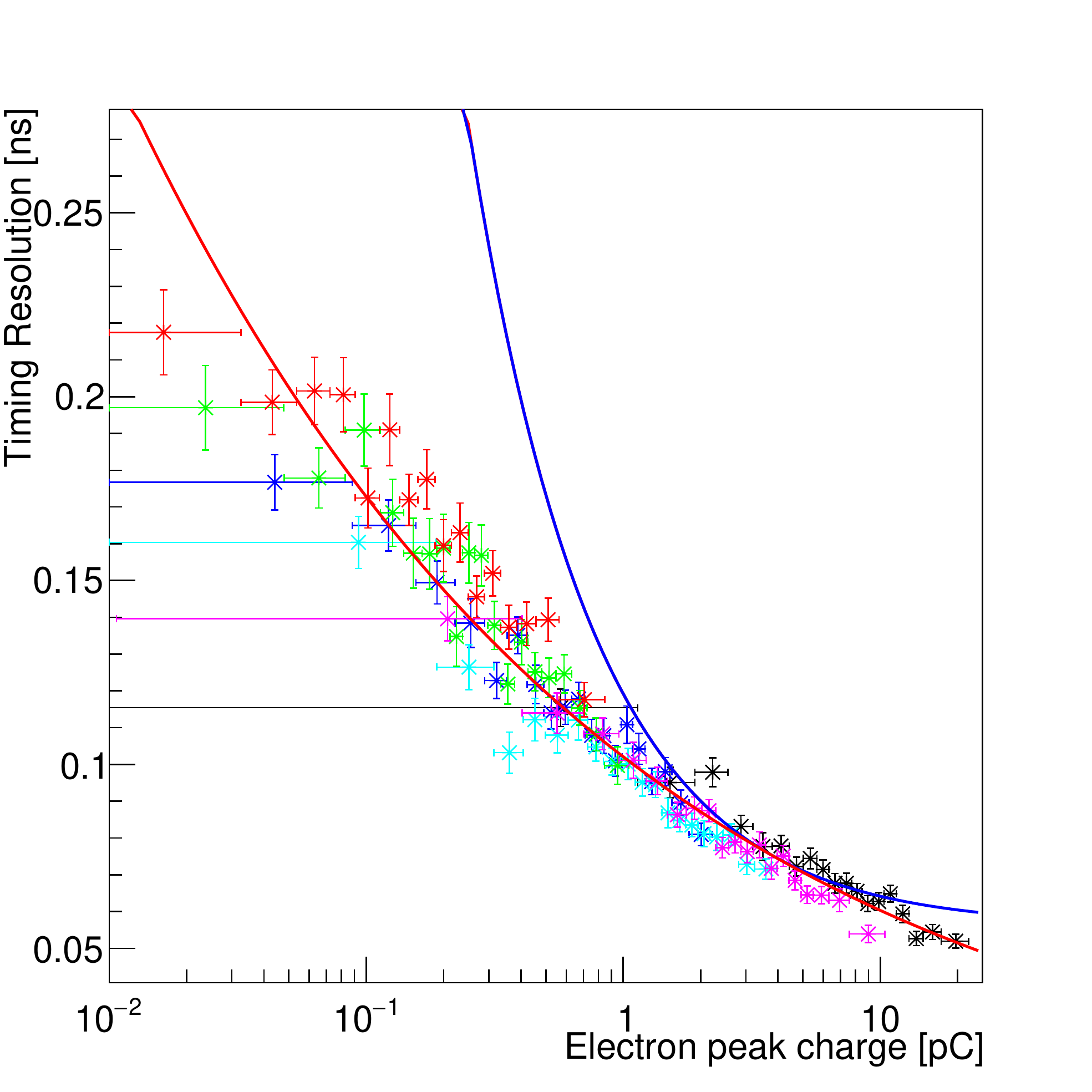}
        \caption{}
    \end{subfigure}
    \caption[Time resolution with respect to electron peak charge for experiment and simulation without noise.]{Both figures show the time resolution with respect to the electron peak charge for different drift voltages and an anode voltage of $450\,V$, with COMPASS gas filling.
    Figure (a) corresponds to the experimental data, whilst (b) corresponds to the simulated data.
    Solid line in (a), and the red line in (b), indicate the corresponding parameterizations through a constant term and a power law, both common to all drift voltages (Eq. \ref{eq-power}).
    The blue line in (b) is the parameterization of the experimental data in (a).
    }  
    \label{simu:fig:timres}
\end{figure}

The dependence of the time resolution on the electron peak charge is shown in Figure \ref{simu:fig:timres} for all different drift voltages and for both the experimental and simulated data.
Experimental data are presented in Figure \ref{simu:fig:timres}\,a, while the simulation prediction is presented in Figure \ref{simu:fig:timres}\,b.
In both experiment and simulation, and independently of the drift voltage, the time resolution is found to follow the same curve on electron peak charge, which means that the simulation supports the experimental observation that the resolution depends on the electron peak size almost independently of the drift voltage.
Again, the time resolution is parameterized with the sum of a power law and a constant term (Equation \ref{eq-power}). 
This fit is shown with the black line in Figure \ref{simu:fig:timres}\,a and the red line in Fig. \ref{simu:fig:timres}\,b. 
The blue line in Fig. \ref{simu:fig:timres}\,b corresponds to the parameterization of the resolution with respect to the charge found in the experimental data.
Comparing the blue line with the prediction of the simulation in Figure \ref{simu:fig:timres}\,b that, even though, the parameterizations agree in a region of the electron peak charge, they diverge especially in small electron peak charges.

\begin{figure}[H]
    \centering
    \begin{subfigure}[h]{0.49\textwidth}
        \includegraphics[width=0.95\textwidth]{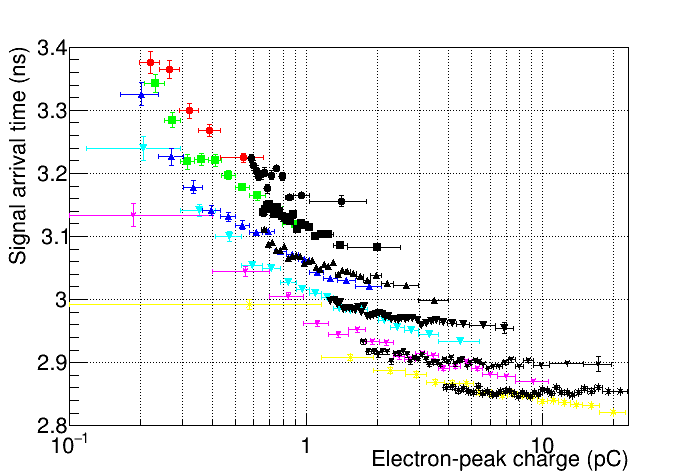}
        \caption{}
    \end{subfigure}
    \begin{subfigure}[h]{0.49\textwidth}
        \includegraphics[width=0.95\textwidth]{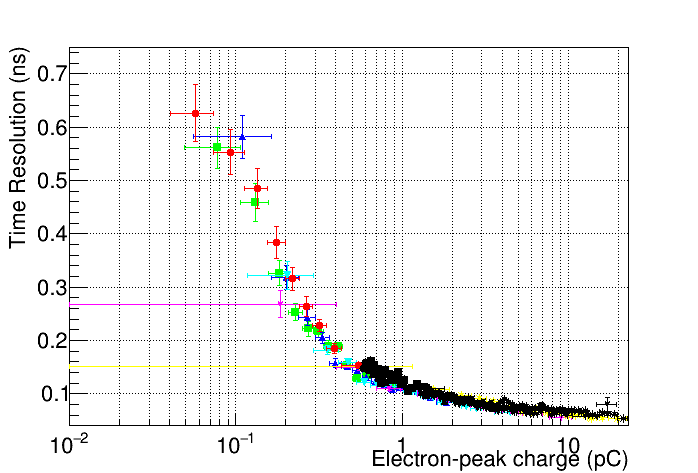}
        \caption{}
    \end{subfigure}
    \caption[Mean signal arrival time and time resolution with respect to the electron peak charge for simulation with noise.]{
        (a) Mean SAT as a function of the electron peak charge.
        (b) Time resolution as a function of the electron peak charge.
        In both figures black points correspond to experimental data while colored correspond to simulated points.
        The gas used is the COMPASS gas with an anode voltage of $450\,V$ and for drift voltages of (red) $300\,V$, (green) $325\,V$, (blue) $350\,V$, (cyan) $375\,V$, (magenta) $400\,V$ and (yellow) $425\,V$.
    }  
    \label{simu:fig:incnoise}
\end{figure}

The reason that the experimental and simulated points exhibit such a different functional dependence is primarily because of the electronic noise.
Indeed, by including the very simple case of Gaussian electronic noise, a very similar dependence both on the mean SAT and the time resolution is achieved.
The Gaussian noise is implemented on a waveform-by-waveform basis.
For each point of the waveform a random number is drawn from a Gaussian, with mean equal to $0$ and a standard deviation equal to a noise RMS $\sigma_n$, and is added on the waveform point.
In this case, the noise RMS was chosen to be equal to $\sigma_n = 2.5\, mV$. 
Figures \ref{simu:fig:incnoise} show the comparison between simulated (colored) and experimental (black) points when including this electronic noise.
Figure \ref{simu:fig:incnoise}\,a corresponds to the mean SAT as a function of the electron peak charge while Figure \ref{simu:fig:incnoise}\,b to the time resolution again versus the electron peak charge.  
The gas used is the COMPASS gas with an anode voltage of $450\,V$ and for drift voltages of (red) $300\,V$, (green) $325\,V$, (blue) $350\,V$, (cyan) $375\,V$, (magenta) $400\,V$ and (yellow) $425\,V$.

\subsection{Dependence of Mean SAT and the Electron Peak Shape}

\begin{figure}[H]
      \centering
      \begin{subfigure}[h]{0.49\textwidth}
          \includegraphics[width=0.95\textwidth]{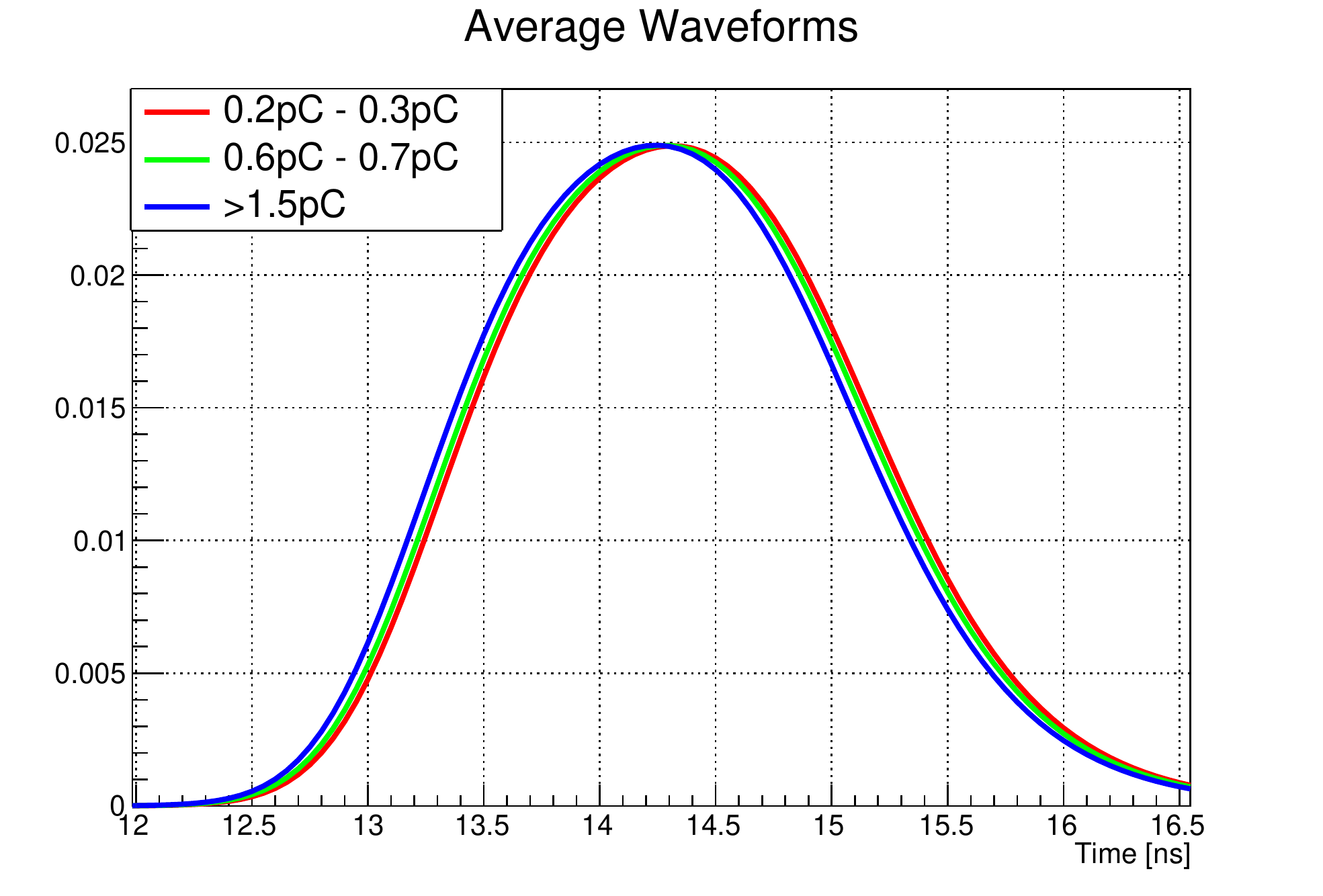}
          \caption{}
      \end{subfigure}
      \begin{subfigure}[h]{0.49\textwidth}
           \includegraphics[width=0.95\textwidth]{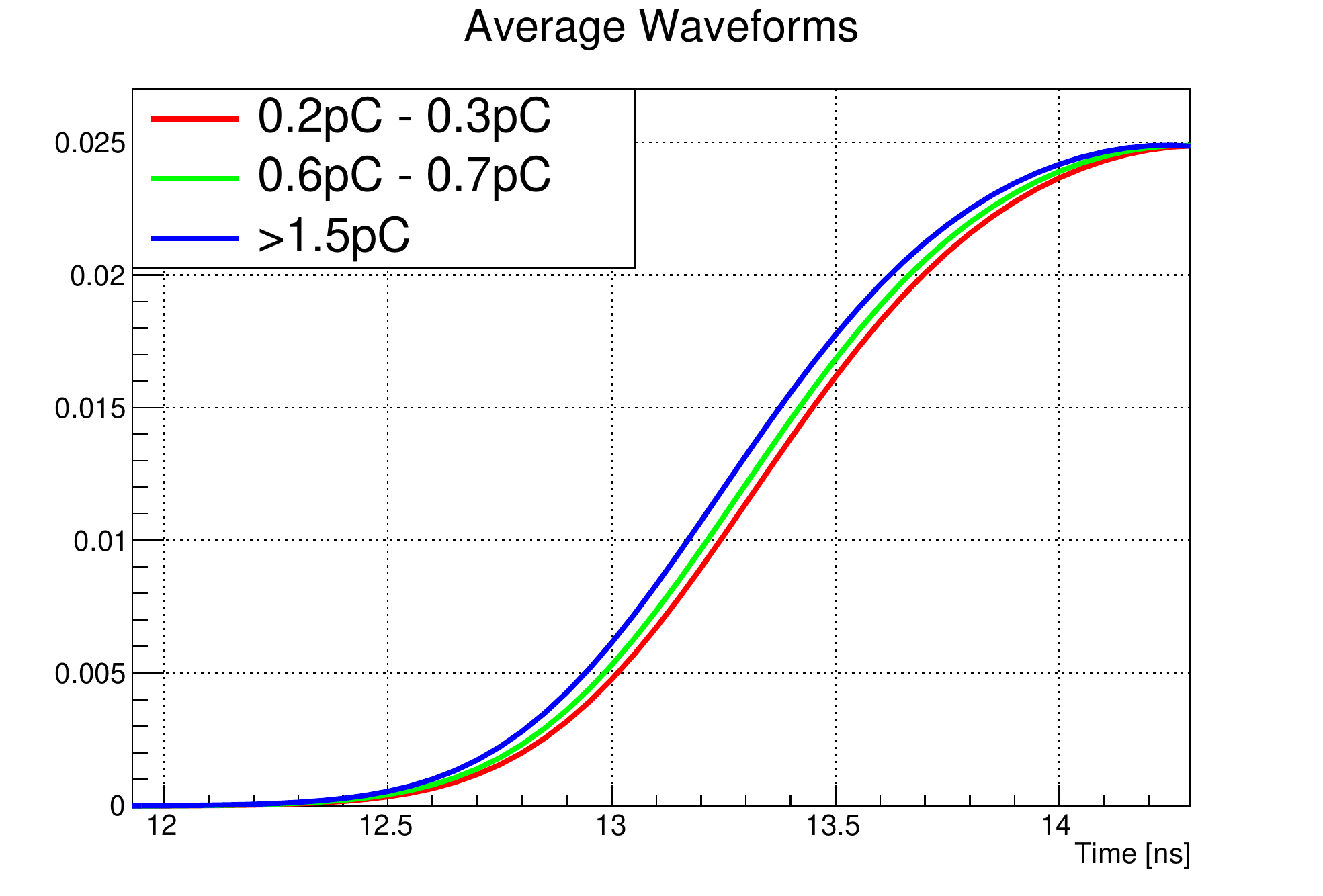}
           \caption{}
      \end{subfigure}
      \caption[Average waveforms in simulation.]{Both figures show the average simulated waveforms in different bins of electron peak charge, denoted by the color code.
         In (a), the whole electron peak is shown, whilst in (b) the focus is on the leading edge.}
      \label{simu:fig:average}
 \end{figure}

It was argued that the dependence of the mean SAT on the size of the electron peak is not a weakness of the constant fraction discrimination technique.
By averaging experimental waveforms in different narrow bins of the electron peak's size, a constant pulse shape is observed independent of the electron peak's size.
In addition to this, the average pulse is shifted in time with respect to an average pulse in a different narrow bin of the electron peak's size.
Exactly the same effects are reproduced in the simulation.
By averaging simulated waveforms in different narrow bins of the electron peak's size, as shown in Figure \ref{simu:fig:average}, the same behaviour is observed as can be seen by comparing Figure \ref{simu:fig:average} and Figure \ref{aver-comp}.

Since the simulation accurately predicts the same characteristics on the mean SAT, it will be explored in the next section so as to associate the origin of this timing effect with a mechanism that takes place in a microscopic level.

 \subsection{Correlation between Microscopic and Macroscopic Time}\label{sec:sim:corr}

 Since the signals' timing characteristics are reproduced almost quantitatively by the simulation, the mechanism that produces these timing characteristics can be investigated using the simulation.
 The first step towards this investigation is to associate the macroscopic variables (electron peak charge, waveform's timing) to microscopic ones. 
The fact that the electron peak charge is proportional to the number of electrons produced is trivial.
Roughly, the number of secondary electrons in the pre-amplification region is proportional to the electron peak charge, subject to a small smearing from the amplification region.
The microscopic variable corresponding to the waveform's timing is not as obvious.

Assuming that the small-gap amplification region is very fast, i.e. the time spread introduced in this region is negligible, then the signal arrival time should depend on the time of the electrons in the pre-amplification avalanche passing through the mesh.
As discussed in the previous Section and as shown in Figure \ref{simu:fig:example2}, both the mean SAT and the spread of the timing distribution depends on the number of electrons passing through the micro-mesh.

\begin{figure}[H]
    \centering
    \begin{subfigure}[h]{0.49\textwidth}
        \includegraphics[width=0.95\textwidth]{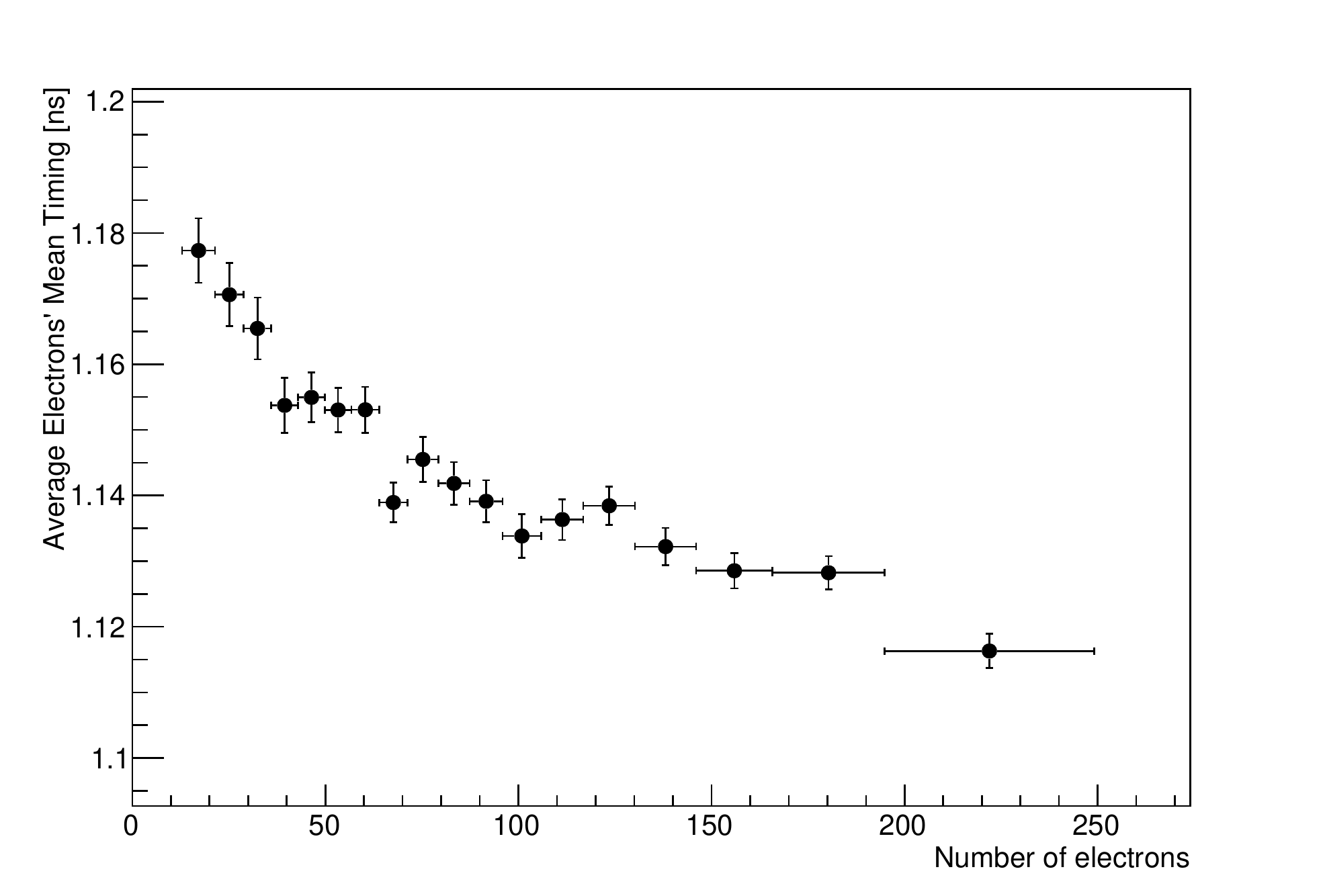}
        \caption{}
    \end{subfigure}
    \begin{subfigure}[h]{0.49\textwidth}
        \includegraphics[width=0.95\textwidth]{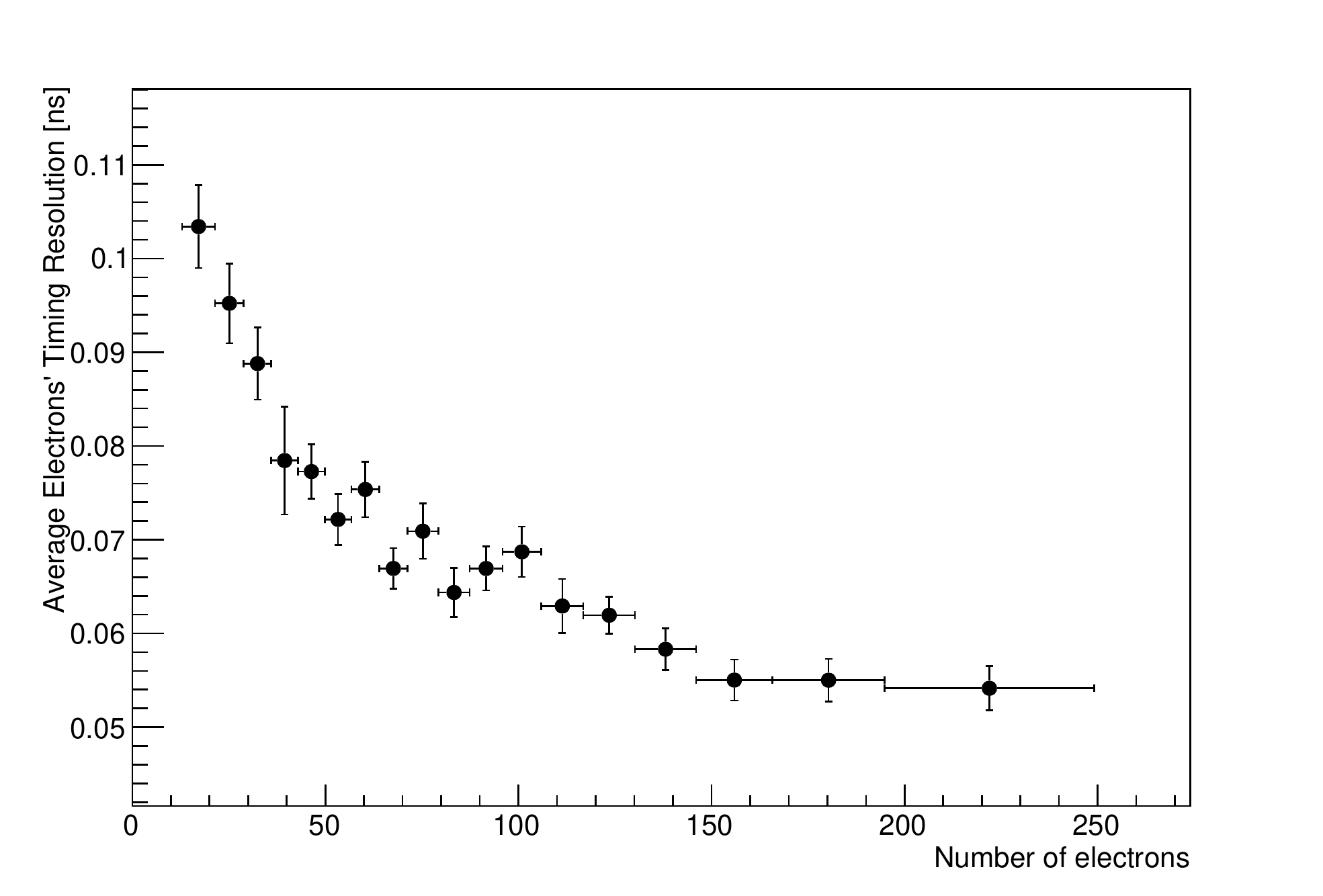}
        \caption{}
    \end{subfigure}
    \caption[Average of mean time of all electrons and time resolution with respect to number of electrons.]{(a) Average of the mean time of all electrons passing through the mesh with respect to the number of these electrons. (b) Time resolution of the mean time of all electrons passing through the mesh with respect to the number of these electrons. These points correspond to an anode and drift voltage of $450\,V$ and $425\,V$, respectively.}  
    \label{simu:fig:slewres}
\end{figure}

In the simulation, apart from the characteristics that would be observed in the experiment, all of the microscopic history of each electron is available.
The analysis proceeded in terms of the secondary number of electrons ,$N_p$, in the pre-amplification avalanche.
For each simulated event, the average time $\bar{\tau}$ of all electrons when passing through the mesh is calculated.
The distribution of $\bar{\tau}$ is then investigated with respect to the number of electrons passing through the mesh.
The a) mean and b) sigma values of the $\bar{\tau}$ distributions with respect to the related $N_p$ are presented in Figure \ref{simu:fig:slewres}, where the horizontal bars denote the size of the bin.
The mean values exhibits a similar dependence on the $N_p$ as the SAT dependence on the electron peak size, while the sigma values depend on the $N_p$ with a similar way as the resolution depends on the electron peak size. 

\begin{figure}[H]
    \centering
    \begin{subfigure}[h]{0.49\textwidth}
        \includegraphics[width=0.95\textwidth]{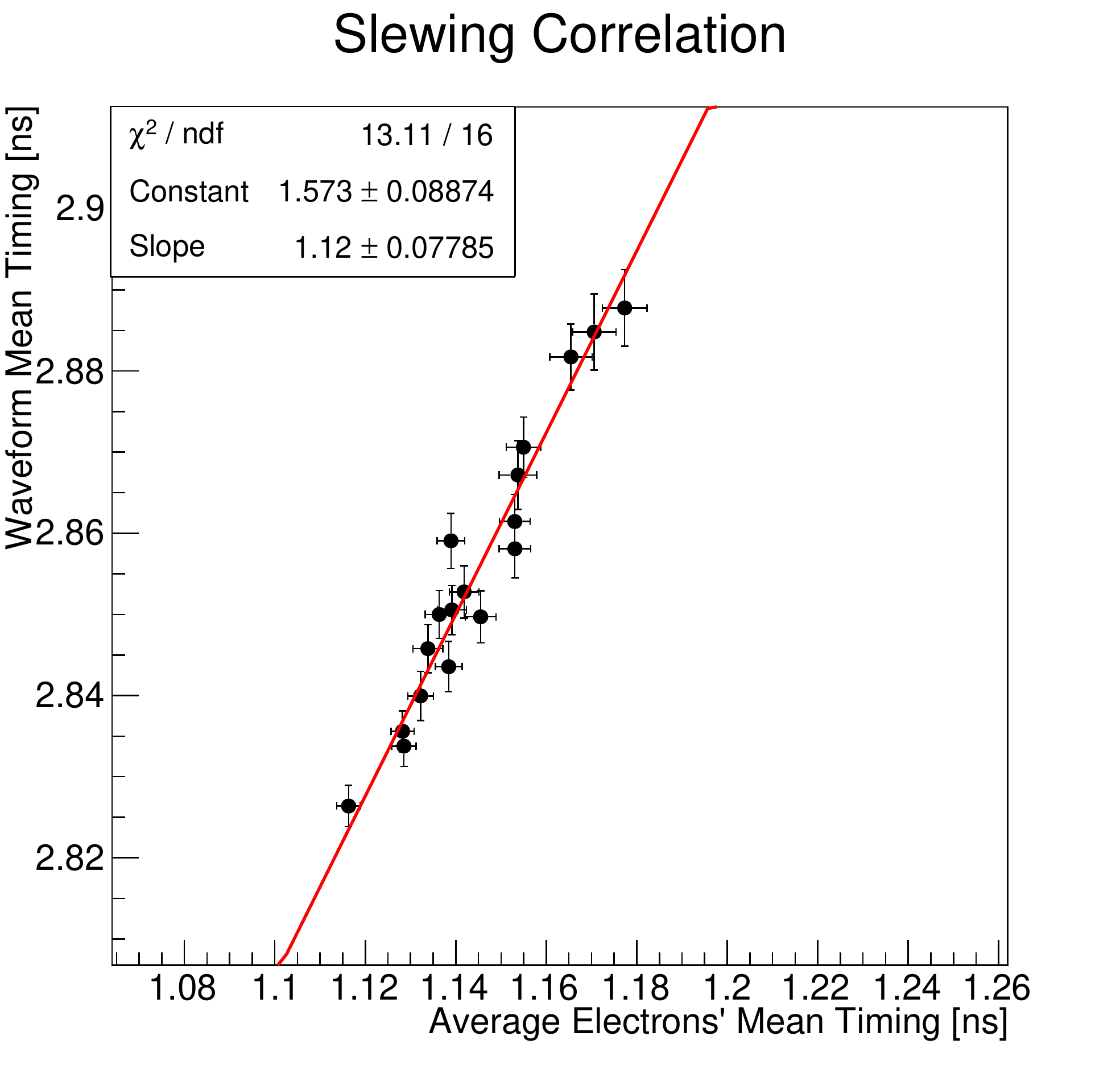}
        \caption{}
    \end{subfigure}
    \begin{subfigure}[h]{0.49\textwidth}
        \includegraphics[width=0.95\textwidth]{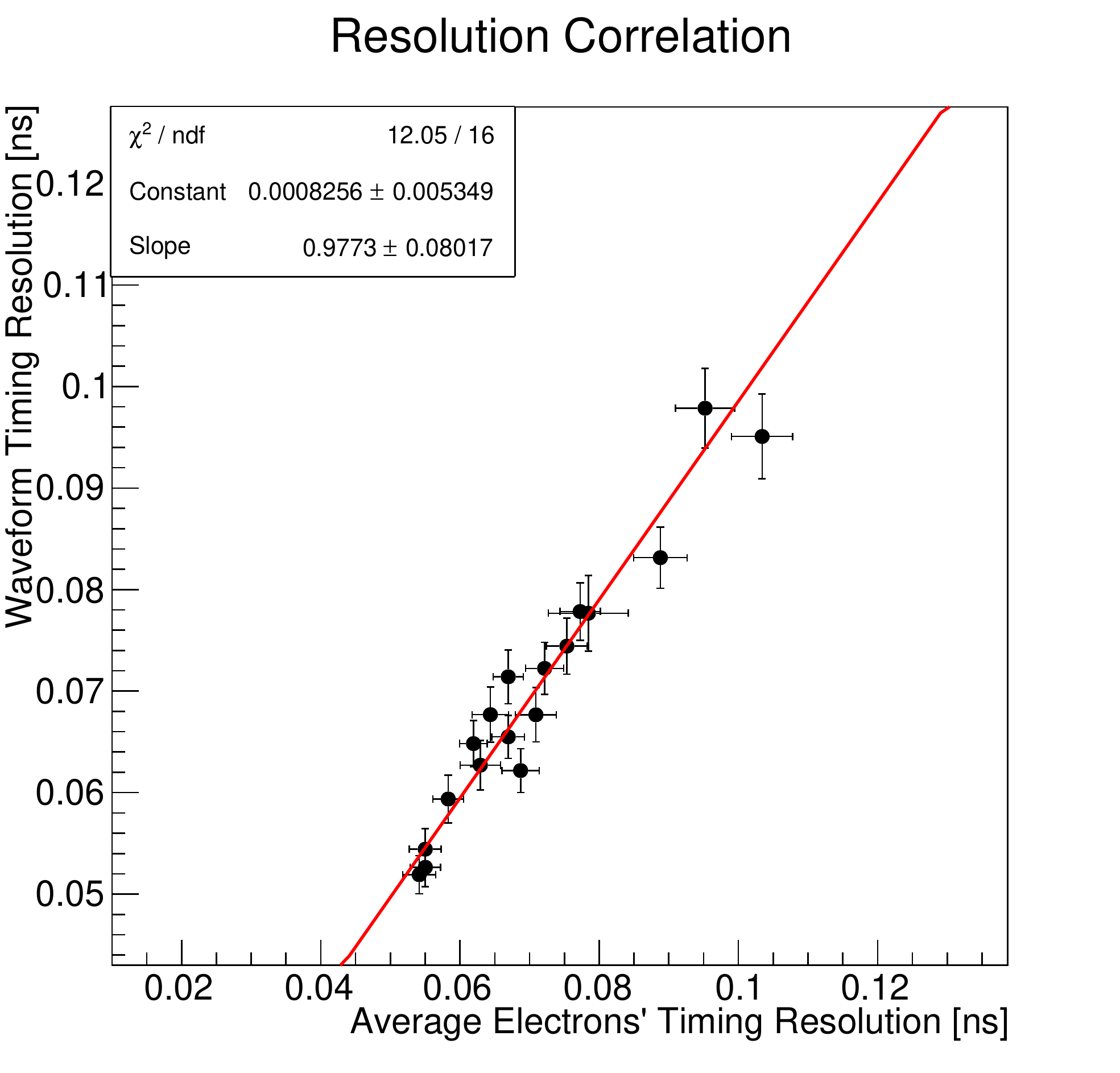}
        \caption{}
    \end{subfigure}
    \caption[Correlation between microscopic and macroscopic times.]{(a) Mean waveform timing with respect to the average of the mean time of all electrons passing through the mesh.
    (b) Time resolution of the waveform with respect to the time resolution of the mean time of all electrons passing through the mesh. 
    Each point is calculated in the same bin of electron peak charge/number of electrons according to their proportionality constant. 
    These points correspond to an anode and drift voltage of $450\,V$ and $425\,V$, respectively.}  
    \label{simu:fig:corr}
\end{figure}

Nevertheless, this qualitative comparison is not enough to associate the above dependencies to the waveforms' timing characteristics.
To explore this indication, the one-to-one correspondence between the arrival time of the signals and $\bar{\tau}$ is investigated.
The correspondence between the arrival time of the electronic signal and the $\bar{\tau}$ is probed through the relation between the mean and sigma values of the electron peak pulses' arrival time distribution and the mean and sigma values of the $\bar{\tau}$ distribution, respectively, when both distributions are related to the same events.
The relation of the means of the above distributions evaluated in bins of the charge of their observed signals is shown in Figure \ref{simu:fig:corr}\,a.
A linear fit on these points shows a slope equal to $1.12 \pm 0.08$, which is consistent with one, which means that indeed these two physical variables are one-to-one correlated.
The constant term is not (and should not necessarily be) equal to zero, because $\bar{\tau}$ expresses a time that is at a different stage of the signal evolution than the time expressed by the arrival of the electronic signal.
In other words, $\bar{\tau}$ corresponds to the average time of arrival on the mesh by definition while the timing of the waveform corresponds to the time when the signal arrives on the readout electronics, with respect to the arrival of the signal on the readout electronics of the timing reference device. 
The correlation between the spreads of the two distributions is shown in Figure \ref{simu:fig:corr} while the line expresses the linear fit whose slope equals to $0.98 \pm 0.08$ and whose constant term equals to $8\cdot 10^{-4} \pm 5 \cdot 10^{-3}$, which are consistent with one and zero, respectively.

The same analysis as the one shown in Figs. \ref{simu:fig:slewres} and \ref{simu:fig:corr}, which corresponds to anode and drift voltage settings $450\,V$ and $425\,V$, respectively, has been repeated for simulated sets of events with anode voltage of $450\,V$ and drift voltages of $300\,V,325\,V,350\,V,375\,V,400\,V$, and the same one-to-one correspondence was found.
In the rest of this study, we will follow the strong indications described above and we will use on an event by event basis the average of the times when the pre-amplification electrons pass through the mesh to represent, in a microscopic level, the e-peak arrival time.

\section{Microscopic Investigation}\label{mech:ch}

The mean time at which the pre-amplification electrons, produced by a single photoelectron, are passing through the mesh behaves the same way as the timing of the waveform. 
A detailed investigation on a microscopic level is performed in this section for the purpose of apprehending the mechanism(s) responsible for the observed timing effects of the mean SAT and time resolution on the size of the electron peak.
It is reminded that the processes in the amplification region as well as the electrical impulse response are ignored in this investigation and the search is focused only in the pre-amplification region.
This is done exactly because the microscopic variable associated to timing of the waveform is the mean time of arrival of the pre-amplification avalanche electrons on the micro-mesh, a variable that is independent of the amplification region and the electronics.
It is in fact indirectly dependent on the amplification region in that the voltage applied there affects the transparency of the micro-mesh, but this is taken into account.

\begin{figure}[H]
    \centering
    \includegraphics[width=0.45\textwidth]{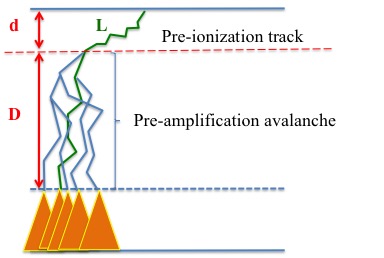}
    \caption[Basic stages of the pre-amplification process.]{Basic stages of the pre-amplification process. The initial photoelectron, beginning at the cathode, scatters and drifts towards the mesh/anode until a secondary electron is produced. At that point, the pre-amplification avalanche begins its exponential development.}  
    \label{mech:fig:diag}
\end{figure}

The basic stages of the signal development are illustrated in the diagram of Figure \ref{mech:fig:diag}.
It is assumed that the pre-amplification avalanche is initiated at the point, in space and time, at which the first ionization happens.

In the context of this investigation and in the pre-amplification region, the variables of interest are:

\begin{enumerate}
    \item The initial spatial and temporal coordinates of the photoelectron. (By default the photoelectron begins at $z=328\,\mu m, t = 0\,ns$.)
    \item The multiplicity of the secondary electrons, produced in the pre-amplification avalanche, reaching the plane just above the mesh at $z=146\,\mu m$.
    \item The mean time of all electrons in the pre-amplificiation avalanche reaching the plane just above the mesh at $z=146\,\mu m$.
    \item The spatial and temporal coordinates at which the first ionization happens and the first secondary electron is produced.
    \item The number of scatterings that the photoelectron undergoes before it produces the first secondary electron.
\end{enumerate}

In this part of the analysis, instead of referring to the plane lying in the middle of the mesh, the plane just above the mesh is used.
This serves the purpose of bypassing the effects of the mesh transparency, making the investigation simpler.
It has already been shown that the transmission through the micro-mesh does not degrade significantly the timing characteristics of the signal.
Therefore, all the arrival times will hereafter refer to the plane just above the micro-mesh unless otherwise noted.
Furthermore, unless otherwise noted, all results in this section correspond to a drift voltage of $350\,V$ and an anode voltage of $450\,V$.
Finally, the term ``event'' refers to the whole process in which a single photoelectron, from the moment of its creation, produces a signal.

\begin{figure}[H]
    \centering
    \includegraphics[width=0.55\textwidth]{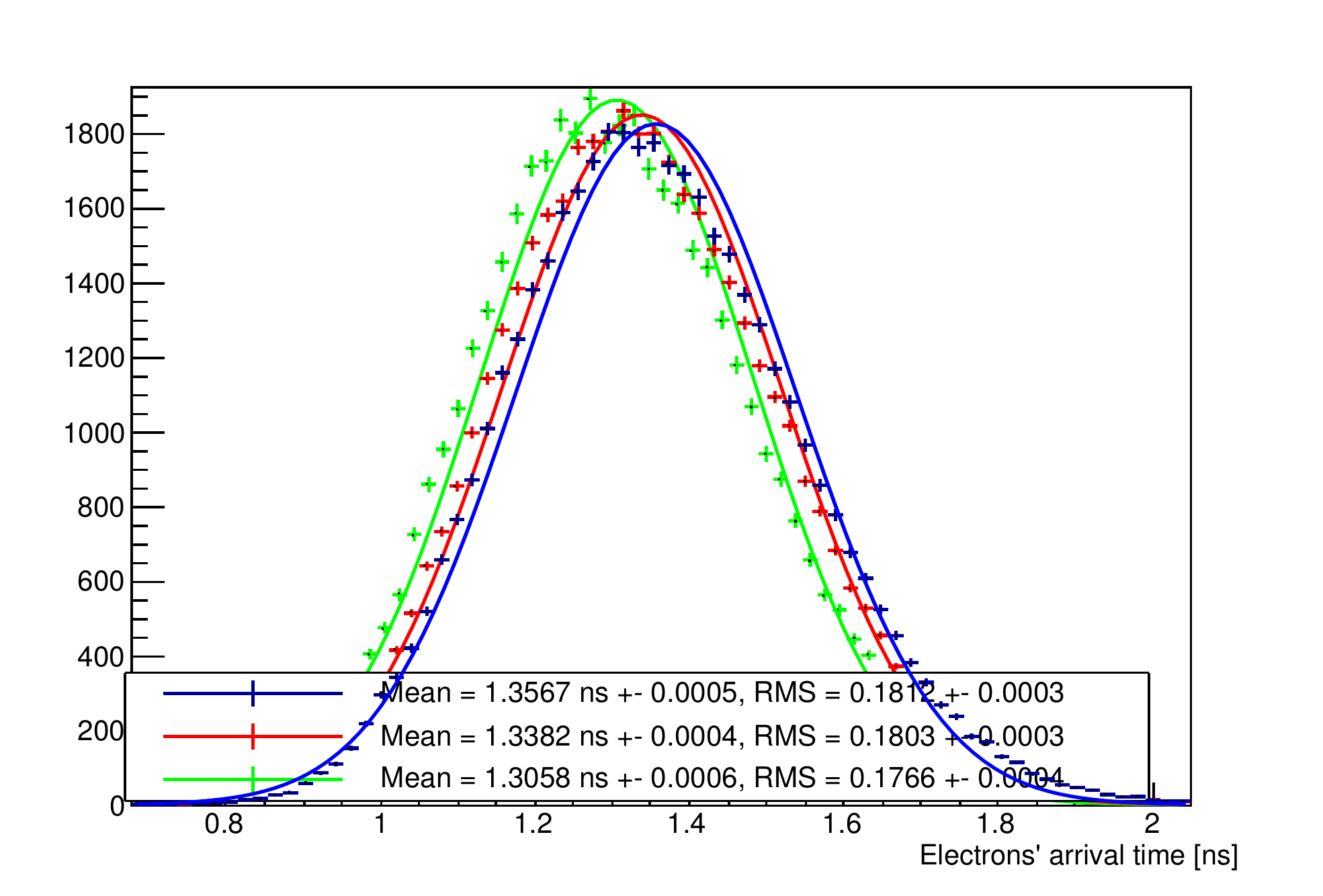}
    \caption[Distributions of arrival times of electrons.]{Distribution of the arrival times of electrons when their number is in the range of (blue) $[5,10]$, (red) $[10,15]$, (green) $[30,35]$. The heights of the distributions have been scaled for visual purposes.}  
    \label{mech:fig:simdist}
\end{figure}

A key observation, which has already been made and is mentioned in previous sections is the fact that the distribution of all the electrons' arrival times has a mean value that is changing with the number of the electrons transmitting through the micro-mesh.
It is explicitly shown in Figure \ref{mech:fig:simdist}, where events in three different bins of number of pre-amplification electrons have been selected. 
Blue corresponds to electrons' arrival times when their number is in the range of $[5,10]$ electrons, red corresponds to the range of $[10,15]$, and green to the range of $[30,35]$.
The fact that the mean arrival time is dependent on the number of electrons, is far from trivial and has no obvious explanation.
It is because of this fact that the mean SAT depends on the number of secondary electrons and requires an explanation.

\subsection{Single Electron Tracking}

\begin{figure}[H]
    \centering
    \begin{subfigure}[h]{0.42\textwidth}
        \includegraphics[width=0.95\textwidth]{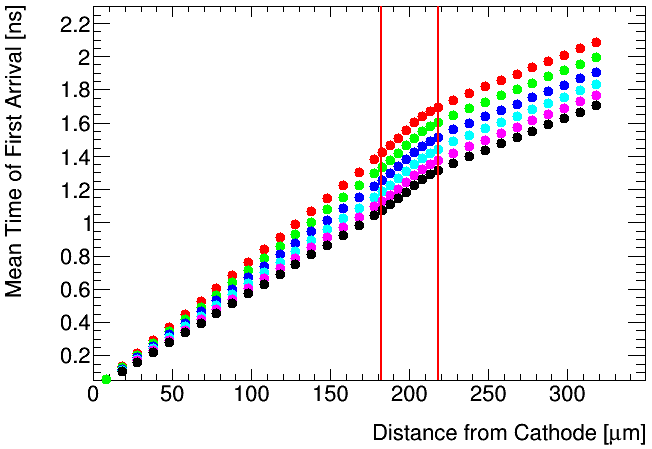}
        \caption{}
    \end{subfigure}
    \begin{subfigure}[h]{0.42\textwidth}
        \includegraphics[width=0.95\textwidth]{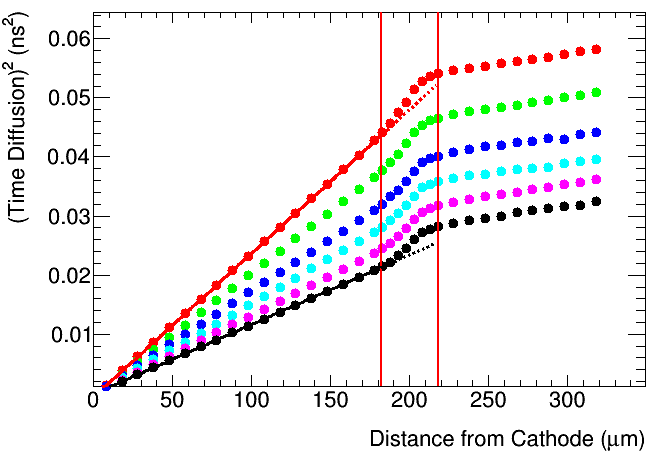}
        \caption{}
    \end{subfigure}
    \caption[Mean and variance of first passage time with respect to distance from the cathode.]{(a) Mean first passage time of a single electron with respect to the distance from the cathode.
    (b) Variance of first passage time a single electron with respect to the distance from the cathode.
    The COMPASS gas mixture is simulated with anode voltage of $450\,V$ and drift voltage of (red) $300\,V$, (green) $325\,V$, (blue) $350\,V$, (cyan) $375\,V$, (magenta) $400\,V$ and (black) $425\,V$.
    The vertical red lines denote the boundaries of the micro-mesh.
}  
    \label{mech:fig:diffuse}
\end{figure}

At a first step, a single electron is simulated in the gas volume beginning from the cathode, allowing it to produce secondary electrons which are however not tracked.
The volume is separated in different levels of $z$ with a distance of $10\mu m$ between each level.
The first passage time of each electron is recorded in each of the levels.
Then, the mean and the variance of these times is calculated and are presented in Figure \ref{mech:fig:diffuse}.
Figure \ref{mech:fig:diffuse}\,a corresponds to the mean time of arrival on each of the levels versus the distance of the level from the Cathode, while Figure \ref{mech:fig:diffuse}\,b corresponds to the variance of the time of arrival versus the distance.
The gas mixture is the COMPASS one with anode voltage of $450\,V$ and drift voltage of (red) $300\,V$, (green) $325\,V$, (blue) $350\,V$, (cyan) $375\,V$, (magenta) $400\,V$ and (black) $425\,V$ with red lines corresponding to the boundaries of the grounded micro-mesh.
From linear fits in the pre-amplfication region (left from the red lines), the drift velocities and the longitudinal diffusion coefficients can be estimated.
These are tabulated in Table \ref{tab:drift:diff}.
It is clear that the mesh introduces a variance that is small compared to the variance that would have been acquired if there was no mesh.
This is shown from the difference between the points on the right red line of Figure \ref{mech:fig:diffuse}\,b and the dashed line which the extrapolation of the linear fit.
Furthermore, the anode region introduces a diffusion that is between $40\,ps$ and $51\,ps$ according to the drift voltage.
Considering that this diffusion is added in quadrature to the one from the pre-amplification region on each electron combined with the fact that there is a very big number of electrons passing through the mesh, then the extra time spread is truly negligible.

\begin{table}[h]
\centering
\caption{Drift Velocity and Longitudinal Diffusion Coefficient of single electrons drifting for different drift voltages.}
\begin{tabular}{ccp{3.5cm}}\hline
        Drift Voltage $(V)$ & Drift Velocity $\mu m/ns$ & Longitudinal Diffusion Coeff. $(\mu m)^2/ns$ \\ \hline
        $300$ & $128.41 \pm 0.09$ & $257.35 \pm 0.64$ \\
        $325$ & $136.94 \pm 0.10$ & $266.42 \pm 0.67$ \\
        $350$ & $145.37 \pm 0.10$ & $268.28 \pm 0.69$ \\
        $375$ & $153.75 \pm 0.11$ & $278.64 \pm 0.72$ \\
        $400$ & $161.92 \pm 0.11$ & $285.66 \pm 0.70$ \\
        $425$ & $170.23 \pm 0.12$ & $290.47 \pm 0.74$ \\ \hline
    \end{tabular}
\label{tab:drift:diff}
\end{table}

\subsection{Avalanche Development}

In the electron avalanche multiplication, each electron has a constant probability per unit of length to ionize the gas again, producing another secondary electron.
Therefore, the differential equation that describes this multiplication in terms of the distance travelled by the collective motion of the electrons, is written as:

\begin{equation}
    \label{mech:eq:town}
    \frac{dN}{N} = (\alpha - \eta)dz
\end{equation}
where $z$ is distance travelled by the avalanche, $N=N(z)$ is the mean number of electrons in the avalanche at $z$, $\alpha$ is the first Townsend coefficient and $\eta$ is the attachment coefficient.
The difference between the first Townsend coefficient and the attachment coefficient is the effective first Townsend coefficient $\bar{\alpha} = \alpha - \eta$.
The solution of the differential equation is:

\begin{equation}
    \label{mech:eq:townsol}
    N(z) = N(0)e^{\bar{\alpha} z}
\end{equation}

This corresponds to mean number of electrons, the number of electrons is, however, a random number that follows a certain distribution.
Although the P\'olya distribution (Equation \ref{eq:polya})
is not the true distribution which the number of electrons follow, it is a very good approximation.
As such, the distribution of the number of electrons will be treated as if it is a P\'olya distribution.
The mean of the P\'olya distribution depends on the travelling distance and is equal to $N(z)$.

In the model shown in Fig. \ref{mech:fig:simdist}, the photoelectron before the ionization performs a Brownian motion that is prone to statistical fluctuations.
On the other hand, the avalanche is a collective statistical phenomenon of many electrons whose mean values (e.g. speed, mean arrival time, etc.) is less sensitive to statistical fluctuations.
To help in the microscopic analysis, a new variable is introduced which is the "pre-amplification avalanche length" ($D$ in Fig. \ref{mech:fig:simdist}) and is the distance between the point in space where the first ionization happened and the plane just above the mesh.
Using the pre-amplification avalanche length $D$ instead of $z$ does not bias the proportionality of Equation \ref{mech:eq:townsol} because the pre-amplification avalanche length is equivalent to the distance which two avalanches travelled, that began at the same point, but without the prominent fluctuations of the primary electron.
The complementary variable $d$ in Figure \ref{mech:fig:simdist} represents the depth in the pre-amplfication region that the photoelectron penetrates before it starts the multiplication.

\begin{figure}[H]
    \centering
    \begin{subfigure}[h]{0.42\textwidth}
        \includegraphics[width=0.95\textwidth]{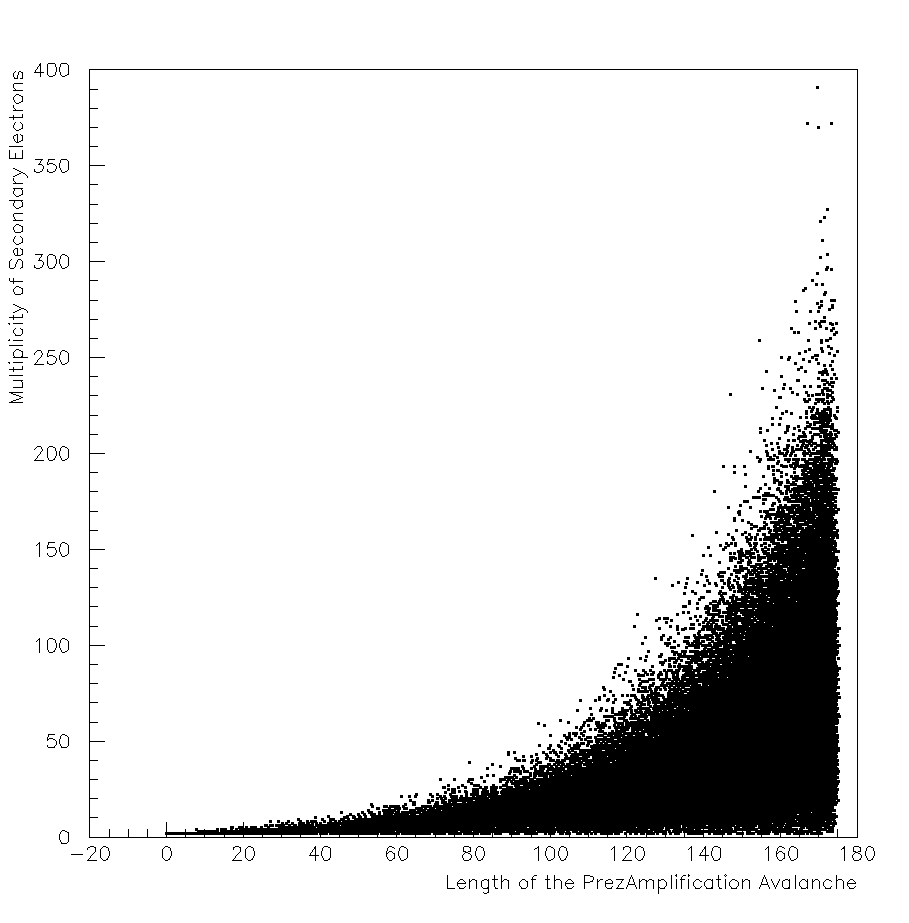}
        \caption{}
    \end{subfigure}
    \begin{subfigure}[h]{0.42\textwidth}
        \includegraphics[width=0.95\textwidth]{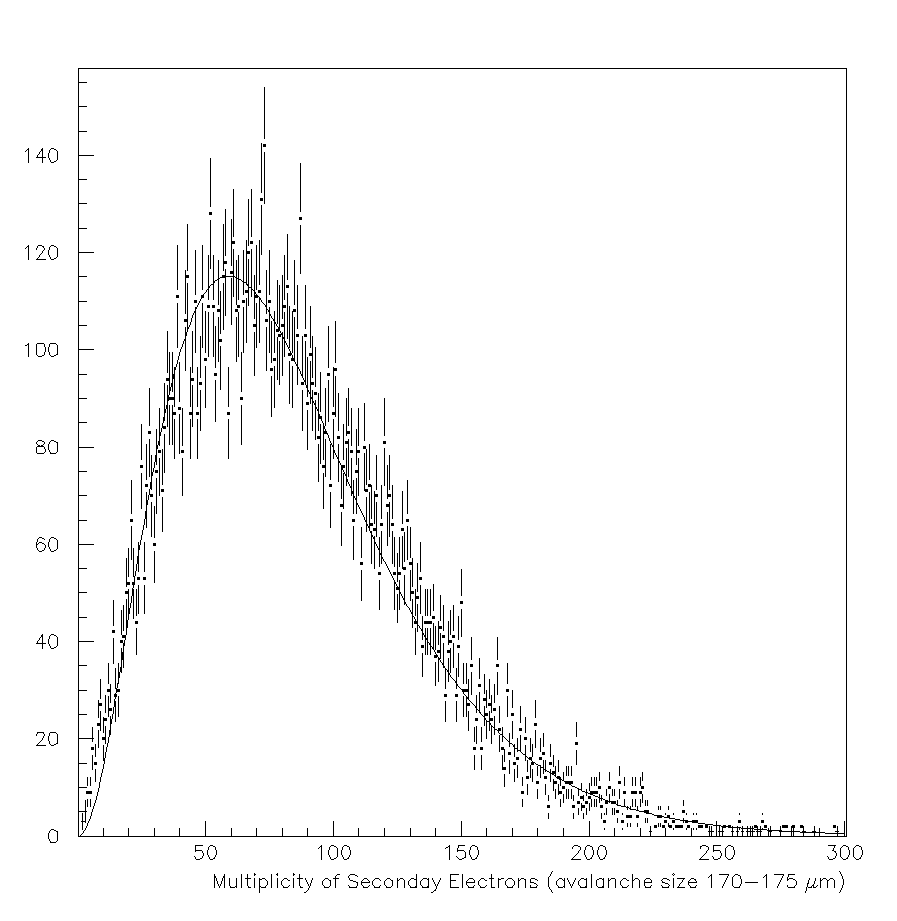}
        \caption{}
    \end{subfigure}
    \caption[Multiplicty of secondary electrons and length of pre-amplification avalanche.]{(a) Scatter plot of the multiplicity of secondary electrons in the pre-amplification avalanche versus its length.
    (b) Distribution of the multiplicity of secondary electrons in a bin of the avalanche's length $(170\,\mu m, 175\,\mu m)$.
    The Polya fit is shown with the solid line.}  
    \label{mech:fig:scatpol}
\end{figure}

\begin{figure}[H]
    \centering
    \begin{subfigure}[h]{0.42\textwidth}
        \includegraphics[width=0.95\textwidth]{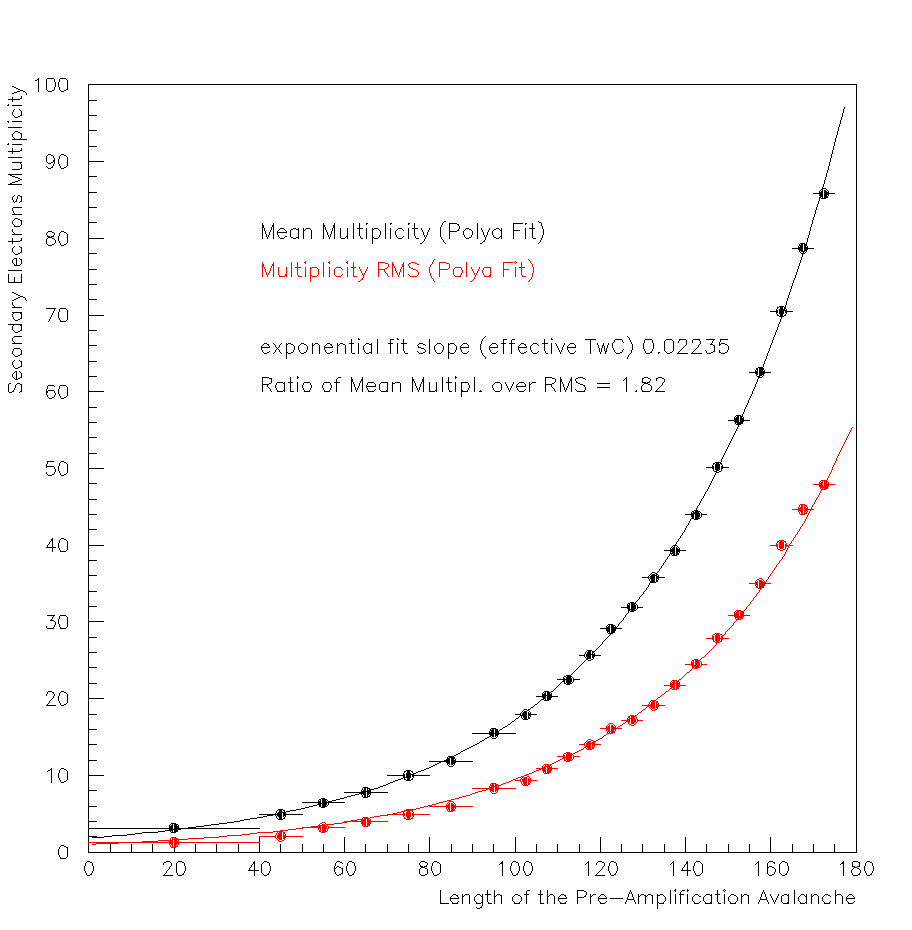}
        \caption{}
    \end{subfigure}
    \begin{subfigure}[h]{0.52\textwidth}
        \includegraphics[width=0.95\textwidth]{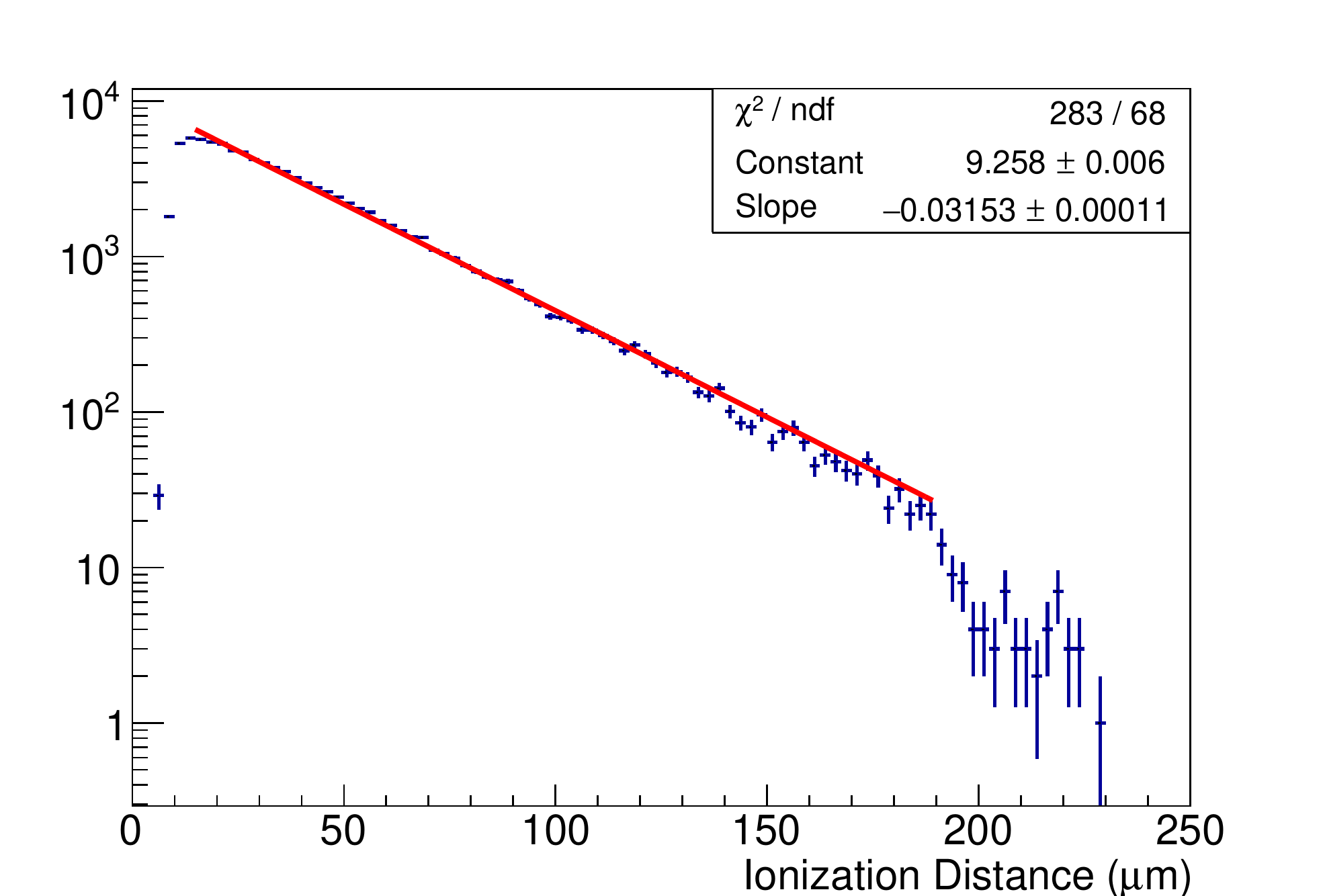}
        \caption{}
    \end{subfigure}
    \caption[Mean and RMS of secondary electrons with respect to avalanche length and distribution of ionization distance.]{(a) Mean and RMS parameters of Polya fits as a function of the length of the pre-amplification avalanche with black and red, respectively. Townsend coefficient is estimated at $\alpha = 0.022\,\mu m^{-1}$ and the ratio between Mean and RMS parameters constant and equal to $R = 1.82$.
    (b) Distribution of the ionization distance. The first Townsend coefficient is estimated at $\alpha = 0.032\,\mu m^{-1}$.
    All estimations are based in exponential fits illustrated with solid lines.}  
    \label{mech:fig:expo}
\end{figure}

A scatter plot of the number of electrons versus the pre-amplification avalanche length $D$ is presented in Figure \ref{mech:fig:scatpol}\,a.
Figure \ref{mech:fig:scatpol}\,b shows the distribution of the number of secondary electrons for avalanches with a length in the range $[170\,\mu m, 175\,\mu m]$, i.e. for very long avalanches, and the solid line represents a fit with a P\'olya distribution.
Many P\'olya fits are realized in the many bins of $D$, where the bins span across the whole range of the avalance length spectrum, and the P\'olya parameters, Mean and RMS, which are estimated by these P\'olya fits are shown with black and red points, respectively, in Figure \ref{mech:fig:expo}\,a.
The Mean number of secondary electrons, as a function of $D$ is fit using Equation \ref{mech:eq:townsol}.
The slope of this fit corresponds to an effective first Townsend coefficient equal to $\bar{\alpha} = 0.022\,\mu m^{-1}$.
The RMS multiplicity is also fit with an exponential whose slope is found to have the same value as $\bar{\alpha}$, when found by fitting the Mean values.
Because of this, the ratio between the P\'olyas' Mean and RMS values is found constant at $1.82$ for all the available avalanche lengths.

The direct way to approximate the first Townsend coefficient is through the ionization's distance distribution $L$ (as noted in Figure \ref{mech:fig:diag}), i.e. the distribution of the distance between the cathode and the point at which the first ionization happened.
Figure \ref{mech:fig:expo} shows this distribution in which the solid line represents the exponential fit.
The first Townsend coefficient is approximated through this fit at $\alpha = 0.032\,\mu m^{-1}$.
Obviously, this value is quite different from the one found by fitting multiplicities, and significant attachment losses are expected.
Attachment phenomena supress the multiplicity increase in the avalanche phenomena, with result the value found by fitting the multiplicities is an effective value that includes electron losses due to attachment, recombination, etc.

\vspace{-0.5cm}

\vspace{-0.3cm}

\subsection{Temporal Development}

\begin{figure}[H]
    \centering
    \begin{subfigure}[h]{0.45\textwidth}
    \includegraphics[width=1.\textwidth]{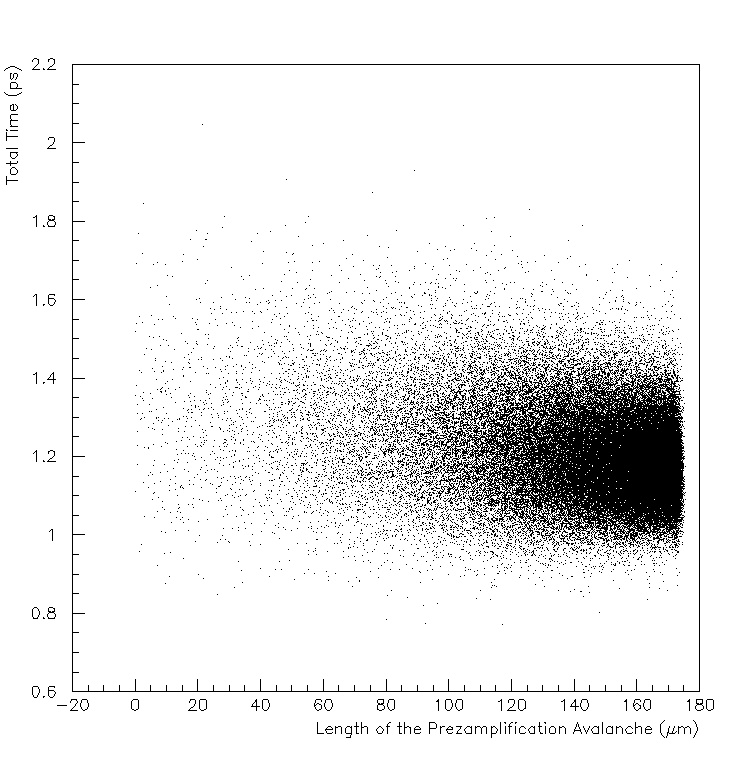}
        \caption{}
    \end{subfigure}
    \begin{subfigure}[h]{0.45\textwidth}
        \includegraphics[width=1.\textwidth]{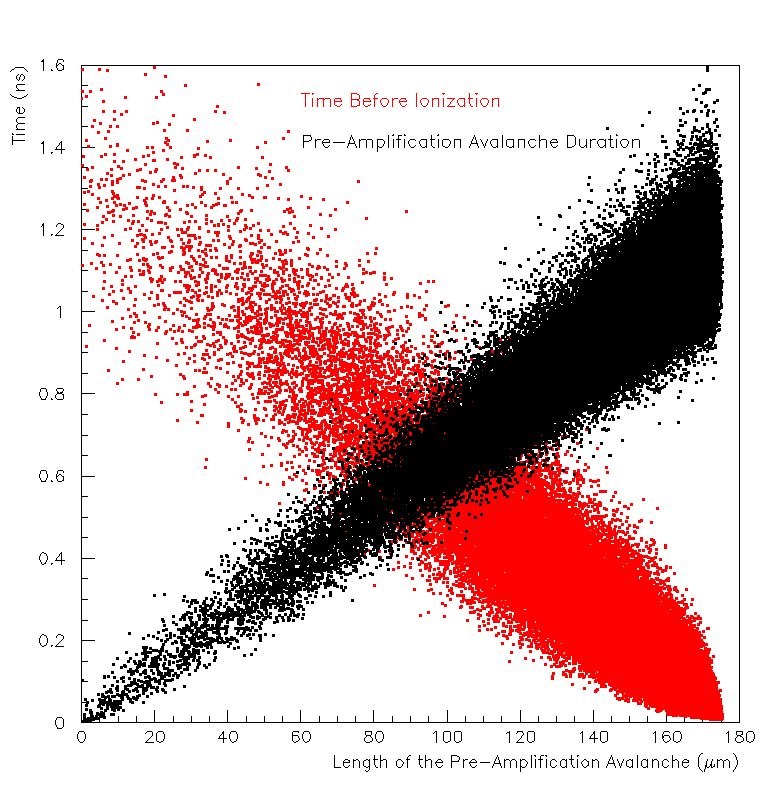}
        \caption{}
    \end{subfigure}
    \caption[Mean arrival time of electrons and time of first ionization.]{a) Mean arrival time of electrons in a pre-amplficiation avalanche on the micro-mesh versus the pre-amplificiaton avalanche length.
        b) The time of first ionization is shown with red points while the time it takes for an avalanche to reach the mesh (mean time of electrons) from the moment of first ionization is shown with black points, both versus the length of the pre-amplification avalanche.  
    }  
    \label{mech:fig:noth}
\end{figure}

A scatter plot of the arrival time (mean time of all electrons arriving on the mesh) versus the pre-amplification avalanche length is illustrated in Figure \ref{mech:fig:noth}\,a.
The dependencies are hard to guess from this scatter plot and for this reason, the events are categorized in small bins of the length of the pre-amplification avalanche length, and the distribution of the arrival time is studied in each of the bins.
This procedure is identical to the one described in Section \ref{sec:tim}
followed to find the mean and spread of the SAT.
The arrival time, $\bar{\tau}$, of the pre-amplification electrons in a bin of avalanche lengths is approximately distributed with a Gaussian shape.
In each of the bins, the means and the standard deviations are estimated with Gaussian fits.

The scatter plot of Figure \ref{mech:fig:noth}\,b, presents the time interval between the production of the photoelectron and the first ionization $\tau_e$ (with red points), and the time duration of the pre-amplification avalanche $\tau_a$ (with black points), as a function of the length of the pre-amplification avalanche.
The distributions of $\tau_e$ and $\tau_a$ variables have been examined in bins of the pre-amplification avalanche length, whose Mean and RMS values have been estimated.

\begin{figure}[H]
    \centering
    \begin{subfigure}[h]{0.41\textwidth}
        \includegraphics[width=0.95\textwidth]{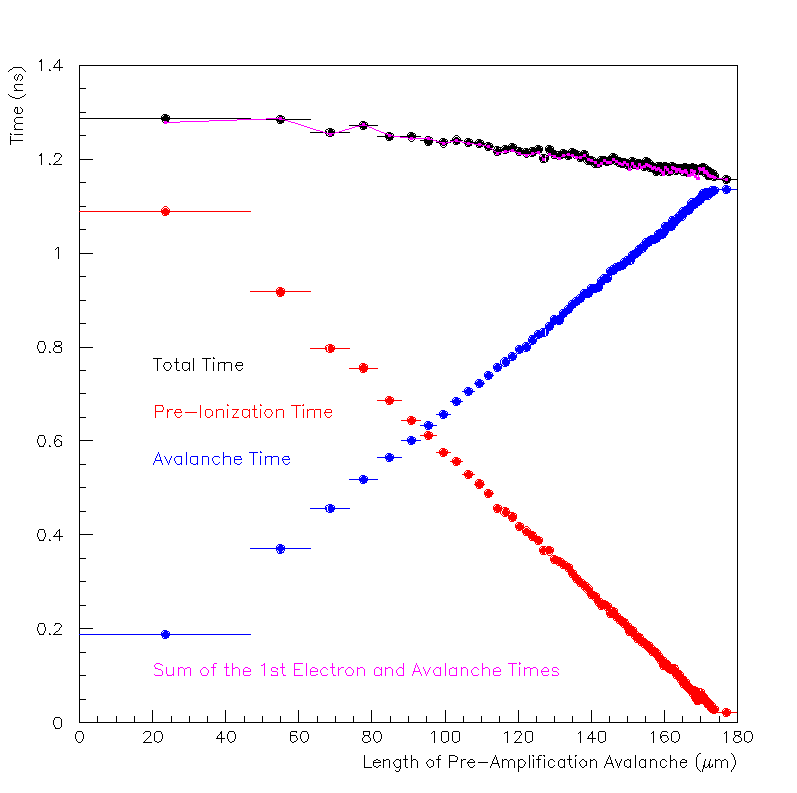}
        \caption{}
    \end{subfigure}
    \begin{subfigure}[h]{0.41\textwidth}
        \includegraphics[width=0.95\textwidth]{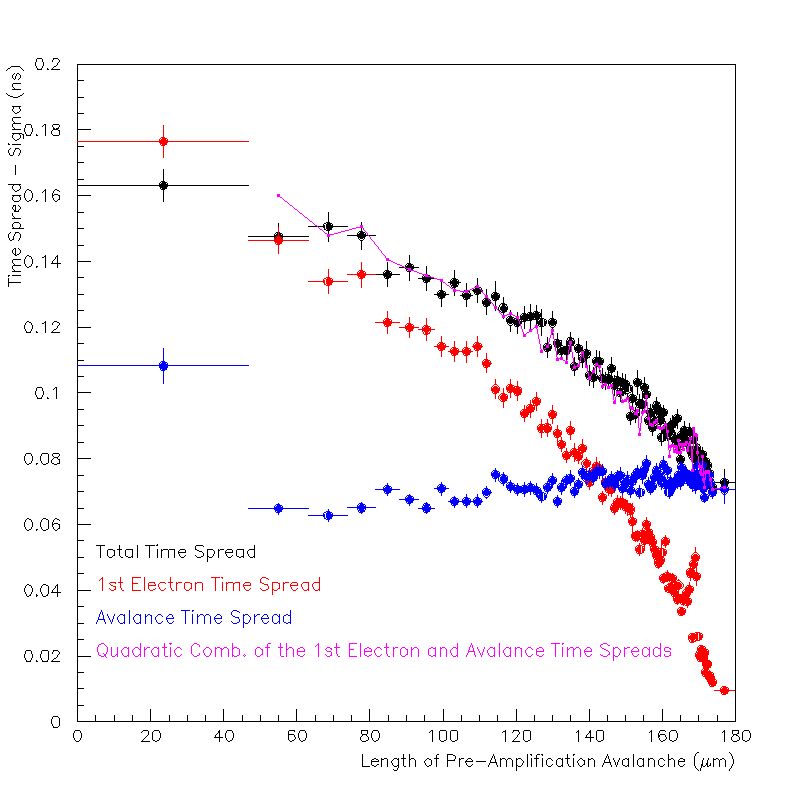}
        \caption{}
    \end{subfigure}
    \caption[Mean and spread of arrivl time as a function of pre-amplification avalanche length.]{(a) Mean arrival time and (b) Arrival time spread as a function of the pre-amplificiaton avalanche length. On both figures, red shows the timing characteristics of the photoelectron until it ionizes. Blue is the characteristics of the avalanche. Black is of the total phenomenon and magenta is the combination of the two effects.}  
    \label{mech:fig:slew}
\end{figure}

In Figure \ref{mech:fig:slew}\,b the standard deviation values of the $\bar{\tau}\, \tau_e, \tau_a$ distributions in bins of the pre-amplification avalanche lengths are shown versus the corresponding avalanche length.
The ionization time spread $\tau_e$ is shown with red, the spread of the time duration of the avalanche $\tau_a$ (from its creation until its arrival on the mesh) is shown with blue, and the standard deviation of $\bar{\tau}$ is presented with $\bar{\tau}$, i.e. the total time from the creation of the photoelectron until the avalanche reaches the mesh.
The magenta points show the square roots of the quadratic sums of the related red and blue points.
Naturally, the sum of the variances of the two effects (random walk of the photoelectron before ionization and avalanche formation) gives the same result as the direct evaluation of the variance of $\bar{\tau}$.
Accordingly, the mean values of the distributions of $\bar{\tau}\, \tau_e, \tau_a$ in bins of the pre-amplification avalanche length as functions 
of the pre-amplification avalanche length are shown in Figure \ref{mech:fig:slew}\,a, with the same color code as Figure \ref{mech:fig:slew}\,b.
Obviously, the sum of the red and blue points, shown with magenta, coincide with the mean values that correspond to the total time $\bar{tau}$.

It is natural that as the pre-amplification avalanche length increases, the time spent by the initial photoelectron before ionization decreases while the time duration of the avalanche increases.
Even though the sum of the vertical distance components of the two phenomena is constant and equal to the drift gap, the sum of the two times $\tau_e + \tau_a$ clearly depends on the pre-amplification avalanche length.
This means that, effectively, the photoelectron and the avalanche do not ``drift'' with the same velocities.
Linear fits are realized on the two mean times and slopes corresponding to different drift velocities are found.
This hints that the primary photoelectron drifts with a ``slow'' but constant drift velocity before ionization while the avalanche drifts with a faster yet again constant drift velocity.

The phenomenological explanation of this is that before ionizing, the primary electron cannot have interacted with the gas through the channel of ionization.
After the first ionization, the electrons are ``free'' to interact through this channel.
The channel of ionization is an inelastic process and when it happens, it damps the energy of the electron.
Therefore, by ``turning on'' another possible inelastic process, an effect similar (but much less pronounced) to adding a quencher happens, i.e. electrons with lower energy participate in f    ewer elastic scatterings and the electric field spends less time recovering a forward direction motion.

\begin{figure}[H]
    \centering
        \includegraphics[width=0.65\textwidth]{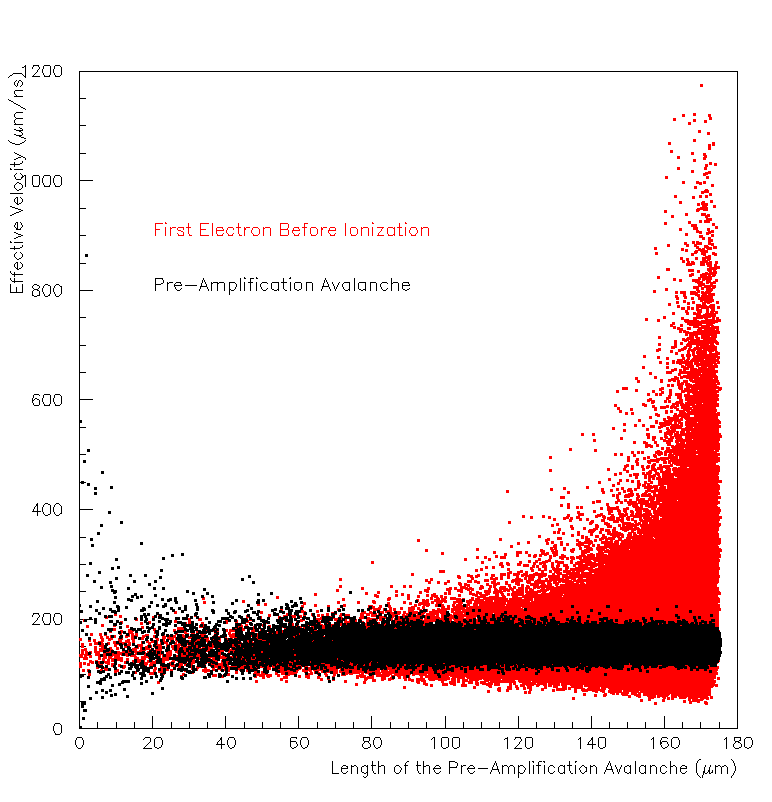}
        \caption[Effective vertical velocity as a function of pre-amplification avalanche length.]{Scatter plot effective vertical velocity as a function of the pre-amplificiaton avalanche length.
    Red shows the speed of the photoelectron until it ionizes.
    Black shows the time and speed of the avalanche.
    }  
    \label{mech:fig:rb}
\end{figure}

The effective vertical speed versus the pre-amplification avalanche length is shown in the scatter plot of Figure \ref{mech:fig:rb}.
The vertical speed is defined as $u_e = \frac{d}{\tau_e}$ and $u_a = \frac{D}{\tau_a}$ for the photoelectron before ionization and the avalanche, respectively, and on an event-by-event basis.
In Figure \ref{mech:fig:rb}, $u_e$ is represented in red and $u_a$ in black.
Obviously the photoelectron before ionization and the avalanche, seem to move with different speeds.
However, this is an artifact of the calculation of the vertical speed.
Essentially, what is measured as vertical speed in this case, is something that is proportional to the inverse of time as the distance is fixed.
It is easy to show through a Taylor expansion around the mean value $\langle \tau \rangle$ that the mean vertical speed is not equal to the drift velocity.

\begin{equation}
    \left< \frac{d}{\tau} \right> = d \left< \frac{1}{\tau} \right> \approx \frac{d}{\left< t \right>} \left( 1 + \frac{\sigma_t^2}{\left< t \right> ^2} \right)
\end{equation}

It is clear that if the variance of time $\sigma_t^2$ and the mean value $\left< t \right>$ both grow linearly with time, and as a result with distance, then the mean speed becomes

\begin{equation}
    \left< \frac{d}{\tau} \right>  \approx v_d \left( 1 + \frac{2D_L}{v_d x} \right)
\end{equation}
where $D_L$ is the longitudinal diffusion coefficient, $v_d$ is the drift velocity and $x$ is the distance from the beginning of the drift.
Obviously, it is clear that the hypothesis of a constant drift velocity supports the fact that the mean vertical speed is increasing with pre-amplification avalanche lengths.
Paradoxically, the primary electron is not drifting with its mean vertical speed.

\begin{figure}[H]
    \centering
    \begin{subfigure}[h]{0.45\textwidth}
        \includegraphics[width=0.95\textwidth]{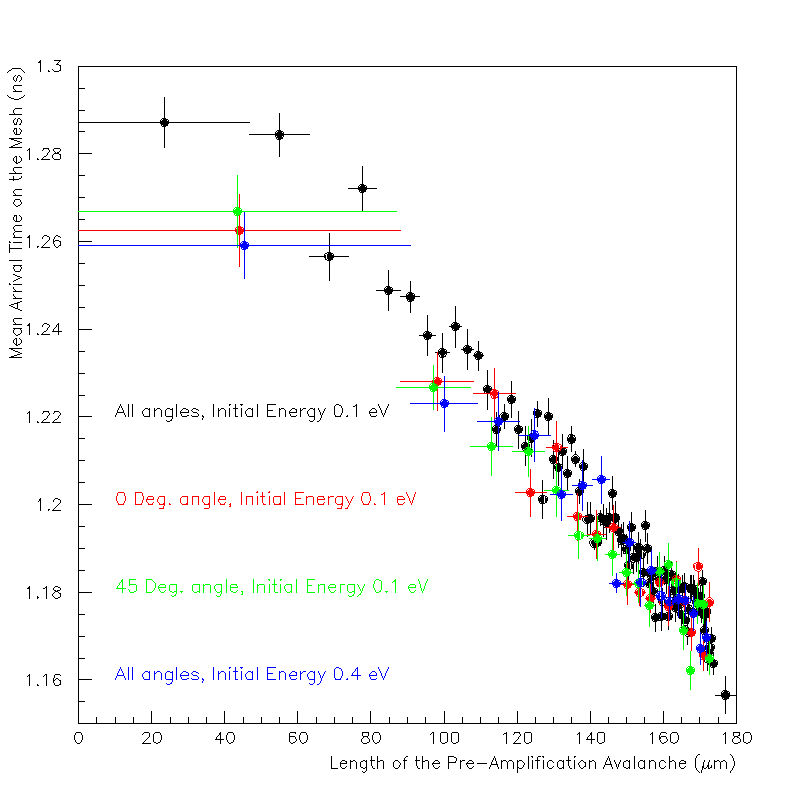}
        \caption{}
    \end{subfigure}
    \begin{subfigure}[h]{0.45\textwidth}
        \includegraphics[width=0.95\textwidth]{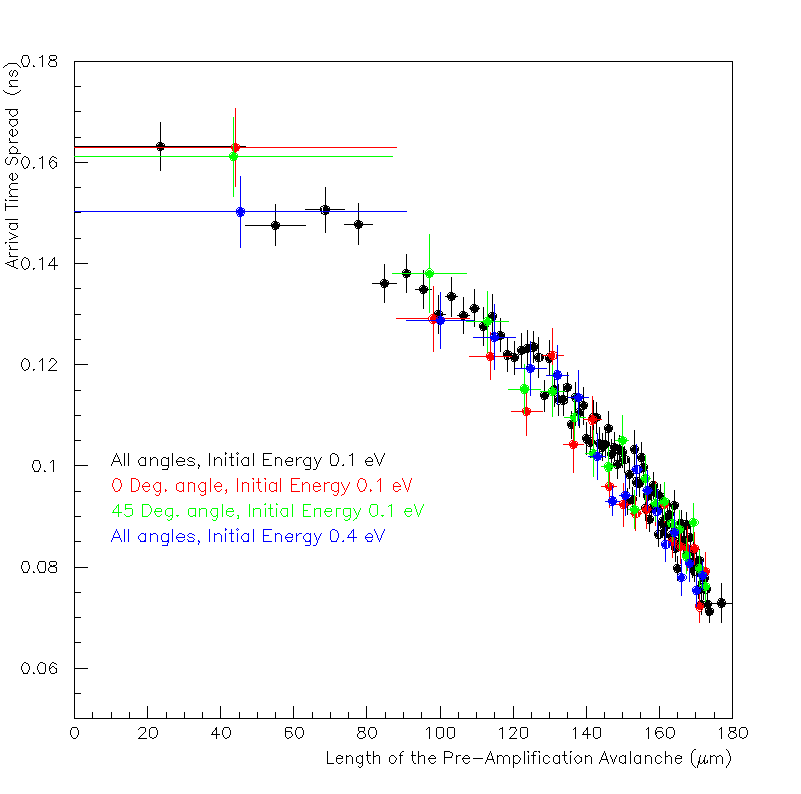}
        \caption{}
    \end{subfigure}
    \caption[Mean and spread of arrival time as a function of avalanche length for different initial conditions.]{(a) Mean arrival time and (b) Arrival time spread as a function of the pre-amplification avalanche length. Initial conditions are (black) isotropic direction with initial energy of $0.1\,eV$, (red) $0^o$ polar angle with initial energy of $0.1\,eV$, (green) $45^o$ polar angle with initial energy of $0.1\,eV$, (blue) isotropic direction with initial energy $0.4\,eV$.  
    }  
    \label{mech:fig:init}
\end{figure}

Before proceeding further, the influence of the photoelectron's initial conditions must be investigated.
The initial direction of motion and energy, are the only initial conditions that can  influence the above time dependencies on the avalanche length.
An electron that has an initial direction perpendicular to the plane of the cathode would on average move faster in the vertical direction.
Furthermore, an electron with a larger initial energy would again move faster vertically.
However, assuming a modest drift voltage of $350\,V$ across a drift gap of $200\,\mu m$, the energy density that an electron acquires is approximately equal to $1.75\,eV\,\mu m^{-1}$.
This energy gain, however, is completely insignificant considering that the mean free path is in the order $\mu ms$.
Indeed, in Figure \ref{mech:fig:init} the independence of the effects with respect to the initial conditions is shown. 
The initial conditions are (black) isotropic direction of the emitted photoelectron from the cathode with an initial energy of $0.1\,eV$, (red) $0^o$ polar angle emission perpendicular to the cathode with an initial energy of $0.1\,eV$, (green) $45^o$ polar emission angle with initial energy of $0.1\,eV$, and (blue) isotropic emission direction with initial energy $0.4\,eV$. 
Both the mean and the standard deviation values of $\bar{\tau}$ are shown in Figures \ref{mech:fig:init}\,a and b, respectively, exhibit the same dependence on the pre-amplification avalanche length for all the different initial conditions.
This fact indicates that the initial conditions do not play any role in the observed timing dependencies.

\subsection{From Microscopic to Observable Effects}

\begin{figure}[H]
    \centering
    \begin{subfigure}[h]{0.49\textwidth}
        \includegraphics[width=1\textwidth]{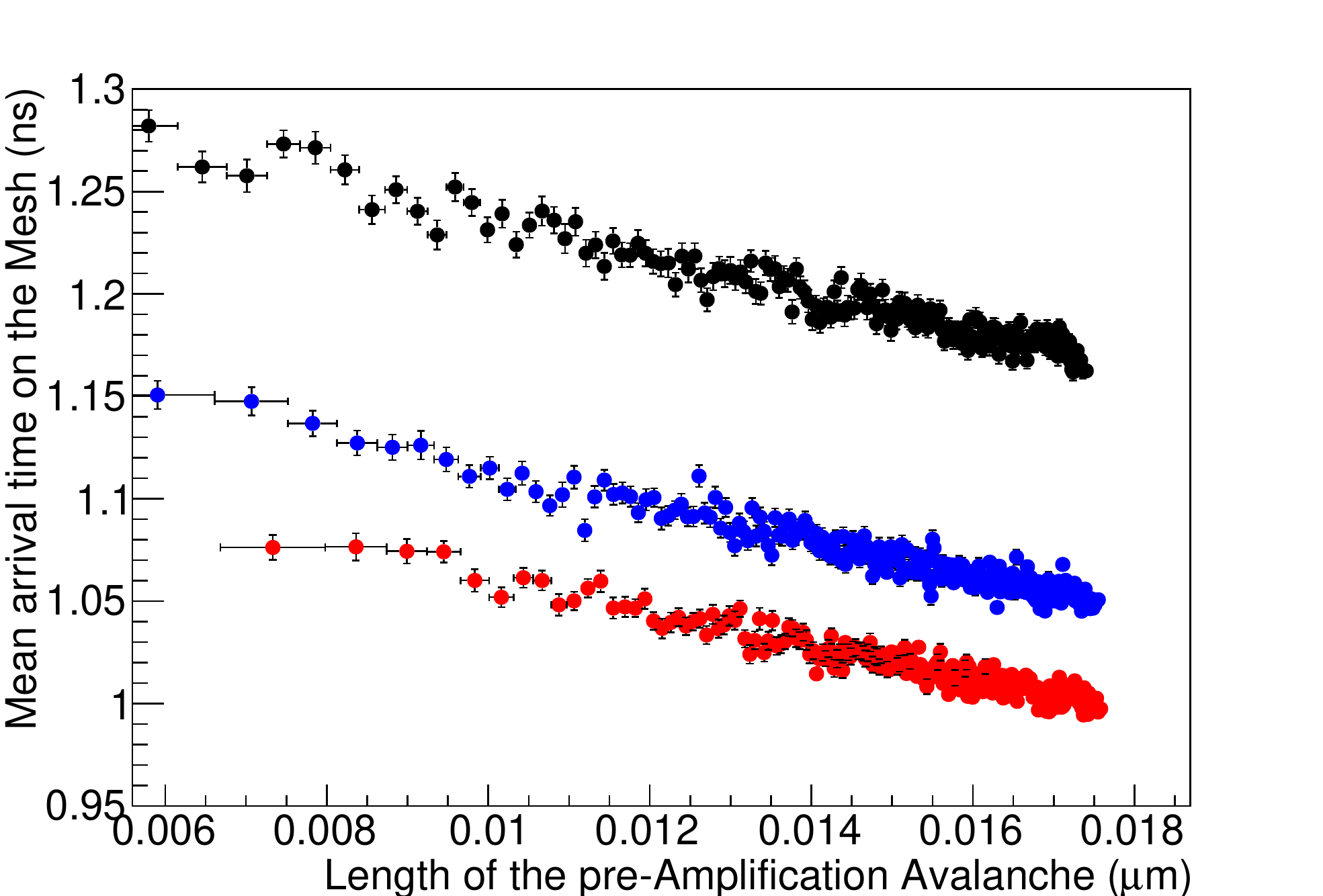}
        \caption{}
    \end{subfigure}
    \begin{subfigure}[h]{0.49\textwidth}
        \includegraphics[width=1\textwidth]{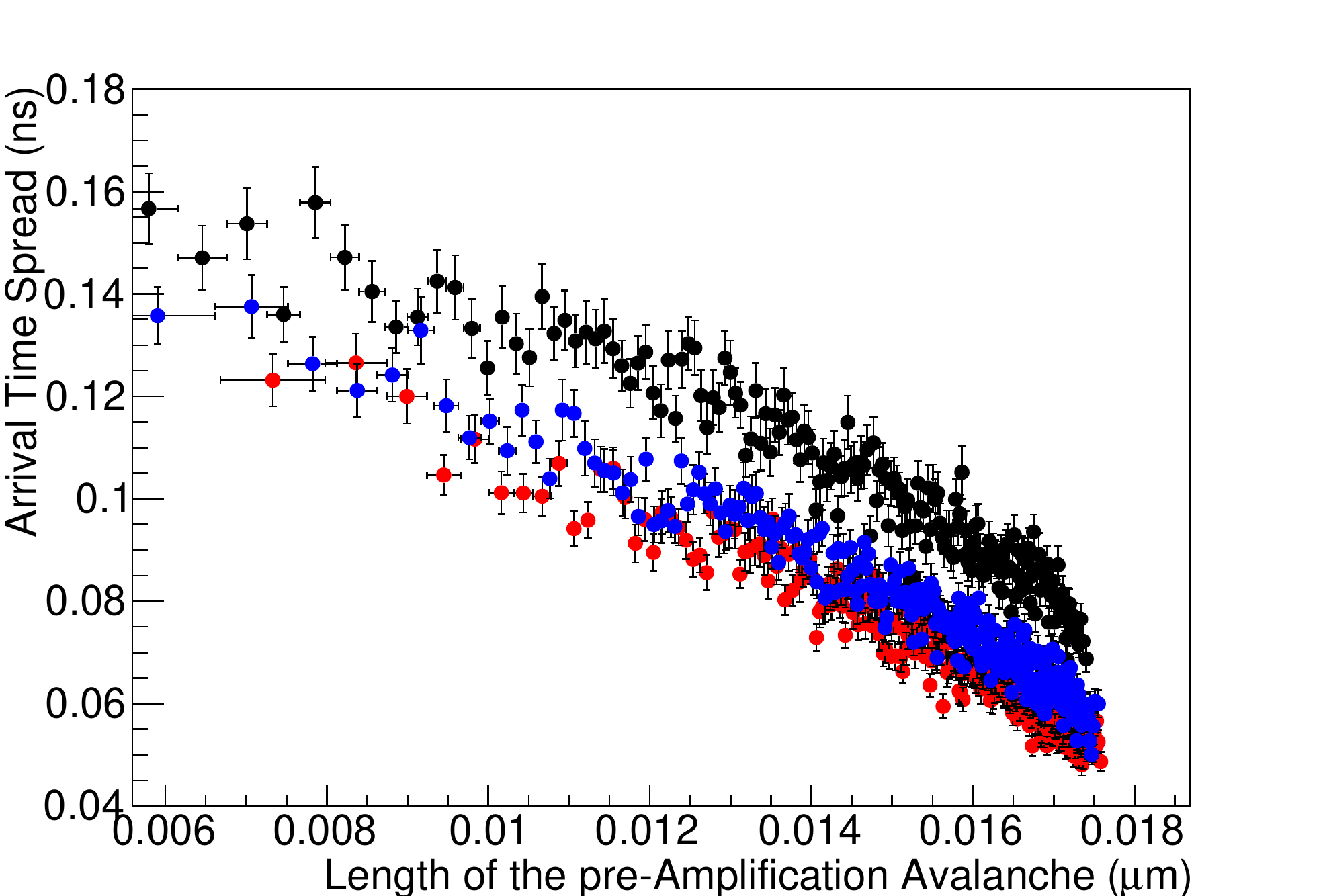}
        \caption{}
    \end{subfigure}
    \caption[Mean and spread of arrival time with respect to pre-amplification avalanche length.]{ (a) Mean and (b) spread of arrival time with respect to the pre-amplification avalanche length. Anode voltage is $450\,$ and drift voltage is (black) $350\,V$, (blue) $400\,V$ and (red) $425\,$.
    }  
    \label{mech:fig:obs}
\end{figure}

The timing characteristics of the signal exist because the statistical properties of the mean time of arrival of (the pre-amplification avalanche) electrons on the micro-mesh depend on the length of the avalanche.
In the previous Sections, support was offered to this claim and the role of the photoelectron before the first ionization on this dependence was highlighted.
Another striking observation in the experimental distributions was the fact that the SAT and the time resolution versus the size of the electron peak are almost independent of the drift field (for the same anode voltage settings).
In this Section, it is investigated whether this observation is supported by the simulation.
Two more simulated sets of events are considered with the same anode voltage of $450\,V$; one with a drift voltage of $400\,V$ and one with $425\,V$, in addition to the one already used with $350\,V$.
The (a) mean and the (b) standard deviation values of the $\bar{\tau}$ distributions in bins of the pre-amplification avalanche length are shown in Figure \ref{mech:fig:obs} as a function the pre-amplification avalanche length. 
Black corresponds to drift voltage of $350\,V$, blue to $400\,V$ and red to $425\,V$.
The anode voltage is at $450\,V$.

\begin{figure}[H]
    \centering
    \begin{subfigure}[h]{0.49\textwidth}
        \includegraphics[width=1.\textwidth]{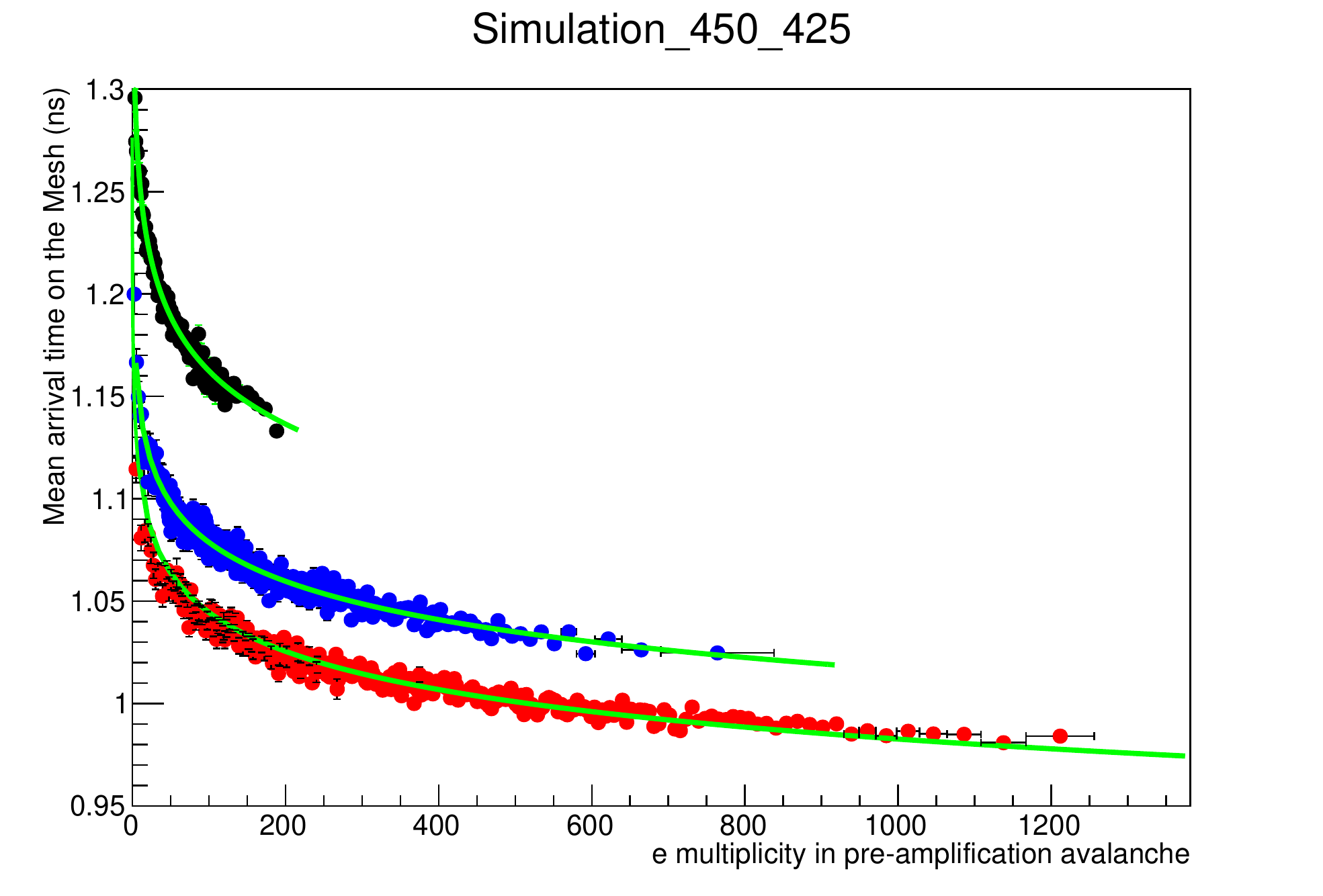}
        \caption{}
    \end{subfigure}
    \begin{subfigure}[h]{0.49\textwidth}
        \includegraphics[width=1.\textwidth]{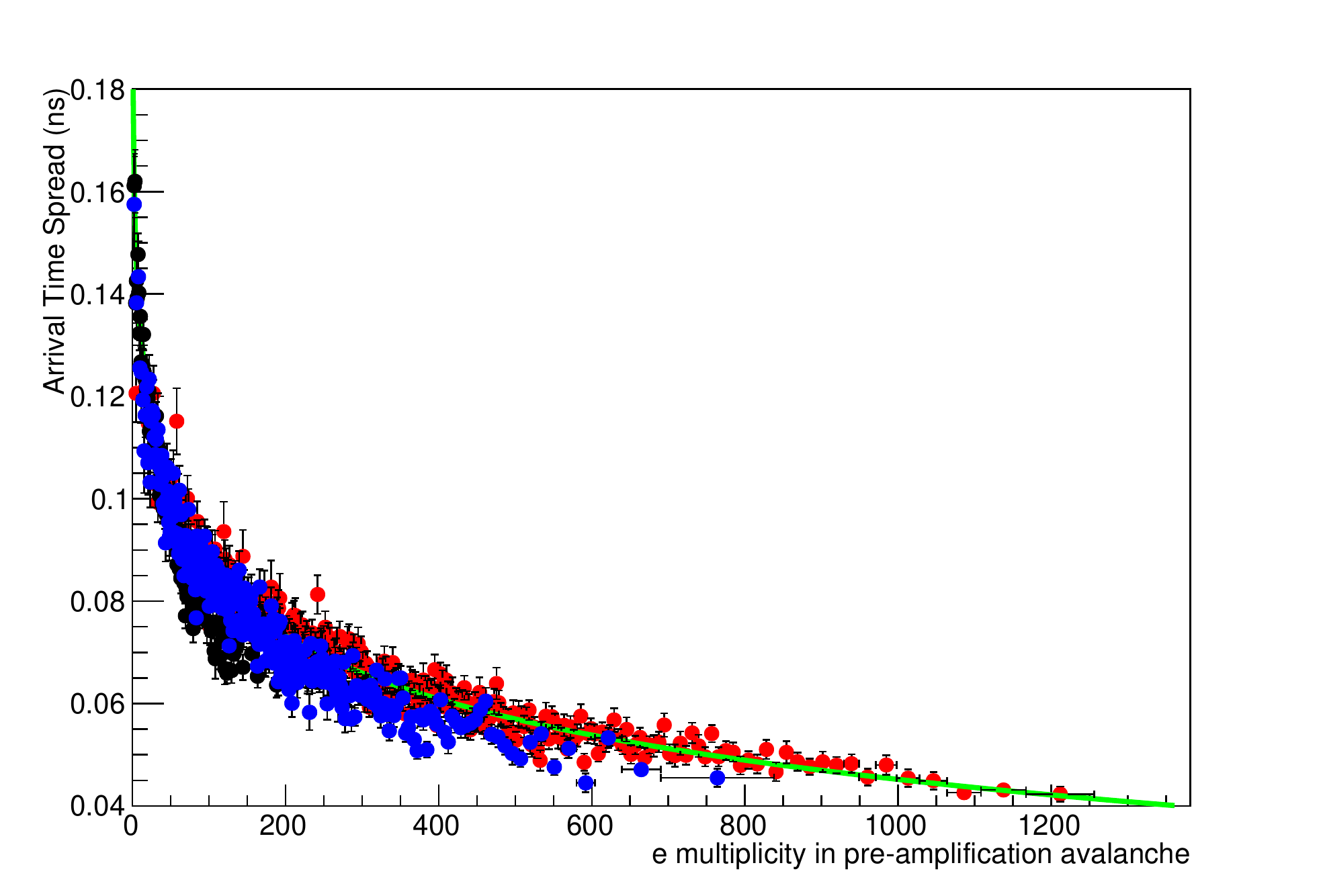}
        \caption{}
    \end{subfigure}
    \caption[Mean and spread of arrival time with respect to pre-amplification avalanche electron multiplicity.]{ (a) Mean and (b) spread of arrival time with respect to the pre-amplification avalanche electron multiplicity. Anode voltage is $450$ and drift voltage is (black) $350\,V$, (blue) $400\,V$ and (red) $425\,V$.
    }  
    \label{mech:fig:2obs}
\end{figure}

The spread of the arrival times, which has a one-to-one correspondence with the time resolution, does not only exhibit a different dependence on the length of the pre-amplification avalanche, but this dependence is also different for different drift voltage settings.
However, the observed effect was that the time resolution shows exactly the same dependence across different drift voltage settings, but on the size of the electron peak.
The electron peak size that is used in the experiment is proportional to the number of electrons. 
These two effects are not yet incompatible because if the drift field is different, then avalanches with the same length will not produce the same number of electrons (on average).
Thus, to check if the simulation is consistent with the experimental observations, the simulation is analyzed now in terms of the number of electrons.
In other words, the $\bar{\tau}$ distributions in bins of the number of electrons in the pre-amplification avalanche are studied,
and the mean and sigma values are defined by Gaussian fits. 
The (a) mean and the (b) standard deviation values of the $\bar{\tau}$ distributions in bins of the number of electrons per avalanche and as a function of the number of electrons per avalanche are shown in Figure \ref{mech:fig:2obs}, while the color code that is followed is the same as in Figure \ref{mech:fig:obs}.

The points are fit with a power law plus a constant term illustrated by the solid line in Figures \ref{mech:fig:2obs}.
As explained in the previous Sections, when the mean values are fit, the constant term takes different values for each drift voltage setting, while the parameters of the power law term are common for all the simulated sets of events.
When the resolution is fit, all of parameters (both constant and power law) are common to all data sets, as is also the case in the experiment.
It is clear from Figure \ref{mech:fig:2obs} that the fact that the time dependence of the timing characteristics on the size of the electron peak is independent of the drift voltage settings, is also supported from the simulation.

\chapter{Analysis of the Muon Beam Data}\label{chap:muons}

The main point of this chapter is the statistical method developed in the analysis of the Testbeam data for the estimation of the mean number of photoelectrons per muon that are extracted from the radiator and the coupled photocathode.
Last but certainly not least, the best time resolution that was achieved is presented.

\section{Calibration with Single Photoelectrons}

Before going on to estimate the mean number of photoelectrons extracted from the photocathode, the single photoelectron response must be learned.
The experimental data which are of interest are the those with anode voltage of $275\,V$, drift voltage of $475\,V$ with the COMPASS gas.

\begin{figure}[H]
    \centering
    \begin{subfigure}[h]{0.49\textwidth}
    \includegraphics[width=0.7\textwidth]{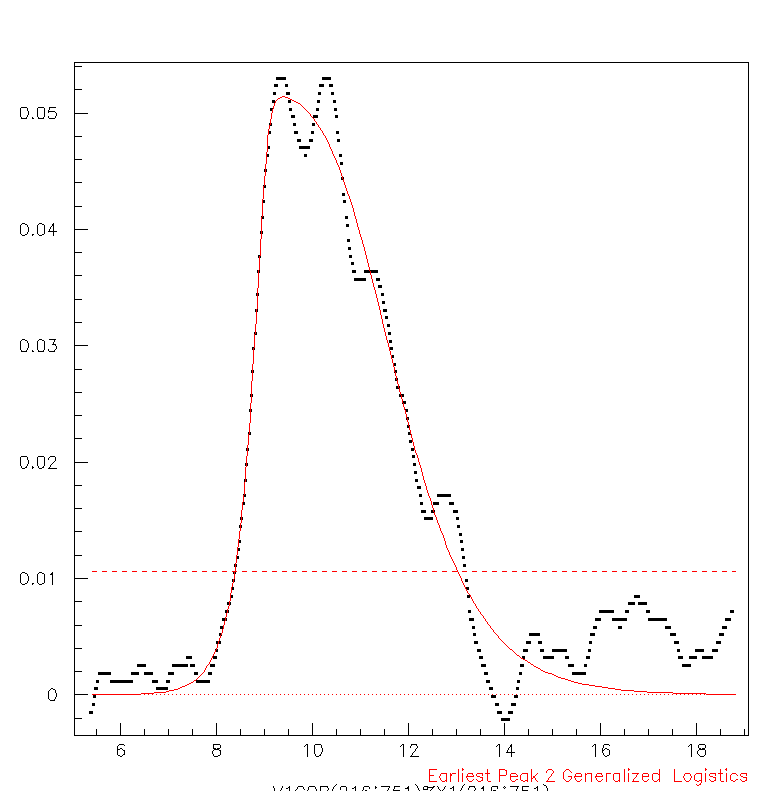}
        \caption{}
    \end{subfigure}
    \begin{subfigure}[h]{0.49\textwidth}
    \includegraphics[width=0.7\textwidth]{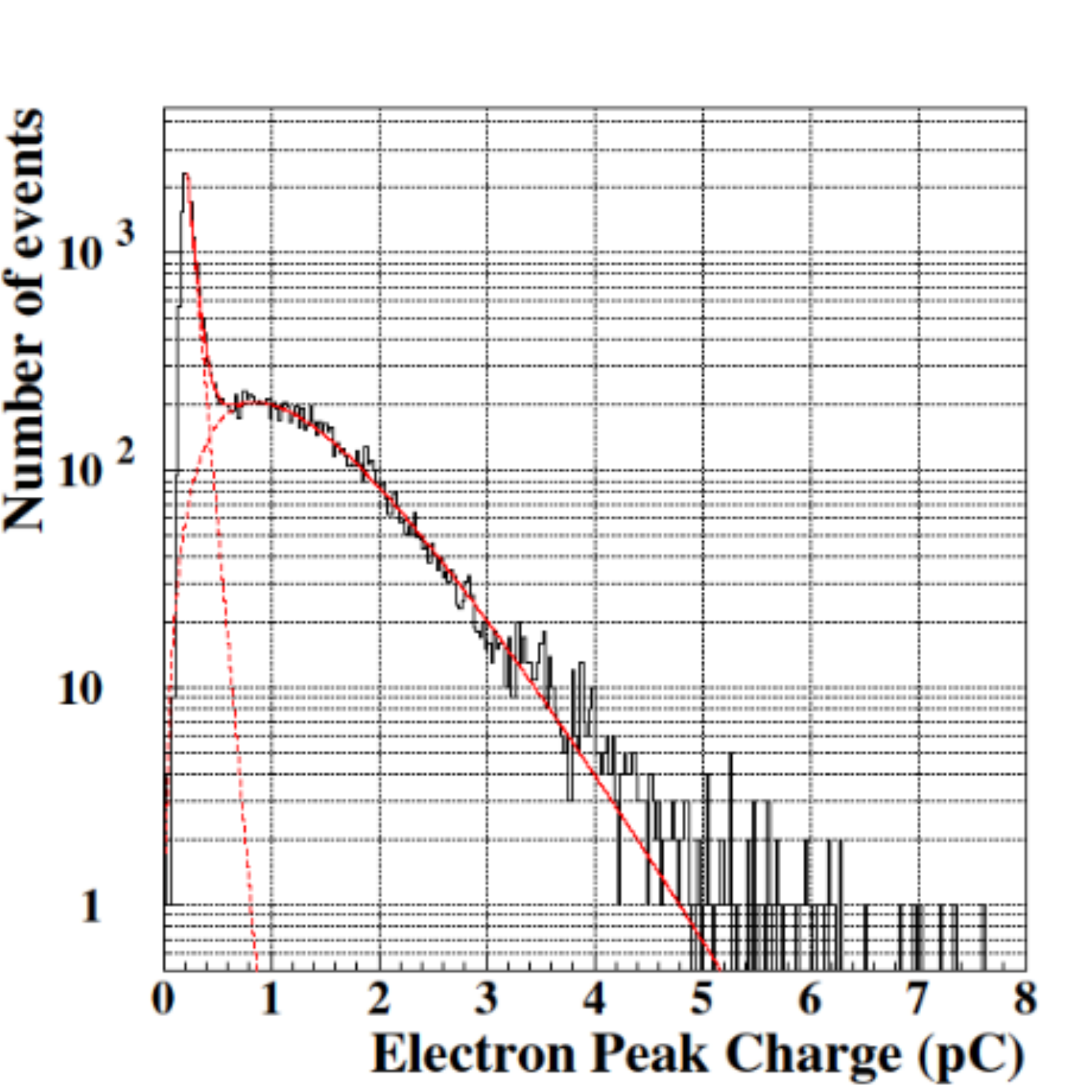}
        \caption{}
    \end{subfigure}
    \caption[Example of noisy waveform and distribution of electron peak charge in calibration data.]{
        Figures show calibration data collected with anode and drift voltages of $275\,V$ and $475\,V$, respectively, and with COMPASS gas.
        (a) Example of noisy waveform shown with black points, where the difference of two generalized logistic functions is fit so as to estimate reliably the charge of the electron peak.
        (b) Distribution of electron peak charge is shown with black, while
        red lines correspond to a global fit of an exponential and a P\'olya distribution.
}
    \label{calib}
\end{figure}

Because the calibration waveforms are of very low amplitudes and are particularly noisy,
the technique used to estimate the electron peak charge is the one involving the difference between two generalized logistic functions is used (Section \ref{section:ept}, Equation \ref{dif-logi}).
A realization of such a fit is presented with red in Figure \ref{calib}\,a, on the waveform's points shown with black. 
This technique allows the reliable estimation of the electron peak charge even in such cases where the noise not negligible.

The distribution of the electron peak charges is shown in Figure \ref{calib}\,b.
The noise is prevalent and is modeled with an exponential while the P\'olya distribution is used to model the spectrum of the electron peak charges (signal).
The red lines in the Figure represent the global fit of the two compononets (noise and signal).
The mean value of the P\'olya fit is found to be $\bar{q}_{SPE} = 1.3246$, while its standard deviation is estimated at $\sigma_q = 0.78072$.
The uncertainties are not presented because the values are heavily correlated and they would be misleading.
However, the uncertainty of the shape of this electron peak charge spectrum is the major contribution of error in the estimation of the mean number of photoelectrons.
This uncertainty is taken into account later by fitting several P\'olya functions in the same spectrum but choosing different bin size or range of fitting, and repeating the estimation.
The statistical combination of the several estimations with their uncertainties give a more realistic estimate. 

\section{Mean Number of Photoelectrons}\label{sec:npe}
Because of the nature of Cherenkov radiation, the photons are not localized in the point that the muon particle is hitting the detector.
Instead, the projection of the Cherenkov cone forms a circle on the plane of the photocathode with a density that is uniform in the polar coordinates $r,\phi$.
The angle between the photons' direction and the muon's direction is roughly $45^o$ degrees, and as the the thickness of the Cherenkov radiator is $3\,mm$, 
the radius of the circle defined by the projection of the Cherenkov cone has a radius also of $3\,mm$.

\begin{figure}[H]
    \centering
    \hspace*{3cm}
    \includegraphics[width=0.7\textwidth]{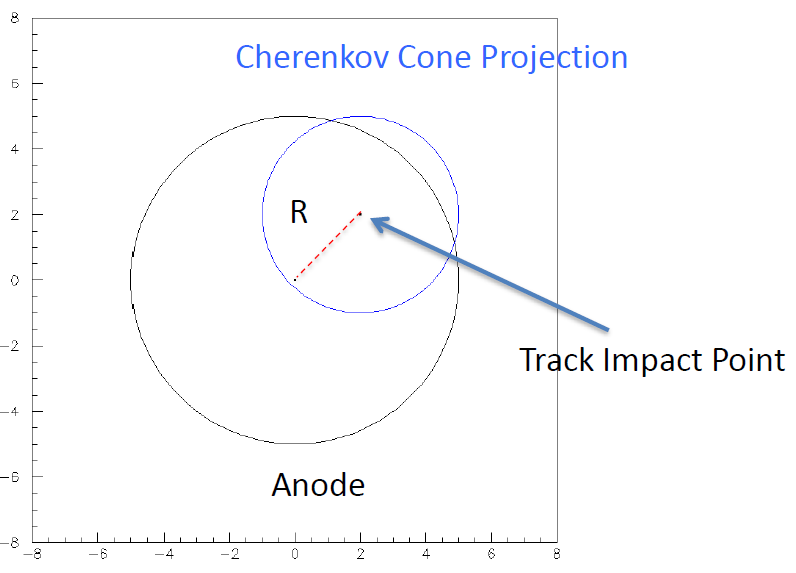}
    \caption[Example of the circle defined by the Cherenkov photons in the passage of muon through the radiator.]{
        Example of the circle defined by the Cherenkov photons in the passage of muon through the radiator.
}
    \label{cher-circle}
\end{figure}

The active area of the detector, projected on the photocathode, is $5\,mm$.
Therefore, if the muon that is passing through the detector is not passing very close to center of the active area, then there will be photons that will escape.
This mechanism is illustrated in Figure \ref{cher-circle}.
It is expected that the number of photoelectrons per muon is constant for track impact parameters (radial distances between MIP and center of detector) up to $2\,mm$.
After that, the percentage will drop until an impact parameter of $8\,mm$, after which the percentage will stay at $0$.

\begin{figure}[H]
    \centering
    \begin{subfigure}[h]{0.49\textwidth}
        \includegraphics[width=0.9\textwidth]{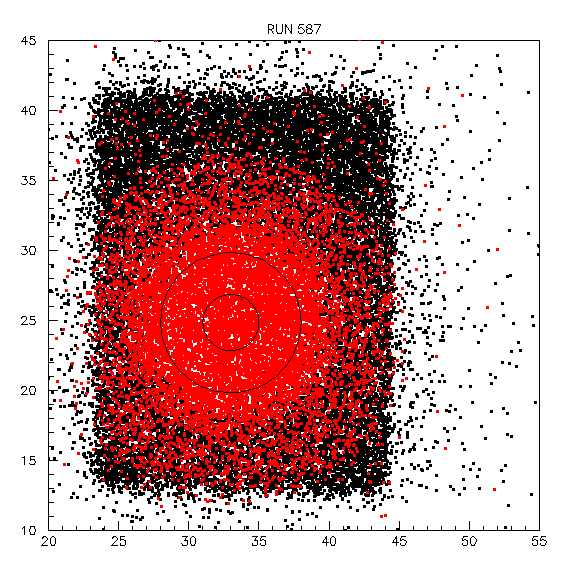}
        \caption{}
    \end{subfigure}
    \begin{subfigure}[h]{0.49\textwidth}
        \includegraphics[width=0.9\textwidth]{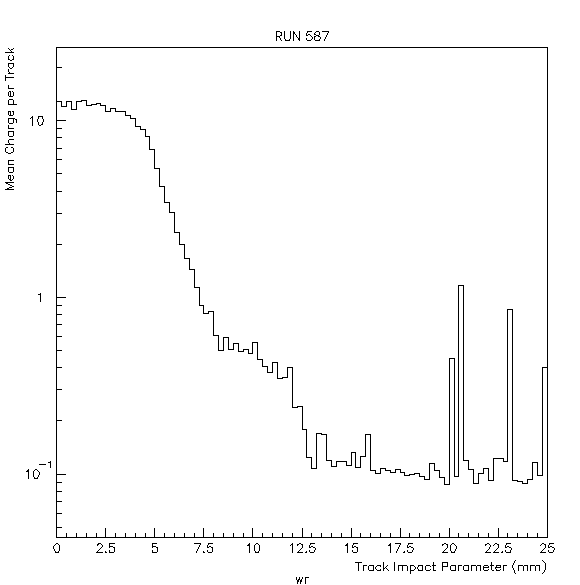}
        \caption{}
    \end{subfigure}
    \caption[Track impact points and mean charge per track.]{
        (a) Scatter plot of track impact points on the plane of the photocathode. Black points are tracks accepted by the trigger setup while red ones also produce a signal in the PICOSEC detector.
        Outer circle represents the boundary of the detector's active area and the inner circle represents the area of impact parameters for which all photons end up in the active area.
        (b) Mean charge per track as a function of the track impact parameter. Blue line corresponds to the inner circle of (a), red to the outer circle of (a), while green denotes the distance after which no photons should have been deposited in the photocathode.
    }
    \label{charge-impact}
\end{figure}

A quantity that is proportional to the number of photoelectrons extracted from the photocathode is the mean charge per muon track.
By looking how the mean charge per track changes with the impact parameter of the track, the described behaviour of the number of photoelectrons is expected.
The distribution of tracks is shown in Figure \ref{charge-impact}\,a, where black corresponds to all of the tracks for which the trigger setup accepted the event,
while red points correspond to only the tracks which produced a signal in the PICOSEC detector.
The inner solid circle is the circle inside which tracks will deposit all photoelectrons inside the outer solid circle which is the boundary of the active area of the detector.
The mean charge per track versus the impact parameter is shown in Figure \ref{charge-impact}\,b.
The blue line denotes the impact parameters up to which all photons are deposited on the photocathode.
The red line denotes the boundary of the detector's boundary while the green line denotes the impact parameter after which there should be no photoelectron detected.

\begin{figure}[H]
    \centering
    \begin{subfigure}[h]{0.49\textwidth}
        \includegraphics[width=0.9\textwidth]{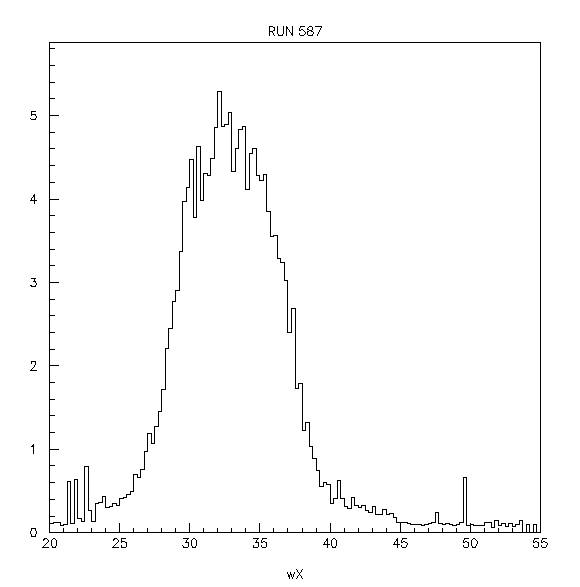}
        \caption{}
    \end{subfigure}
    \begin{subfigure}[h]{0.49\textwidth}
        \includegraphics[width=0.9\textwidth]{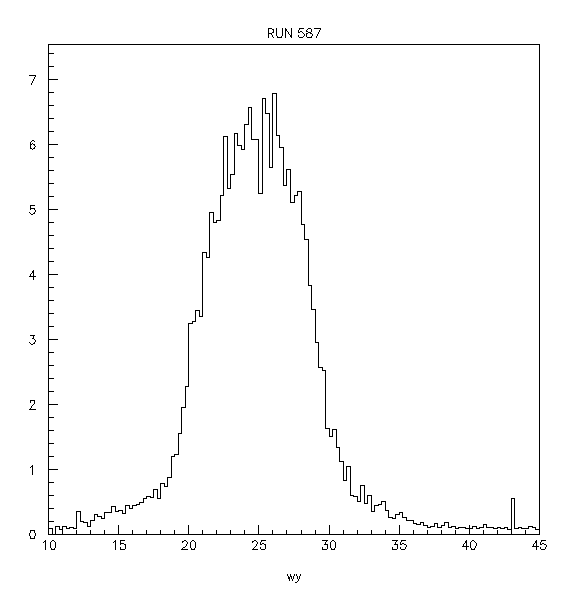}
        \caption{}
    \end{subfigure}
    \caption[Distributions of mean charge per track.]{
        Distribution of the mean charge per track versus the (a) horizontal coordinate and (b) the vertical coordinate.
    }
    \label{tracks}
\end{figure}

The center of the detector was found by projecting the mean charge per muon in the horizontal axis (Figure \ref{tracks}\,a) and the vertical axis (Figure \ref{tracks}\,b), and then finding the mean value of the projected distributions.
Obviously, there is a structure in Figure \ref{charge-impact}\,b that behaves unexpectedly.
For impact parameters in the range $[8\,mm,\ 12.5\,mm]$, there is a bump where according to the above mechanism it should have been equal to zero (or at least equal to charge of the noise).

\begin{figure}[H]
    \centering
    \hspace*{3cm}
    \includegraphics[width=0.7\textwidth]{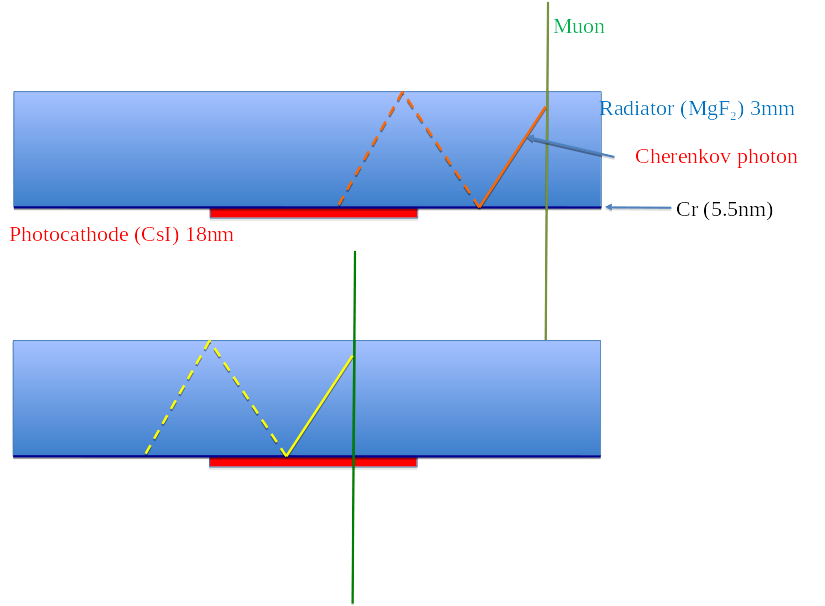}
    \caption[Examples of Cherenkov photon reflection.]{
        Examples of Cherenkov photon reflection. (top) A muon passing far away from the detector can still deposit photons on the photocathode.
        (bottom) A muon passing through the center of the detector will not deposit all of the electrons.
}
    \label{refl}
\end{figure}

The simplest possible explanation is that in the boundary between the metallic thin surface and the Cherenkov radiatior, the photons can be reflected.
Two examples of this mechanism are illustrated in Figure \ref{refl}.
Not only can photons be detected from events with muons passing far away, but some of the photons of events with muons passing through the center of the detector will escape.

\begin{figure}[H]
    \centering
    \begin{subfigure}[h]{0.49\textwidth}
        \includegraphics[width=0.9\textwidth]{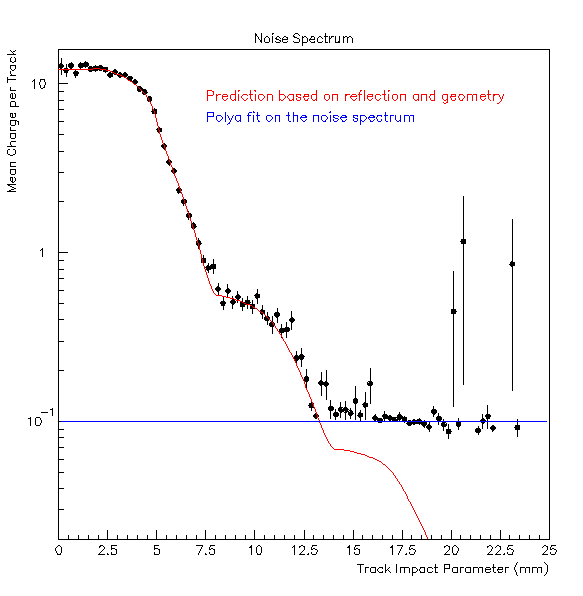}
        \caption{}
    \end{subfigure}
    \begin{subfigure}[h]{0.49\textwidth}
        \includegraphics[width=0.9\textwidth]{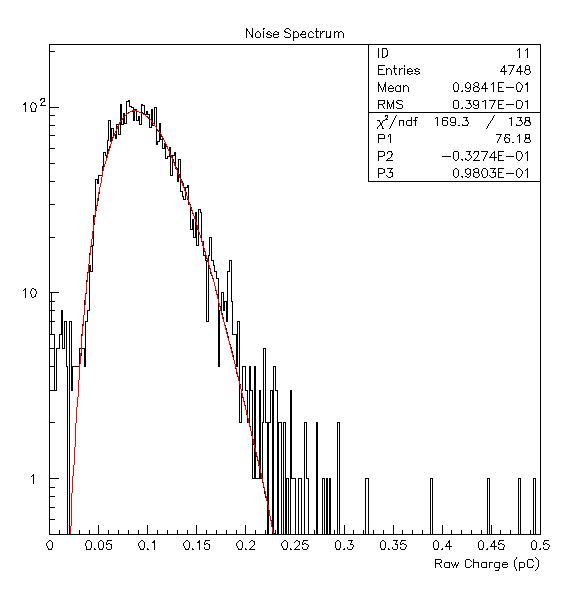}
        \caption{}
    \end{subfigure}
    \caption[Mean charge per muon and mean charge of noise events.]{
        (a) Mean charge per muon as a function of the impact parameter.
        The prediction of the model assuming a reflection probability of $22\%$ is shown with red while with blue is shown the mean charge of the noise spectrum.
        (b) Distribution of charge where noise is mimicking the electron peak. Solid line corresponds to a P\'olya fit.
        The data were collected with anode voltage of $375\,V$ and drift voltage of $375\,V$ using the COMPASS gas.
    }
    \label{model}
\end{figure}

To estimate the mean number of photoelectrons a likelihood method is developed.
Firstly, the probability distributions of each detection stage must be identified.

\begin{enumerate}
    \item A muon passes through the Cherenkov radiator and $n$ photons are produced. This number of photons $n$ is a random variable which follows the Poisson distribution $Poisson(n;\mu)$:
        \begin{equation}
            Poisson(n;\mu) = \frac{\mu^n e^{-\mu}}{n!}
        \end{equation}
        where $\mu$ is the mean number of photons that are produced by an incident muon.
    \item Each of the $n$ produced photons has a probability to either interact in the active area of the photocathode or to escape the active area.
        This probability $\epsilon (r)$ depends on the impact parameter $r$ of the muon.
        Of the $n$ photons produced, $k$ of them will be deposited on the photocathode according to the Binomial distribution:
        \begin{equation}
            Binomial(k;n,\epsilon(r)) = \frac{n!}{k!(n-k)!} \left(\epsilon\left(r\right) \right)^k\left(1-\epsilon\left(r\right)\right)^k
        \end{equation}
    \item The $k$ photoelectrons which synchronously start the signal formation in the detector will ultimately produce a signal with charge $q$ according the N-P\'olya distribution defined in Equation \ref{npolya}, or if $k=0$ a noise charge $q$ will be produced with a PDF $g(q)$.
        The parameters of the N-P\'olya distribution that must be known a priori are the parameters of the P\'olya distribution that describe the spectrum of a single photoelectron response,i.e. the mean electron peak charge $\bar{q}_{SPE}$ and the $\theta_{SPE}$ (or RMS) parameter of Single PhotoElectrons.
        These are available through the Calibration Data.
    \item However, because the observable variable is the charge $q$, all possible $k$ electrons producing an electron peak charge equal to $q$ must be considered.
        Thus, the probability to observe an electron peak charge equal to $q$ is:
        \begin{equation}
            P(q;\mu,r,\bar{q}_{SPE},\theta_{SPE}) = \sum_{k=0}^\infty f(k;\mu,r) \cdot G(q;\bar{q}_{SPE},\theta_{SPE},k)
        \end{equation}
        where $f(k;\mu,r)$ is the probability to have $k$ synchronous photoelectrons given that the impact parameter is equal to $r$ and
        \begin{equation}
            G(q;\bar{q}_{SPE},\theta_{SPE},k) = 
            \begin{cases} 
                Polya_N(q;\bar{q}_{SPE},\theta_{SPE},k) & k > 0 \\
                g(q) & k = 0
            \end{cases}
        \end{equation}
\end{enumerate}

The probability $f(k;\mu,r)$ is the infinite sum:

\begin{equation}
    f(k;\mu,r) = \sum_{n=0}^\infty Binomial(k;n,\epsilon(r)\cdot Poisson(n;\mu)
\end{equation}

It is easy to show that because $k$ is a non-negative integer then $f(k;\mu,r)$ is a Poisson distribution:

\begin{equation}
    f(k;\mu,r) = Poisson\left(k;\mu\cdot \epsilon(r)\right)
\end{equation}

To find the function that expresses the ``geometrical efficieny'' $\epsilon(r)$, a toy Monte-Carlo simulation is employed.
The toy Monte-Carlo is in fact very simple; photon coordinates are sampled with a uniform random radial distance from the impact parameter and a uniform random polar angle with respect to the impact parameter.
Then, with a certain probability (a free parameter of the Monte-Carlo) the photon's radial distance from the impact parameter is increased by $6\,mm$ (length derived according to classical optics).
This probability expresses the probability for the photon to be reflected up and down the radiator.
If the photon does get reflected in the simulation, a random number is again sampled to find whether it is reflected again, and the process repeats until it does not get reflected.
It is assumed that the photons cannot be absorbed. 
Then, the percentage of photons that are deposited inside the active area are counted to find the geometrical efficieny $\epsilon(r)$.

In Figure \ref{model}\,a the mean charge per track is shown versus the charge of the electron peak where the red line corresponds to the above described model with a reflection probability of $22\%$ and the blue line corresponds to the mean charge of the noise spectrum.
The mean charge of the noise spectrum was estimated through a P\'olya fit on events that were out of time with regular muon events, shown in Figure \ref{model}\,b.
The agreement between the model and the experimental data is excellent.

Notice that in the toy Monte-Carlo, the quantum efficiency of the photocathode is not included.
However, this quantum efficiency can be ``bypassed'' by normalizing the geometrical efficiency such that $\epsilon(0) = 1$.
In that case $\mu$ is no longer the mean number of photons that are produced by a single muon.
Instead $\mu$ is the mean number of photoelectrons that are extracted from the photocathode 
locally\footnote{Notice that some photons that would originally be deposited on the photocathode get reflected and are lost.
If the detector was infinite, all of the photons would eventually be deposited on the photocathode but they would be converted into electrons in a different place and in a later time.
If the detector had a segmented readout they would be easily distinguishable.},
given that the muon passes through the center of the detector.

Finally, the probability to detect an electron peak charge of $q$, given that the impact parameter is $r$, is:

\begin{multline}
    P(q;\mu,r,\bar{q}_{SPE},\theta_{SPE}) = \\
   = \sum_{k=1}^\infty Poisson\left(k;\mu\cdot \epsilon(r)\right) \cdot Polya_N(q;\bar{q}_{SPE},\theta_{SPE},k)
\end{multline}

However, the center of the detector may not always be known and may even be misaligned with respect to the scintillator trigger, as is usually the case with the small trigger.
The tracking detectors provide horizontal and vertical coordinates $x,\ y$.
In that case the impact parameter will be equal to $r = \sqrt{(x-\delta x)^2 + (y-\delta y)^2}$.
Where $\delta x$ and $\delta y$ are the misalignment coordinates of the PICOSEC detector.
For a set of N muon track measurements $\left\{(q_1,x_1,y_1),(q_2,x_2,y_2,),...,(q_N,x_N,y_N)\right\}$, the likelihood is defined:

\begin{multline}
    L\left(q_1,q_2,...,q_N,x_1,x_2,x_N,...,y_1,y_2,,,,.y_N; \mu, \delta x, \delta y\right) =\\ 
    = \prod_{i=1}^N \left(
    \sum_{k=0}^\infty Poisson\left(k;\mu\cdot \epsilon(r_i)\right) \cdot G(q_i;\bar{q}_{SPE},\theta_{SPE},k)
\right)
\end{multline}

where
\begin{multline}
    G(q;\bar{q}_{SPE},\theta_{SPE},k) = \\
    =
    \begin{cases} 
        \frac{ (\theta_{SPE}+1)^{k(\theta_{SPE}+1)}}{\bar{q}_{SPE}\Gamma(k(\theta_{SPE}+1))} \left( \frac{q}{\bar{q}_{SPE}}\right)^{k(\theta_{SPE}+1)-1} e^{-(\theta_{SPE}+1)q/\bar{q}_{SPE}} & k > 0 \\
        g(q) & k = 0
    \end{cases}
\end{multline}
\begin{equation}
    Poisson(k;\mu\cdot\epsilon(r)) = \frac{ (\mu\cdot\epsilon(r))^k e^{-\mu\cdot\epsilon(r)}}{k!}
\end{equation}
\begin{equation}
r_i = \sqrt{(x_i-\delta x)^2 + (y_i-\delta y)^2}
\end{equation}
and the parameters $\bar{q}_{SPE},\theta_{SPE}$ are the parameters of the P\'olya distribution describing the single photoelectron response and must be estimated from calibration data, while the geometrical efficiency $\epsilon(r)$ must be simulated assuming a model.

\begin{figure}[H]
    \centering
    \begin{subfigure}[h]{0.49\textwidth}
        \includegraphics[width=0.9\textwidth]{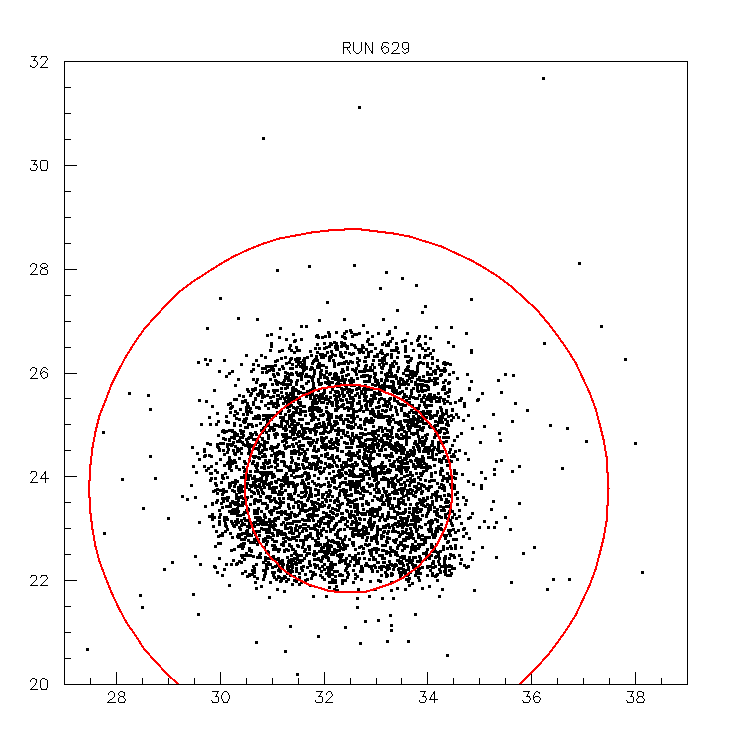}
        \caption{}
    \end{subfigure}
    \begin{subfigure}[h]{0.49\textwidth}
        \includegraphics[width=0.9\textwidth]{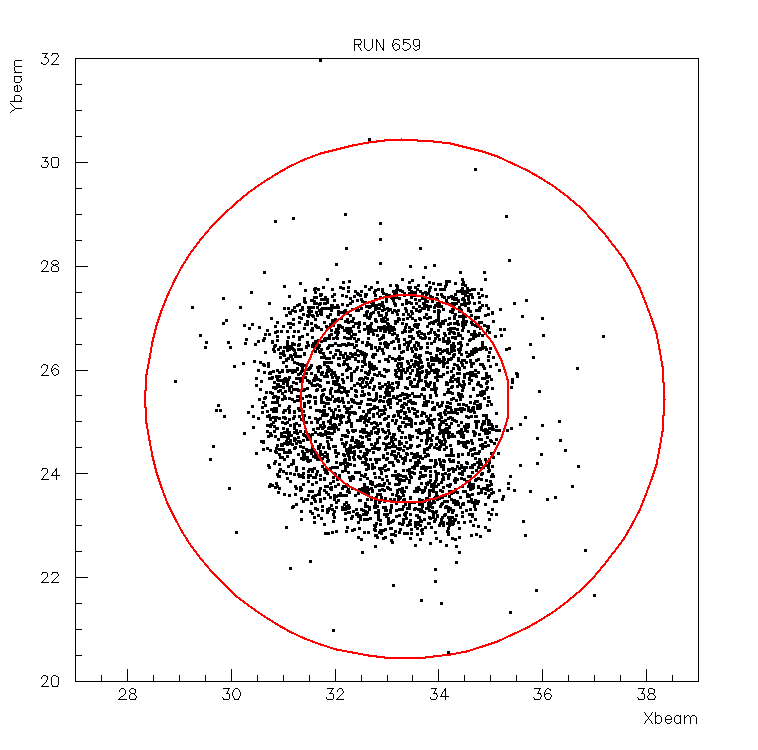}
        \caption{}
    \end{subfigure}
    \caption[Track impact points with data collected with the small trigger.]{
        Track impact points on the $x-y$ plane for two different datasets.
        In both datasets, the anode voltage is set to $275\,V$, the drift voltage to $475\,V$.
        The inner circle denotes the area in which all incident tracks deposit the maximum number of photons on the photocathode.
        The outer circle denotes the boundaries of the photocathode.
    }
    \label{align}
\end{figure}

This likelihood is numerically maximized to find estimations for parameters $\mu$, $\delta x$ and $\delta y$.
Two examples where this alignment was necessary to find the center of the detector are presented in Figures \ref{align} where the track impact points are shown on the $x-y$ plane.
Tracks hitting inside the inner circle deposit the maximum number of photons on the photocathode while the outer circle denotes the boundaries of the photocathode.
Both datasets are collected with an anode voltage of $275\,V$, a drift voltage of $475\,V$.

\begin{figure}[H]
    \centering
    \includegraphics[width=0.5\textwidth]{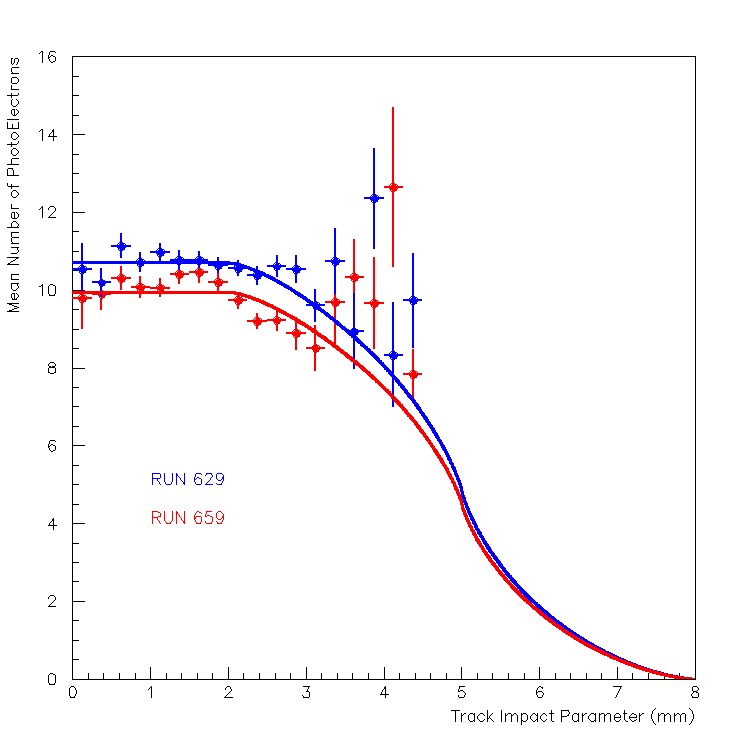}
    \caption[Mean number of photoelectrons as a function of track impact parameter.]{
        Mean number of photoelectrons versus the track impact parameter for the two datasets presented in Figure \ref{align}.
        Solid lines represent the prediction of the model described in the Section.
        (blue) $\mu = 10.7\pm 0.5$.
        (red) $\mu = 10 \pm 0.5$.
}
    \label{photo}
\end{figure}

The mean number of photoelectron versus the impact parameter is shown for the same two datasets in Figures \ref{photo}\,a and b.
The solid lines correspond to the model's prediction.
The model provides a decent description but the experimental data are not enough to investigate the dependence on the impact parameter.
It is clear that although the operational settings were identical, the mean number of photoelectron is different.
This could be happening for a number of reasons like variations in environment conditions (pressure,temperature) but the most likely scenario is that the photocathode was damaged because of the large ion backflow from the amplification region to the pre-amplification region (because of large drift voltage).

\begin{figure}[H]
    \centering
    \begin{subfigure}[h]{0.49\textwidth}
        \includegraphics[width=0.9\textwidth]{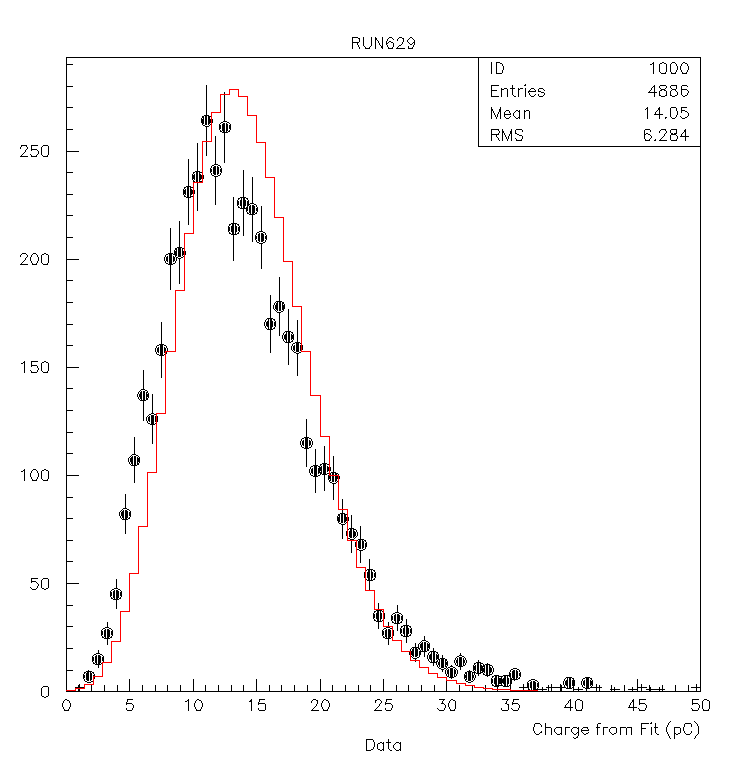}
        \caption{}
    \end{subfigure}
    \begin{subfigure}[h]{0.49\textwidth}
        \includegraphics[width=0.9\textwidth]{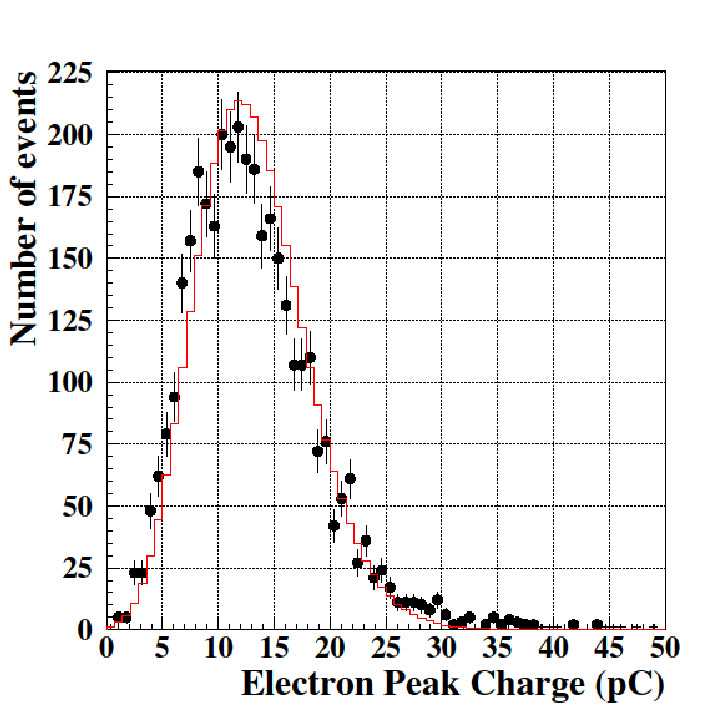}
        \caption{}
    \end{subfigure}
    \caption[Distributions of electron peak charge and prediction.]{
        Distributions of electron peak charges are shown with black points for the same two datasets presented in Figures \ref{align}, \ref{photo}.
        Solid red lines represent the prediction of the model after the maximum likelihood estimation.
        (a) $\mu = 10.7\pm 0.5$.
        (b) $\mu = 10 \pm 0.5$.
    }
    \label{mean-npe}
\end{figure}

Figures \ref{mean-npe}\,a and b show the distributions of the electron peak charge with black points along with the predictions of the model depicted with the solid red lines.
The estimated mean number of photoelectrons is (blue) $\mu =10.7\pm 0.5$ and (red) $\mu = 10 \pm 0.5$.
The combined estimation of the mean number of photoelectrons is 
\begin{equation}
\mu = 10.4 \pm 0.4
\end{equation}

\section{Timing Results}

\begin{figure}[H]
    \centering
    \includegraphics[width=0.7\textwidth]{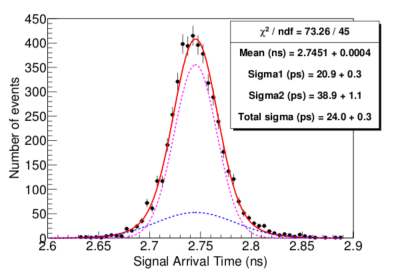}
    \caption[Distribution of Signal Arrival Time for muons.]{
        Distribution of SAT for events collected with muons at anode voltage of $275\,V$ and a drift voltage of $475\,V$ using the COMPASS gas.
        The solid lines represents a fit of a double Gaussian with the dashed lines depicting the two individual Gaussians.
        The standard deviation is $\sigma_t = 24.0\pm 0.3\,ps$.
}
    \label{final}
\end{figure}

In the same two datasets for which the mean number of photoelectrons was measured, the timing results should be presented.
Because of the very large drift voltage, the mean SAT does not seem to depend on the electron peak charge and because of the limited number of events a small dependence is not easy to find.
The distribution of the SAT is shown in Figure \ref{final} with black points.
The resulting histogram is fit with a double Gaussian distribution with common means, shown with the solid line.
The dashed lines correspond to the two Gaussian components of the double Gaussian.
The standard deviation of the double Gaussian distribution is estimated at
\begin{equation}
\sigma_t = 24.0\pm0.3\,ps
\end{equation}
This result is the highlight of the PICOSEC detector proving an unprecedented timing precision for gas-filled detectors and with a single measurement of a MIP. 

\begin{figure}[H]
    \centering
    \begin{subfigure}[h]{0.49\textwidth}
        \includegraphics[width=0.9\textwidth]{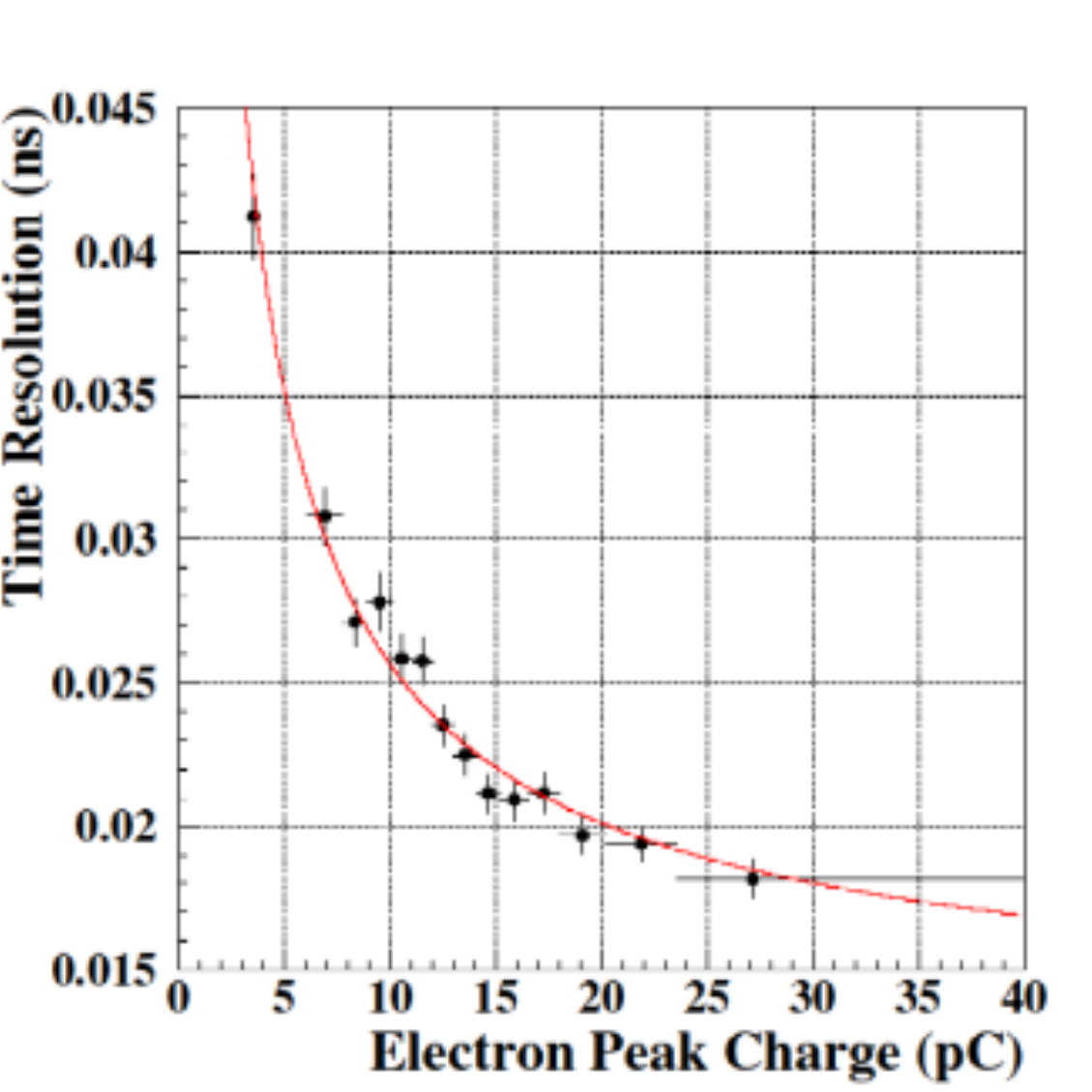}
        \caption{}
    \end{subfigure}
    \begin{subfigure}[h]{0.49\textwidth}
        \includegraphics[width=0.9\textwidth]{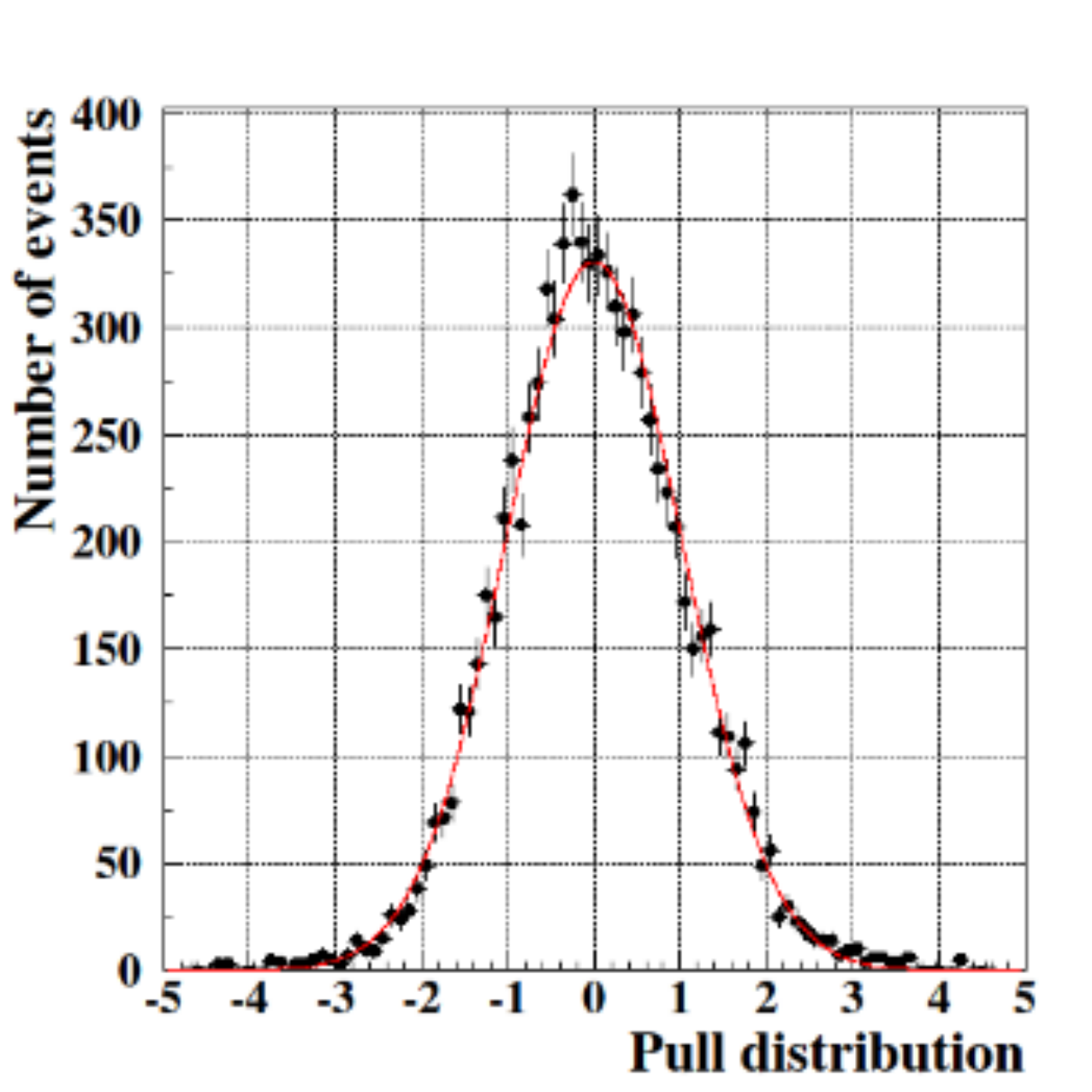}
        \caption{}
    \end{subfigure}
    \caption[Time resolution versus electron peak charge and pull distribution.]{
        (a) Time resolution versus the electron peak charge. Red line corresponds to a fit using Equation \ref{eq-power}.
        (b) Pull distribution using the parameterization of (a). Red line represents a Gaussian fit with parameters $\mu = -0.0004\pm 0.011$ and $\sigma = 1.02\pm 0.01$, consistent with the standard normal distribution.
    }
    \label{finalpull}
\end{figure}

The fact the distribution could not be fit with a single gaussian is an indication that the different events do not all exhibit the same time resolution.
Any number (larger than $1$) number of Gaussians could have been used to fit the disitribution.
Two Gaussians are sufficient to describe the experimental data.
To test the consistency of the result, the time resolution as a function of the electron peak charge is found and presented in Figure \ref{finalpull}\,a.
As usually, the time resolution becomes better for larger electron peak sizes and a fit is realized using the power law and a constant term of Equation \ref{eq-power}.
The fit parameters provide a parameterization of the time resolution in the form:
\begin{equation}
    R(q) = \frac{0.08}{q^{0.785}}+0.012
\end{equation}
Using this parameterization, the pull distribution is made, i.e. from each event's SAT the mean SAT is subtracted and is then divided by the corresponding time resolution $R(q)$ according to the electron peak charge $q$ of the event.
This is Pull distribution which is presented in Figure \ref{finalpull}\,b.
The solid line represent a Gaussian fit whose parameters are found to be equal to $\mu = -0.0004\pm 0.011$ and $\sigma = 1.02\pm 0.01$.
The result is consistent with a standard normal distribution and the statistical test is passed.

\chapter{Conclusion}

In the context of this thesis, an easy-to-use signal processing software package was 
developed to apply the techniques described in Section \ref{sec:sp} among others.
Combined with the application of the statistical techniques mention in Sections \ref{sec:tim} and \ref{sec:tot},
the best time resolution that could be achieved with the PICOSEC detector on a single photoelectron response is around $76\,ps$ by operating the detector either with CF$_4$ gas filling or with the COMPASS gas mixture.
The dependence of the total time resolution (Figure \ref{tot_res}) with the drift voltage hinted that the drift voltage could be furtherly increased to achieve even better time resolution.

Indeed, this was tried in a later Testbeam campaign whose best results are presented in Chapter \ref{chap:muons}.
The best time resolution was achieved by operating the detector with the COMPASS gas mixture with an anode voltage of $275\,V$ and a drift voltage of $475\,V$.
At these operational conditions a mean number of photoelectron per muon equal to $10.4\pm 0.4$ is extracted and the time resolution is found equal to $24 \pm 0.3\,ps$.
To find the mean number of photoelectrons another software package was developed that applies the maximum likelihood technique shown in Section \ref{sec:npe}.

In the analysis of the Laser Test, three peculiar results were observed:
\begin{enumerate}
\item There exists a dependence of the mean SAT on the electron peak size which is similar to the ``time walk'' effect.
\item The dependence of the mean SAT on the electron peak size changes only through a constant term between different drift voltages and the same anode voltage. 
The functional form of it stays the same.
\item The dependence of the time resolution on the electron peak size does not change for different drift voltages.
\end{enumerate}
To address these issues a simulation based on Garfield++ was made.
In order to verify the simulation the unknown pre-amplifier's response was needed.
To overcome this problem, a statistical technique was developed, in a rigorous manner, to take this response into account.

After verifying the simulation, a microscopic investigation was launched where it was found that the macroscopic electron peak charge and SAT were associated directly to the (microscopic) number of electrons that were passing through the mesh and the (microscopic) mean time of all electrons on the moment of passing through the mesh.
With this correspondence, it was found that the first issue was caused by the fact that electrons are moving with different drift velocities before and after the first ionization in the pre-amplification avalanche.
The phenomenological explanation of this is that before ionizing, the primary electron cannot have interacted with the gas through the channel of ionization.
After the first ionization, the electrons are ``free'' to interact through this channel.
The channel of ionization is an inelastic process and when it happens, it damps the energy of the electron.
Therefore, by ``turning on'' another possible inelastic process, an effect similar (but much less pronounced) to adding a quencher happens, i.e. electrons with lower energy participate in fewer elastic scatterings and the electric field spends less time recovering a forward direction motion.

Although the other two issues are shown to be reproduced by the simulation, there is no explanation yet.
This part is the subject of a future work.

This work has equipped the collaboration with a better understanding of the detector's inner workings.
This understanding can and will be used to achieve better results with the PICOSEC and design novel experimental applications of the timing information that it can provide.

\appendix
\chapter{Time of First Passage in Brownian Motion with Drift}\label{app-igaus}

Consider a 1-dimensional diffusion process where the diffusing particle starts at a time $t = 0$ at position $x = 0$.
After time $t$, the Probability Density Function (PDF) to find the particle at position $x$ will be given by the Gaussian distribution:

\begin{equation} 
    G'(x;t) = \frac{1}{\sqrt{2\pi \sigma^2(t)}}e ^{-\frac{x^2}{2\sigma^2(t)}}
\end{equation}

where $\sigma^2(t)$ grows linearly with time and is associated with a diffusion coefficient $D_L$ such that

\begin{equation}
    \sigma^2(t) = 2 D_L t
\end{equation}

Suppose that this diffusion process is biased through a force field 
and the particle is drifting towards positive $x$ with a constant drift velocity $v_d$.

The PDF then to find the particle at position $x$ will be given by another Gaussian whose mean value has been shifted by $\mu(t) = v_d t$:

\begin{equation} 
G(x;t) = \frac{1}{\sqrt{2\pi \sigma^2(t)}} 
    e^{-\frac{\left(x-\mu(t)\right)^2}{2\sigma^2(t)}}
\end{equation}

We are interested in the PDF of the arrival time of the particle at a constant level at $x = a$.
We begin by evaluating the probability that the particle has already passed from the position $x = a$.
This is given by the integral of $G(x;t)$ from the position $a$ until infinity.

\begin{equation}
    C(t) = \int_a^\infty  G(x;t) dx
\end{equation}

which is rewritten as

\begin{equation}
    C(t) = \frac{1}{\sqrt{\pi}}  \int_{\left(\frac{a -\mu(t)}{\sqrt{2}\sigma(t)}\right)}^\infty  e^{ -\left( \frac{a - \mu(t)}{\sqrt{2}\sigma(t)} \right)^2} d\left(\frac{x -\mu(t)}{\sqrt{2}\sigma(t)}\right)
\end{equation}

or with a change of variable $u = \frac{x -\mu(t)}{\sqrt{2}\sigma(t)}$:

\begin{equation}
    C(t) = \frac{-1}{\sqrt{\pi}} \int_\infty^{\left(\frac{a -\mu(t)}{\sqrt{2}\sigma(t)}\right)}  e^{ -u^2 } du
\end{equation}

It is easy to realize that this integral corresponds to the Cumulative Density Function (CDF) 
of the PDF we are in search of (of the time of arrival of electrons on a fixed level $x=a$).
Therefore, we find the PDF $P(t)$ by differentiating the CDF with respect to the time of arrival $t$.

\begin{equation}
    P(t) = \frac{dC(t)}{dt} = \frac{-1}{\sqrt{\pi}} \frac{d}{dt}\left(\int_\infty^{\left(\frac{a -\mu(t)}{\sqrt{2}\sigma(t)}\right)}  e^{ -u^2 } du\right)
\end{equation}

or

\begin{equation}
    P(t) = \frac{-1}{\sqrt{\pi}} \frac{d}{du}\left(\int_\infty^{\left(\frac{a -\mu(t)}{\sqrt{2}\sigma(t)}\right)}  e^{ -u^2 } du\right)\frac{du}{dt}
\end{equation}

By the fundamental theorem of calculus, this PDF reduces to:

\begin{equation}\label{a1:1}
    P(t) = \frac{-1}{\sqrt{\pi}}   \exp\left( -\left(\frac{a -\mu(t)}{\sqrt{2}\sigma(t)}\right)^2 \right) 
    \frac{du}{dt}
\end{equation}

By substituting $\mu(t) = v_d t$ and $\sigma(t) = \sqrt{2 D_L t}$ into $u$ it is straightforward to show that

\begin{equation}
    \frac{du}{dt} = - \sqrt{\frac{a^2}{4D_Lt^3}}
\end{equation}

and Equation \ref{a1:1} becomes

\begin{equation}
    P(t) = \sqrt{\frac{a^2}{2D_L}}\cdot\sqrt{\frac{1}{2\pi t^3}}   \exp\left( -\frac{1}{2D_L}\frac{\left(a - v_d t\right)^2}{t} \right) 
\end{equation}

or

\begin{equation}
    P(t) = \sqrt{\frac{a^2}{2D_L}}\cdot\sqrt{\frac{1}{2\pi t^3}}   \exp\left( -\frac{1}{2} \cdot \frac{a^2}{2D_L} \cdot \frac{v_d^2}{a^2}\cdot\frac{1}{t}\cdot\left(\frac{a}{v_d} - t\right)^2 \right) 
\end{equation}

In other words the PDF $P(t)$ is an Inverse Gaussian Distribution with mean $\mu = \frac{a}{v_d}$ and a shape parameter $\lambda = \frac{a^2}{2D_L}$. 
The time spread of the first passage is equal to 
\begin{equation}
    \sigma_t = \sqrt{\frac{2D_La}{v_d^3}}
\end{equation}

\bibliography{pico}
\bibliographystyle{ieeetr}

\end{document}